\documentclass[phd,tocprelim,cornellheadings,12pt]{cornell}
\usepackage{graphicx}
\usepackage{graphics}
\usepackage{moreverb}
\usepackage{subfigure}
\usepackage{epsfig}
\usepackage{hangcaption}

\usepackage{palatino}

\usepackage{amsmath}
\usepackage{amssymb}
\usepackage{amsfonts}
\usepackage{color}
\usepackage[round]{natbib}
\usepackage{lscape}
\usepackage[mediumspace,mediumqspace,Gray,squaren]{SIunits}

\usepackage{yfonts}

\usepackage[numbered]{mcode}

\usepackage{wasysym}

\usepackage{booktabs}
\usepackage{multirow}

\usepackage{mathrsfs}

\usepackage{paralist}

\usepackage{framed}

\usepackage{ifpdf}

\renewcommand{\caption}[1]{\singlespacing\hangcaption{#1}\normalspacing}

\newcommand{\myrlambda}{480}

\title {Explorations into the inertial and integral scales of 
homogeneous axisymmetric turbulence}
\author {Kelken Chang}
\conferraldate {January}{2012}
\degreefield {Ph.D.}
\copyrightholder{Kelken Chang}
\copyrightyear{2012}

\begin{document}

\maketitle

\makecopyright

\begin{abstract}

A flow generator is described in which homogeneous 
axisymmetric turbulent air flows with varying and fully 
controllable degrees of anisotropy, 
including the much studied isotropic case, are generated 
by the combined agitations produced by 32 acoustic mixers 
focusing at the center of the system.  
The axisymmetric turbulence in a central
volume of the size of the inertial scale is shown
to have negligible mean and shear.  
The Taylor Reynolds number is about \myrlambda.  

The influence of large scale anisotropy on the turbulence
is examined from three aspects, namely the velocity structure 
functions, the velocity correlation functions, and the integral lengths.  
The directional dependence of two different second order transverse 
structure functions, in which one of them has separations stretched 
along the axis of symmetry of the turbulence and the other one 
normal to it, is studied.  
It is shown that the inertial range scaling exponents, 
determined using the extended-self-similarity procedure, 
and the Kolmogorov constants of the two structure functions 
are unaffected by the direction in which the structure functions
are measured.  

As an extension, because of its relevance to the study of
intermittency, the directional dependence of transverse structure 
functions of the fourth to the sixth order is studied.  
Despite some issues with measurement noise and statistical
convergence, some indications are found that anisotropy
in the velocity field intensifies the asymmetry of the probability
density of the velocity increments.  
In addition, some evidence is found that the inertial range 
scaling exponents of the fourth, fifth, and sixth order are
independent of the anisotropy.  

Finally, it is found that, except in the isotropic case, the 
second order transverse velocity correlation functions 
deviate from each other at the large scale with increasing
anisotropy.  
A self-similarity argument similar to one found in the
study of critical phenomena is proposed.  
It is shown that the argument leads to a power-law relationship 
between the large scale velocity fluctuation and the 
correlation length, with an exponent that depends on the 
inertial range scaling exponent of the turbulence.  
The data collapse predicted by the self-similarity hypothesis 
is verified.  
It is demonstrated that the value of the power-law exponent
is consistent with the value of the inertial range scaling exponent.  
\end{abstract}

\begin{biosketch}
Kelken Chang was born in Muar, Malaysia in September, 
1978, the fourth child of five of Chia-Ming and Kim-Leu Chang.  
Chia-Ming Chang was a laboratory technician who was working in the Lee
Rubber Company in Medan, Indonesia at the time his second son was
born.  
In fact the original family name had been Teo, the family being
ethnic Chinese from Fujian, China, but Chia-Ming Teo had changed his
name to Chia-Ming Chang in 1965 when he took up Singaporean
citizenship.  
Kelken's mother was the fifth child of eight of a well-to-do chinese
trader and she was ten years younger than her husband.  
Kelken was educated at Muar High School, a former Government English
School established in 1902 by the British government during the
colonization period.  
He went on to tertiary education at the National University of
Singapore (NUS) and earned a Bachelor of Science degree in Physics
with first class honors in 2003.  
He spent the first half of his graduate career at Cornell in Ithaca,
New York, from 2003 to 2006, and afterwards at the Max-Planck-Institut
f\"{u}r Dynamik und Selbstorganisation in G\"{o}ttingen, Germany.  
In his spare time, he enjoys swimming in the lakes, reading poetry,
and if he finds a good victim, playing practical jokes.
\end{biosketch}

\begin{dedication}
\begin{quote}
\hfill To my family and friends.
\end{quote}
\end{dedication}

\begin{acknowledgements}
Many outstanding individuals have made my graduate experience
one of the highlights of my life so far.  
I thank my advisor Eberhard Bodenschatz, the intrepid explorer whose
undiminishable energy and contagious enthusiasm for science has 
provided much of the impetus for this dissertation.  
Eberhard provided me the freedom to grow as an independent thinker.  
He is also very kind and understanding when experiments did not go 
as well as planned.  

I am beholden to my co-worker Greg Bewley for sharing with 
me his technical expertise and numerous invaluable ideas.  
His work attitude is very instructive and decisive every time in 
pushing the project a step forward.  
I appreciate the patience and rigor shown by both Greg and 
Eberhard in weeding out the waffle of my thoughts.  

I am grateful to Erich Mueller and Itai Cohen for agreeing to sit 
on my special committee and for providing the many levels of 
administrative support throughout the entire stage of my studies.  
Outside of classroom, Erich has taught me the value of scientific 
collaboration, a value that has also been reprised by many 
Cornell professors.  
I thank David Cassel, Veit Elser, Henry Tye, and Michelle Wang for 
the advice they have given me at various stages of my graduate 
studies.  

I am indebted to Sned for teaching me the rudiments of machining 
and for running a superb student machine shop; 
Andreas Kopp, Dr. Artur Kubitzek, Ortwin Kurre, Gerhard Nolte, 
and Andreas Renner for their invaluable technical assistance; 
Angela Meister for her meticulous handling of administrative 
issues at the institute; 
and Katharina Schneider for understandingly hearing out my 
many grouses -- our daily banter at the lunch table has been 
a bulwark against the monotony of laboratory work.  

I owe a debt of gratitude to my inspiring office mates from Cornell:
Albert, Amgad, Dario, and Haitao; and from the Max Planck Institute 
for Dynamics and Self-Organization: 
Azam, Christian, Eva, Ewe Wei, Fabio, Gabriel A., Gabriel S., 
Haitao, Hengdong, Holger, Jens, Marco, Mathieu, Matthias, Mireia,
Noriko, Robert, Shinji, Stephan, Toni, Vladimir, and Walter.  
Foremost among them, I wish to express my gratitude to 
Mathieu Gibert for graciously making his apartment 
available at critical stages of the writing of this thesis.  
I enjoyed the genuine camaraderie of these wonderful individuals, 
with whom I have had some of the most engaging conversations, 
sometimes discussing far into the night in some pub in G\"{o}ttingen
and finally thinking that we had solved our respective problems 
with alcoholically induced elation.  

I have received financial support from the National Science Foundation,  
through grants PHY-9988755 and PHY-0216406, for my studies at Cornell 
and from the Max Planck Society and Deutsche 
Forschungsgemeinschaft (German Science Foundation), through 
the grant XU91/3-1, for my studies in G\"{o}ttingen, Germany.  

I thank my family for their unfaltering support that has carried me
through the ups and downs of graduate school, 
and my friends for riding with me through thick and thin wood.  
\end{acknowledgements}

\contentspage

\tablelistpage

\figurelistpage

\normalspacing
\setcounter{page}{1}
\pagenumbering{arabic}
\pagestyle{cornell}
\addtolength{\parskip}{0.5\baselineskip}

\chapter{Introduction}

Homogeneous and axisymmetric turbulence, a simple and unique class
of anisotropic turbulence, has eluded detailed investigations due to
the lack of theoretical descriptions and flow generators whose large
scale forcing can be tuned systematically.  
The intent of the present experimental study is to expound on some ideas for making quantitative
measurements of homogeneous and axisymmetric turbulent flows, in hopes
that a better way of characterizing such a flow may be found in the
future.  
In his valuable review article on experimental methods in turbulence 
research for the {\it Handbuch der Physik}, \cite{corrsin:1963b} states that
\begin{quote}
As in other areas of science, the goals of experiment are, loosely, of
two kinds,
\begin{inparaenum}[\itshape a\upshape)]
\item exploratory and
\item to confirm or disprove theories.
\end{inparaenum}
The former puts a premium on imagination, with accuracy often
secondary; the latter requires accuracies at least sufficient to 
distinguish among competing theories.  In exploratory measurements
of statistical properties of turbulent flows 20\% accuracy is sometimes
satisfactory.  The more permanent crucial data, especially where
signal-to-noise ratio is large, are taken with perhaps 2 to 10\% uncertainty.
\end{quote}
It is in this spirit that we would like to invite the reader to join in our
exploration of this fascinating field of anisotropic turbulence.  
As a prelude to the experimental study, let us begin with a short 
theoretical description of turbulence.  

\section{Kolmogorov theory}
\label{sec:kolmogorov_theory}

The equation of motion of an unforced incompressible fluid, 
the Navier-Stokes equation (in nondimensional form)
\begin{equation}
\label{eq:navier_stokes}
\dfrac{\partial \boldsymbol{u}}{\partial t} + \boldsymbol{u} \cdot
\nabla \boldsymbol{u} = - \nabla p + \dfrac{1}{{\rm Re}}
\, \nabla^2 \boldsymbol{u} \,,
\end{equation}
is a nonlinear partial differential equation.  
Here, $\boldsymbol{u} (\boldsymbol{x}, t)$ and $p (\boldsymbol{x}, t)$ 
are the dimensionless fluid velocity and pressure, respectively, which
are functions of space, $\boldsymbol{x}$, and time, $t$.  
The dimensionless parameter ${\rm Re}$ is the Reynolds number.  
Suppose the fluid has a well-defined characteristic length scale, 
$\mathscr{L}$, and a characteristic velocity scale, $\mathscr{U}$, 
the Reynolds number is defined as
\begin{equation}
{\rm Re} = \dfrac{\mathscr{U} \mathscr{L}}{\nu} \,,
\label{eq:reynolds_number}
\end{equation}
where $\nu$ is the kinematic viscosity of the fluid.  
At low Reynolds numbers, the inertial forces (the terms on the left hand
side of the equation of motion) are negligible and solutions to the
Navier-Stokes equation exist.  
The flows are smooth and laminar, like honey flowing down a plate.  
At high Reynolds numbers, the flows become turbulent and no analytical
solution satisfying realistic boundary and initial conditions has yet
been found.  
A complete process in solving a physical problem usually involves
finding the right governing equations, followed by finding the solutions
to these equations.  
In turbulence, we have only the equations and the solutions are
still missing.  
\cite{nelkin:1992}, \cite{warhaft:2002a}, and chapter 41 of
\cite{feynman:1963} give concise introductions to the turbulence
problem.  
For a slightly idiosyncratic review of the milestones of turbulence, 
the reader could refer to \cite{lumley:2001}.  

For simplicity, research in turbulence has focused on idealized 
turbulent flows detached from the influence of externally imposed 
boundary.  
As a result, there are no apparent velocity and length scales 
with which one could use to define the Reynolds number.  
For this reason, the Reynolds number is constructed from 
the statistics of the turbulence.  
The characteristic velocity scale is taken as the root-mean-square 
(RMS) of the velocity fluctuations
\begin{equation}
u^{\prime} = \bigg[ \dfrac{1}{3} \, \sum_{i=1}^3 u_i^2 \bigg]^{1/2} \, , 
\end{equation}
and the characteristic
length scale is taken as the integral length, $L$, which is the integral of 
the velocity autocorrelation function (see section~\ref{sec:integralscale}).  
This yields a Reynolds number 
\begin{equation}
\mathrm{Re}_L = \dfrac{u^{\prime} L}{\nu} \, ,
\label{eq:Re_L}
\end{equation}
intrinsic to the turbulence and independent of the external 
geometry of the flow.  
More frequently, the Taylor scale Reynolds number \cite[][]{taylor:1935}
\begin{equation}
R_{\lambda} = \dfrac{u^{\prime} \, \lambda}{\nu} \, ,
\end{equation}
based on a different intrinsic length scale is chosen to describe the
turbulence level.  
The Taylor scale, $\lambda$, is the separation at which
the parabolic expansion of the velocity autocorrelation function near 
the origin becomes zero, and, in the case of isotropic turbulence (to
be described shortly), is given by \cite[][]{taylor:1935}
\begin{equation}
\lambda = \sqrt{\dfrac{15 \, \nu \, u^{\prime 2}}{\epsilon}} \,,
\end{equation}
where $\epsilon$ is the energy dissipation rate per unit mass of the fluid.  
The Taylor scale Reynolds number is related to $\mathrm{Re}_L$ by
\cite[][]{taylor:1935}
\begin{equation}
R_{\lambda} \approx (15 \, {\rm Re}_L)^{1/2} \,.
\label{eq:Rlambda_Re}
\end{equation}
It is routinely employed in the literature and is useful for
comparison to the plethora of theoretical, numerical, and experimental
results.  

Taking the divergence of equation~\ref{eq:navier_stokes} and
using the incompressibility condition, $\nabla \cdot \boldsymbol{u} = 0$, 
yields a Poisson equation for the pressure field
\begin{equation}
\label{eq:pressure_poisson}
\nabla^2 p = 
- \nabla \cdot (\boldsymbol{u} \cdot \nabla) \, \boldsymbol{u} \,.
\end{equation}
Equation~\ref{eq:pressure_poisson} shows two features
of the pressure field.  
Firstly, the pressure term in the Navier-Stokes equation
is a nonlinear term of the same order as the advective term, 
$(\boldsymbol{u} \cdot \nabla) \, \boldsymbol{u} $.  
Secondly, the pressure at a given point is determined by the 
velocity field everywhere in space at the same time.  
Equation~\ref{eq:pressure_poisson} thus highlights the 
crux of the turbulence problem -- the Navier-Stokes equation
is nonlinear, spatially nonlocal, and contains no small 
parameter in the theory \cite[][]{warhaft:2002a}.  
Any small perturbative expansion around the linear part of the
equation will be amplified by the nonlinear terms and eventually
diverge.  

To make progress, we turn away from the equation of motion and
construct phenomenological models based on physical assumptions.  
A recurring theme in many branches of physics is the search for
universality.  
Physical systems exhibiting universality display macroscopic phenomena
that are independent of the microscopic details of the systems.  
The emergence of simple continuum laws, like the Navier-Stokes 
equation, from complex underlying microscopic interactions would 
not have been possible had the fluid motions depended in great 
detail on the atomic interactions in the fluid molecules.  
In turbulence, there is an additional layer to the universality of the
physical phenomena we are seeking.  
The universality must not only be unaffected by 
the quantum-mechanical properties of the fluid, but it must also 
be impervious to the influence of the initial and boundary conditions 
of the flow.  
This is the vision of the forerunners of turbulence theorists
\cite[see][]{taylor:1935, kolmogorov:1941, prandtl:1945, 
heisenberg:1948, weizsaecker:1948, onsager:1949}.  
These related works, built upon Richardson's idea of the energy 
cascade \cite[][]{richardson:1922}, consider localized turbulent 
motions (eddies) as the putative mediators of energy.  
Turbulence is viewed as a hierarchy of eddies spanning over a 
wide range of length scales.  
Each eddy of size $\ell $ has a characteristic velocity $u(\ell)$ and
timescale $\tau (\ell)$.  
Larger eddies may contain smaller eddies.  
The largest eddies, of typical size $L_{0}$, are vigorously created by
the forcing mechanisms of turbulence.  
Feeding on the energy they received from the forcing, they become
unstable and break up into successively smaller unstable eddies.  
As the cascade proceeds, energy is being passed from the larger eddies
to the smaller ones, until the effect of viscosity on eddies of a
particular size becomes significant, and energy is dissipated as
heat.  

\cite{kolmogorov:1941}, building on the foundation of statistically
homogeneous isotropic turbulence laid down by \cite{taylor:1935}, 
is credited for quantifying Richardson's energy cascade and for 
introducing the concept of universality in turbulence.  
The tenets of homogeneous isotropic turbulence put forth by 
\cite{kolmogorov:1941} can be summarized as follows.  
The presentation has been abridged from \cite{pope:2000}.  
\newcounter{qcounter}
\begin{list}{(\alph{qcounter})}
{
\usecounter{qcounter}
\setlength\listparindent{0.25in}
}
\item For very high Reynolds number turbulent flows, the 
small-scale\footnote{In the language of turbulence, the term
`scale' is suggestive of a characteristic length scale for the 
size of an eddy.} turbulent motions are statistically 
isotropic.  

This hypothesis is usually referred to as the postulate of 
local isotropy.  
A turbulent flow is locally isotropic if it is locally homogeneous and 
if it is invariant with respect to rotations and reflections of the 
coordinate axes.

\item The statistical properties of the small scales 
($\ell \ll L_{0}$) is governed universally by the energy dissipation rate
per unit mass, $\epsilon$, and the fluid viscosity, $\nu$.  

The range of scales $\ell \ll L_{0}$  is known as the universal
equilibrium range, see figure~\ref{fig:eddy_size_scale}.  
The motion of eddies in the energy-containing range, $\ell \sim L_{0}$, 
is shaped by the initial and boundary conditions, and is therefore 
anisotropic and non-universal.  

Kolmogorov gave a prescription for calculating the length, time, and
velocity scales of the smallest eddies
\begin{align}
\label{eq:kolmogorov_length}
\eta &= \nu^{3/4} \, \epsilon^{-1/4} \, , \\
\label{eq:kolmogorov_time}
\tau_{\eta} &= \nu^{1/2} \, \epsilon^{-1/2} \,, \\
\label{eq:kolmogorov_velocity}
u_{\eta} &= \nu^{1/4} \, \epsilon^{1/4} \,.
\end{align}
They are appropriately referred to as the Kolmogorov length, time, and
velocity scales.  
The Reynolds number formed by these scales $u_{\eta} \, \eta / \nu$ is
unity, the smallness of which indicates that the motion of eddies at
scales $\ell < \eta$ is laminar.  

\item The statistical properties of those scales in the inertial range
  is determined by $\epsilon$ only.  

The universal equilibrium range is split into the
inertial range ($\eta \ll \ell \ll L_{0}$) and the dissipation range
($\ell < \eta$), see figure~\ref{fig:eddy_size_scale}.  
Eddy motion in the inertial range is neither affected by molecular
viscosity nor the forcing mechanism.  
\begin{figure}
\begin{center}
\includegraphics[scale=0.6]{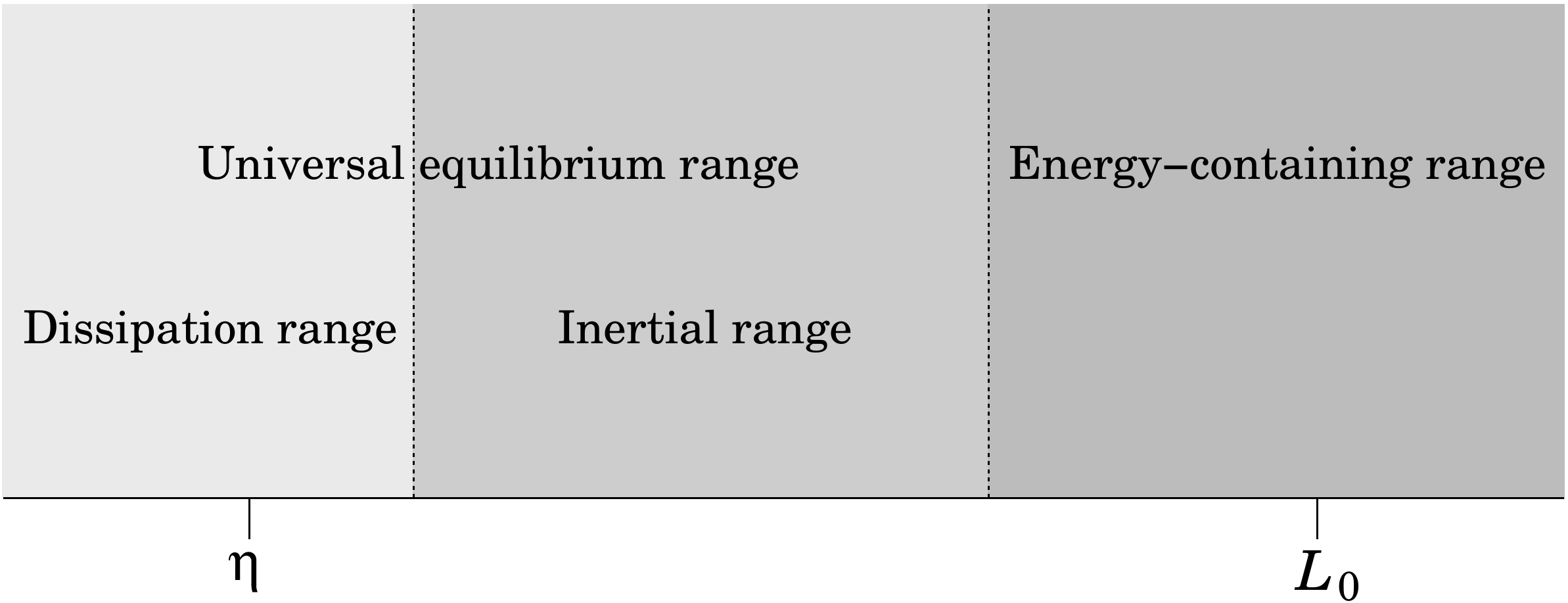}
\caption{The figure shows the range of eddy sizes $\ell$ from the
  largest ($L_{0}$) to the smallest ($\eta$) at very high Reynolds
  number.  The meaning of each length scale is explained in the text.}
\label{fig:eddy_size_scale}
\end{center}
\end{figure}

Recasting this hypothesis in the cascade picture, this means that the
rate of energy input ($\epsilon$) received by the largest eddies is
equal to the rate of energy transferred in succession from the largest
eddies to the smallest eddies.  
This rate, in turn, is equal to the energy dissipated by the smallest
eddies.  
An implicit assumption in this premise is that $\epsilon$ is
independent of $\nu$.  
Experimental evidence for this assumption has been collected by
\cite{sreenivasan:1984} and, more recently, by \cite{pearson:2002}.  
\end{list}

The validity of Kolmogorov's hypotheses is constantly under debate
\cite[e.g.][]{frisch:1995}.  
Its limitation to describing only theoretically ideal homogeneous
and isotropic turbulence has not only very little relevance to
anisotropic and non-infinite Reynolds number turbulence occurring 
in geophysical environment and in the laboratory, 
but the practical aspect of these hypotheses is also in question.  
For example, it is not clear to which statistical quantities they can 
be applied \cite[see][]{nelkin:1994}.  
There is, however, considerable predictive power in them and a myriad
of the existing work on turbulence rests on them.  
As an illustration, consider the second order velocity structure function
\begin{equation}
D_{ij} (\boldsymbol{r}) = \langle \delta u_i (\boldsymbol{r}, \boldsymbol{x}) 
\, \delta u_j (\boldsymbol{r}, \boldsymbol{x}) \rangle \,.
\end{equation}
$\delta u_i (\boldsymbol{r}, \boldsymbol{x})$
is the velocity increment
\begin{equation}
\delta u_i (\boldsymbol{r}, \boldsymbol{x}) = 
u_i^{\prime} (\boldsymbol{r} + \boldsymbol{x}) 
- u_i^{\prime} (\boldsymbol{x}) \,,
\end{equation}
where $u_i^{\prime}$ is the fluctuation in the $i$-th velocity component.  
$\boldsymbol{r}$ is the spatial increment and $\boldsymbol{x}$ is an
arbitrary reference point in space.  
The symbol $\langle \cdots \rangle$ is an ensemble average
in theoretical analysis, and a time average in experiment.  
The commonly studied one-dimensional structure functions are the 
projections onto Cartesian $\boldsymbol{r}$-axes of 
$D_{ij} (\boldsymbol{r})$.  
These functions are important because they are experimentally measurable.  
For simplicity, the coordinate system is chosen such that the separation $\boldsymbol{r}$ is in the $x_1$ direction.  
The two other mutually orthogonal directions $x_2$ and $x_3$ are then 
normal to $\boldsymbol{r}$.  
For homogeneous and isotropic turbulence, $D_{ij} (\boldsymbol{r})$ 
is related to the longitudinal structure function
\begin{equation}
D_{11} (r) = \langle (u_1^{\prime} (r,0,0) 
- u_1^{\prime} (0,0,0))^{2} \rangle \, , 
\end{equation}
and the transverse structure functions
\begin{gather}
D_{22} (r) = \langle (u_2^{\prime} (r,0,0) 
- u_2^{\prime} (0,0,0))^{2} \rangle \, , \\
D_{33} (r) = \langle (u_3^{\prime} (r,0,0) 
- u_3^{\prime} (0,0,0))^{2} \rangle \, ,
\end{gather}
by the equation
\begin{equation}
D_{ij} (\boldsymbol{r}) = D_{22} (r) \, \delta_{ij} + [D_{11} (r) -
D_{22} (r)] \, \dfrac{r_i \, r_j}{r^2} \,,
\end{equation}
where $D_{22} (r) = D_{33} (r)$ by symmetry.  
See e.g. \cite{pope:2000} for a derivation.  

For an incompressible fluid, the transverse structure
functions are related to the longitudinal structure function
\begin{equation}
\label{eq:sf_incompressibility}
D_{22} (r) = D_{33} (r) = \biggl( 1 + \dfrac{r}{2} \, \dfrac{\partial }{\partial r} \biggr) D_{11} (r)\,,
\end{equation}
so that only one scalar function, either $D_{11} (r)$ or $D_{22} (r)$, 
is needed for describing structure functions with distances 
measured along arbitrary direction.  

In the inertial range, $\eta \ll r \ll L_0$, an application of 
dimensional analysis and the universality hypothesis leads to 
the scaling for the structure functions at sufficiently high Reynolds
numbers \cite[][]{kolmogorov:1941}
\begin{gather}
\label{eq:dll_k41}
D_{11} (r) = C_2 \, (\epsilon \, r)^{2/3} \, , \\
\label{eq:dnn_k41}
 D_{22} (r) = D_{33} (r) = \dfrac{4}{3} \, C_2 \, (\epsilon \, r)^{2/3} \,,
\end{gather}
where $C_2$ is presumed to be a universal scaling constant, 
known as the Kolmogorov constant.  
These power laws are the first nontrivial predictions of Kolmogorov's
theory\footnote{An experimental discovery of the law of pair
  dispersion by \cite{richardson:1926} predates Kolmogorov's
  theoretical discovery.}.  
(See, however, \cite{frisch:1995}, chapter 6, for a modern derivation
assuming only the symmetries of the Navier-Stokes equation, i.e. time and
space invariances, rotational invariance, Galilean invariance, and
scale invariance.)  

The power-law behavior (equations~\ref{eq:dll_k41} and \ref{eq:dnn_k41}) 
has its Fourier space counterpart.  
This can be seen by applying the Wiener-Khinchin theorem 
(e.g. equation~$4.52$ of \cite{frisch:1995}), which states that the 
correlation function
\begin{equation}
\label{eq:correlation_function}
R_{ij} (\boldsymbol{r}) = \langle u^{\prime}_{i} (\boldsymbol{r}+\boldsymbol{x}) \, u^{\prime}_{j} (\boldsymbol{x}) \rangle \, ,
\end{equation}
and the three-dimensional velocity spectrum tensor
\begin{equation}
\Phi_{ij} (\boldsymbol{k}) = \langle \tilde{u}_{i} (\boldsymbol{k}) \, 
\tilde{u}_{j}^{*} (\boldsymbol{k}) \rangle \,,
\end{equation}
are Fourier transforms of each other
\begin{gather}
R_{ij} (\boldsymbol{r}) = \oint  \mathrm{d} \boldsymbol{k} \, \Phi_{ij} (\boldsymbol{k}) \, \mathrm{e}^{i \boldsymbol{k} \cdot \boldsymbol{r}} \,, \\
\Phi_{ij} (\boldsymbol{k}) = \dfrac{1}{(2 \, \pi)^3} \, \oint \mathrm{d} \boldsymbol{r} \, R_{ij} (\boldsymbol{r}) \, \mathrm{e}^{-i\boldsymbol{k} \cdot \boldsymbol{r}}\,.
\end{gather}
Here, the Fourier representation of $\tilde{u}_i (\boldsymbol{k})$ 
was introduced to turbulence theory by \cite{heisenberg:1948} 
and it is written as
\begin{equation}
\tilde{u}_i (\boldsymbol{k}) = \dfrac{1}{(2 \, \pi)^3} \, \oint \mathrm{d} \boldsymbol{r} \, u_i^{\prime} (\boldsymbol{r}) \, 
\mathrm{e}^{-i\boldsymbol{k} \cdot \boldsymbol{r}}\,.
\end{equation}
The asterisk in $\tilde{u}_{i}^{*} (\boldsymbol{k})$ denotes complex 
conjugate and reality of $u_i^{\prime} (\boldsymbol{r})$ implies
$\tilde{u}_{i}^{*} (\boldsymbol{k}) = \tilde{u}_{i} (\boldsymbol{-k})$.

Just as with the structure function $D_{ij} (r)$, $\Phi_{ij} (\boldsymbol{k})$ is usually projected onto Cartesian $\boldsymbol{k}$-axes to yield the experimentally accessible one-dimensional spectra
\begin{equation}
E_{ij} (k_1) = 2 \, \int_{-\infty}^{\infty} \mathrm{d} k_2 \int_{-\infty}^{\infty} \mathrm{d} k_3 \, \Phi_{ij} (\boldsymbol{k}) \,.
\end{equation}
Using the Wiener-Khinchin theorem, \cite{taylor:1938} showed 
that the one-dimensional spectra and the correlation functions 
form a Fourier transform pair
\begin{gather}
\label{eq:rtoe}
E_{ij} (k_1) = \dfrac{1}{\pi} \, \int_{-\infty}^{\infty} 
R_{ij} (r) \, \mathrm{e}^{-i k_1 r} \, \mathrm{d} r \, \\
\label{eq:etor}
R_{ij} (r) = \dfrac{1}{2} \, \int_{-\infty}^{\infty} E_{ij} (k_1) \, 
\mathrm{e}^{i k_1 r} \, \mathrm{d}k_1 \,.
\end{gather}
In the above, $\boldsymbol{r}$ is taken as $r \, \boldsymbol{e}_1$.  
The normalization factor is chosen so that when $r=0$, equation~\ref{eq:etor} gives, for example when $i=j=1$
\begin{equation}
\label{eq:r11_e11}
R_{11} (0) = \langle u_1^{\prime 2} \rangle 
= \int_{0}^{\infty} E_{11} (k_1) \, \mathrm{d} k_1 \,.
\end{equation}
Here, $E_{11} (k_1)$ is extended to negative frequencies by 
$E_{11} (-k_1) = E_{11} (k_1)$.  
If the turbulence is locally isotropic\footnote{If the turbulence is locally isotropic, then $\langle u^{\prime}_i (\boldsymbol{r}) \, 
u^{\prime}_j (\boldsymbol{r}) \rangle = 
\langle u^{\prime}_i (0) \, u^{\prime}_j (0) \rangle 
= R_{ij} (0)$, and also 
$R_{ij} (-\boldsymbol{r}) = R_{ij} (\boldsymbol{r})$}, the correlation 
function $R_{ij} (r)$ is then linked to the structure function $D_{ij} (r)$ 
by
\begin{equation}
\label{eq:sf_correlation}
D_{ij} (r) = 2 \, ( R_{ij} (0) - R_{ij} (r) )\,,
\end{equation}
which, when equation~\ref{eq:etor} is substituted into the above, yields
\begin{equation}
\label{eq:sf_spectrum_relation}
D_{ij} (r) = \int_{-\infty}^{\infty} (1 - \mathrm{e}^{i k_1 r}) \, E_{ij} (k_1) \, \mathrm{d} k_1 \,.
\end{equation}

In two short notes that predated his prediction for the structure function, Kolmogorov presented relevant mathematical description for turbulence energy spectra and their connection to structure functions \cite[see][]{kolmogorov:1940a, kolmogorov:1940b}.  
He considered the case when the spectra follow a power law, 
taking the longitudinal case, $i=j=1$, as an example
\begin{equation}
E_{11} (k_1) = B \, |k_1|^{-(\gamma + 1)} \,,
\end{equation}
where $B$ is a positive constant.  
A substitution of the above into equation~\ref{eq:sf_spectrum_relation}
with a change of variables, $x=k_1 r$, yields the following power law 
for the longitudinal structure function
\begin{equation}
D_{11} (r) = B \, A_{\gamma} \, |r|^{\gamma} \,,
\end{equation}
with the dimensionless integral
\begin{equation}
A_{\gamma} = \int_{-\infty}^{\infty} (1 - \mathrm{e}^{i x}) \, 
|x|^{-(\gamma + 1)} \, \mathrm{d} x \,.
\end{equation}
Convergence of the structure function limits the value of $\gamma$ 
to $0 < \gamma < 2$ (see e.g. page~90 of \cite{monin:1975}), 
in which case the integral can be evaluated analytically to give 
(for a derivation, see appendix~\ref{app:agamma})
\begin{equation}
A_{\gamma} = \dfrac{\pi}{\Gamma (\gamma + 1) \, 
\sin (\gamma \, \pi / 2)} \quad , \quad (0 < \gamma < 2) \,.
\end{equation}
Here, $\Gamma (x)$ is the gamma function.  
The conclusion to be drawn from this analysis is that if the spectrum 
exhibits a power-law behavior $E_{11} (k_1) = B |k_1|^{-(\gamma + 1)}$, 
then there is a corresponding power-law behavior $D_{11} (r) = B \, A_\gamma \, |r|^{\gamma}$ for the structure function.  

\cite{kolmogorov:1941} and \cite{obukhov:1941a, obukhov:1941b}
made the first attempt at determining the spectral power-law exponent
by assuming that the energy spectrum $E_{11} (k_1)$ has a universal 
form uniquely determined by the energy dissipation rate $\epsilon$.  
A straightforward dimensional analysis yields
\begin{equation}
\label{eq:k41_long_spectrum}
E_{11} (k_1) = C_k \, \epsilon^{2/3} k_1^{-5/3} \,,
\end{equation}
where $C_k$ is a dimensionless universal constant.  
It is related to the Kolmogorov constant, $C_2$, through 
(equation 21.25 of \cite{monin:1975})
\begin{equation}
\dfrac{C_2}{C_k} = A_{2/3} 
= \tfrac{3}{2} \, \Gamma (\tfrac{1}{3}) \approx 4 \,.
\end{equation}
Just as in the case with the transverse structure functions, the transverse spectra are also related to $E_{11} (k_1)$ through
\begin{equation}
\label{eq:k41_trans_spectra}
E_{22} (k_1) = E_{33} (k_1) 
= \dfrac{1}{2} \, \biggl(1 - k_1 \, \dfrac{\partial}{\partial k_1} \biggr) 
\, E_{11} (k_1) = \dfrac{4}{3} \, C_k \, \epsilon^{2/3} \, k_1^{-5/3} \,.
\end{equation}
Equations~\ref{eq:k41_long_spectrum} and \ref{eq:k41_trans_spectra}
are the famous one-dimensional Kolmogorov $-\tfrac{5}{3}$ spectra.  

\section{Verification of Kolmogorov theory}
\label{sec:k41_verification}

As a preliminary to the investigation of their angular dependence
in chapter~\ref{chap:universality}, we review, in the following, 
efforts made in establishing the exact value of the scaling exponents, 
the Kolmogorov constants, $C_2$ and $C_k$, and their Reynolds 
number trends.  

\begin{figure}
\begin{center}
\includegraphics[scale=0.5]{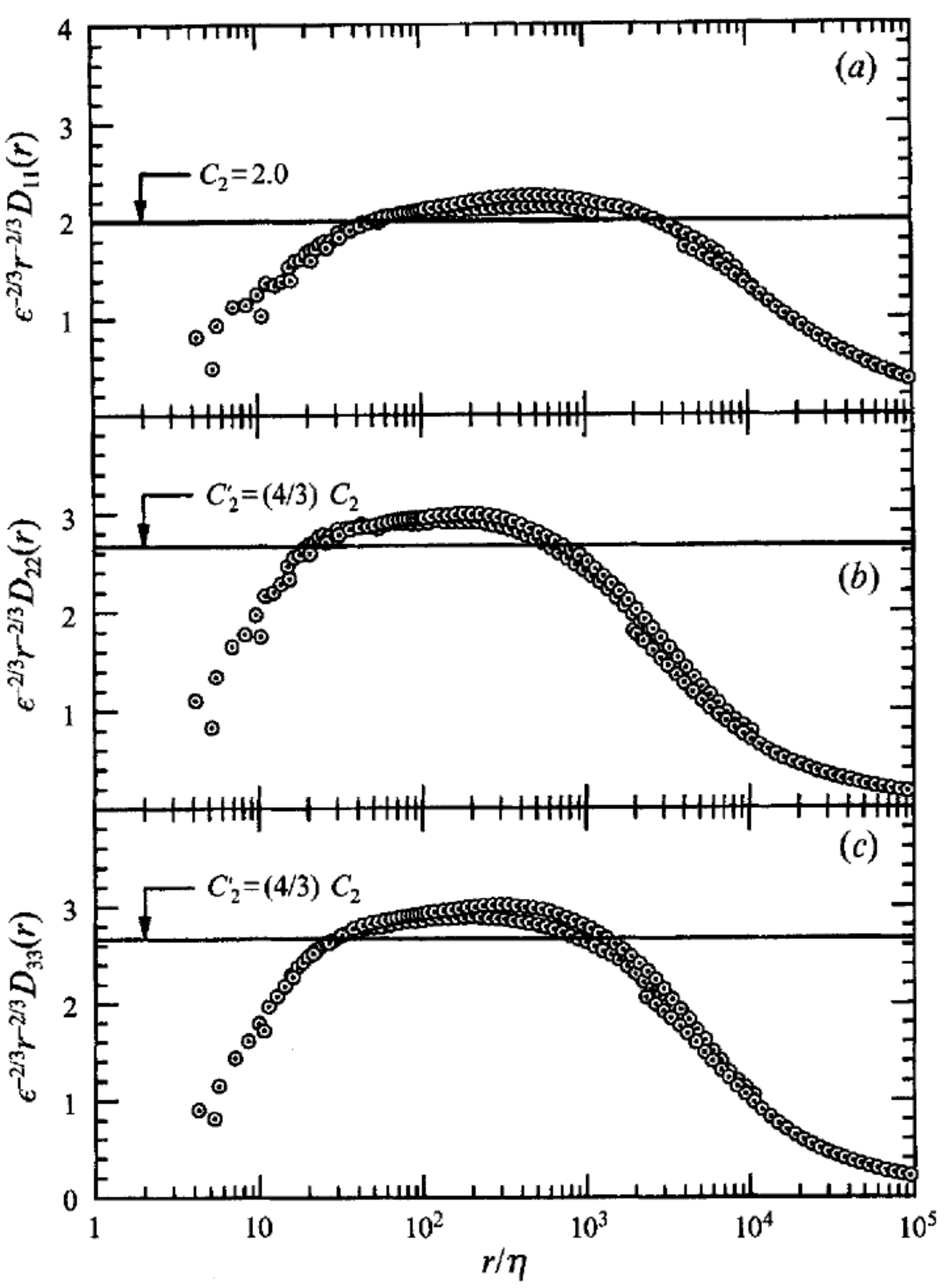}
\caption{The second-order longitudinal ($D_{11} (r)$) and transverse
  ($D_{22} (r)$ and $D_{33} (r)$) velocity structure functions
  measured in a turbulent boundary layer with $R_{\lambda} = 1450$.
  When compensated appropriately, the structure functions within the
  inertial range of scales are a measure of the value of the
  Kolmogorov constant.  Reprinted with permission
  from \cite{saddoughi:1994}: {\it J. Fluid Mech.}, 268:333-372,
  Copyright (1994) with permission from Cambridge University Press.}
\label{fig:saddoughi1994}
\end{center}
\end{figure}

Work on verification of the two-thirds law, equations~\ref{eq:dll_k41} and
\ref{eq:dnn_k41}, as well as estimation of $C_2$, up to the early 1970s 
has been reviewed by \cite{monin:1975} (see page~461).  
Kolmogorov made the first attempt to compare his theoretical 
prediction for the structure functions to the data obtained by 
\cite{dryden:1937} with hot-wire anemometry, in which
Taylor's frozen-turbulence hypothesis \cite[][]{taylor:1938} has been 
invoked to surrogate temporal velocity fluctuations for spatial velocity 
fluctuations (see the sidebar Taylor's Hypothesis).  
The measurements were obtained from a wind tunnel grid turbulence, 
\makebox[\textwidth][s]{whose Reynolds numbers ($R_{\lambda}$) 
we estimated to be in the range from $15$ to $140$.}
\begin{framed}
\noindent
{\sc Taylor's Hypothesis:} When studying a multiple-scale problem like turbulence, it is difficult to
obtain an instantaneous snapshot of all the eddies formed within
the turbulence.  
It is easier, and cheaper, to make measurements at one point 
over a long period of time.  
To study the turbulence from a continuous record of measurements 
from a single point, we need to assume that the advective velocity, 
$\langle U \rangle$, is uniform and the turbulence intensity, 
$u^{\prime}/\langle U \rangle$, is low.  
In other words, the turbulence is frozen, or evolves on a time scale 
much larger than the advective one, so that the conversion from 
time to space, $x = \langle U \rangle \, t$, can be made.  
For a reference on Taylor's hypothesis, see \cite{frisch:1995}, page 58.  
\end{framed}
\noindent
Kolmogorov suggested that $C_2 = 3/2$ \cite[][]{kolmogorov:1941b}.  
(See \cite{kolmogorov:1991} for an English translation of this work.)  
Shortly thereafter, a study carried out in the atmospheric surface layer 
by means of hot-wire anemometry was undertaken by \cite{obukhov:1942}, but the study was halted by World War II.  
Obukhov later resumed the study using the hot wire anemometer 
data of G\"{o}decke \cite[][]{obukhov:1949a, goedecke:1935}.  
Around the same time, many atmospheric surface layer studies 
employing hot wire anemometry were initiated in different 
parts of the world; for example \cite{townsend:1948} in the UK; 
\cite{ maccready:1953} and \cite{cramer:1959} in the USA; 
\cite{ shiotani:1955} in Japan; and \cite{taylor:1955} in Australia; 
after Kolmogorov's work received significant exposure due to 
\cite{batchelor:1946b, batchelor:1947}.  
The 1970s saw the emergence of precise hot-wire measurements.  
Measurements of \cite{vanatta:1970} in the atmospheric boundary
layer over the ocean showed that both $r^{2/3}$ and $r^{0.722}$ 
power laws fit well to their second order structure functions data, 
and that the value of $C_2$ is $2.3$ (error not reported).  
\cite{comtebellot:1971} measured longitudinal and transverse
velocity autocorrelations but did not report on measurements of
the structure functions.  
\cite{anselmet:1984} investigated structure functions up to the
18th order on the axis of turbulent jets, at $R_{\lambda}$ ranging 
from $536$ to $852$, and in a turbulent duct flow, at $R_{\lambda}=515$.  
They found that $C_2 = 2.2$ and $D_{11} (r) \sim r^{0.71}$ in the inertial
range.  
An experimental study by \cite{saddoughi:1994}, which later 
became a benchmark experiment in the field, 
was performed on the test-section ceiling of the $80 \times
120$~foot wind tunnel at NASA Ames Research Center.  
The flows are boundary layer flows at $R_{\lambda}$ ranging
from 500 to 1450.  
The Reynolds numbers achieved in this experiment are superior 
to those in many laboratory flows.  
It has only been surpassed by a few flows generated in 
high-Reynolds-number facilities, including the ONERA wind tunnel 
with $R_{\lambda} = 3374$ \cite[e.g.][]{gledzer:1996}, 
the Princeton University turbulent pipe \cite[e.g.][]{zagarola:1998} 
with a $R_{\lambda}$ of $2.3 \times 10^4$ (estimated from 
equation~\ref{eq:Rlambda_Re} where $\mathrm{Re}_L$ is based
on the average velocity and the pipe diameter), 
and the wind tunnel of the Central Aerohydrodynamic Institute in 
Moscow, Russia, with a $R_{\lambda}$ of $3200$ \cite[][]{praskovsky:1994}.  
\cite{saddoughi:1994} observed approximately one decade of
$r^{2/3}$ scaling (reproduced in figure~\ref{fig:saddoughi1994}).  
The value of the Kolmogorov constant reported is $2.0 \pm 0.1$.  
Subsequent measurements of the Kolmogorov constant by
\cite{anfossi:2000} in the atmospheric surface layer ($C_2 = 1.7$), 
\cite{degrazia:2008} and \cite{welter:2009} in the wind tunnel
and atmospheric boundary layers ($C_2 = 2.47 \pm 0.30$) at 
$R_{\lambda}$ in the range between $373$ and $2\times 10^4$ 
are marginally consistent with the value reported by 
\cite{saddoughi:1994}.  
The discrepancy between the values of the Kolmogorov constant 
measured in the laboratory and in the atmosphere may arise due 
to the uncertainty in atmospheric flow conditions, for example variability 
of wind speed and direction.  
We note that measurements from the controlled laboratory studies in 
\cite{degrazia:2008} and \cite{welter:2009} give $C_2 = 2.34 \pm 0.26$, 
in closer agreement with the value reported by \cite{saddoughi:1994}.  

Many of the past studies however, have been devoted to verifying
the spectra (equations~\ref{eq:k41_long_spectrum} 
and \ref{eq:k41_trans_spectra}).  
Experimental studies prior to the early 1970s have been 
reviewed by \cite{monin:1975} (see page~467), and subsequent 
studies up to the early 1990s have been extensively reviewed 
by \cite{sreenivasan:1995}.  
Data from numerical simulations have been collected by
\cite{yeung:1997}.  
We review in the following a few milestone experiments and
supplement the list with several recent ones.  

In a pioneering tidal-channel experiment at $R_{\lambda}$ 
ranging from $3000$ to $18000$, \cite{grant:1962}\footnote{These 
estimates for the Reynolds numbers were obtained 
by \cite{sreenivasan:1995}.  See the footnote therein for details.} 
observed more than two decades of $-5/3$ scaling and measured 
a value of $0.47 \pm 0.02$ for $C_k$ from data obtained with 
hot-film anemometry, which uses the same principle as hot-wire 
anemometry but the probe is less likely to be contaminated by dirt 
in water \footnote{\cite{kraichnan:1966} has analyzed the same 
data and observed that the estimate would be higher by more than 
10\% if one looked, instead, for flat region in plots of $\epsilon^{-2/3} 
\, k_1^{5/3} \, E_{11} (k_1)$ vs $k_1$.}.  
The hot-wire measurements of \cite{saddoughi:1994} show slightly less 
than a decade of $-5/3$ scaling, and in that range, $C_k = 0.49$.  
The reduced scaling range in this study might be an effect of the 
Reynolds number; we note that their Reynolds number is an order of
magnitude smaller than that of \cite{grant:1962}.  
The data of over 100 spectra, mostly collected with hot-wire technique, 
from various geophysical and laboratory flows collected in 
\cite{sreenivasan:1995} support a universal value of $C_k$, 
approximately $0.53 \pm 0.055$, beyond $R_{\lambda}$ of about $50$, 
as can also be inferred from recent experiments of \cite{welter:2009}.  
Similar conclusions can be drawn from numerical simulation studies
\cite[e.g.][]{yeung:1997, gotoh:2002, kaneda:2003}.  
On the other hand, these findings seem to be at odds with
the outcome uncovered in a few experiments, which suggest
$C_k$ follows a power law of the form $C_k \propto R_{\lambda}^{-\mu}$.  
The value of the exponent, $\mu$, is $0.094 \pm 0.004$ in 
the atmospheric surface layer \cite[][]{praskovsky:1994}, $2/3$ in 
a wind tunnel with specially designed active grid 
\cite[][]{mydlarski:1998}, and $0.16$ in a hydrodynamic 
simulation forced by Taylor-Green vortex \cite[][]{mininni:2008};  
whereas theory predicts $C_k \propto (\ln R_{\lambda})^{-1}$ 
\cite[][]{barenblatt:1995}.  

These conflicts seem to have been resolved in a recent study 
by \cite{donzis:2010}.  
The apparent Reynolds-number dependence is ascribable to 
inconsistent procedures followed in the estimation of $C_k$.  
A typical procedure for estimating $C_k$ is to plot the 
compensated spectrum, $\epsilon^{-2/3} \, k_1^{5/3} \, E_{11} (k_1)$, 
and seek a plateau, whose height in the inertial range is taken 
as $C_k$.  
This procedure is, however, complicated by the absence of 
a distinct plateau in the inertial range, and the existence of a 
spectral bump at the high-wavenumber end (the near-dissipation 
range), known as the bottleneck, whose effect is most prominent 
in the three-dimensional spectrum.  
The occurrence of the spectral bump was first captured in an 
analysis of \cite{qian:1984}.  
The physical mechanism leading to its appearance has been 
attempted by various workers \cite[e.g.][]{yakhot:1993, 
falkovich:1994, martinez:1997, kurien:2004, 
verma:2007, bershadskii:2008, frisch:2008}.  
\cite{donzis:2010} showed that the height of this bump in the three-dimensional spectrum varies as $R_{\lambda}^{-0.04}$, 
while the height and the location of the local minimum preceding the 
bump remain practically constant.  
\cite{donzis:2010} suggest evaluating $C_k$ from the height
of the local minimum before the bump, as this leads to 
a Kolmogorov constant independent of $R_{\lambda}$.  
It remains for future investigations to elucidate the underlying mechanism
of this operation.  

Since the spectra are uniquely related to the structure functions, 
one may expect that the bottleneck effect in wavenumber space 
will have its physical space correspondence near the transition 
between the dissipation and the inertial ranges.  
\cite{dobler:2003} argued that localized features in wavenumber 
space will become sufficiently non-local in physical 
space that they are practically undetectable.  
\cite{lohse:1995}, on the other hand, argued that a rapid transition 
between these ranges in the physical space is the manifestation 
of the spectral bump.  
The findings of \cite{donzis:2010} confirm this belief.
To assess the effect of the Reynolds number on the transition region,
they studied the Batchelor interpolation formula \cite[][]{batchelor:1951}
\begin{equation}
\label{eq:batchelor_interpolation}
\dfrac{D (r)}{u^{2}_{\eta}} = K \, 
\dfrac{(r/\eta)^2}{[1+(c_B \, r / \eta)^q]^{(2-\zeta_2)/q}} \,,
\end{equation}
with $K=1/15$ for the longitudinal and $2/15$ for the transverse
structure functions, $u_\eta$ is the Kolmogorov velocity 
(see equation~\ref{eq:kolmogorov_velocity}), and $\zeta_2 = 0.67$ 
is the inertial range scaling exponent.  
$c_B$ and $q$ are fit parameters.  
They found $c_B = 0.076$ and $0.102$ for longitudinal and transverse
structure functions, independent of $R_{\lambda}$, and the
power laws $q \propto R_{\lambda}^{-1.06}$ and 
$q \propto R_{\lambda}^{-0.92}$ for the longitudinal and transverse 
structure functions, respectively.  
The decrease of $q$ with $R_{\lambda}$ confirms the belief that the
transition at high Reynolds numbers is smoother, which is the 
manifestation of a flatter spectral bump.  
Such a subtle Reynolds number dependence is currently beyond 
the resolution limit of experiments and may have escaped 
definitive detection.  
Notwithstanding this subtle effect, at least the contamination by a 
'physical' bump in the determination of $C_2$ can be obviated in 
the physical space\footnote{The typical procedure in physical space 
is to plot, for example, $(\epsilon \, r)^{-2/3} \, D_{11} (r)$ vs $r$ and 
seek a plateau whose height is taken as the value of $C_2$.}.  
Between the approach in spectral space and that in physical space, the approach in physical space remains the better one.  

\section{Relation to other inertial-range constants}

The following discussion is intended to illustrate the interrelations 
between the Kolmogorov constant and the scaling constants of 
the Lagrangian structure functions, the Richardson-Obukhov constant
for turbulent relative dispersion, and the Obukhov-Corrsin constant 
for passive scalar structure function.  
The scalar-valuedness of the Kolmogorov constant to be established 
in chapter~\ref{chap:universality} may be taken as a constraint
on the angular dependence of the various constants.  

It is often the case that analyzing the same physical problem 
from a different perspective may reveal unexpected results 
and advance our understanding of the problem.  
The structure functions introduced in the previous section are 
functions of separations fixed in space.  
Analyzing fluid motion at fixed spatial locations is known as
the Eulerian description.  
To inspect the turbulence problem in a different way,
researchers have devised the Lagrangian description, where an 
observer moves along with the flow \cite[e.g.][]{toschi:2009}.  
The basic tools in the Lagrangian description of turbulence are
the Lagrangian structure functions, defined as
\begin{equation}
D_p^L (\tau) = \langle \delta u_{i} (\tau) \, \delta u_{j} (\tau) \rangle \,,
\end{equation}
where the velocity increments, $\delta u_i (\tau) = 
u_i^{\prime} (t + \tau) - u_i^{\prime} (t)$, 
are taken along the trajectory of an individual fluid particle.  

Obukhov and Landau \footnote{The scaling form for the 
Lagrangian velocity structure function was independently 
obtained by Obukhov and Landau shortly after the publication 
of Kolmogorov's 1941 paper.  See \cite{monin:1975}, page 359, 
for a historical account.} applied Kolmogorov's theory to
derive scaling forms for the Lagrangian velocity structure 
functions valid in the inertial range
\begin{equation}
\langle \delta u_i (\tau) \, \delta u_j (\tau)\rangle 
= C_0 \, \epsilon \, \tau \, \delta_{ij} \, ,
\end{equation}
where $i$ is any Cartesian component $x$, $y$, or $z$; $\epsilon$
is the energy dissipation rate per unit mass; and $\delta_{ij}$ is the
Kronecker delta.  
See e.g. \cite{monin:1975}, page~358 for a derivation.  
The constant $C_0$, presumed to be universal, is known as the 
Lagrangian velocity structure function constant.  
It is an important parameter in stochastic models of turbulent transport
and dispersion \cite[e.g.][]{rodean:1991, sawford:1991, weinman:2000}.  
Quality measurements of $C_0$ are scarce and its value is very uncertain,
partly because Lagrangian experiments, where the trajectories
of fluid particles are followed in both time and space 
\cite[e.g.][]{laporta:2000}, have historically been very difficult.  
Reported values of $C_0$ range from $1.0$ to $7.0$ \cite[][]{poggi:2008}.  
\cite{inoue:1951} applied the Kolmogorov theory
to the Lagrangian velocity spectrum $\phi_{ij}^{L} (\omega)$, defined 
as the Fourier transform of the Lagrangian velocity autocorrelation tensor, 
and obtained $\phi_{ij} (\omega) = B \, \epsilon \, \omega^{-2} \, 
\delta_{ij}$, where $\omega$ is the angular frequency.  
$B$ and $C_0$ are simply related by $C_0 / B = A_1 = \pi$, see e.g. \cite{monin:1975}, page 361.  
Under the assumption that the velocity fluctuations integrated from Eulerian and Lagrangian spectra are equal \cite[][]{lumley:1957, lumley:1962}, 
\cite{corrsin:1963} approximated the Eulerian and Lagrangian spectra 
by their inertial-range forms between appropriate wave number and
frequency limits and obtained $B / C_{k}^{3/2} = \mathrm{constant}$, 
where $B$ is the scaling constant of the Lagrangian velocity 
spectrum.  
The same power law has recently been derived by 
\cite{franzese:2007} based on a statistical diffusion theory 
of relative dispersion of fluid particles.  

The Lagrangian velocity structure function constant is, quite 
astonishingly, also related to the Richardson-Obukhov constant 
governing the turbulent relative dispersion from a point source
in isotropic turbulence.  
\cite{obukhov:1941b}, refining the empirical finding of 
\cite{richardson:1926}, applied Kolmogorov's dimensional reasoning
to the mean square separation distance between two particles 
$\langle \boldsymbol{r}^2 (t) \rangle$ and derived an expression 
valid in the inertial range of scales
\begin{equation}
\langle \boldsymbol{r}^2 (t) \rangle = g \, \epsilon \, t^3 \,,
\end{equation}
where $\epsilon$ is the energy dissipation rate.  
The scaling constant $g$ is known as the Richardson-Obukhov constant.  
See e.g. \cite{ouellette:2006b} for a modern derivation.  
Experimental and numerical studies seem to be converging
on a value in the range $0.5  \leqslant g \leqslant 0.6$ 
\cite[][]{salazar:2009}.  
The interest in the law of two-particle dispersion is immense 
because of its utility in describing transport and mixing processes in environmental and engineering problems.  
Several investigators \cite[see][]{lin:1960, novikov:1963, ivanov:1963} 
exploited the assumption that the accelerations of fluid particles are uncorrelated at high Reynolds number to connect two-particle 
statistics with Lagrangian one-particle statistics.  
All their analyses yield the result $g = 2 \, C_0$.  
\cite{borgas:1991} showed that taking into account the two-particle acceleration correlation provides a value of $g$ smaller than 
$2 \, C_0$.  
The recent work of \cite{franzese:2007} based on a statistical diffusion
theory of relative dispersion provides an estimate $g \approx C_0 / 11$.  

The relative motion between particles is, remarkably, connected to 
the concentration profile of a substance, the so-called passive scalar.  
The substance could be a contaminant, like smoke dispersing in air; 
it could also be odor, like the smell of perfume traveling in air; 
or heat, as in the case when a weakly heated object is cooled in 
the flow.  
In his seminal paper on turbulent relative dispersion, \cite{richardson:1926}
introduced the distance-neighbor function $q(\boldsymbol{r}, t)$, which
is the equal-time probability density function for the separation 
$\boldsymbol{r}$ between fluid particle pairs randomly chosen from
a scalar field $\theta^{\prime} (\boldsymbol{r}, t)$ passively 
advected by the turbulence.  
They are related by \cite[e.g.][]{ott:2000}
\begin{equation}
\label{eq:q_theta}
q (\boldsymbol{r}, t) = \int \langle \theta^{\prime} (\boldsymbol{x}, t) \,
\theta^{\prime} (\boldsymbol{x}+\boldsymbol{r}, t) \rangle \, 
\mathrm{d} \boldsymbol{x} \,.
\end{equation}
Assuming homogeneity, isotropy, and stationarity, we also have by 
definition
\begin{equation}
\label{eq:rsqr_q}
\langle \boldsymbol{r}^2 (t) \rangle = 
\int_{0}^{\infty} 4 \, \pi \, s^4 \, q (s, t) \, \mathrm{d} s \,.
\end{equation}
Equations~\ref{eq:q_theta} and \ref{eq:rsqr_q} provide a formal
link between the mean square separation distance between 
particle pairs and the autocorrelation of the passive scalar fluctuations.  
Assuming the classical inertial range scaling for $q$, \cite{thomson:1996}
showed that the connection between the scalar dissipation
rate and relative dispersion is given by
\begin{equation}
\dfrac{\partial \langle \theta^{\prime 2} \rangle}{\partial t}
= - \dfrac{3}{2} \, \langle \theta^{\prime 2} \rangle \, 
\dfrac{\partial \ln \langle \boldsymbol{r}^2 \rangle}{\partial t} \,.  
\end{equation}
\cite{obukhov:1949b} and \cite{corrsin:1951} independently 
extended Kolmogorov's theory to the passive scalar and derived
for the second order structure function of scalar increments
\begin{equation}
\langle (\theta^{\prime} (\boldsymbol{r}, t) - 
\theta^{\prime} (0, t))^2 \rangle 
= C_{\theta} \, \epsilon_{\theta} \, \epsilon^{-1/3} \, r^{2/3} \,,
\end{equation}
where $\epsilon$ is the energy dissipation rate, 
$\epsilon_{\theta} = 2 \, \kappa \, \langle (\nabla 
\theta^{\prime})^2 \rangle$ is the rate of dissipation of concentration fluctuations, and $\kappa$ is the diffusivity.  
$C_{\theta}$ is known as the Obukhov-Corrsin constant.  
Experimental values of $C_{\theta}$ lie mostly in the range
$1.21 \leqslant C_{\theta} \leqslant 2.01$ \cite[][]{sreenivasan:1996}.  
\cite{thomson:1996}, applying the two-particle statistical theory of \cite{batchelor:1952}, obtained a result relating the Obukhov-Corrsin constant
to the Richardson-Obukhov constant
\begin{equation}
g = \tfrac{64}{243} \, \alpha_1^3 \, C_{\theta}^{-3} \,,
\end{equation}
for some constant $\alpha_1$ that shapes the spreading of 
the \cite{richardson:1926} distance-neighbor function.  

While many of the inertial range constants still elude precise
theoretical modeling and experimental quantification, 
we believe that a complete theory of turbulence should
and must unveil the interrelations between them.  
Here, we make no attempt at seeking such a fundamental
relation, but we verify in chapter~\ref{chap:universality} the scalar 
invariance of $C_2$, which is inherently implied in all theoretical 
investigations that invoked the local isotropy hypothesis, as well as 
in many single-point measurements that assumed Taylor's hypothesis.  

\section{Higher-order structure functions}
\label{sec:higher_order_sf}

In preparation for the examination of the influence of anisotropy
on the scaling exponents of structure functions of order greater 
than 2 in chapter~\ref{chap:higherorderstats}, we introduce
in the following the higher-order structure functions and their
role in the study of intermittency.  

According to \cite{landau:1959}, the arguments leading
to Kolmogorov's prediction presented in the preceding sections 
do not take into account the possible 'bursty' nature of the energy 
dissipation rate.  
A distinctive feature of turbulence is its abrupt and extremely
intense fluctuations in the observables.  
Landau's main objection is that the $\epsilon$ in Kolmogorov's
theory is a mean taken over time.  
He argued that since $\epsilon$ fluctuates in time, 
$\langle \epsilon^n \rangle \neq \langle \epsilon \rangle^n$, 
except when $n=1$, and therefore the mean rate of energy 
$\langle \epsilon \rangle$ is insufficient to describe turbulence.  
This phenomenon is known as intermittency.  
(See \cite{frisch:1995}, 6.4, for a historical account of Landau's
objection.)

It has now become common in studies of intermittency
to look at higher-order velocity structure functions.  
Kolmogorov's arguments can be naturally extended to moments
of velocity increments, $\delta u (r) = u^{\prime} (x+r) - 
u^{\prime} (x)$, to order higher than two.  
For example, the $p$th-order longitudinal structure function, 
$D_{p}^{L} (r) = \langle (\delta u_L (r))^p \rangle$, 
where the subscript $L$ indicates that the direction of velocity 
is along the separation vector, scales in the inertial range as
\begin{equation}
\label{eq:k41_moments}
D_{p}^{L} (r) = C_p \, (\epsilon r)^{\zeta_p^L} \,,
\end{equation}
with universal scaling exponents given by the Kolmogorov prediction 
\begin{equation}
\label{eq:zetap_k41}
\zeta_p^L = p/3 \,.
\end{equation}
The values of $\zeta_p^{L}$ given by the above equation for order 
$1$ to $6$ are listed in table~\ref{table:she_leveque_exponents}.  
The constants $C_p$s are presumed to be universal.  
To date, there exist no measurements that display unambiguous 
inertial range scaling over an enormously wide scaling 
range \cite[][]{sreenivasan:1998}.  
Thus, several self-consistent procedures have been devised for
better estimation of the scaling exponents.  
The Batchelor interpolation formula 
(equation~\ref{eq:batchelor_interpolation}), based on matched
asymptotics, is useful but involves a very elaborate nonlinear 
fitting procedure \cite[see e.g.][]{kurien:2000b}.  
Another simple and intriguing procedure has been proposed
by \cite{benzi:1993}.  
They observed that structure functions plotted against any other
structure function of a given order, rather than against the 
separation, generally yields a measurably improved inertial
scaling range at low Reynolds numbers.  
This procedure is known as the extended-self-similarity (ESS) method.  
The reason for its success in Navier-Stokes turbulence is unclear, but
recent progress in Burgers turbulence explains this 
transformation of variables from physical to structure function 
space as a method to deplete the subdominant inertial and 
dissipative range contributions that mask power-law scaling
\cite[][]{chakraborty:2010}.  

In practice, the third-order longitudinal structure function,
$\langle (\delta u_L (r))^3 \rangle$, is chosen as the 
surrogate for separation.  
The foundation for this choice lies in an exact result obtained by
\cite{kolmogorov:1941b} from the Navier-Stokes equation for 
freely decaying isotropic turbulence, and it remains the only 
exact result derived from the equations of motion.  
Starting with the K\'{a}rm\'{a}n-Harwarth equation 
\cite[][]{karman:1938}, Kolmogorov rigorously showed that
\begin{equation}
D_{3}^{L} (r) = - \dfrac{4}{5} \, \epsilon \, r  + 6 \, \nu \, \dfrac{\mathrm{d} D_{2}^{L} (r)}{\mathrm{d} r} \,.
\end{equation}
In the inertial range, this reduces to the celebrated Kolmogorov's 
four-fifths law \cite[][]{kolmogorov:1941b}
\begin{equation}
D_3^L (r) = - \dfrac{4}{5} \, \epsilon \, r \,.
\end{equation}
Thus, relating the above to equation~\ref{eq:k41_moments}, the 
results $C_3 = -4/5$ and $\zeta_3^L = 1$ are both exact and nontrivial.  
To be precise, ESS can only be used to determine relative scaling
exponents.  
Since $\zeta_3^L$ is exactly $1$, the relative scaling exponents
may be directly related to the true scaling exponents.  

Another variant of ESS, which gives even better scaling, modifies 
structure functions of order $p$ by taking the moments of the absolute 
values of the velocity increments, $\langle |\delta u_L|^p \rangle$, 
the so-called generalized structure functions \cite[][]{vainshtein:1994}.   
Strictly speaking, the four-fifths law only applies to 
$\langle (\delta u_L)^3 \rangle$, with no absolute value.  
\cite{sreenivasan:1996b} have found small differences in the
scaling of $\langle (\delta u)^{p}\rangle$ and 
$\langle |\delta u|^p \rangle$.  
The difference is small, however, and is very difficult to observe 
experimentally.  
One then replaces the scaling laws in 
equation~\ref{eq:k41_moments} by
\begin{equation}
\langle |\delta u_L|^p \rangle 
\propto \langle |\delta u_L|^3 \rangle^{\zeta_p^L} \,.
\end{equation}
Plotting the structure functions relative to $\langle |\delta u_L|^3 \rangle$
shows cleaner inertial range scaling behavior than plotting them
against the separations.  
Because of its utility in the longitudinal structure functions, the ESS 
technique has been extended to the transverse case.  
The transverse ESS scaling laws
\begin{equation}
\label{eq:sf_trans_ESS}
\langle |\delta u_T|^p \rangle 
\propto \langle |\delta u_T|^3 \rangle^{\zeta_p^T} \,,
\end{equation}
where the subscript $T$ denotes velocity direction perpendicular
to the separation vector, and $\zeta_{p}^{T} = p/3$ is the inertial 
range scaling exponent given by the Kolmogorov prediction, 
$D_{p}^{T} (r) = C_p \, (\epsilon \, r)^{\zeta_p^T}$.  
This version of ESS is less justified, but is widely in use 
\cite[e.g.][]{herweijer:1995, camussi:1996, kahalerras:1996, 
noullez:1997, vandewater:1999, zhou:2000, pearson:2001, zhou:2001}, 
although defining the transverse structure functions in this way 
always yields values for the odd moments, even when they do not 
exist for homogeneous isotropic turbulence \cite[][]{shen:2002}.  
We will see in chapter~\ref{chap:universality} that at moderate 
Reynolds numbers, the inertial range is too narrow to fit a power law 
in $r$, and we will make use of ESS to determine $\zeta_p$.  

\section{Anomalous transverse scaling exponents}
\label{sec:anomalous_exponents}

As we have seen in the previous section, in order to verify the 
Kolmogorov prediction for the higher-order scaling exponents
(equation~\ref{eq:zetap_k41}), we naturally need to measure
the higher-order structure functions.  
Measurements of the longitudinal structure functions up to
the fourth order were first obtained by \cite{vanatta:1970}, 
and subsequently, up to the 18th order, by \cite{anselmet:1984}.  
The longitudinal scaling exponents, $\zeta_p^L$, from both 
measurements show departure from the expression \ref{eq:zetap_k41} 
(the difference is less than $5\%$ at fourth order, and increases 
monotonically up to $30\%$ at 12th order), a phenomenon known 
as anomalous scaling, but it is well predicted by the She-Leveque 
model \cite[][]{she:1994} using a hierarchical model for the energy 
dissipation rate, which gives for the scaling exponent
\begin{equation}
\label{eq:zetap_she_leveque}
\zeta_p^L = \dfrac{p}{9} + 2 \, 
\bigg[ 1 - \bigg(\dfrac{2}{3}\bigg)^{p/3} \bigg] \,.
\end{equation}
The values of $\zeta_p^L$ given by the above equation for $p$ up to 
the sixth order are listed in table~\ref{table:she_leveque_exponents}.  
\begin{table}
\begin{center}
\begin{tabular}{ll*{6}{c}}
\toprule
\addlinespace[8pt]
\multirow{2}{*}{$\zeta_p$} & \multirow{2}{*}{Source} & 
\multicolumn{6}{c}{Order $p$} \\
\addlinespace[5pt]
\cmidrule(l){3-8}
\addlinespace[5pt]
& & $1$ & $2$ & $3$ & $4$ & $5$ & $6$ \\
\addlinespace[5pt]
\midrule
\addlinespace[8pt]
\multirow{3}{*}{$\zeta_p^L$} & \cite{kolmogorov:1941} &
$0.333$ & $0.667$ & $1$ & $1.333$ & $1.667$ & $2$ \\
\addlinespace[8pt]
& \cite{she:1994} & 
$0.364$ & $0.696$ & $1$ & $1.280$ & $1.538$ & $1.778$ \\
\addlinespace[8pt]
& \cite{dhruva:1997} &
$0.366$ & $0.700$ & $1$ & $1.266$ & $1.493$ & $1.692$ \\
\addlinespace[5pt]
\midrule
\addlinespace[8pt]
$\zeta_p^T$ & \cite{dhruva:1997} &
$0.359$ & $0.680$ & $0.960$ & $1.200$ & $1.402$ & $1.567$ \\
\addlinespace[5pt]
\bottomrule
\end{tabular}
\caption{The inertial range scaling exponents of the longitudinal 
and transverse structure functions.  
The first two rows are the longitudinal scaling
exponents, $\zeta_p^L$, predicted by \cite{kolmogorov:1941} 
(see equation~\ref{eq:zetap_k41}) and \cite{she:1994} 
(see equation~\ref{eq:zetap_she_leveque}).  
The third and fourth rows show the longitudinal and transverse 
exponents measured by \cite{dhruva:1997} using the ESS method 
(see section~\ref{sec:higher_order_sf}) in atmospheric turbulence 
at $R_{\lambda}$ between $10^4$ and $1.5 \times 10^4$.}
\label{table:she_leveque_exponents}
\end{center}
\end{table}

It was thought that the transverse scaling exponent, $\zeta_p^T$ 
(defined in equation~\ref{eq:sf_trans_ESS}), is equal to $\zeta_p^L$.  
This is true only for $p=2$ because of the incompressibility condition
(see equation~\ref{eq:sf_incompressibility}).  
For $p\neq 2$, the transverse and longitudinal scaling exponents 
are not related by any constraint.  
In fact, measurements of \cite{herweijer:1995} in shear flows at 
$R_{\lambda} = 340-810$ did suggest that the difference 
$\zeta_p^T - p/3$ is larger than $\zeta_p^L - p/3$ for $p \geqslant 4$, 
implying that $\zeta_p^T < \zeta_p^L$ for $p \geqslant 4$.  
The difference persists even at $R_{\lambda}$ up to $10^4$, as shown by
the hot-wire measurements of \cite{dhruva:1997} in the atmospheric 
surface layer, reproduced in table~\ref{table:she_leveque_exponents}.  
The same result is found in other experiments \cite[][]{camussi:1997, 
antonia:1999, zhou:2000, romano:2001, hao:2008} 
and numerical simulations \cite[][]{chen:1997, boratav:1997, 
grossmann:1997, gotoh:2002}.  
A number of experiments, however, suggest that the two exponents
are equal \cite[][]{camussi:1996, noullez:1997, kahalerras:1998, he:1999}.  
A few authors \cite[][]{pearson:2001, zhou:2001, antonia:2002, 
shen:2002, zhou:2005} report both $\zeta_p^T \sim \zeta_p^L$ and 
$\zeta_p^T < \zeta_p^L$ in their experiments.  

Table~\ref{table:zeta_truthtable} is an assessment of the issue.  
We would like to caution the reader of a too simplistic 
interpretation of table~\ref{table:zeta_truthtable}.  
The flow conditions, measurement techniques, and the definitions of 
$\zeta_p^L$ and $\zeta_p^T$ vary from experiment to experiment.  
We note that in most experiments, Taylor's hypothesis has been invoked 
to surrogate temporal series of measurements for spatial ones (see the
sidebar Taylor's Hypothesis in section~\ref{sec:k41_verification}).  
On the other hand, \cite{lin:1953} has shown that the hypothesis
breaks down in the case of flows containing high shear.  
Correction to Taylor's hypothesis based on spectral information
gathered in only one direction and on a local derivative in the
remaining direction has been proposed by \cite{delalamo:2009}.  
We note that in some cases when the scaling exponents do agree, 
Taylor's hypothesis has not been invoked \cite[e.g.][]{noullez:1997,
shen:2002, zhou:2005} or a modified version based on the instantaneous
velocity $u(t)$, instead of the mean velocity $\langle U \rangle$, 
has been used to obtain the spatial separation $\Delta r = u(t) \, \Delta t$ 
\cite[e.g.][]{kahalerras:1998}.  
Because many authors report only structure functions constructed
from single-point measurements of the velocity signals, we have 
deliberately limited the scope of comparison in 
table~\ref{table:zeta_truthtable} to those that report two-point
measurements that truly scan the physical space.  
\begin{table}
\begin{center}
\begin{tabular}{llccl}
\toprule
\addlinespace[8pt]
Flow & TH & $R_{\lambda}$ & Scaling exponent & Source\\
\addlinespace[5pt]
\midrule
\addlinespace[8pt]
Shearless & No & $100-300$ & $\zeta_p^T \lesssim \zeta_p^L$ & 
\cite{zhou:2005}\\
\addlinespace[8pt]
Shearless & No & $863$ & $\zeta_p^T \sim \zeta_p^L$ & 
\cite{shen:2002} \\
\addlinespace[8pt]
Shearless & Yes & $100-300$ & $\zeta_p^T < \zeta_p^L$ & \cite{zhou:2005} \\
\addlinespace[8pt]
Shearless & Yes & $863$ & $\zeta_p^T < \zeta_p^L$ & \cite{shen:2002} \\
\addlinespace[8pt]
Shearless & Yes & $10^4-1.5\times 10^4$ & $\zeta_p^T \lesssim \zeta_p^L$ & \cite{dhruva:1997} \\
\addlinespace[8pt]
Sheared & No & $254$ & $\zeta_p^T < \zeta_p^L$ & \cite{shen:2002} \\
\addlinespace[8pt]
Sheared & No & $875$ & $\zeta_p^T \sim \zeta_p^L$ & \cite{shen:2002} \\
\addlinespace[8pt]
Sheared & Yes & $254$ & ESS not applicable & \cite{shen:2002} \\
\addlinespace[8pt]
Sheared & Yes & $875$ & $\zeta_p^T < \zeta_p^L$ & \cite{shen:2002} \\
\addlinespace[5pt]
\bottomrule
\end{tabular}
\caption{The equality between longitudinal and transverse scaling 
exponents for sheared and unsheared turbulence and their trends with
Reynolds number.  
The acronym TH denotes the use of Taylor's hypothesis in the experiments.}
\label{table:zeta_truthtable}
\end{center}
\end{table}

Our second observation is that there is no consensus among researchers
on the choice of the third order structure function for the ESS method.  
While all researchers agree on using the ESS scaling 
$\langle |\delta u_L|^p \rangle 
\propto \langle |\delta u_L |^3 \rangle^{\zeta_p^L}$ for 
$\zeta_p^L$, the ESS scalings
$\langle |\delta u_T|^p \rangle 
\propto \langle |\delta u_L |^3 \rangle^{\zeta_p^T}$ and
$\langle |\delta u_T|^p \rangle 
\propto \langle |\delta u_T |^3 \rangle^{\zeta_p^T}$ 
have been proposed for $\zeta_p^T$.  
We observe in the data of \cite{pearson:2001, zhou:2001, zhou:2005} 
that ESS scaling based on $\langle |\delta u_L |^3 \rangle$ tends to
produce a wider scatter and a value for $\zeta_p^T$ that is smaller 
than the one obtained with $\langle |\delta u_T |^3 \rangle$.  
We think that the uncertainty in the reported values of $\zeta_p^T$
partly stems from the inconsistency in the procedure used in obtaining
the transverse scaling exponent and our attempt here is to test
the directional dependence of $\langle |\delta u_T|^p \rangle \propto 
\langle |\delta u_T |^3 \rangle^{\zeta_p^T}$.  

The difference between the longitudinal and transverse scaling
exponents remains one of the unresolved issues of turbulence.  
The reasons for this inequality have been proposed by various
authors.  
For example, \cite{romano:2001} and \cite{zhou:2001} list the 
following as plausible explanations:
\begin{inparaenum}[\itshape a\upshape)]
\item the anisotropy of the large scale, 
\item the Reynolds number effect, 
\item the differences in initial and boundary conditions,
\item different intermittency.  
\item the effect of Taylor's hypothesis.  
\end{inparaenum}
Among these possible causes, which may be related, 
\cite{romano:2001} have singled out the anisotropy of the flow 
as the most likely source for the observed inequality.  

A complementary approach has been developed to extract
the anisotropic contributions to the scaling of structure functions 
\cite[][]{biferale:2005}.  
This approach uses the irreducible representation of the SO(3)
symmetry group.  
Instead of projecting the structure functions onto the
longitudinal and transverse components in the traditional
way, we expand the structure functions in terms of
spherical harmonics\footnote{Following the usual convention, 
the orbital angular momentum is represented by $\ell$.  
Its projection onto the $z$ axis is represented by $m$.}, 
$Y_{\ell}^m$, sidestepping the questions about the 
differences between longitudinal and transverse scaling 
exponents.  
Because the isotropic ($\ell = 0$) and anisotropic parts 
($\ell = 1, 2, \ldots$) of the structure 
functions are naturally decoupled in this representation,
one could ask how does the value of the anisotropic scaling exponent
compare to that of the isotropic one?  
The first-order anisotropic scaling exponent has been calculated
using perturbation theory \cite[][]{grossmann:1994, falkovich:1995}
and is found to be higher than the isotropic one.  
Borrowing results from passive scalar turbulence \cite[][]{fairhall:1996}, 
it is then conjectured that all higher order anisotropic scaling exponents
are positive, greater than the isotropic scaling exponent, and increase
with increasing order \cite[][]{lvov:1996}.  
If this is true, then these nondecreasing anisotropic exponents would 
neatly explain the diminishing anisotropic scaling contributions with decreasing scale.  
Initial tests have been conducted in atmospheric boundary layer 
flows \cite[][]{arad:1998, kurien:2000a, kurien:2000b} and the
results suggest a hierarchy of increasingly larger anisotropic scaling 
exponents with increasing order.  
Succeeding measurements in laboratory homogeneous shear 
flows \cite[][]{warhaft:2002b}, however, have cast doubt on the 
conclusion reached by the above authors and more refined 
experimental analyses have found the results to depend on the 
geometrical configuration of the measurement probes 
\cite[][]{staicu:2003b}.  
The SO(3) decomposition is in its infancy.  
The technique clearly awaits a more impartial analysis and more
refined experimental techniques\footnote{The well-crafted 
argument that the method might improve at higher 
Reynolds numbers is problematic because it is precisely
at moderate Reynolds numbers that the method makes 
many of its claims \cite[e.g.][]{arad:1999a, biferale:2001}.}.  

\section{Anisotropic turbulence}
\label{sec:anisotropic_turbulence}

As discussed in section~\ref{sec:kolmogorov_theory}, 
Kolmogorov's theory presumes that, at sufficiently high Reynolds 
numbers, the small scales of the turbulence are statistically 
homogeneous, isotropic, and free of externally imposed boundary.  
Few geophysical and laboratory flows fulfill these requirements.  
Field measurements in the atmospheric boundary layer 
show that atmospheric flows attain the highest ever Reynolds 
number ($R_{\lambda}$) achievable on Earth, typically of the 
order $10^{4}$ \cite[e.g.][]{dhruva:1997}.  
Complex terrestrial terrain and planetary dynamics, however, 
may influence the energy-containing scales of turbulent 
motion.  
Consequently, one would expect motion at these scales to be
anisotropic.  
Table~\ref{table:anisotropy_geophysical} offers a few glimpses of
the degree of anisotropy in nature.  
And how does man-made turbulence compare with nature?
Our reaction to table~\ref{table:anisotropy_lab} is one of amazement.    
Despite the moderate Reynolds number, laboratory flows 
created under different conditions possess approximately
the same degree of anisotropy as nature.

\begin{table}
\begin{center}
\footnotesize
\begin{tabular}{llll}
\toprule
\addlinespace[8pt]
Flow & Anisotropy & Direction & Source \\
\addlinespace[5pt]
\midrule
\addlinespace[8pt]
Urban surface layer & $1.9:1.5:1$ & streamwise/spanwise/vertical & \cite{roth:2000} \\
\addlinespace[8pt]
Rural surface layer & $2:1.5:1$ & streamwise/spanwise/vertical & \cite{counihan:1975} \\
\addlinespace[8pt]
Marine surface layer & $1.83:2.6:1$ & streamwise/spanwise/vertical & \cite{friehe:1991} \\
\addlinespace[8pt]
Mixing layer & $1.3$ & streamwise/vertical & \cite{finnigan:2000} \\
\addlinespace[8pt]
Canopy & $1.7$ & streamwise/vertical & \cite{finnigan:2000} \\
\addlinespace[8pt]
Sea surface current & $0.44-2.18$ & zonal/meridional & \cite{ducet:2000} \\
\addlinespace[8pt]
Martian boundary layer & $1.06$ & zonal/meridional & \cite{sullivan:2000} \\
\addlinespace[5pt]
\bottomrule
\end{tabular}
\normalsize
\caption{The landscape of anisotropy for geophysical flows.  
Anisotropy refers to the ratio of streamwise to spanwise to 
vertical, streamwise to vertical, or zonal to meridional RMS 
velocity fluctuations, depending on the context.  In 
\cite{sullivan:2000} the RMS velocity fluctuation data are 
unavailable.  The anisotropy quoted is for the mean wind speed.}
\label{table:anisotropy_geophysical}
\end{center}
\end{table}
What causes this anisotropy in the energy-containing
scales of turbulence?  
Is it a reflection of the asymmetry in the mechanism that
agitates the fluid?  
There is ample evidence to support a causal relationship between 
asymmetry and anisotropy, but the relationship has not been 
tested experimentally.  
This evidence includes flows where asymmetry and anisotropy
are present at the same time, such as in 
turbulent pipe flows \cite[][]{pearson:2001}, 
turbulent jets \cite[e.g.][]{romano:2001}, 
counter-rotating von K\'{a}rm\'{a}n flows \cite[e.g.][]{voth:2002}, 
and wind tunnels with specially designed shear generators 
\cite[e.g.][]{shen:2000, isaza:2009}.  
Turbulence produced by computer simulation can also be 
forced in an asymmetric way, and \cite{yeung:1991} studied 
the influence of this asymmetry.  
One exception is the turbulence produced by a grid in a wind
tunnel with a specially designed contraction, where the
axis of the tunnel introduces a clear asymmetry,
yet the turbulence produced is nearly locally isotropic, but
decaying \cite[][]{comtebellot:1966}.  

There is a pattern in the relationship between asymmetry
of the forcing and anisotropy of the turbulence.  
Machines with a single axis of symmetry, such as the
von K\'{a}rm\'{a}n flow generators \cite[e.g.][]{voth:2002}
and wind tunnels \cite[e.g.][]{comtebellot:1966}, produced 
axisymmetric turbulence.  
Machines with more axes, such as the one developed by
\cite{hwang:2004}, produced isotropic turbulence.  
A careful study by \cite{zimmermann:2010} showed that
six axes were sufficient to produce turbulence
without a preferred direction.  
The machine to be described in this thesis has 16 axes.  
\begin{landscape}
\begin{table}
\begin{center}
\begin{tabular}{llrlll}
\toprule
\addlinespace[8pt]
Flow & Measurement & $R_{\lambda}$ & 
$u^{\prime} / v^{\prime}$ & Direction & Source \\
& location& & & & \\
\addlinespace[5pt]
\midrule
\addlinespace[8pt]
Grid turbulence & $45 \, d$ & $669$ & $1.23$ & streamwise/spanwise & \cite{kistler:1966} \\
\addlinespace[8pt]
Grid (uncontracted) & $20 \, d - 200 \, d$ &
$37-72$ & $1.05-1.25$ & streamwise/spanwise & \cite{comtebellot:1966} \\
\addlinespace[8pt]
Grid (contracted) & $20 \, d - 200 \, d$ &
$37-72$ & $0.96 - 1.07$ & streamwise/spanwise & \cite{comtebellot:1966} \\
\addlinespace[8pt]
Oscillating grid & center plane & 
$31$ & $1.05$ & vertical/horizontal & \cite{shy:1997} \\
\addlinespace[8pt]
Cylinder wake & $75 \, d$ centerline& 
$320$ & $1.17$ & streamwise/spanwise & \cite{hao:2008} \\
\addlinespace[8pt]
Pipe flow & $100 \, d$ centerline & 
$326$ & $1.67$ & streamwise/spanwise & \cite{pearson:2001} \\
\addlinespace[8pt]
Jet & $40 \, d$ centerline &
$500$ & $1.32$ & streamwise/spanwise & \cite{romano:2001} \\
\addlinespace[8pt]
von K\'{a}rm\'{a}n flow & center & 
$970$ & $1.47$ & radial/axial & \cite{voth:2002} \\
\addlinespace[8pt]
Wind tunnel shear flow & $6 \, d$ & 
$974$ & $1.53$ & streamwise/spanwise & \cite{shen:2000} \\
\addlinespace[8pt]
Plane jet & $50 \, d$ & 
$1107$ & $1.13$ & streamwise/spanwise & \cite{pearson:2001} \\
\addlinespace[8pt]
Boundary layer & $0.37 \, d$ & 
$1450$ & $1.26$ & streamwise/spanwise & \cite{saddoughi:1994} \\
\addlinespace[5pt]
\bottomrule
\end{tabular}
\caption{A bird's eye view of the anisotropy observed in laboratory flows.  
The symbol $d$ refers to distance between grid center lines, 
cylinder diameter, pipe diameter, jet nozzle diameter, tunnel width, 
jet width, or the boundary layer thickness, depending on the context.  
$u^{\prime}/v^{\prime}$ refers to the ratio of streamwise to spanwise, 
vertical to horizontal, or radial to axial RMS velocity fluctuations, 
depending on the context.  \cite{shy:1997} did not specify the 
Reynolds number, it is estimated from equation~\ref{eq:Rlambda_Re}.}
\label{table:anisotropy_lab}
\end{center}
\end{table}
\end{landscape}

\cite{batchelor:1946a} initiated the study of anisotropic 
turbulence by considering turbulence with reduced symmetry, 
namely symmetry about an axis, drawing motivation from 
wind-tunnel and pipe flows.  
Applying the invariant theory of \cite{robertson:1940}, 
\cite{batchelor:1946a} showed that the correlation tensor, 
$R_{ij} (\boldsymbol{r}) = \langle u^{\prime}_{i} (\boldsymbol{r}
+\boldsymbol{x}) \, u^{\prime}_{j} (\boldsymbol{x}) \rangle$, 
can be described by a single scalar function in systems of high 
symmetry (isotropic), but in axisymmetric turbulence additional 
functions are required.  
\cite{chandrasekhar:1950} showed that it can be written 
in terms of only two independent scalar functions, which are 
not directly measurable.  
For two special forms of this tensor, which represent possibly 
the simplest kinds of non-isotropic turbulence, \cite{sreenivasan:1978} 
showed that it is sufficient to give two functions of a single variable 
or even one function of a single variable with an additional parameter.  
An experimentally-oriented formulation developed by 
\cite{lindborg:1995} re-expressed the correlation tensor in terms of 
6 measurable two-parameter scalar functions, but because of 
reflectional symmetry only two of them are independent.  

As the rank of the tensor increases, so does the number of 
independent functions we need to describe the tensor.  
Simple symmetry arguments, however, may reduce the number
of terms and greatly simplify the expression.  
Let us consider $R_{ij} (x_k)$ where $i,j,k$ are cyclic permutations of 
the indices $1, 2, 3$.  
Reflectional symmetries of the Navier-Stokes equation about the 
$x-y$, $y-z$, and $x-z$ planes yield \cite[e.g.][]{lindborg:1995}
\begin{equation}
R_{ij} (x_k) = 0 \,,
\end{equation}
for all permutations of $(i,j,k) = (1,2,3)$.  
This is nothing but a consequence of the conservation of parity.  

In principle, to fully describe the axisymmetric turbulence,  
all 6 functions would need to be measured.  
Here, we ingeniously designed our flow apparatus to
allow rotations of the anisotropic large scale forcing, instead of the
measurement apparatus, in order to measure parts of two 
out of these six functions with separations that extended radially 
from the center of the turbulent region.  
In the same spirit as the previous sections, we shall examine
in chapter~\ref{chap:integralscale} the directional dependence of 
these correlations in anisotropic turbulence.  

\section{Overview}

The central idea that interweaves all the chapters in this thesis
is anisotropic turbulence.  
From this idea, we develop three interrelated sub-themes.  
First, we discuss the generation of anisotropic turbulence in the 
laboratory.  
In chapter~\ref{chap:apparatus}, we introduce the turbulence 
generation apparatus and describe the accompanying flow 
measurement technique.  
Following this, in chapter~\ref{chap:procedure}, we explain the 
coordinate system we have used to present our measurements, 
and describe the measurement protocol and methods.  
In chapter~\ref{chap:flows}, we examine the performance of the 
turbulence generator.  
We discuss three aspects of the large scale velocity fluctuations: 
the uniformity in the anisotropy, axisymmetry, and homogeneity.  
Combining these and the auxiliary measurements of the mean flow, 
the velocity fluctuations mixed correlation, and the Reynolds
stress, we evaluate to what degree the turbulence in the fluid is 
homogeneous and axisymmetric.  

The second idea concerns the influence of large scale anisotropy
on the inertial scales of turbulence.  
This influence is traditionally studied through the structure functions.  
In chapter~\ref{chap:universality}, we review the existing works
on anisotropy in structure functions.  
We examine an aspect that has been overlooked in the literature,
namely the behavior of second-order transverse structure 
functions in different directions of the flow.  
We found that the two structure functions become equal at 
the inertial scales.  
This contrasts with existing results obtained in shear flows and 
suggests that anisotropy produced in the fluctuations and
that produced by shear are likely to be fundamentally different.  
Our measurements suggest that the Kolmogorov constant is 
independent of the direction in which it is measured.  
This we exploit to determine the parameters of the turbulence.  
Following this, in chapter~\ref{chap:higherorderstats}, we 
examine structure functions of order four, five, and six.  
We illustrate the difficulty in obtaining precise measurements
of these structure functions with measurements of their
probability density functions.  
We demonstrate the possible influence anisotropy might
have on the shape of the probability density functions.  

The third idea covers the influence of large scale anisotropy
on the integral scales of turbulence.  
This is studied through the correlation functions.  
In chapter~\ref{chap:integralscale}, we devote ourselves to the analysis 
of transverse correlation functions measured in different directions.  
We observed that they deviate from each other at the large scale.  
We propose a scaling form similar to one that is studied
in problems dealing with critical phenomena, and combine it 
with Kolmogorov-type inertial range scaling to show that the 
data collapse when appropriately scaled.  
We then derive a power-law relationship between the integral 
length and the large scale velocity fluctuation and demonstrate
that the power-law exponent is indeed linked to the inertial range
scaling exponent.  

Finally, in chapter~\ref{chap:summary}, we summarize the results 
presented in this thesis and discuss several shortcomings of our 
arguments, as well as future measurements that may complement 
and extend the results of this thesis.  
\chapter{The flow apparatus and the measurement technique}
\label{chap:apparatus}

The goal of the present experimental study is to generate a closer
approximation to tunable and anisotropic, but unsheared, turbulent 
flows than those obtained by \cite{shen:2002}, which we 
accomplished through the construction of a flow apparatus 
described hereunder.  
The apparatus makes systematic studies of anisotropic turbulence 
feasible, and a direct comparison of anisotropic turbulence with 
isotropic turbulence in a single apparatus possible.  
The flows generated by this apparatus are characterized using laser
Doppler velocimetry (LDV).  
We review the LDV measurement technique and discuss some of the 
essential features, only to a degree sufficient for the understanding of 
this thesis.  

\section{The flow apparatus}

We now describe our turbulence chamber, the loudspeaker turbulence
generator and the loudspeaker amplifier unit.  
We review in some detail the algorithm we used for driving the 
turbulence generators.  

\subsection{The chamber}
\label{subsec:soccerball}

\begin{figure}
\begin{center}
\includegraphics[width=12cm]{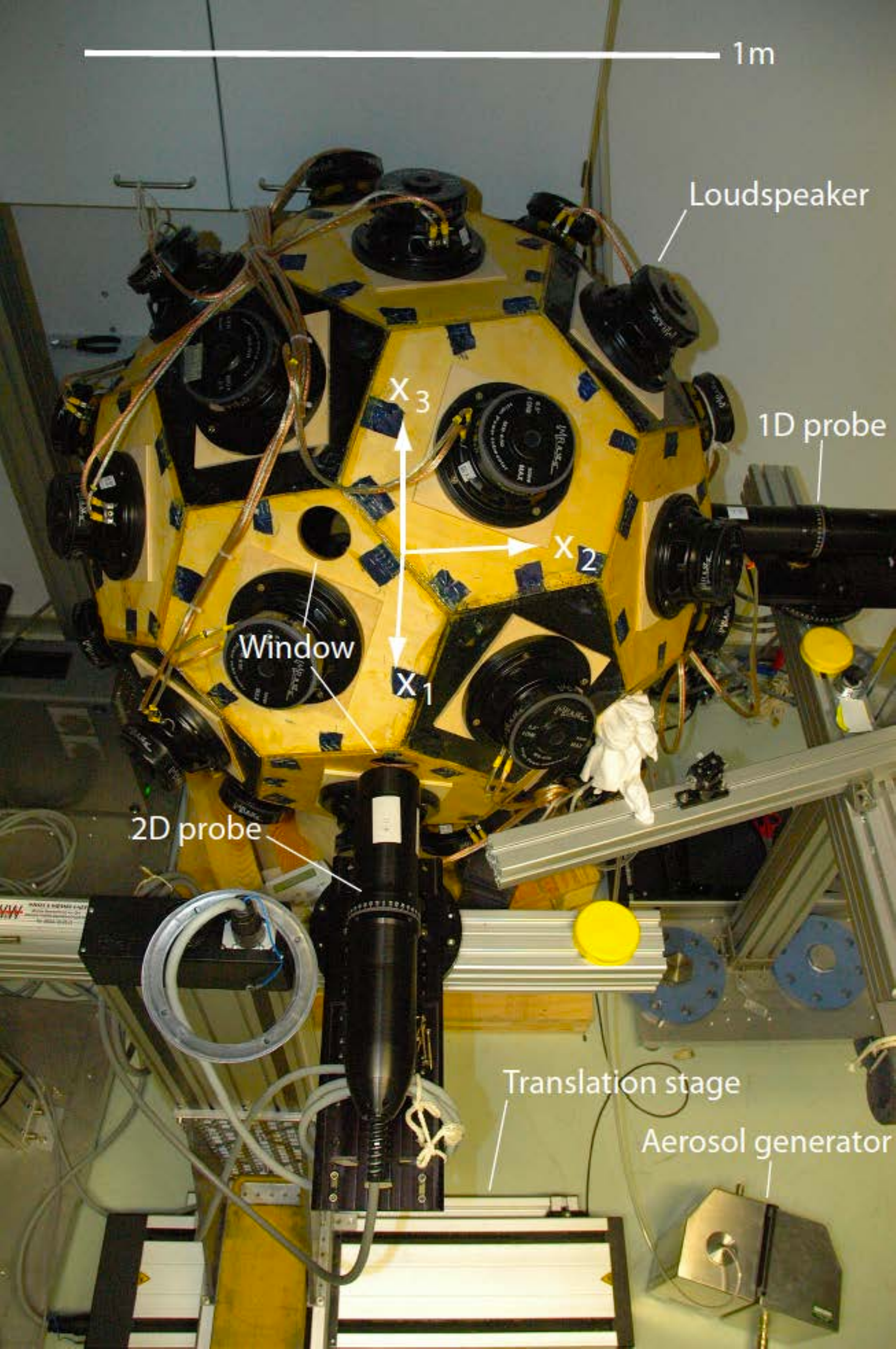}
\caption{The `soccer ball' chamber.  Also shown is the layout of
  the Laser Doppler velocimetry probes in the experiment.  A
  two-component probe aligned along the $x_1$-axis measured the
  components of velocity in the $x_2$ and $x_3$ directions, and an
  additional one-component probe aligned along the $x_2$ axis
  measured velocities in the $x_3$ direction.}
\label{fig:soccerball}
\end{center}
\end{figure}
\begin{figure}
\begin{center}
\includegraphics[width=14cm]{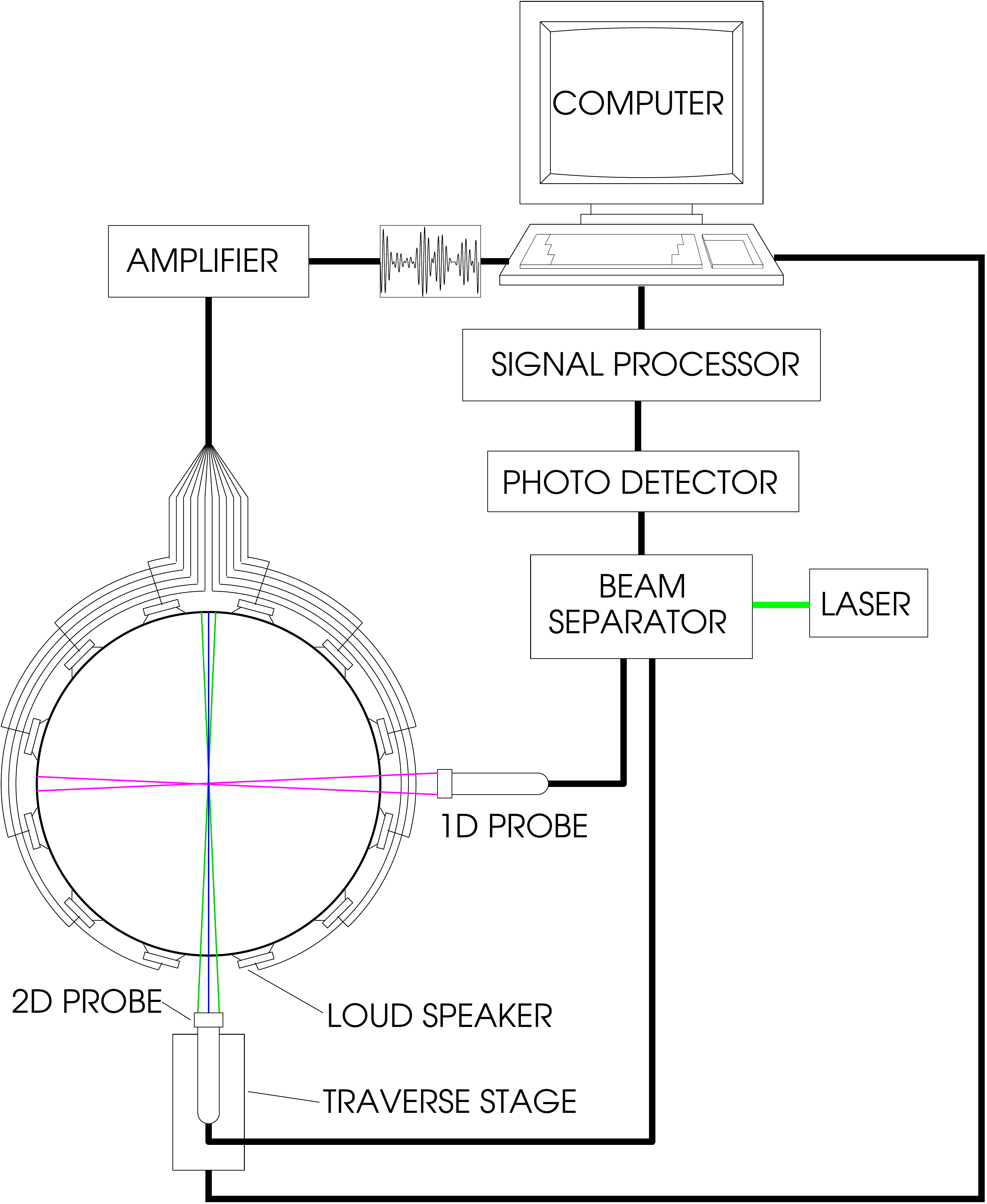}
\caption{An overview of the experiment showing the `soccer ball', the
  amplifier unit, and the measurement apparatus.  
  The amplifier received signals from the multichannel high-speed
  voltage output device on the computer through a BNC connector block,
  and sent the signals to the loudspeaker array on the soccer ball.
  A laser Doppler velocimetry (LDV) system collected velocity signals
  and saved the data on a computer (not shown).}
\label{fig:expschematic}
\end{center}
\end{figure}
The geometry of our flow chamber was inspired by the apparatus
developed by \cite{zimmermann:2010} at the Max Planck
Institute for Dynamics and Self-Organization in G\"{o}ttingen, Germany.  
As shown in figure~\ref{fig:soccerball}, our turbulence chamber
had the shape of a truncated icosahedron and it had a diameter
of 99~cm.  
It was made of wood with twelve regular pentagonal and twenty 
regular hexagonal faces joined together with nylon straps and glue.  
The inner and outer surfaces were coated with several layers of lacquer 
to make the chamber water repellant.  
In the center of each face, a circular hole was cut for the jet 
generator.  
In addition, further circular holes adjacent to the jet generators
were cut out for optical access.  
As the shape resembles that of a soccer ball, we also refer to
the flow apparatus as the `soccer ball' chamber.  

\subsection{The loudspeaker}
\label{subsec:loudspeaker}

\begin{figure}
\begin{center}
\includegraphics[width=6cm]{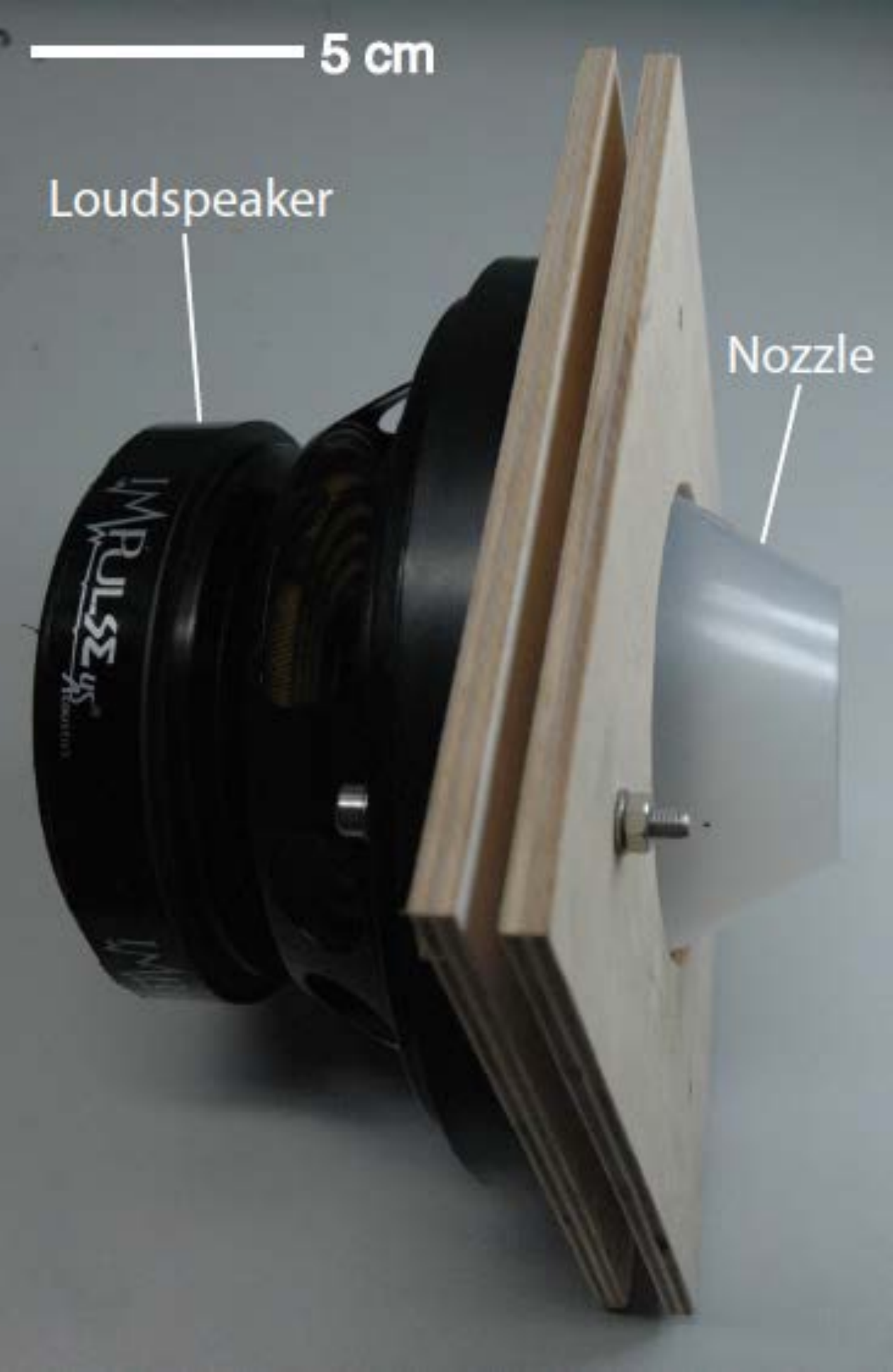}
\caption{A loudspeaker with a conical nozzle.}
\label{fig:loudspeaker}
\end{center}
\end{figure}

As depicted in figure~\ref{fig:loudspeaker}, the turbulence
mixers were jets, or jet-like winds that are produced by 
powerful acoustic sources in air, developed for flow control
applications \cite[][]{glezer:2002}.  
A 150~W loudspeaker with a diameter of 16.5~cm and a uniform
frequency response between 50 and 3000~Hz, was mounted on
each face of the soccer ball and pointed towards the center
of the chamber.  
The loudspeakers pushed air in and out through conical nozzles
of opening angle 30$^{\circ}$, length 4.3~cm and orifice diameter
5~cm.  
Each nozzle was held between a face of the soccer ball and a square
wooden plate.  
By using a sinusoidal driving of the loudspeaker, we generated
a pulsating turbulent jet, a phenomena known as acoustic 
streaming \cite[][]{lighthill:1978}.  
When backing away from the orifice, the loudspeaker 
diaphragm ingests air from all directions, while it blows air
out in a single direction when moving toward the orifice.  
As the jet travels downstream and reaches the center of 
the soccer ball, it interacts with other jets created by other
loudspeakers.  
These interactions produce a turbulent region that is more
intense than the turbulence due to a single turbulent jet.  
The sound level inside the soccer ball was typically 135~dB, 
which corresponds to 0.2\% of the measured turbulent
kinetic energy.  
Thus, we expect the sound to have a negligible effect
on the turbulence and the measurements of velocity.  

The flow chamber was similar to the ones described by 
\cite{hwang:2004, webster:2004, warnaars:2006, lu:2008, goepfert:2010}.  
The chamber by Hwang and Eaton was a 410 $\times$ 410 $\times$ 410
mm$^{3}$ cubical Plexiglas box, from which the corners were cut
off to make the internal volume of the chamber closer to 
spherical.  
A loudspeaker-driven jet was mounted at each vertex, and 
aimed toward the center of the chamber.  
The loudspeakers were driven with sine waves, each with a random
frequency and phase, to discourage the formation of standing waves or
periodic structure 
\makebox[\textwidth][s]{inside the chamber; the power spectra were 
uniform in the range from 90 Hz} 
\begin{landscape}
\begin{table}
\begin{center}
\begin{tabular}{cllllll}
\toprule
\addlinespace[5pt]
& Hwang \& Eaton & Webster {\it et al.} & Warnaars {\it
  et al.} & Lu {\it et al.} & Goepfert {\it et al.} & Present work \\
& (2004) & (2004) & (2006) & (2008) & (2010) & 
\\
\addlinespace[5pt]
\midrule
\addlinespace[8pt]
Geometry & Truncated & Truncated & Rectangular & Truncated
& Octahedron & Truncated \\
& cube & cube & & cube & & icosahedron \\
\addlinespace[8pt]
Flow medium & Air & Saltwater & Water & Air & Air & Air \\
\addlinespace[8pt]
$\epsilon$ (m$^2$/s$^3$) & $11$ & $2.5 \times 10^{-5}$ & $1.25 \times
10^{-6}$ & $1.4$ & $5.8$ & $6.71$ \\
\addlinespace[8pt]
$\eta$ ($\micro$m) & 130 & 450 & 950 & 220 & 155 & 150 \\
\addlinespace[8pt]
$\tau_\eta$ (ms) & 1.2 & 200 & 1000 & 3.3 & 1.6 & 1.5 \\
\addlinespace[8pt]
$R_{\lambda}$ & 220 & 68 & 5.89 & 260 & 250 & \myrlambda \\
\addlinespace[5pt]
\bottomrule
\end{tabular}
\caption{The chamber geometry and flow
  statistics from various speaker-driven flow chambers.  The symbols
  from top to bottom are: energy dissipation rate ($\epsilon$),
  Kolmogorov length scale ($\eta$), Kolmogorov time scale
  ($\tau_\eta$), Taylor micro-scale Reynolds number ($R_\lambda$).
  For each study, only the case that gave the highest Reynolds number
  is reported.}
\label{table:flowchambers}
\end{center}
\end{table}
\end{landscape}
\noindent
to 110 Hz.  
The approach by Hwang and Eaton had inspired the design of flow
apparatuses by various workers interested in the generation of
homogeneous and isotropic turbulence; see \cite{webster:2004,
warnaars:2006, lu:2008, goepfert:2010}.  
We compare the flow parameters in table~\ref{table:flowchambers}.  

\subsection{The driving algorithm}
\label{subsec:algorithm}

Here, we drove the loudspeakers with signals chosen to generate
a flow with the desired anisotropy and low mean velocity.  
We achieved this by modulating the amplitude of a sine wave with
independent noise \cite[][]{fox:1988} of correlation time $0.1$ seconds, 
which was approximately equal to the large-scale eddy turn over
time, $\mathscr{L}/u^{\prime}$, $\mathscr{L}$ being a characteristic
length scale describing the large-scale motions of the flow 
(see section~\ref{sec:integralscale} for a definition), and
$u^{\prime}$ being the RMS velocity fluctuations.  
This condition ensured that fluctuations in the energy input
rate to the turbulence occurred on time scales that were equal 
to or faster than the turbulence decay time, so that the turbulence
was in a steady state.  
In any case, we found that the statistical properties of the turbulence 
were insensitive to the correlation time.

At any given time $t$, the voltage for the $i$-th
speaker was an amplitude-modulated sine signal
\begin{equation}
a_i = \delta_i \, \sin (2 \, \pi \, f \, t) \,,
\end{equation}
where $\delta_i$ is the modulating noise
and $f$ is the base frequency of the sine wave.  
We drove the loudspeakers in phase and 50~Hz produced the
strongest jet.  
The exponentially correlated noise $\delta_i$ was calculated 
according to the algorithm given by \cite{fox:1988}.  
We first set the step size $\Delta t$, and the parameters 
$D$ and $\lambda$.  
At each time step, we picked two random numbers $p$ and $q$, 
uniformly distributed on $[0,1]$, and computed the following
\begin{align}
&E = \exp (- \lambda \, \Delta t) \,, \\
&h = \bigl[- 2 \, D \, \lambda \, (1-E^2) \, \ln (p) \bigr]^{1/2} \,
\cos (2 \, \pi \, q) \,, \\
&\delta_i (t + \Delta t) = E \, \delta_i (t) + h \,.
\end{align}
It follows that $\delta_i$ is an exponentially correlated noise
with the properties \cite[][]{fox:1988}
\begin{gather}
\langle \delta_i (t) \rangle = 0 \,, \\
\label{eq:noise_correlation}
\overline{\langle \delta_i (t) \, \delta_i (s) \rangle} = D
\, \lambda \, \exp (- \lambda |t - s|) \,.
\end{gather}
It can be seen from \ref{eq:noise_correlation} that 
$\lambda^{-1}$ is the correlation time for the noise.  

\begin{figure}
\begin{center}
\subfigure[]{
\includegraphics[scale=0.65]{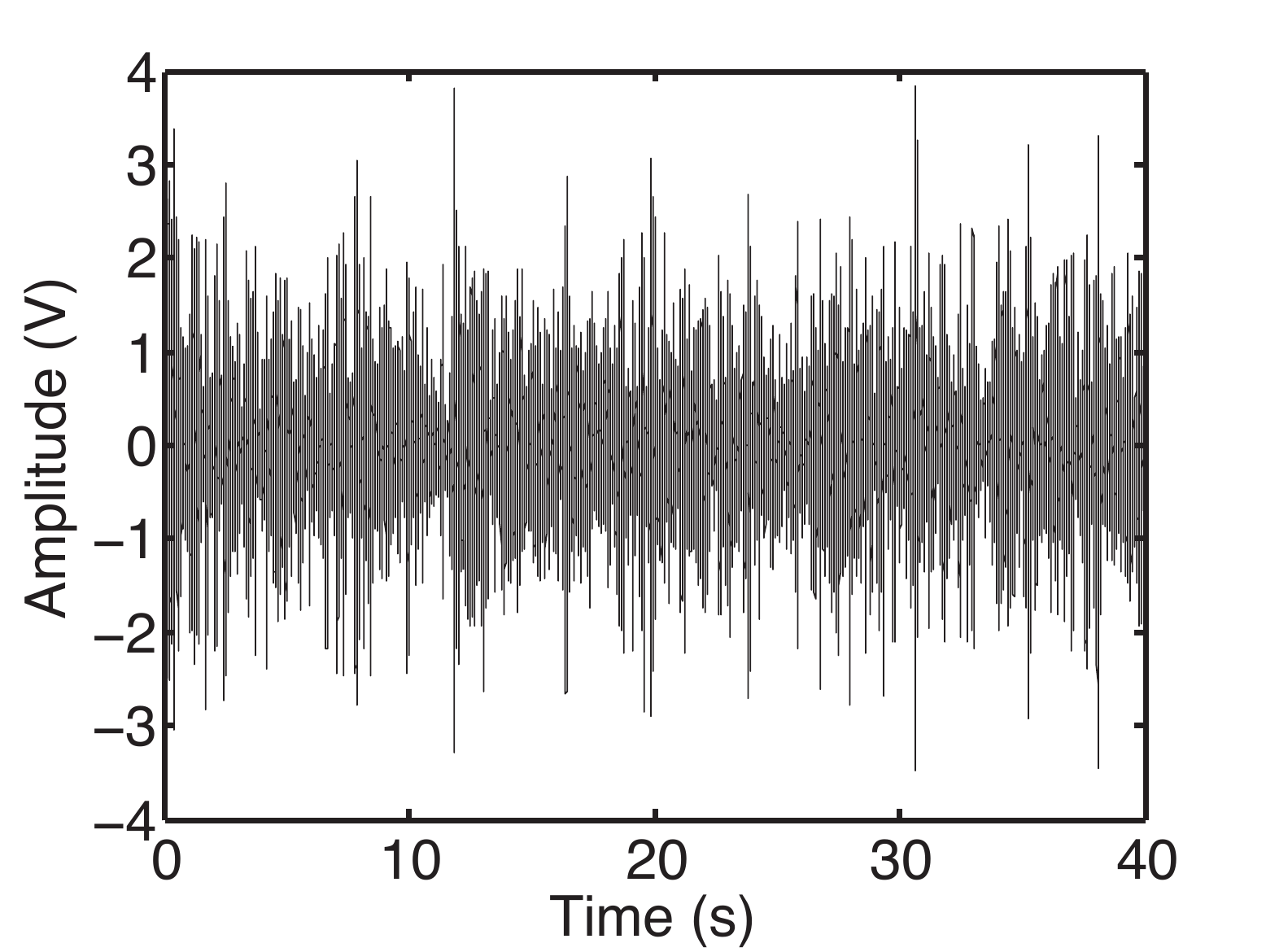}
}
\subfigure[]{
\includegraphics[scale=0.65]{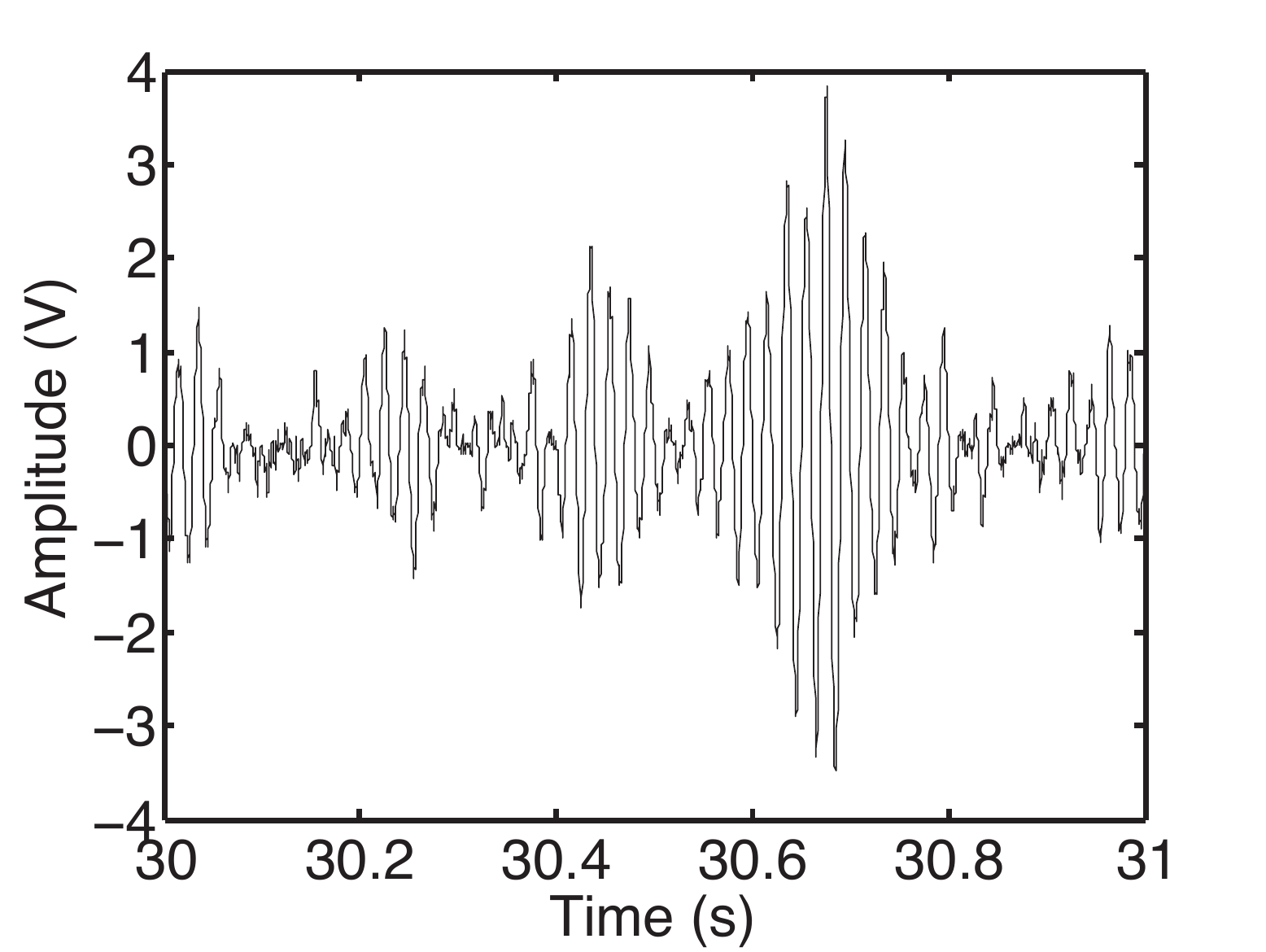}
}
\caption{(a) A 40-second time series of the voltage applied to 
the loudspeakers.  (b) A part of the time series, zoomed in to reveal
the 50~Hz carrier wave.}
\label{fig:fox_noise}
\end{center}
\end{figure}

Figure~\ref{fig:fox_noise}(a) shows a typical time series
of the voltages generated at a sampling frequency 
$1/\Delta t = 3000$ Hz.  
A {\sc Matlab}$\textsuperscript{\textregistered}$ script to 
generate this is given in appendix~\ref{app:foxnoise}.  
We had set $\lambda = 10$~s$^{-1}$, $D = 0.1$~s, 
and the initial condition $\delta_i (t = 0) = 0$.  
The sequence was about $40$~s long and the RMS voltage was 1~V.  
Zooming into the sequence, figure~\ref{fig:fox_noise}(b) shows
the fine structure within 1~s of the signal.  
Because the correlation time of the noise is five times larger than
the period of the carrier wave, the voltage amplitude varies slowly
with time.  

\begin{figure}
\begin{center}
\subfigure[]{
\includegraphics[scale=0.5]{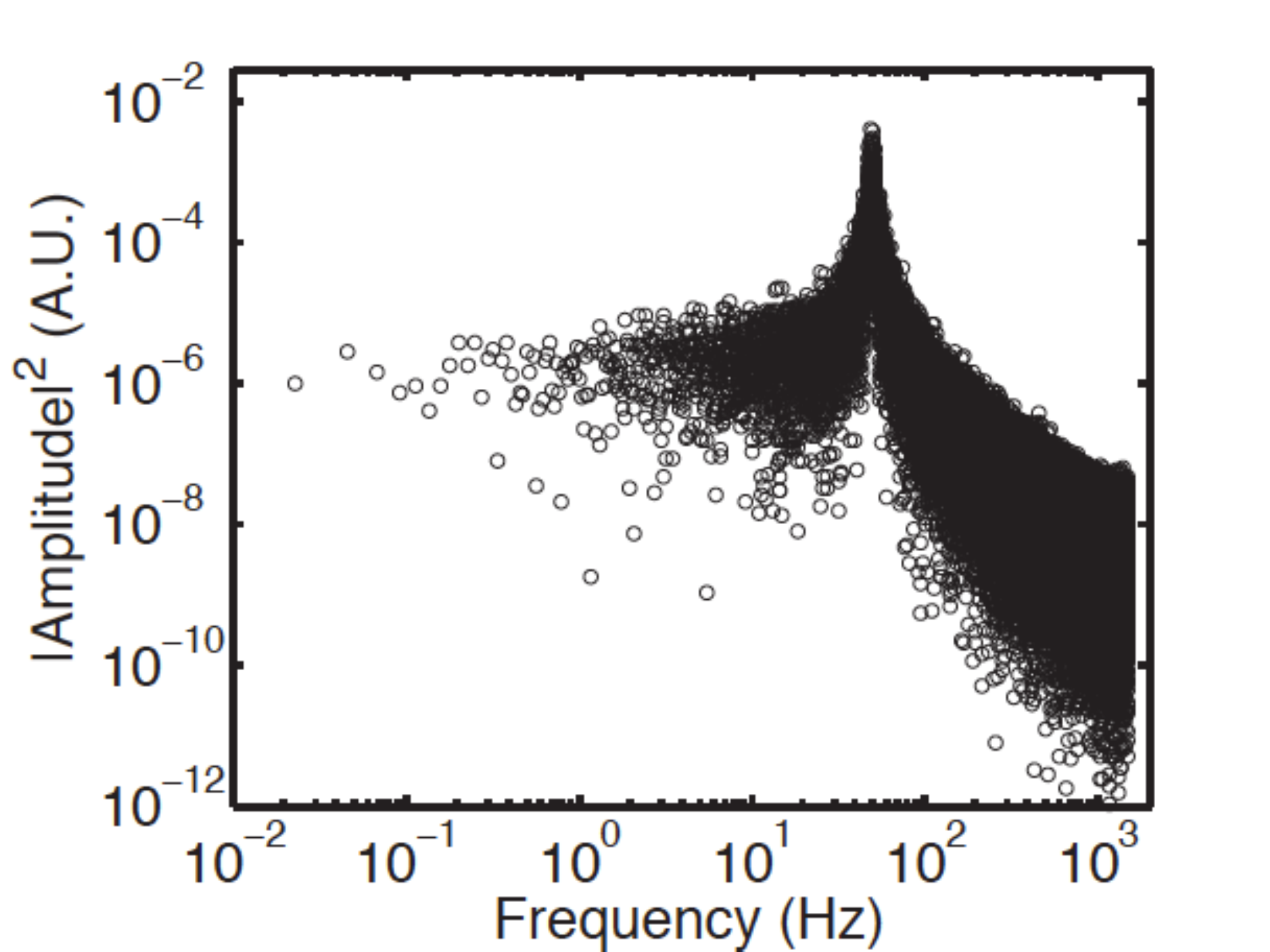}
}
\subfigure[]{
\includegraphics[scale=0.5]{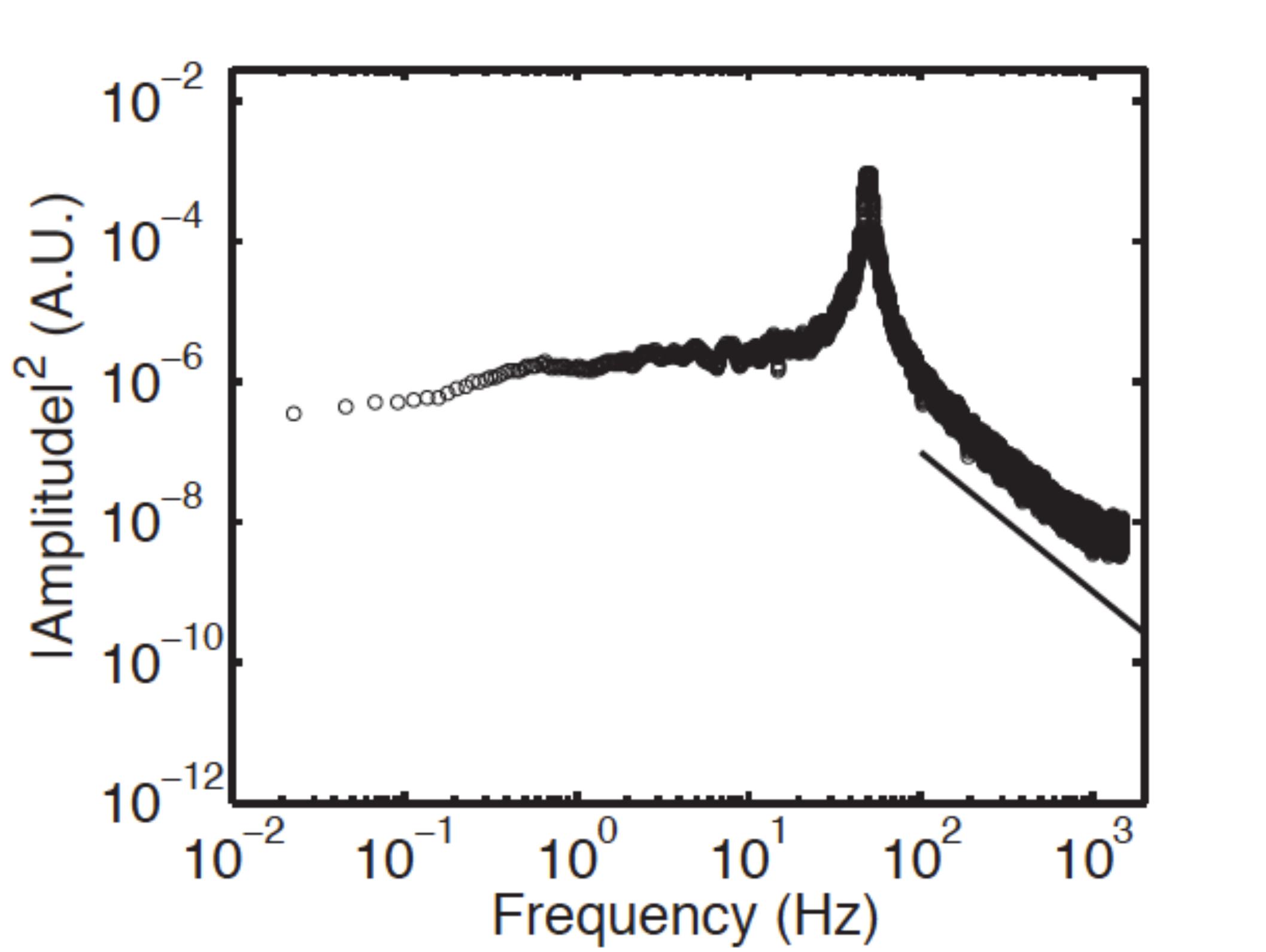}
}
\caption{The spectrum of the voltage applied to the loudspeakers
(a) before filtering and (b) after filtering.  The solid line in (b) is 
proportional to a power law with an exponent of $-2$.}
\label{fig:fox_fft}
\end{center}
\end{figure}

Figure~\ref{fig:fox_fft}(a) shows the unfiltered power spectrum of 
the voltage signal.  
The trend being masked by noise, in figure~\ref{fig:fox_fft}(b) we filter 
the spectrum with a running average to show the intended data.  
The spectrum was uniform up to about $30$~Hz, 
at which point the spectrum increased steadily with increasing frequency 
and reached a peak at the carrier frequency of $50$~Hz, after 
which the spectrum fell off approximately inversely proportional 
to the square of the frequency.  


Finally, to avoid mean flows, the amplitudes of each driving signal
are constrained so that they sum to zero.  
This is ensured by sending 
\begin{equation}
a_i - \frac{1}{32} \, \sum_{j=1}^{32} a_j \,,
\end{equation}
to the $i$-th loudspeaker.  
If the response of the loudspeakers to the driving signal is linear,
this adjustment maintains a constant volume of air in the flow 
apparatus.  
It also reduces the amplitude of the sound generated by the 
loudspeakers, and minimizes the amount of air exchanged 
between the inside of the apparatus and the room.  

The code for driving the loudspeakers was programmed in a 
LabView environment.  
The 32 independent voltage signals were generated with a
National Instruments 32-channel DAQ card (NI PCI-6723) 
on a computer and amplified by an amplifier unit before 
being sent out to the loudspeakers.  

\subsection{The distribution of amplitudes}
\label{subsec:amp_distribution}

\begin{figure}
\begin{center}
\includegraphics[width=6cm]{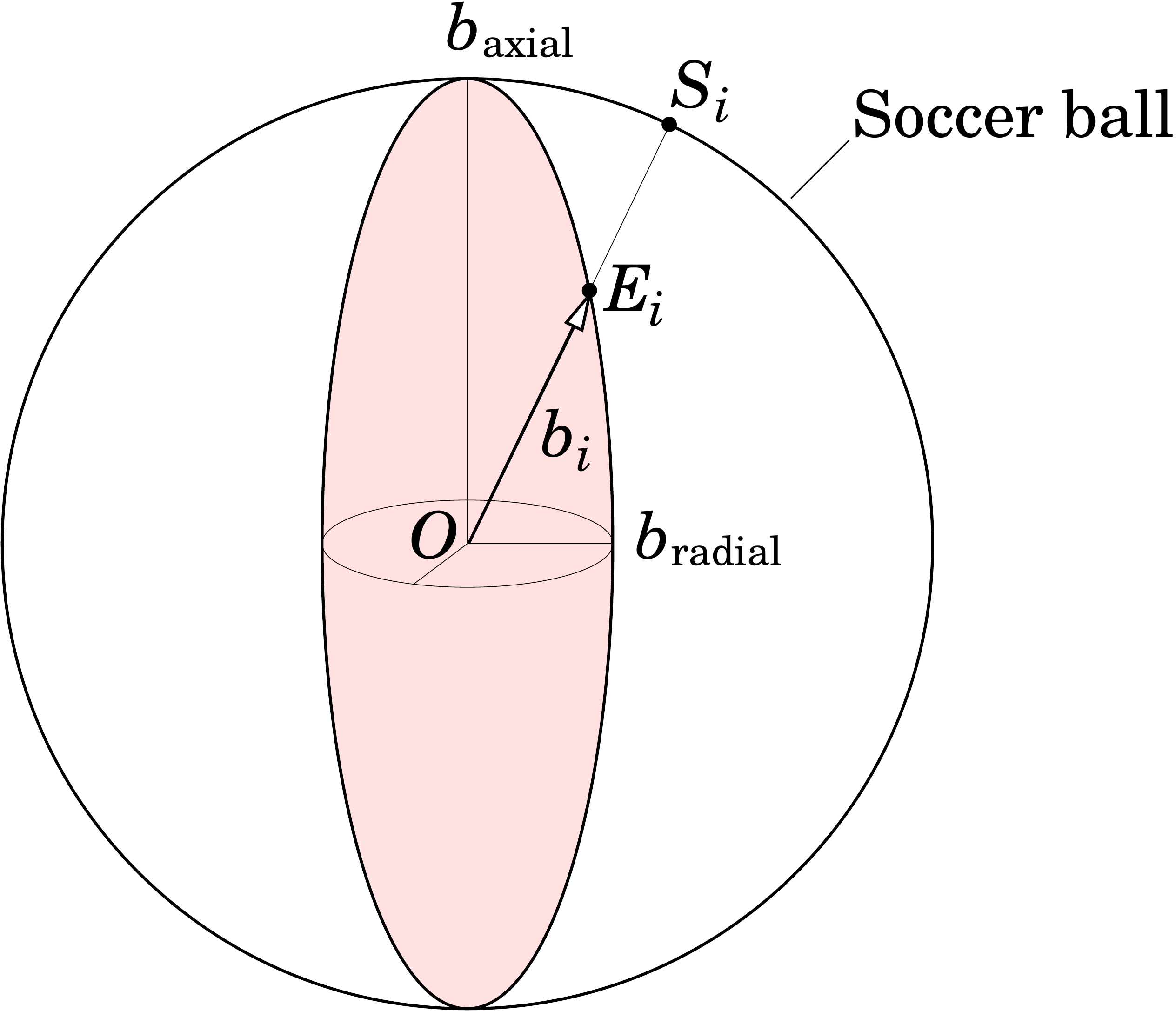}
\caption{A sketch showing the assignment of RMS amplitude for a given
  loudspeaker.  The vector pointing from the center of a
  sphere, $O$, to the center of a loudspeaker, $S_i$, intersects an 
  inscribed ellipsoid at $E_i$.  
  The distance of $E_i$ from the center of the sphere sets the RMS 
  amplitude $b_i$ for the given loudspeaker.}
\label{fig:amp_dist}
\end{center}
\end{figure}

The distribution of loudspeaker amplitudes is shown in 
figure~\ref{fig:amp_dist}.  
By calibrating the RMS amplitude of each loudspeaker 
along the surface of the soccer ball, we were able to select 
the desired anisotropy.  
As we restricted ourselves to cylindrically symmetric forcing, 
the asymmetry can be characterized by the ratio
of the axial amplitude, $b_{\rm axial}$, to the radial
amplitude, $b_{\rm radial}$:
\begin{equation}
\label{eq:amplitude_ratio}
A = \frac{b_{\rm axial}}{b_{\rm radial}} \,.
\end{equation}
Here, the RMS amplitudes are taken as $b_{\rm axial} = 
\langle a_{\rm axial}^2 \rangle^{1/2}$ and 
$b_{\rm radial} = \langle a_{\rm radial}^2 \rangle^{1/2}$, where 
$\langle a_{\rm axial}^2 \rangle$ and $\langle a_{\rm radial}^2 \rangle$
refer to the RMS amplitude of the polar and equatorial loudspeakers, 
respectively.  
A forcing is then described as exhibiting oblate (pancake) asymmetry
when $0<A<1$, spherical symmetry when $A=1$, and prolate (cigar)
asymmetry when $A>1$.  

As we have 32 loudspeakers distributed in an icosahedral 
symmetry, the RMS amplitude for a given loudspeaker was set 
according to the following scheme.  
As shown in figure~\ref{fig:amp_dist}, a vector was drawn
from the center of the sphere to the center of
the loudspeaker.  
We calculated the intersection between the vector 
and the surface of an inscribed ellipsoid, which
had the same center as the sphere.  
The distance between this intersection and the center 
of the soccer ball set the relative RMS amplitude for the 
given loudspeaker.  
Given an asymmetry $A$ and any one of the principal radii, 
$b_{\rm axial}$ and $b_{\rm raial}$, the geometrical shape of 
the ellipsoid is uniquely specified.  
Therefore, the distances may be precalculated once and 
used for setting the RMS amplitudes for all the loudspeakers.

\begin{figure}
\begin{center}
\subfigure[]{
\includegraphics[scale=0.6]{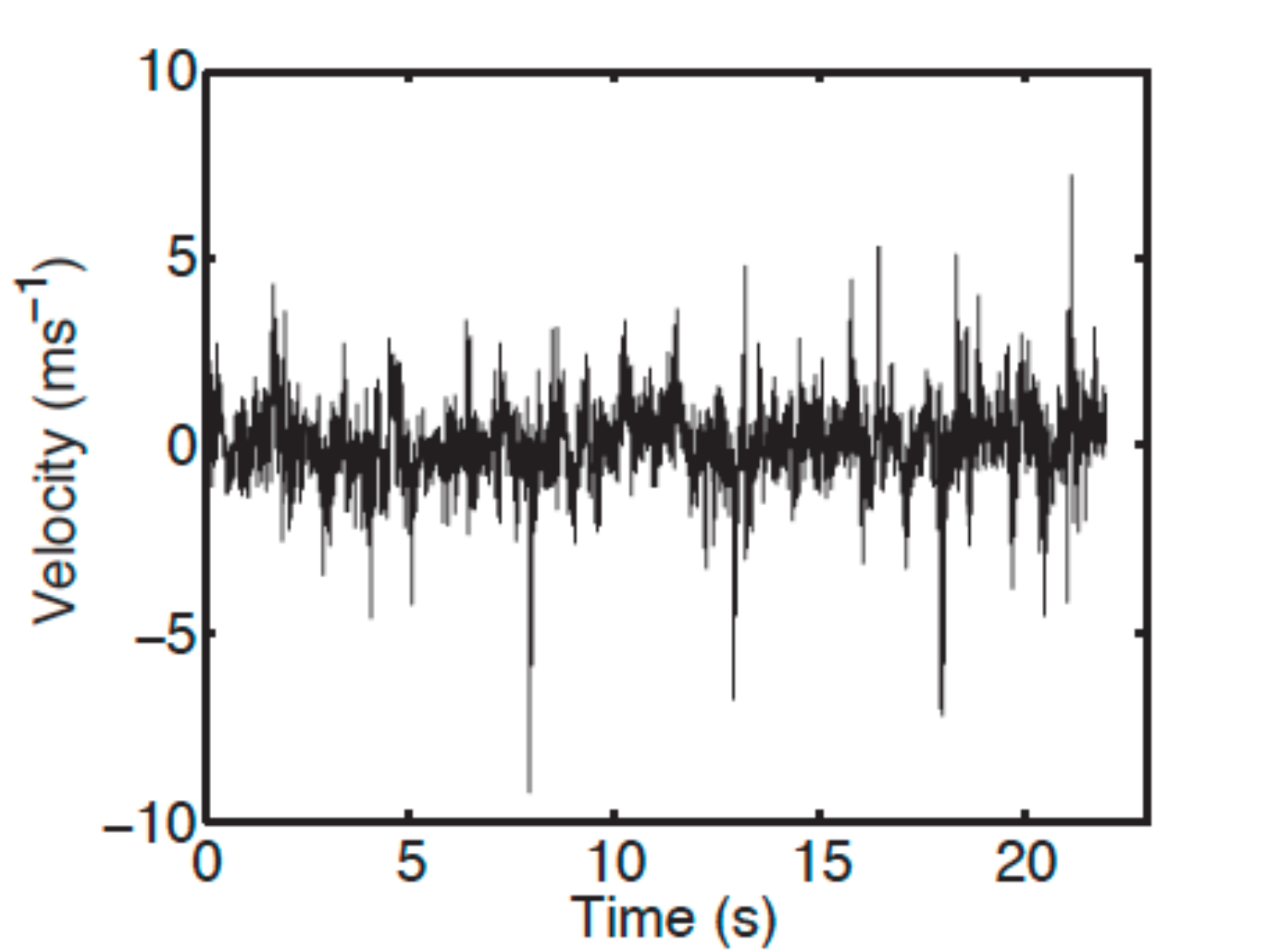}
}
\subfigure[]{
\includegraphics[scale=0.6]{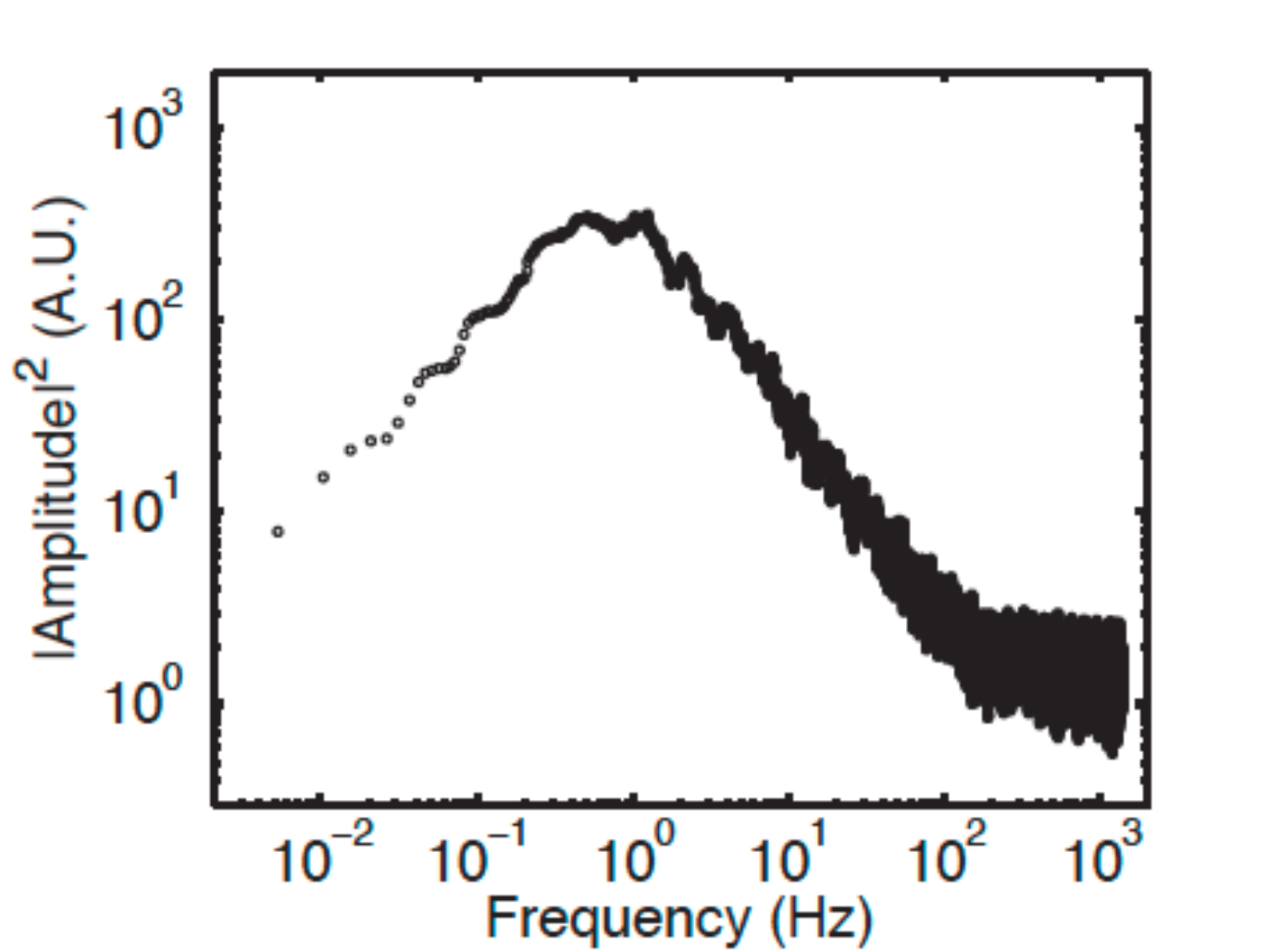}
}
\caption{The figure shows (a) a truncated time series of one 
component of the fluid velocities in response to the forcing and
(b) its spectrum calculated using the Lomb algorithm for
unequally spaced data.}
\label{fig:vel_timeseries_spectrum}
\end{center}
\end{figure}

Figure~\ref{fig:vel_timeseries_spectrum}(a) shows a typical fluid
velocity time series when the 32 loudspeakers were driven 
according to the prescription described above.  
Figure~\ref{fig:vel_timeseries_spectrum}(b) shows the velocity 
spectrum calculated using Lomb algorithm for unequally
spaced data\footnote{The {\sc Matlab}$\textsuperscript{\textregistered}$ 
code by C. Saragiotis for calculating spectrum with Lomb 
algorithm is greatly acknowledged.} 
(see e.g. \cite{press:2007}).  
We did not detect the 50~Hz oscillation in the velocity spectrum.  
We think that any traces of such periodic excitation may have been 
erased as a result of the fluid mechanical interactions between 
multiple turbulent jets.

\subsection{The amplifier}
\label{subsec:amplifier}

Sixteen Raveland XCA 700 two-channel amplifiers, each being
capable of delivering up to 700~Watts of power, 
amplified the digital signals sent through a PCI-to-BNC interface.  
The design and building of the amplifier unit was done by 
Ortwin Kurre and his technical assistants from the electronic 
shop at the Max Planck Institute for Dynamics and Self-Organization 
in Goettingen, Germany.  
\begin{figure}
\begin{center}
\includegraphics[width=8cm]{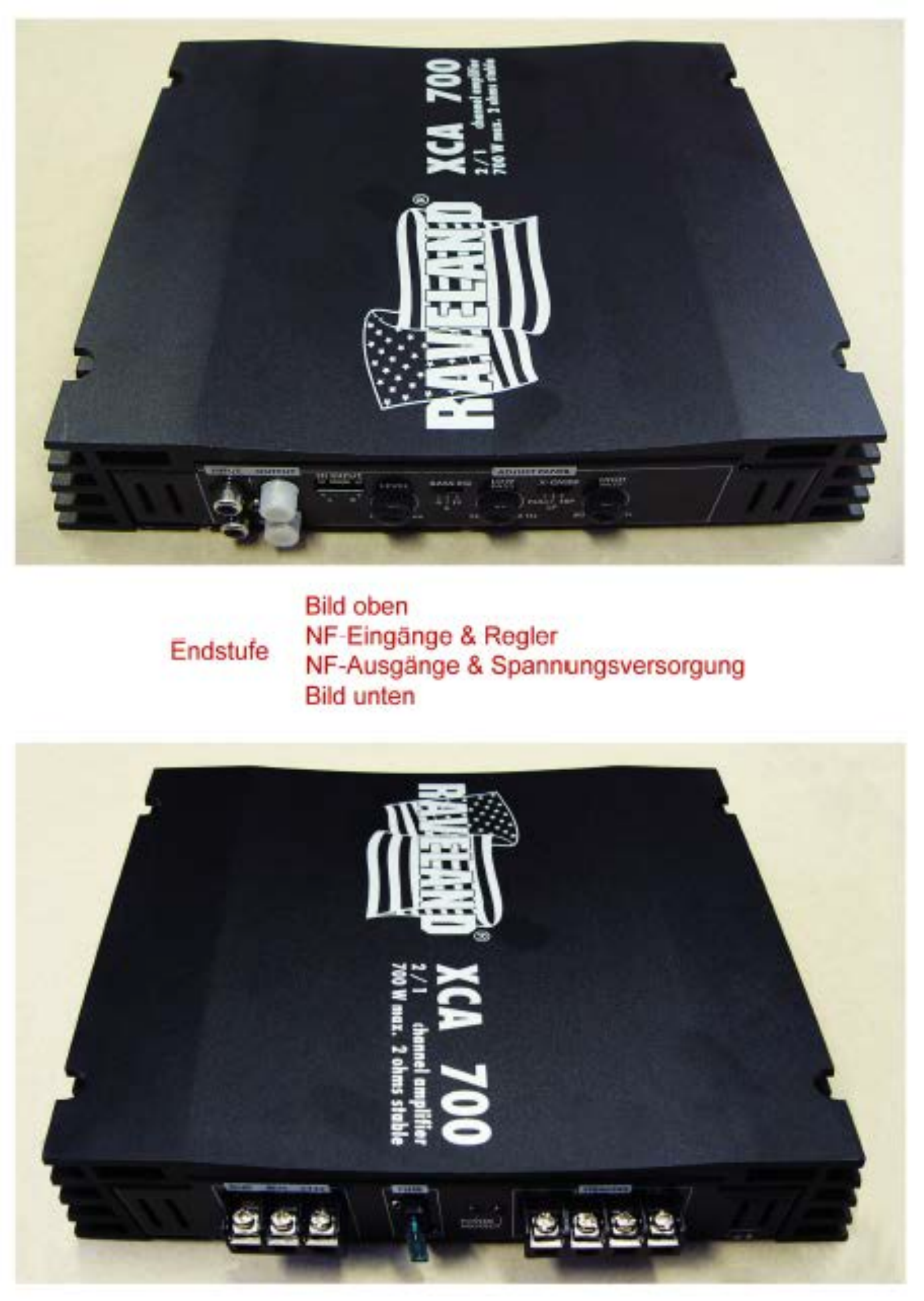}
\caption{The Raveland amplifier.}
\label{fig:raveland}
\end{center}
\end{figure}
\begin{figure}
\begin{center}
\includegraphics[width=12cm]{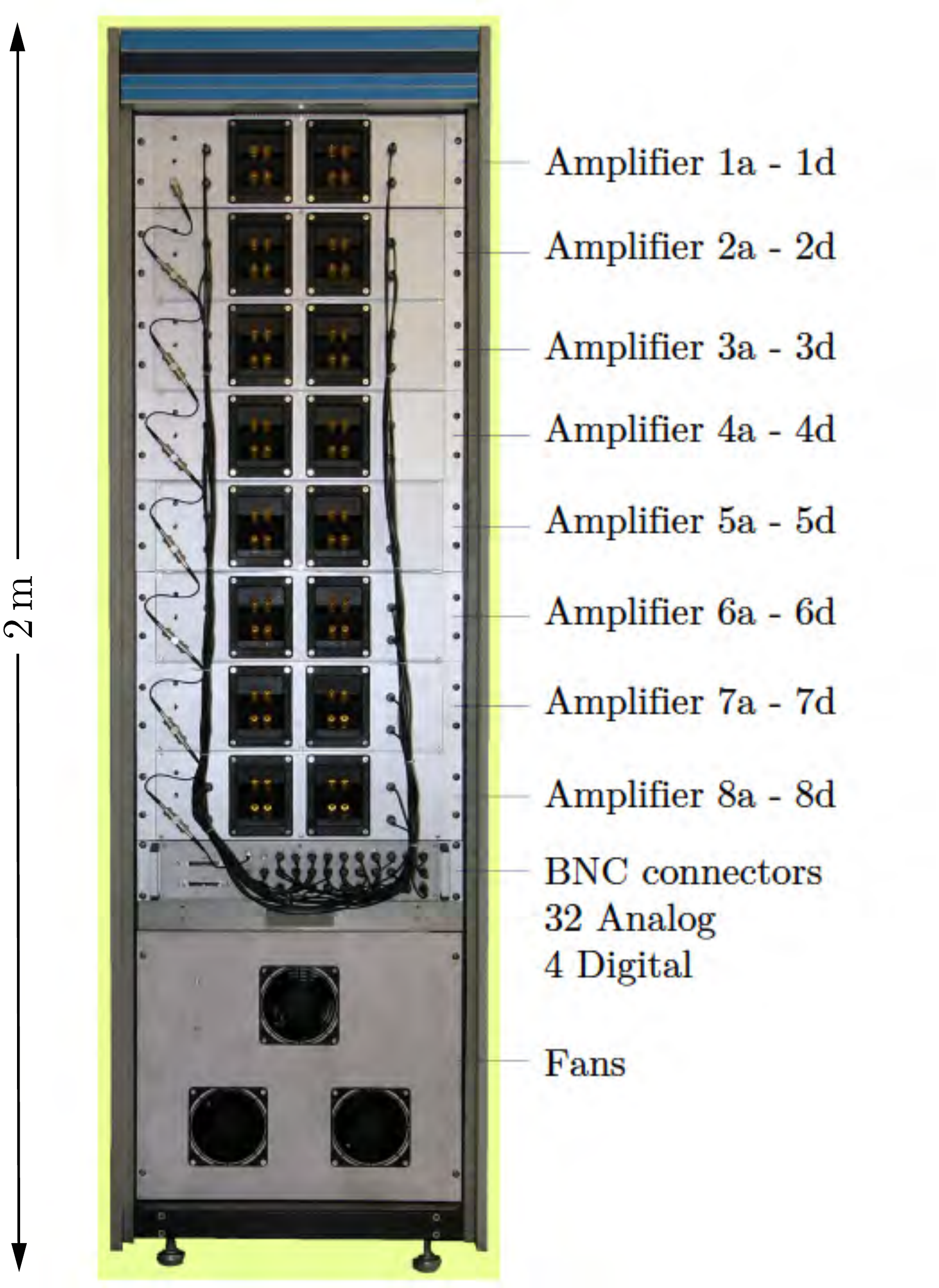}
\caption{The amplifier unit.}
\label{fig:amplifier}
\end{center}
\end{figure}

\section{Laser Doppler Velocimetry}
\label{sec:ldv}

\begin{figure}
\begin{center}
\includegraphics[width=10cm]{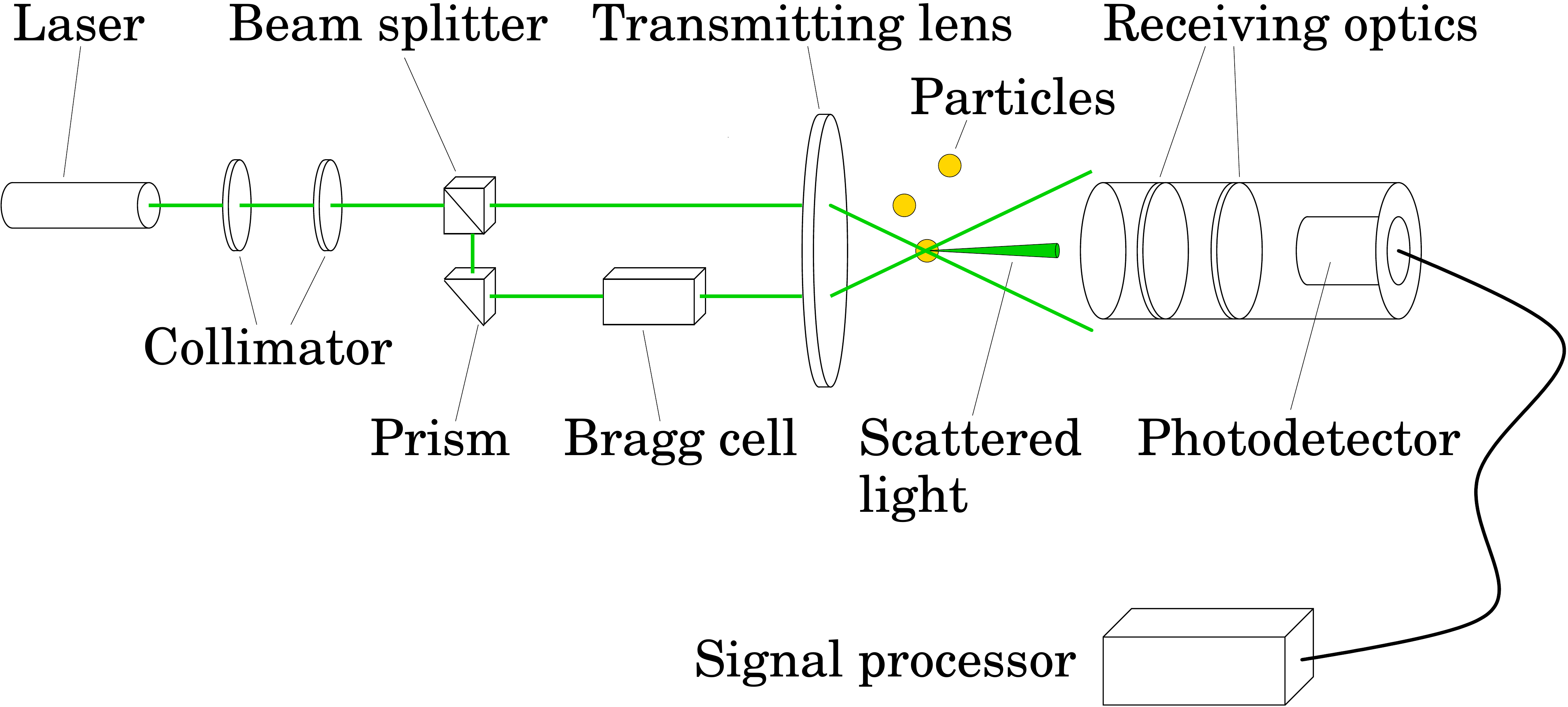}
\caption{A schematic of a laser Doppler velocimeter.}
\label{fig:ldv_schematic}
\end{center}
\end{figure}

Much of the information about the flows we generated inside the soccer
ball was collected using laser Doppler velocimetry (LDV).  
It is a technique for extracting the components
of the velocity of individual particles from the light scattered
by them \cite[e.g.][]{albrecht:2003}.  
A laser light source is an essential part of the technique.  
As shown in figure~\ref{fig:ldv_schematic}, two beams of collimated,
monochromatic, and coherent laser light are focused by a converging 
lens and made to cross at their respective focal point, where they 
interfere and generate a set of straight fringes.  
Focusing is not essential to the explanation that will be made, 
but is only required for making the intersection volume small 
so that only one particle can be inside this volume at any given time.  
As individual particles pass through the fringes in the measurement 
volume, defined by the crossing of the two beams and is 
approximately ellipsoidal in shape, they reflect light from the 
regions of constructive interference into a photo detector.  
The component of velocity perpendicular to the fringes can be 
calculated from the frequency of the signal received at the detector.  
There exist, in the literature, two explanations for the frequency 
dependency of the particle velocity \cite[e.g.][]{albrecht:2003}.  
Both yield the same result but the explanations differ slightly in
their starting point and emphasis.  
We review them in the following sections and discuss the limitations 
of LDV for turbulence measurements in air.  

\subsection{The Doppler model}
\label{subsec:doppler_model}

\begin{figure}
\begin{center}
\includegraphics[scale=0.25]{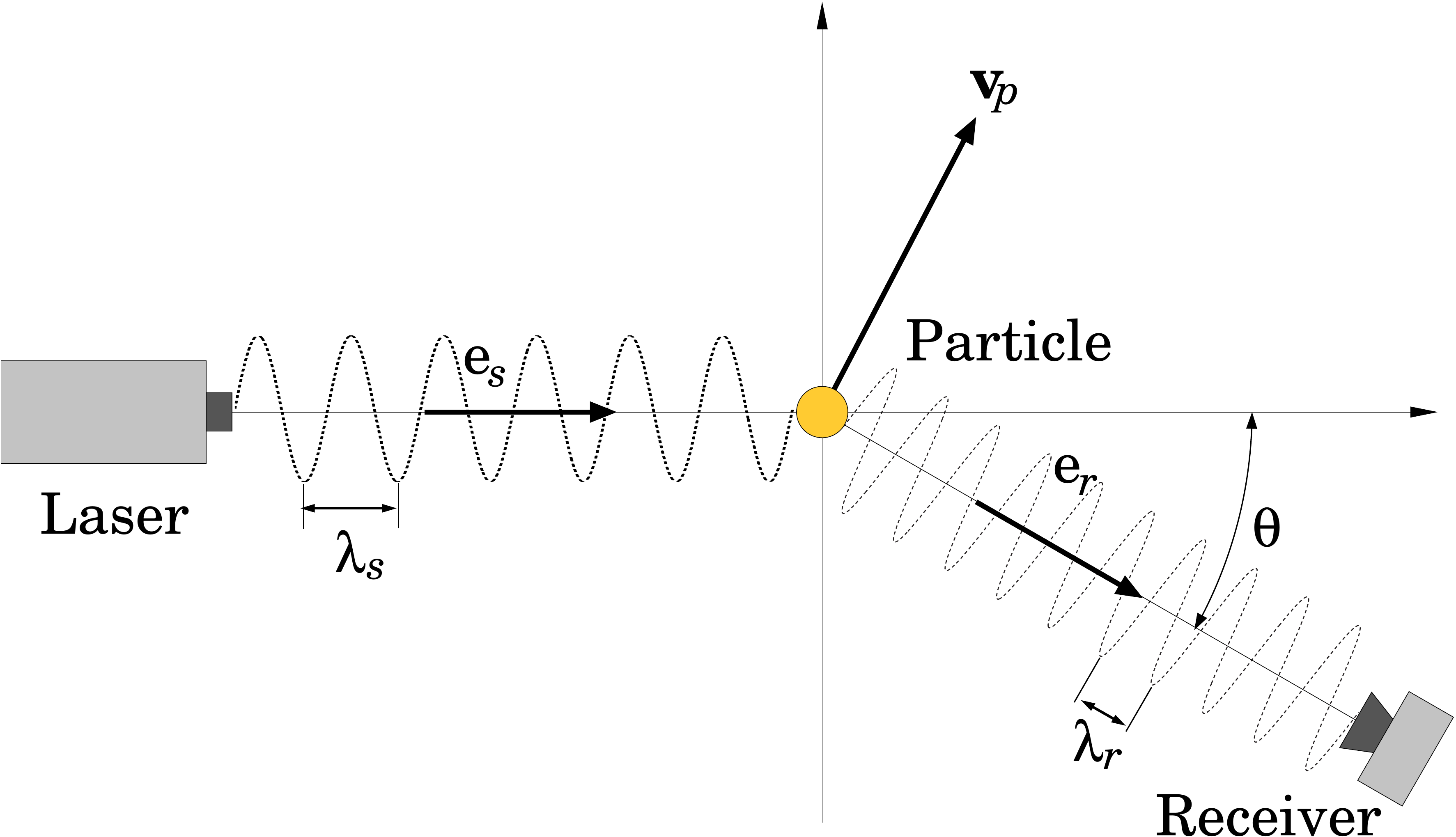}
\caption{The figure illustrates the change in received frequency caused
  by the motion of the particle in a single-beam scattering process.}
\label{fig:doppler_effect}
\end{center}
\end{figure}

The first explanation is based on the Doppler effect.  
Consider the scenario in figure~\ref{fig:doppler_effect}, 
in which a stationary monochromatic light source with 
wavelength $\lambda_s$ and frequency
$\nu_s$ impinges on a particle moving with velocity $\mathbf{v}_p$ in
the flow.  
Denoting the Doppler frequency perceived by the particle as
$\nu_p$, we have
\begin{equation}
\nu_p = \nu_s \, \biggl(1 - \frac{\mathbf{e}_s \cdot \mathbf{v}_p}{c}
\biggr) \,,
\end{equation}
where $\mathbf{e}_s$ is the unit wave vector of the incident beam and
$c$ is the speed of light.
Invoking the Doppler effect a second time,
the frequency received by a stationary detector is
\begin{equation}
\label{eq:doppler_freq}
\nu_r = \frac{\nu_p}{1 - \mathbf{e}_r \cdot \mathbf{v}_p / c}
      = \nu_s \, \biggl(\frac{1 - \mathbf{e}_s \cdot \mathbf{v}_p /
        c}{1 - \mathbf{e}_r \cdot \mathbf{v}_p /c}\biggr) \,,
\end{equation}
where $\mathbf{e}_r$ is the unit wave vector of the scattered beam at
the receiver's end.  
For typical turbulent flows, the particle speed is much less than 
the speed of light.  
Thus, we have for the expression in equation~\ref{eq:doppler_freq} 
the Taylor expansion
\begin{equation}
\nu_r = \nu_s + \frac{\mathbf{v}_p \cdot (\mathbf{e}_r -
  \mathbf{e}_s)}{\lambda_s} \quad (|\mathbf{v}_p| \ll c \,,\, c =
\lambda_s \, \nu_s) \,.
\end{equation}

\begin{figure}
\begin{center}
\includegraphics[scale=0.325]{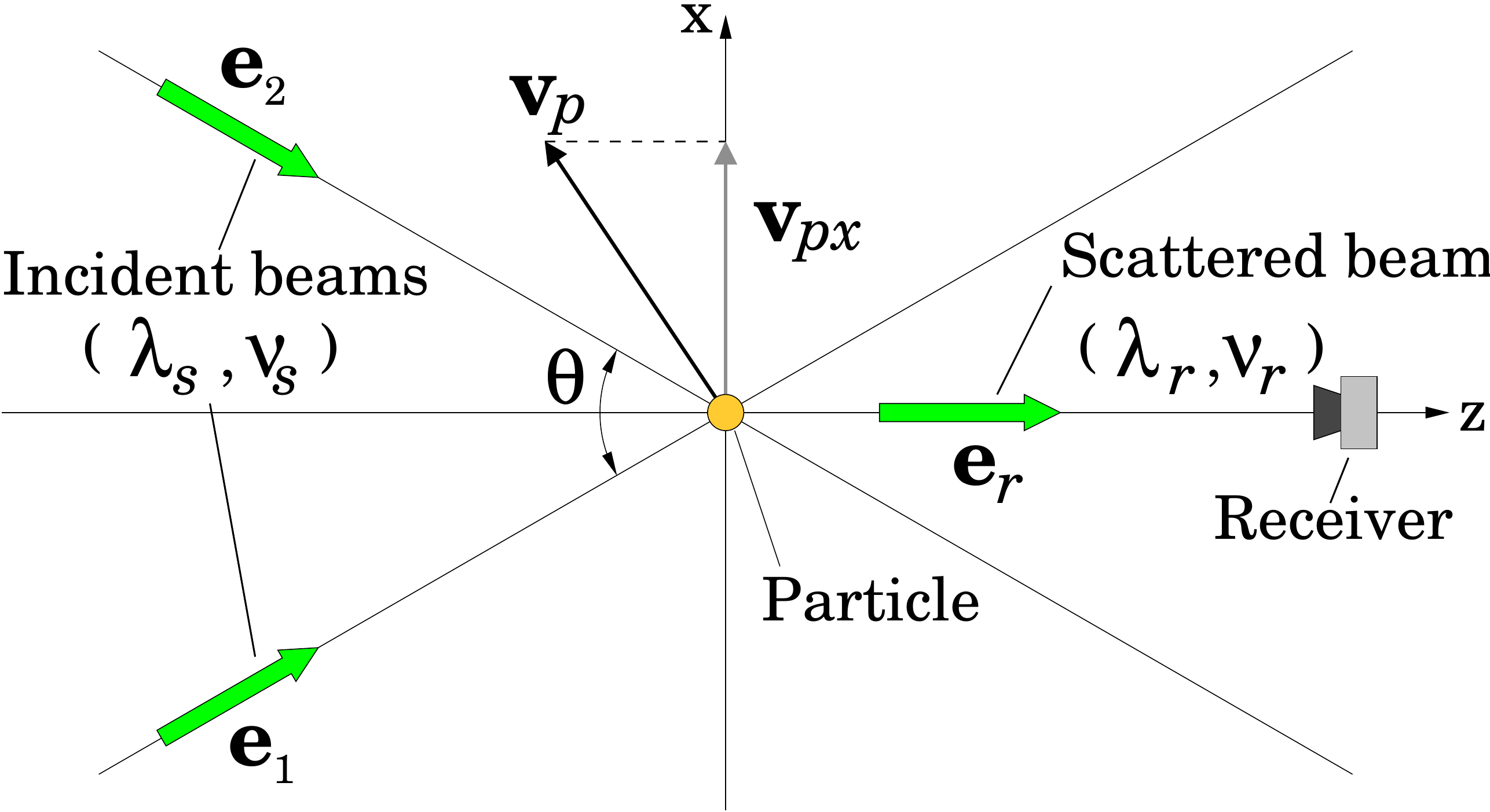}
\caption{The figure illustrates the dual-beam scattering process.}
\label{fig:two_beam_scattering}
\end{center}
\end{figure}

For typical flow velocities (1$\sim$100 m/s), the Doppler shift
frequency is of the order 1 MHz to 100 MHz, and is contained in the
light frequency, which is approximately $10^{14}$ Hz.  
To resolve this small fraction of Doppler shift from the light source is 
technically challenging, if not at all impossible; see \cite{paul:1971}.  
In order to make the Doppler frequency lie in a more manageable 
range, two coherent laser beams of equal intensity and wavelength 
are used instead of one.  
Figure~\ref{fig:two_beam_scattering} shows the dual-beam 
scattering process.
When a particle moves across the measurement volume formed at the
intersection of the two beams, it scatters light from both beams, each
with a frequency
\begin{equation}
\nu_1 = \nu_s + \frac{\mathbf{v}_p \cdot (\mathbf{e}_r -
  \mathbf{e}_1)}{\lambda_s} \quad , \quad
\nu_2 = \nu_s + \frac{\mathbf{v}_p \cdot (\mathbf{e}_r -
  \mathbf{e}_2)}{\lambda_s} \,.
\end{equation}
The light is focused and mixed on the detector to yield the beat
frequency
\begin{equation}
\nu_B = \nu_2 - \nu_1 = \frac{\mathbf{v}_p \cdot (\mathbf{e}_1 -
  \mathbf{e}_2)}{\lambda_s} \,.
\end{equation}
If the angle formed by the intersection of the two beams is denoted by
$\theta$, and the component of particle velocity perpendicular to the
bisector of the beams (the $z$-axis) is denoted by $\mathbf{v}_{px}$,
then the beat frequency can be expressed as
\begin{equation}
\label{eq:beat_freq}
\nu_B = \frac{2 \, \sin (\theta / 2)}{\lambda_s} \, 
|\mathbf{v}_{px}| \,.
\end{equation}
Thus, the frequency difference is linearly proportional to the 
velocity component perpendicular to the bisector
of the two beams.

\subsection{The fringe model}
\label{subsec:fringe_model}

The expression in equation~\ref{eq:beat_freq} is derived largely based
on kinematic consideration and it does not explain the origin of the
beat frequency.  
The fringe model fills this gap of our understanding.  
We illustrate it using the configuration in 
figure~\ref{fig:two_beam_scattering}.  
If two incident beams of equal amplitude $E_0$ and frequency 
$\omega_s$, and linearly polarized perpendicular to the plane 
in which the two beams lie, then the electric field of each beam 
at an arbitrary point $\mathbf{r}$ in space is described by
\begin{align}
\label{eq:electric_field_1}
\mathbf{E}_1 &= \mathbf{e}_y \, E_0 \, \exp\bigl( i [\omega_s \, t -
\mathbf{k}_1 \cdot \mathbf{r} + \phi_1] \bigr) \,, \\
\label{eq:electric_field_2}
\mathbf{E}_2 &= \mathbf{e}_y \, E_0 \, \exp\bigl( i [\omega_s \, t -
\mathbf{k}_2 \cdot \mathbf{r} + \phi_2] \bigr) \,.
\end{align}
$\mathbf{k}_1$ and $\mathbf{k}_2$ are the wave vectors of the two
beams, given by
\begin{equation}
\label{eq:kvectors}
\mathbf{k}_1 = k_s \, \mathbf{e}_1 \quad , \quad \mathbf{k}_2 = k_s \,
\mathbf{e}_2 \quad , \quad k_s = 2 \, \pi / \lambda_s \,.
\end{equation}
$\phi_1$ and $\phi_2$ are the initial phases of the waves at the
origin, respectively.  
At the crossing of the two beams, the two electric fields add 
according to the superposition principle to give an intensity $I$ 
proportional to
\begin{equation}
I \propto |\mathbf{E}_1 + \mathbf{E}_2|^2 \,.
\end{equation}
By substituting the expressions for the electric fields from
equations~\ref{eq:electric_field_1}, \ref{eq:electric_field_2}, 
and \ref{eq:kvectors}, we have for the intensity
\begin{equation}
\label{eq:intensity_fringes}
I \propto 2 \, E_0^2 \, \biggl[ 1 + \cos \biggl( 2 \, \pi \, 
\biggl( \frac{2 \, \sin(\theta/2)}{\lambda_s} \biggr) \, x 
- (\phi_1 - \phi_2) \biggr) \biggr] \,.
\end{equation}
It can be seen that the electromagnetic wave intensity 
varies periodically in $x$, with a fringe spacing given by
\begin{equation}
\Delta x = \frac{\lambda_s}{2 \, \sin (\theta / 2)} \,.
\end{equation}
A particle of diameter $d_p \ll \Delta x$ passing through the
intersection volume samples the local intensity of the interference
pattern and produces a burst of flickering light.  
If $\Delta t$ is the particle's fringe crossing time, then the frequency 
of flickering is given by
\begin{equation}
\label{eq:detector_freq}
\nu_{\rm detector} = \frac{1}{\Delta t} = 
\frac{2 \, \sin (\theta / 2)}{\lambda_s} \, \frac{\Delta x}{\Delta t}
= \frac{2 \, \sin (\theta / 2)}{\lambda_s} \, |\mathbf{v}_{px}| \,,
\end{equation}
which is precisely equation~\ref{eq:beat_freq}.  
The fringe model offers a very physical and intuitive approach for understanding the laser Doppler technique.  
This model is only valid when the particle diameter is less than the 
wavelength of light.  
For particles larger than the wavelength of light, the arguments 
have to be modified in order to take into account the additional 
phase shift due to wave propagation inside the particles.  
This additional phase shift only causes the fringe pattern to be 
shifted in phase and equation~\ref{eq:detector_freq} remains valid.  
The reader is referred to chapter 2 of \cite{albrecht:2003} for a full 
account of the particle finite size effect.  

\subsection{The Bragg cell}
\label{subsec:braggcell}

The reader may note that particles with equal, but opposite velocities,
moving across the measurement volume formed in
the dual beam arrangement will produce the same 
frequency of flickering.  
To resolve the particle velocity direction, a Bragg cell is placed
in one of the beam paths to modify the Doppler frequency.  
It consists of a mechanical oscillator, such as a 
piezoelectric transducer, attached to an optically transparent
material, such as quartz.  
An oscillating electric field drives the transducer to vibrate,
emanating a traveling acoustic wave.  
Under the mechanical pressure of the wave, a periodic
refractive index grating results.  
This traveling periodic grating diffracts the incident light much 
like the atomic planes of a crystal in Bragg diffraction.  
\begin{figure}
\begin{center}
\includegraphics[scale=0.25]{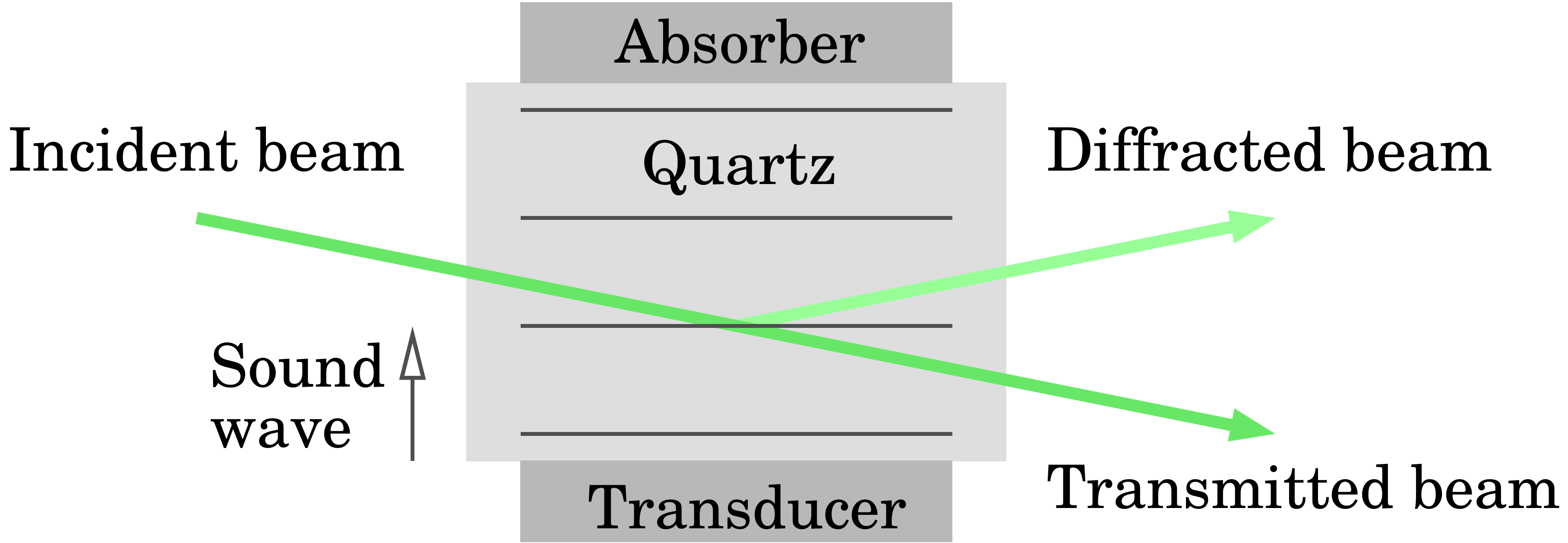}
\caption{A schematic of a Bragg cell.  A transducer generates an
  acoustic wave, resulting in a moving periodic refractive index
  grating that diffracts the incident light similar to in Bragg
  diffraction.  The arrow points in the direction of travel 
  of the acoustic wave.}
\label{fig:braggcell}
\end{center}
\end{figure}

If a Bragg cell is placed in the path of beam 1 in figure~\ref{fig:two_beam_scattering}, the frequency
of the incident beam is shifted by an amount $\nu_{\rm sh}$ to
$\nu_{1} + \nu_{\rm sh}$ when the acoustic wave moves
toward the beam, and $\nu_{1} - \nu_{\rm sh}$ when the
acoustic wave moves away from the beam.  In the fringe model, the
first case corresponds to a movement of the fringes in the $-x$
direction, whereas the second case corresponds to a movement of the
fringes in the $+x$ direction.  The light from beam 1 and
2 is mixed on the detector to yield a beat frequency
\begin{equation}
\label{eq:shift_frequency}
\nu_{B} = 
\begin{cases}
\nu_{\rm sh} + \dfrac{2 \, \sin(\theta / 2)}{\lambda_s}
\mathbf{e}_x \cdot \mathbf{v}_p & \, \text{fringes move in the}
\, -x \,\, \text{direction} \,, \\
\nu_{\rm sh} - \dfrac{2 \, \sin(\theta / 2)}{\lambda_s}
\mathbf{e}_x \cdot \mathbf{v}_p & \, \text{fringes move in the}
\, +x \,\, \text{direction} \,.
\end{cases}
\end{equation}
It is easily seen that a stationary particle will yield a signal
with frequency $\nu_{\mathrm{sh}}$, a movement in the 
direction of the fringes a lower frequency, and a movement
against the fringes a higher frequency.  

\begin{landscape}
\begin{figure}
\begin{center}
\includegraphics[width=24cm]{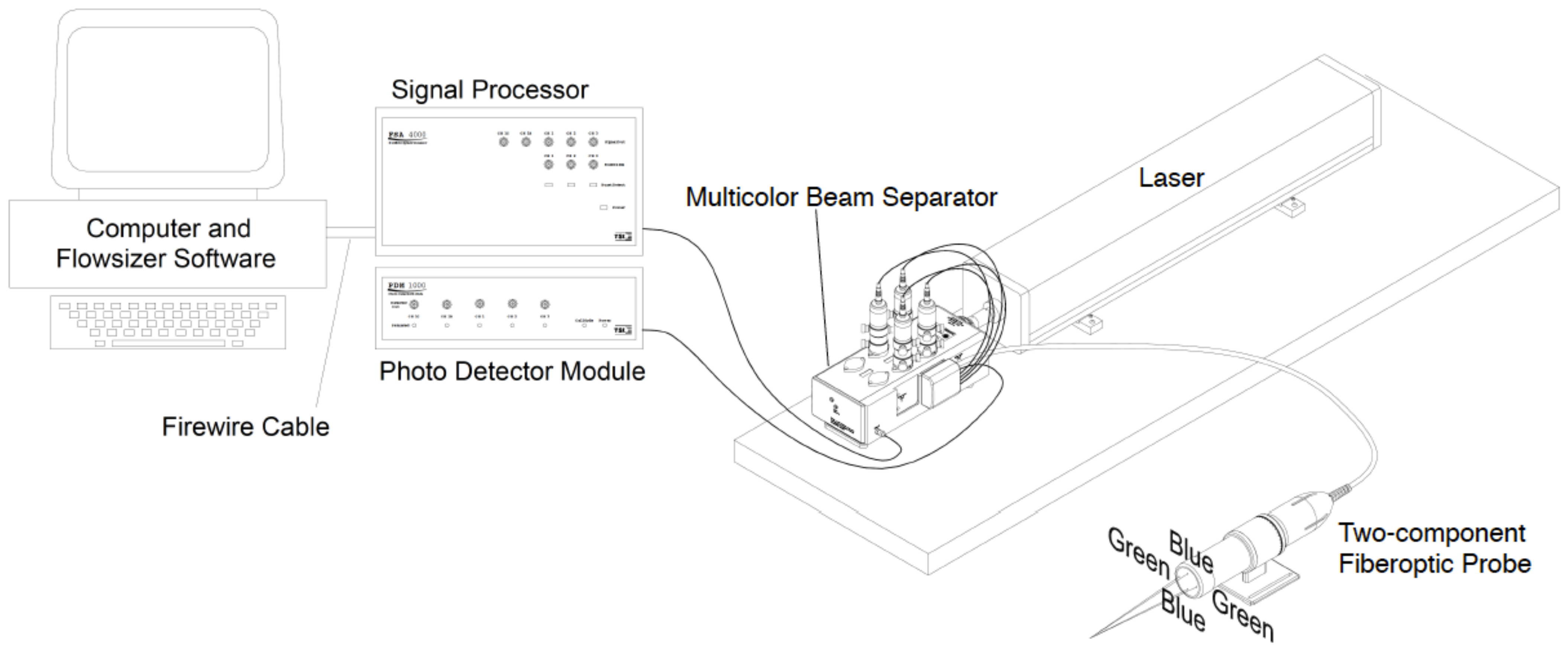}
\caption{An overview of a TSI LDV system.  Figure adapted from
TSI {LDV/PDPA} system installation manual, page 73.}
\label{fig:tsi_ldv}
\end{center}
\end{figure}
\end{landscape}

Figure~\ref{fig:tsi_ldv} shows an overview of the TSI LDV
system used in our investigation.  
An argon ion laser by Coherent (Innova 90C-5) provided a 
multicolored beam.  
A multicolor beam separator (TSI Model 450200),
converted the beam into three beam pairs of
different wavelengths: green, blue, and violet.  
A Bragg cell in the beam separator shifted one in each beam
pair by 40~MHz.  
Two fiberoptic probes focused the light from the beams to form a measurement volume.  
One probe measured two components, and the other one only one
component, of the velocity of individual particles that passed 
\makebox[\textwidth][s]{through the approximately ellipsoidal 
measurement volume, which measured}
\noindent
100~$\micro$m in diameter and 2~mm in length.  
The effect of a finite measurement volume on the measurement 
of particle velocity is discussed in greater detail in 
section~\ref{subsubsec:probevolume}.  
This resolution, however, was sufficient to resolve scales larger 
than the dissipation scale of the turbulence in this experiment.  
A receiving fiber in the probe collected light scattered by particles 
in the volume and sent the light to a photodetector (PDM 1000).  
The photodetector converted the scattered light into electrical signal 
which was then sent to a signal processor (FSA 4000).  
The signal processor extracted Doppler frequency
information from the electrical signals and sent the digitized signal 
to be analyzed by software ({\sc FlowSizer\texttrademark}) on a computer for
data acquisition, yielding a single velocity measurement.  
Data consisted of a stream of velocity samples with irregular time
intervals between them.  
The mean data rate was between 300 and 3000 samples per second, 
depending on the probe.  
For statistical convergence, we typically collected $2\times 10^6$ 
data points per aspect ratio of the forcing and per spatial 
position of the probes.  

\subsection{Data processing}
\label{subsec:dataprocessing}

In LDV measurement, the flow velocity is sampled in time by 
random passages of particles that go through the
measurement volume.
This has two consequences.  
First, the velocity samples are irregularly spaced in time.  
Second, the measurement volume is sampled by high-speed
particles more often than low-speed ones.  
As a consequence, even if the particle spatial distribution is
independent of the velocity field, the rate of particle arrival at the
measurement volume is correlated with the velocity field
\cite[][]{mclaughlin:1973}.  
Without careful attention to this latter fact, statistical average
of flow quantities can be biased.  
Thus, velocity samples must be weighted by appropriate 
statistical factors to yield unbiased statistical averages 
\cite[][]{mclaughlin:1973}.  
The statistical factor we used for correcting single-point statistics
is the arrival time weighting factor \cite[see e.g.][]{adrian:1987}.  
Other methods may also work \cite[see e.g.][]{buchhave:1975,
  buchhave:1979, mcdougall:1980}, but the arrival time method has the
advantage of being computationally efficient and is well-suited for
non-homogeneously seeded flow fields.  

The method works by weighting the observables with the arrival
times between particles.  
For example, the estimator for the mean and the RMS fluctuations 
are 
\begin{align}
\langle v \rangle &= \dfrac{\sum_{i=1}^{N} v_i \,
(t_i - t_{i-1})}{\sum_{i=1}^{N} t_i - t_{i-1}} \,, \\
\langle u^{\prime 2} \rangle &= \dfrac{\sum_{i=1}^{N} (v_i -
\langle v \rangle )^2 \, (t_i - t_{i-1})}{\sum_{i=1}^{N} t_i - t_{i-1}} \,,
\end{align}
where $N$ is the total number of velocity samples.  
As noted by \cite{albrecht:2003}, this method is suitable for moment
estimation of single point statistics but fails for estimation of
correlation functions.  
Therefore, we did not apply this method to our two-point correlation
measurements, and we did not take any measures to correct for possible 
bias in our two-point correlations.  

\begin{figure}
\begin{center}
\includegraphics[width=10cm]{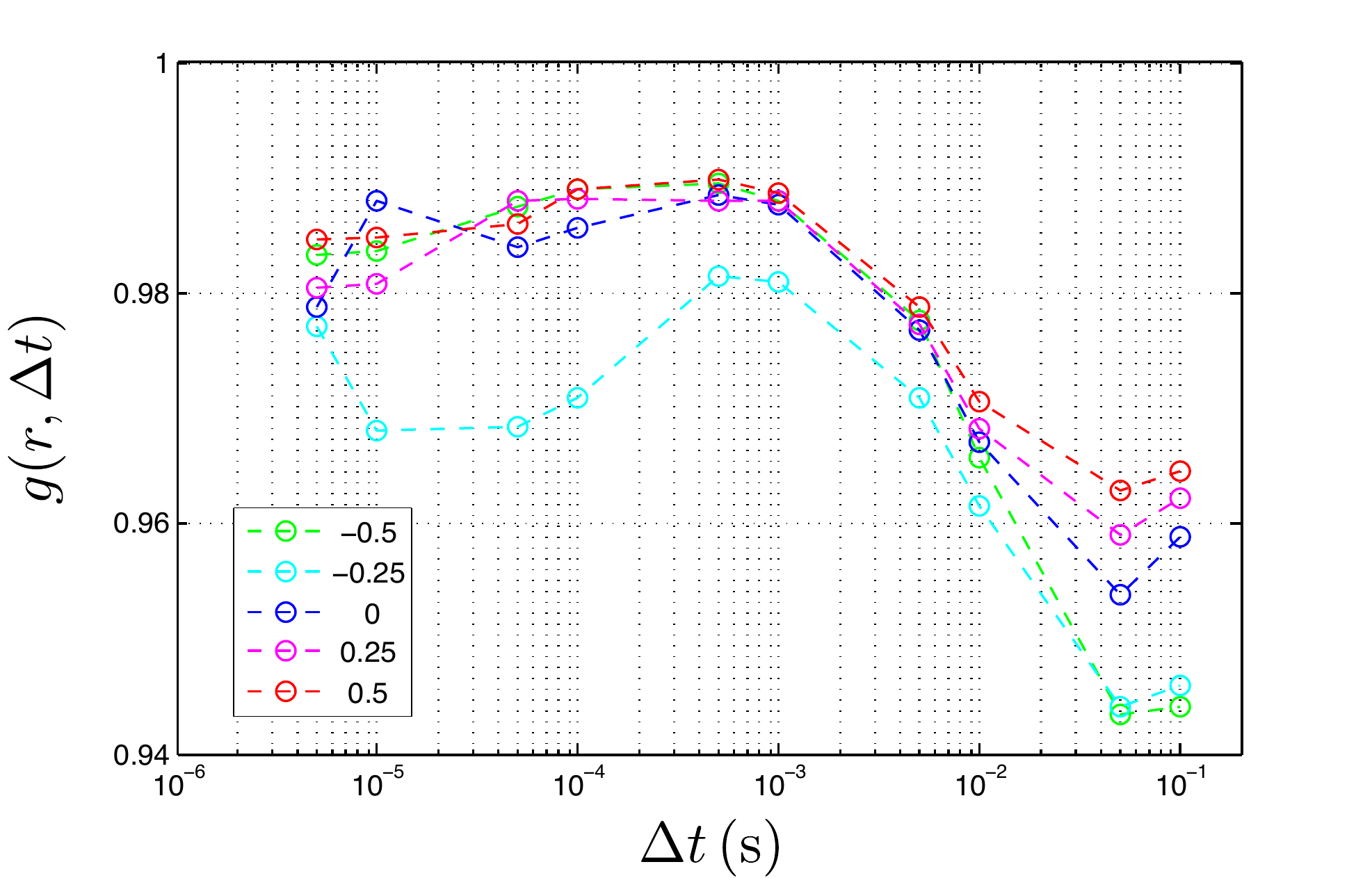}
\caption{The effect of temporal binning on the value of the velocity 
  correlation, $g(r)$, near zero separation.  
  The values in the legend are the separations in millimeters.}
\label{fig:timebin_corr}
\end{center}
\end{figure}
\begin{figure}
\begin{center}
\includegraphics[width=10cm]{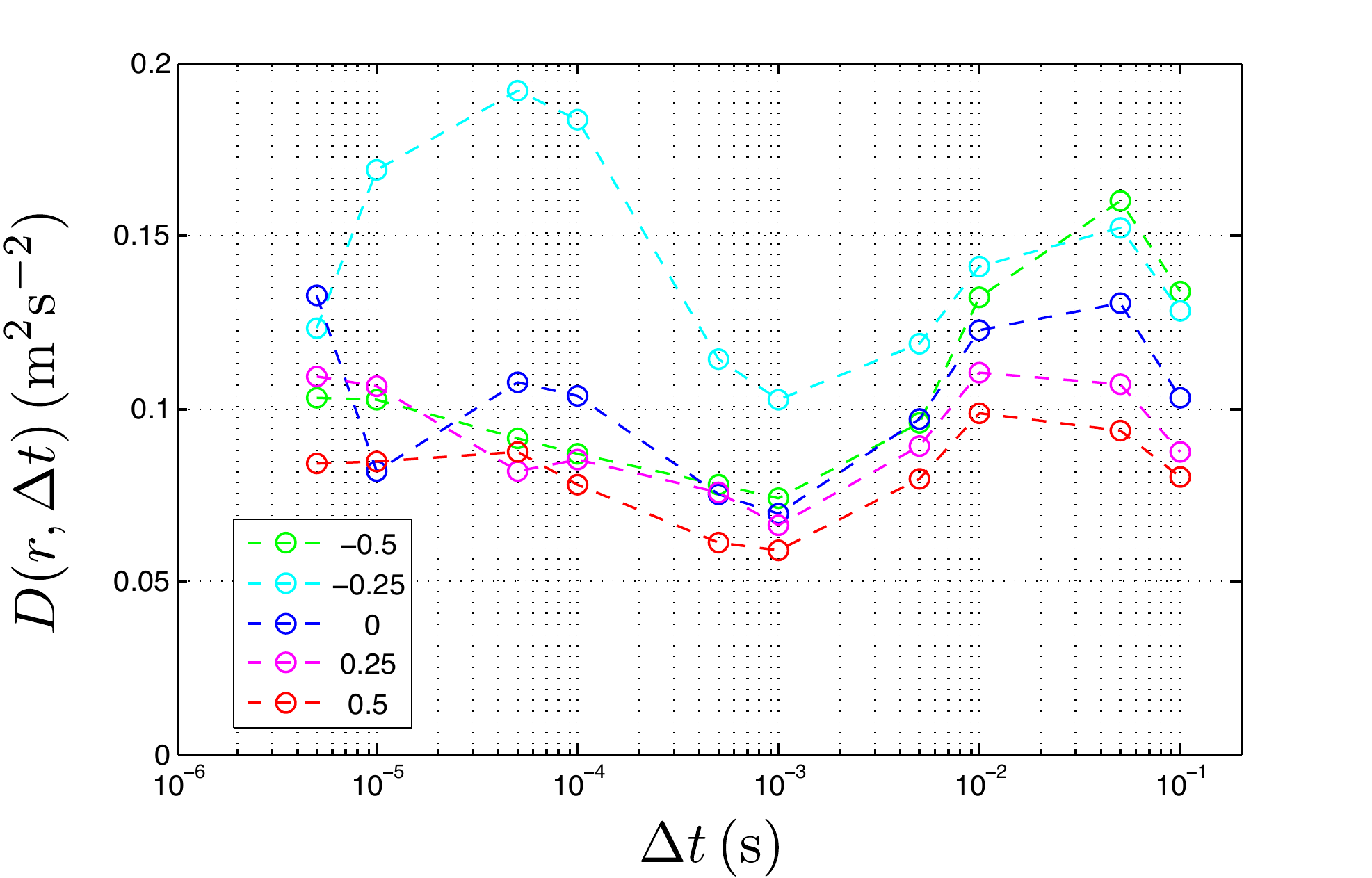}
\caption{The structure function as a function of temporal bin width.  
For legend, see previous figure.}
\label{fig:timebin_struct}
\end{center}
\end{figure}
\begin{figure}
\begin{center}
\includegraphics[width=10cm]{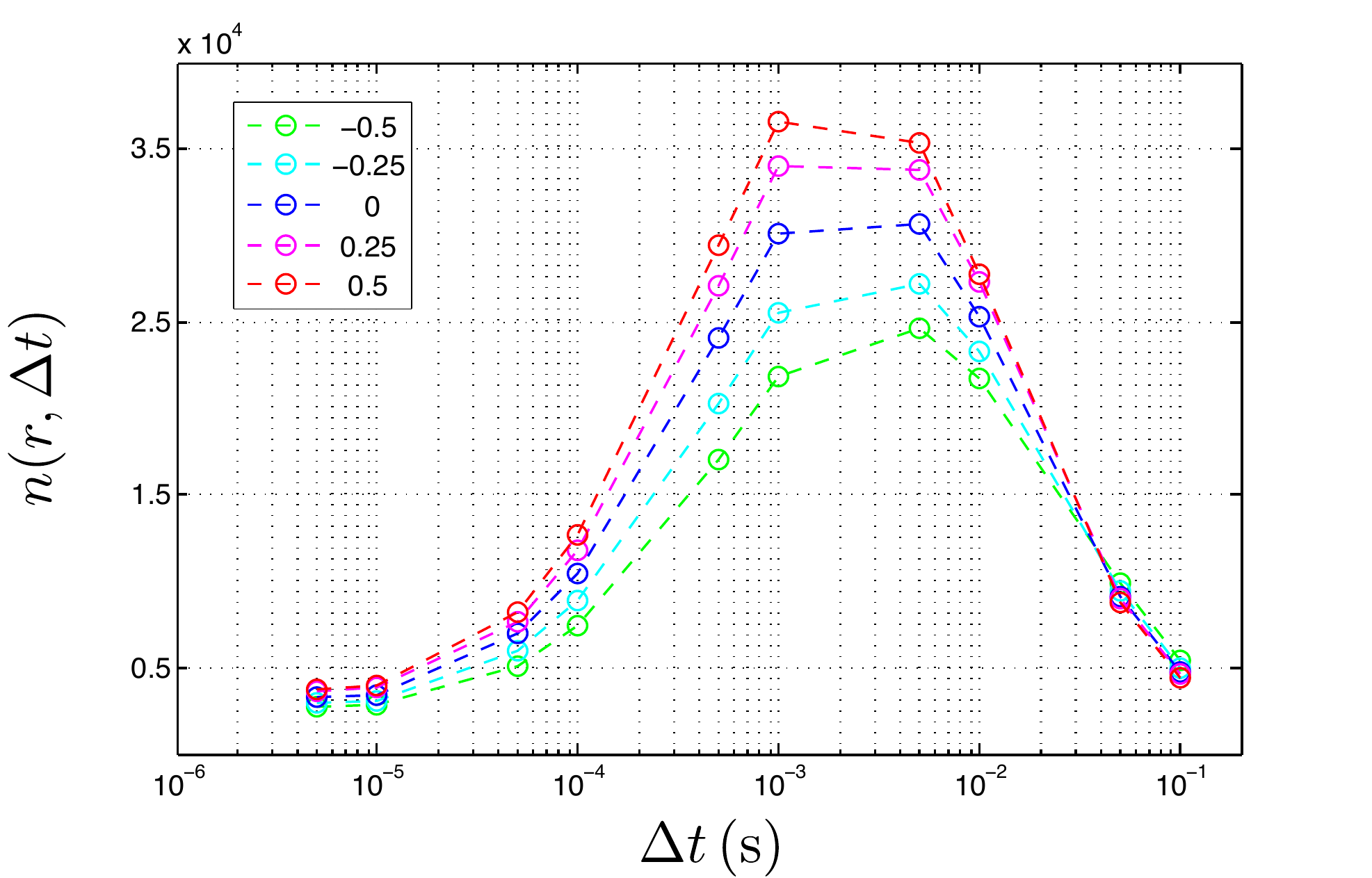}
\caption{The number of samples yielding the velocity 
correlation as a function of temporal bin width.  
For legend, see figure~\ref{fig:timebin_corr}.}
\label{fig:timebin_n}
\end{center}
\end{figure}

Because different probes record slightly different particle arrival times,
in order to calculate two-point statistics, one must look for
coincidences of velocity samples coming from these probes.  
We found these coincidences by discretizing the velocity 
time series using a slotting technique \cite[][]{mayo:1974} and
searched for samples coming from each of the two probes with 
time separations falling within a given temporal bin, $\Delta t$, of
duration $0.5$~ms.  
This duration was chosen because it maximized both the number
of coincidences and the value of the velocity 
autocorrelations (see section~\ref{sec:integralscale})
near zero separation while keeping the value of the structure 
functions near zero separation small (see figures~\ref{fig:timebin_corr}, 
\ref{fig:timebin_struct}, and \ref{fig:timebin_n}).  
We typically collected about $10^4 - 10^5$ velocity samples.

For each of the two probes, an average was taken of velocity signal 
coming from the same probe falling within $\Delta t$.  
The product of all averaged velocity pairs was then added to a sum.  
After processing the entire temporal sequence, the total sum was 
divided by the number of accumulated products, giving us the value of 
the velocity autocorrelation for that given spatial separation.  
We found that so long as $\Delta t$ was smaller than 15~ms, 
the difference in the values of the correlation was less than 
1\%, as shown in figure~\ref{fig:timebin_corr}.  

By definition, the velocity correlation is unity at zero separation.
Noise, however, may contaminate the data and mask the true
value of the correlation at zero separation.  
In order for the correlation to comply with this definition, 
we normalized the velocity correlation with its
value at zero separation.  
The measured structure functions could have been corrected
to yield better accuracy at the smaller separations, but
since the measured structure functions had leveled off at the
low end (see figures~\ref{fig:comp_structure_func} and
\ref{fig:sf_3rd}), and correction of any data is never a satisfying
proposition, no matter how carefully executed, because
in some sense it presumes the answer, therefore no attempt
was made to apply corrections to these measurements, but
they may be required in some future investigation.  
It can certainly be said that the uncorrected data were
internally consistent and the inertial scale measurements were
done with sufficient accuracy that no special correction needed
to be developed.  

\subsection{Limitations of the LDV technique}
%

In addition to the bias discussed in \ref{subsec:dataprocessing},
other physical parameters may limit the accuracy of LDV.  We review
these limitations in the following by dividing them into those arising
from the flow and those arising from measurement instruments.  

\subsubsection{Tracer particles}
\label{subsubsec:tracers}


The motion of an individual spherical particle in a turbulent flow has
been extensively studied \cite[see e.g.][]{corrsin:1956, manton:1977,
  maxey:1983}, and the response of a tracer particle to isotropic
turbulence has been examined by \cite{mei:1996}.  
In making accurate velocity measurements of a fluid with LDV, we
require the velocity of the tracer particle be close to the local,
instantaneous fluid velocity.  
The particle inertial effects are described by the Stokes number
\begin{equation}
\mathrm{St} = \dfrac{\tau_p}{\tau_\eta} \,,
\end{equation}
where $\tau_p$ is the particle response time defined as
\begin{equation}
\tau_p = \dfrac{1}{18} \, \biggl( \dfrac{\varrho_p}{\varrho_f} \biggr) \,
\biggl( \dfrac{D^2}{\nu} \biggr) \,,
\end{equation}
where $\varrho_p$, $\varrho_f$, $D$, and $\nu$ are the particle
density, the density of the fluid, the particle diameter, and 
the fluid viscosity, respectively.  
$\tau_\eta$ is the Kolmogorov time defined in 
equation~\ref{eq:kolmogorov_time}.  
\cite{buchhave:1979} considered the particle-fluid relative velocity,
using a model by \cite{lumley:1976}, and estimated an upper limit of
the particle Stokes number of ${\rm St} \leqslant 1/74$ with the criterion
that the error in the velocity be less than 1\%.
We used oil droplets (density 0.9~g~cm$^{-3}$) as tracer particles in
air (density 1.2~kg~m$^{-3}$), generated by a Palas AGF 10.0 aerosol
generator.  
The droplets had a most probable diameter of about 3~$\micro$m.  
They settled in air at 200~$\micro$m~s$^{-1}$.  
The Stokes number based on the most probably diameter was about 
0.02, the smallness of which qualifies these oil droplets as passive 
tracers \cite[see e.g.][]{bewley:2008}.  

A second problem connected to seeding the fluid with
particles is turbulence modification due to particle-fluid
interaction.  
This is an area of active research \cite[see e.g.][]{hestroni:1989,
  elghobashi:1993}.  
The addition of particles at a volume fraction of as low as $10^{-5}$
to a turbulent flow modifies its motion \cite[][]{elghobashi:1994}.  
For low values of particle volume fraction ($\leqslant 10^{-6}$) the
particles have negligible effect on the flow \cite[][]{elghobashi:1994}.  
The oil droplets used in our experiment were present at a volume
fraction of less than $10^{-7}$.  
This particle concentration was too small to alter the turbulence.

\subsubsection{The effect of refractive index fluctuations}
\label{subsubsec:refractive_index}

Since the laser beams of the LDV probes must propagate through a
medium having random index of refraction fluctuations, the beam paths
will no longer be straight, but will take on random fluctuating
passages through the medium, and the phase and amplitude of the
receiving signal will fluctuate randomly.  
The fluctuations in the index of refraction are primarily due to
pressure fluctuations and temperature fluctuations in the medium.  
\cite{buchhave:1979} estimated that, by considering an isentropic 
variations in the pressure in a compressible flow and assuming a 
quasi-Gaussian relationship between pressure and velocity fluctuations, 
a root-mean-square (RMS) velocity fluctuations of the order of
70~m~s$^{-1}$ are necessary to produce significant changes in the
index of refraction.  
For temperature fluctuations, they obtained an estimate for the rate
of decrease of index of refraction with temperature of the order of $1
\times 10^{-6}$~K$^{-1}$ for air at 273~K.  
Since the typical velocity fluctuations in our experiment was
1~m~s$^{-1}$ ($\ll 70$~m~s$^{-1}$), and the rate of change of index of
refraction with temperature is small, we can neglect the
effect of index-of-refraction fluctuations.  

\subsubsection{The probe volume size}
\label{subsubsec:probevolume}

\begin{figure}
\begin{center}
\includegraphics[scale=0.5]{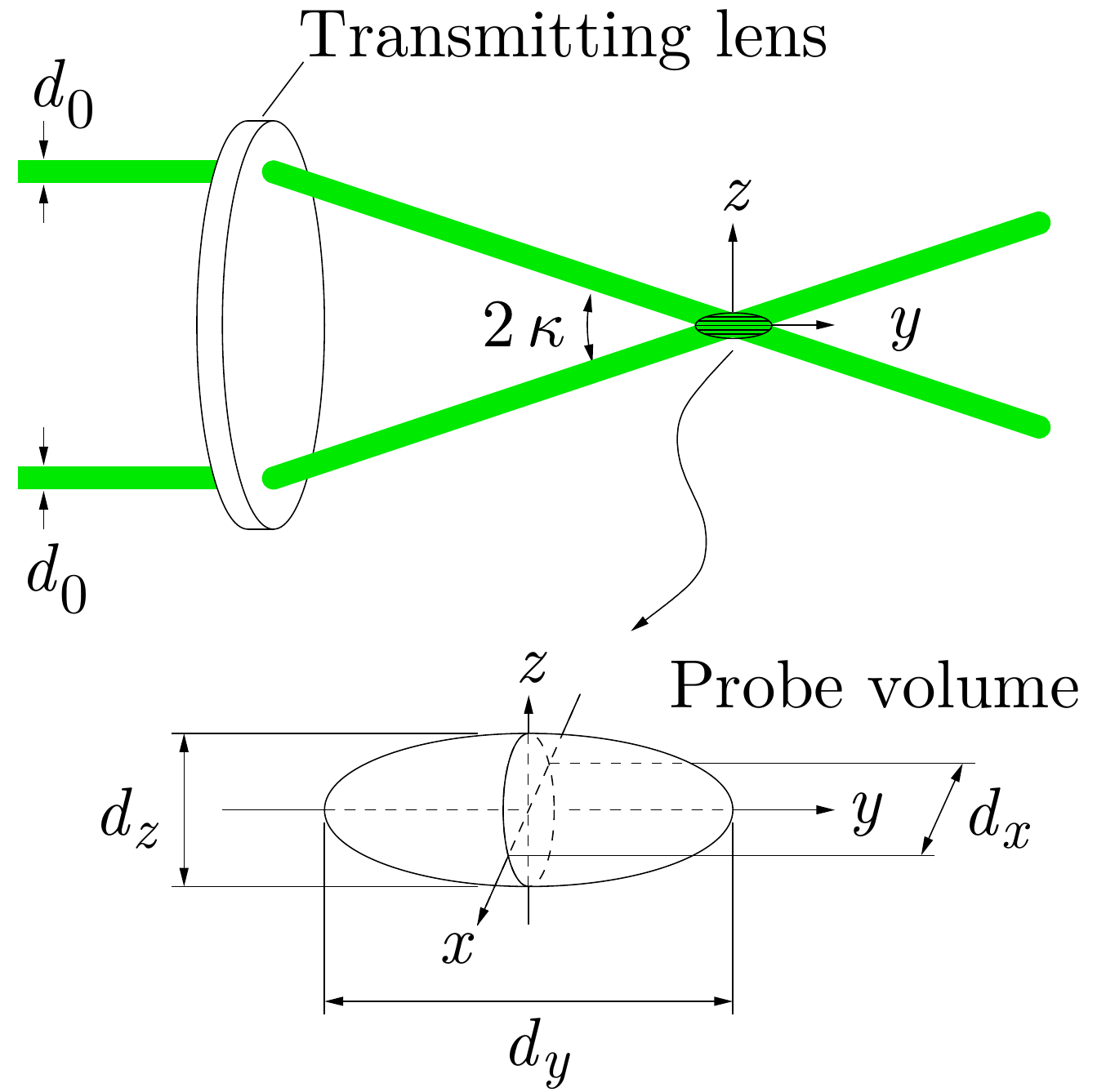}
\caption{The schematic shows the LDV probe volume dimensions.}
\label{fig:probe_volume}
\end{center}
\end{figure}

The probe volume is defined by the surface on which the light
intensity of the fringes is $1/e^2$ of the maximum intensity in the
beam intersection region.  
With Gaussian beams the probe volume is an ellipsoid, and we have the
expression for the probe volume dimensions in terms of the principal
diameters of the ellipsoid in the coordinate system defined in
figure~\ref{fig:probe_volume}:  
\begin{align}
d_x &= \frac{d_{0}}{\cos \kappa} \,, \\
d_y &= d_0 \,, \\
d_z &= \frac{d_0}{\sin \kappa} \,,
\end{align}
where $d_0$, the width of the waist of the focused beam, is 
\begin{equation}
d_0 = \frac{4 \, \lambda \, f}{\pi \, a} \,,
\end{equation}
for $\lambda$, the wavelength of the light, $f$, the focal length of
the transmitting lens, and $a$, the diameter of the beam.  
For our LDV system, the probe volume formed by the two beams of
wavelength $\lambda = 514.5$~nm had the largest dimension.  
Each beam had a diameter of $3.5$~mm, and for our optics, $f =
512$~mm, $\kappa = 2.8^{\circ}$, the probe volume dimensions were
 $d_x \approx d_y \approx 96$~{\micro}m, and $d_z = 1.97$~mm.  
In order to resolve the smallest length scales in the flow, we required
that any of the three probe volume linear dimension be smaller than
the Kolmogorov scale, $\eta$ (see equation~\ref{eq:kolmogorov_length}).  
Our probe volume dimensions were $d_x \approx 0.6 \, \eta$, $d_y \approx
0.6 \, \eta$, and $d_z \approx 12 \, \eta$ (see table~\ref{table:flowstat} 
for values of $\eta$ in the experiment).  
Thus, our optical setup had not met this condition but it had sufficient
dynamic range to resolve the intermediate and the very large scale
motion.  
\chapter{The experimental procedure}
\label{chap:procedure}

Here we describe the coordinate system we used to present our results.  
We describe the measurement protocol that exploited the symmetry
of the flow generation system to rotate the turbulence relative to
our LDV measurement apparatus.  
We discuss the quantity we fixed when comparing flows with 
different values of the forcing anisotropy.

\section{The coordinate system}
\label{sec:coordinate_system}

\begin{figure}
\begin{center}
\includegraphics[width=10cm]{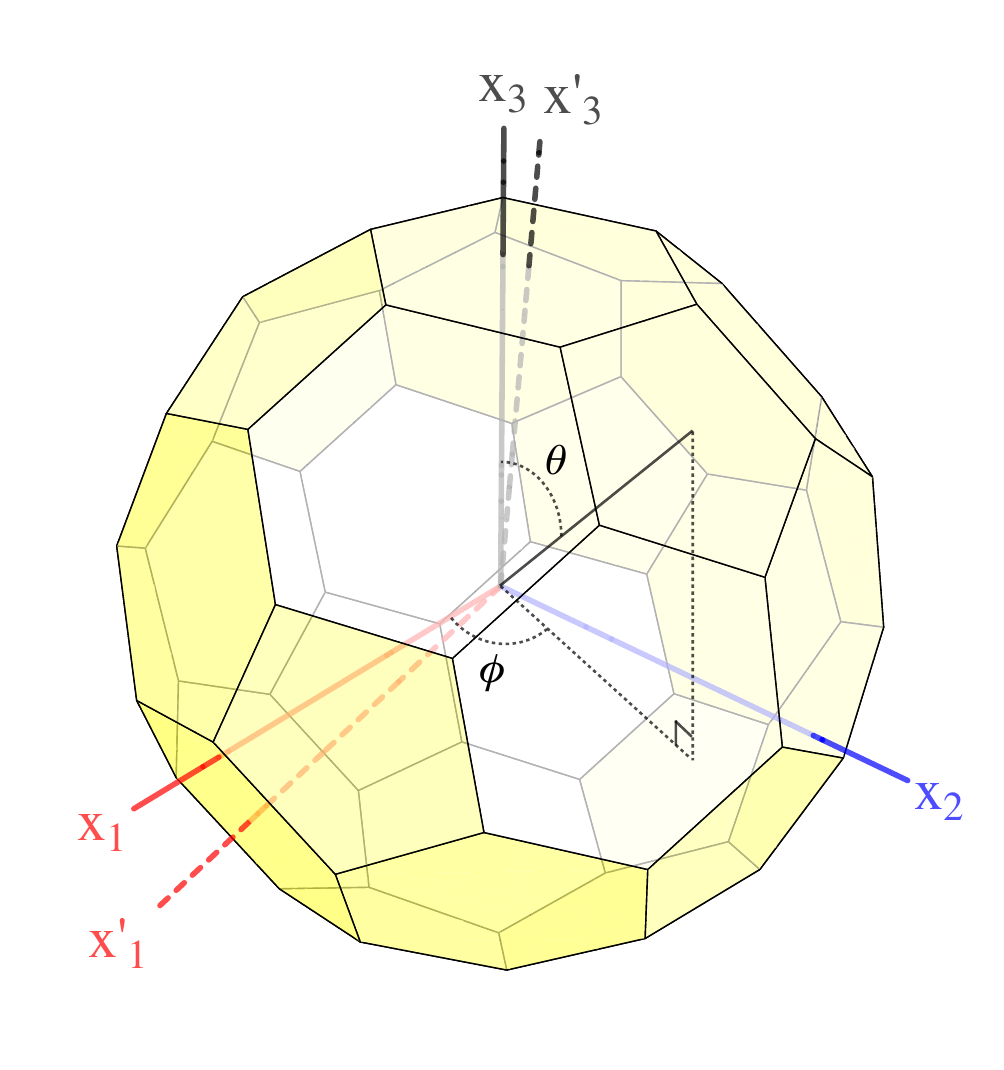}
\caption{The schematic shows the coordinate system in the laboratory 
  frame.  $x_3$ was aligned with the vertical.  
  The axis of symmetry of the forcing lay along 
  either $x^{\prime}_1$ or $x^{\prime}_3$, depending on the 
  configuration.  Our measurements were made at points lay
  along $x_1$.}
\label{fig:lab_frame}
\end{center}
\end{figure}
The turbulence coordinate system was not fixed to the laboratory
frame, $(x_1, x_2, x_3)$, as is shown in figure~\ref{fig:lab_frame}, but
was fixed to the turbulence symmetry axis.  
Because the turbulence was axisymmetric, it had two principal axes; 
one being the axis of symmetry of the forcing, and the other one, 
a radial axis, perpendicular to it.  
However, as described in \ref{sec:measurement_protocol}, our
measurement apparatus measured particle velocities only at points that
lie along a straight line fixed in the laboratory frame.  
In order to sample the statistics of turbulence along both axes, 
we rotated the axis of symmetry of the turbulence by taking advantage 
of the symmetries of the flow apparatus.  

The two orientations of the symmetry axis in the laboratory frame 
are shown as $x_{1}^{\prime}$ and $x_{3}^{\prime}$ in 
figure~\ref{fig:lab_frame}, each of which passes through the centers 
of opposite hexagonal faces of the flow apparatus.  
The $x_1^{\prime}$ axis was nearly parallel to the line along which we 
collected velocity samples, $x_{1}$, and it lay along 
$\theta = 107^{\circ}$ and $\phi = -4^{\circ}$.  
The $x_{3}^{\prime}$ axis lay along $\theta = 4^{\circ}$ and 
$\phi = 180^{\circ}$, and was close to the $x_3$ axis.  
These two axes, $x_1^{\prime}$ and $x_3^{\prime}$, were also 
nearly perpendicular and parallel to the vertical axis, respectively.  
We assume that the two orientations are equivalent because the
acceleration of gravity is negligible relative to particle
accelerations in the turbulent field \cite[see e.g.][]{voth:2002}.  
Because the primed and unprimed coordinate systems are close 
to each other, we do not distinguish between them in the rest of 
this thesis.  
\begin{figure}
\begin{center}
\subfigure[]{\includegraphics[width=7cm]{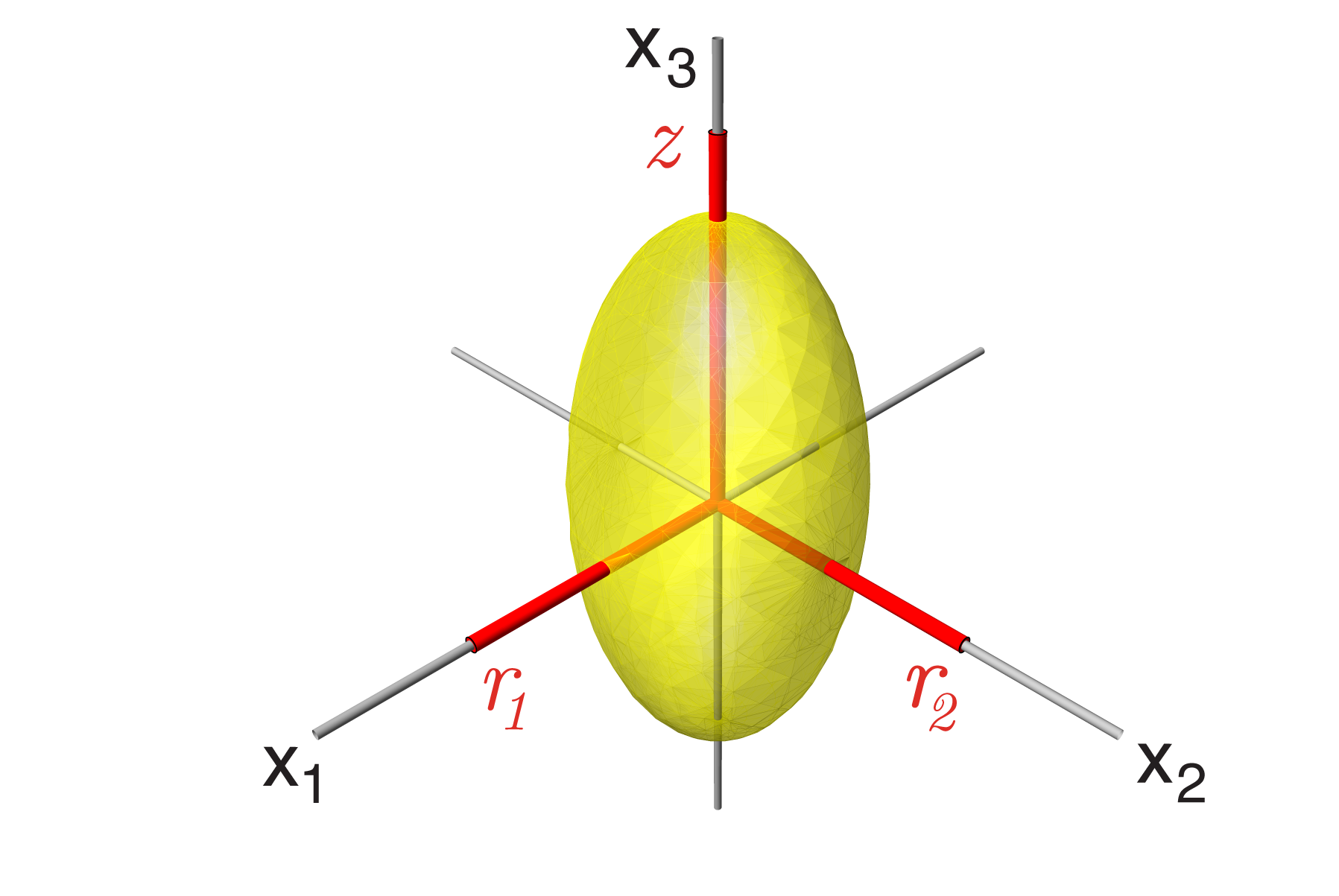}}
\hfill
\subfigure[]{\includegraphics[width=7cm]{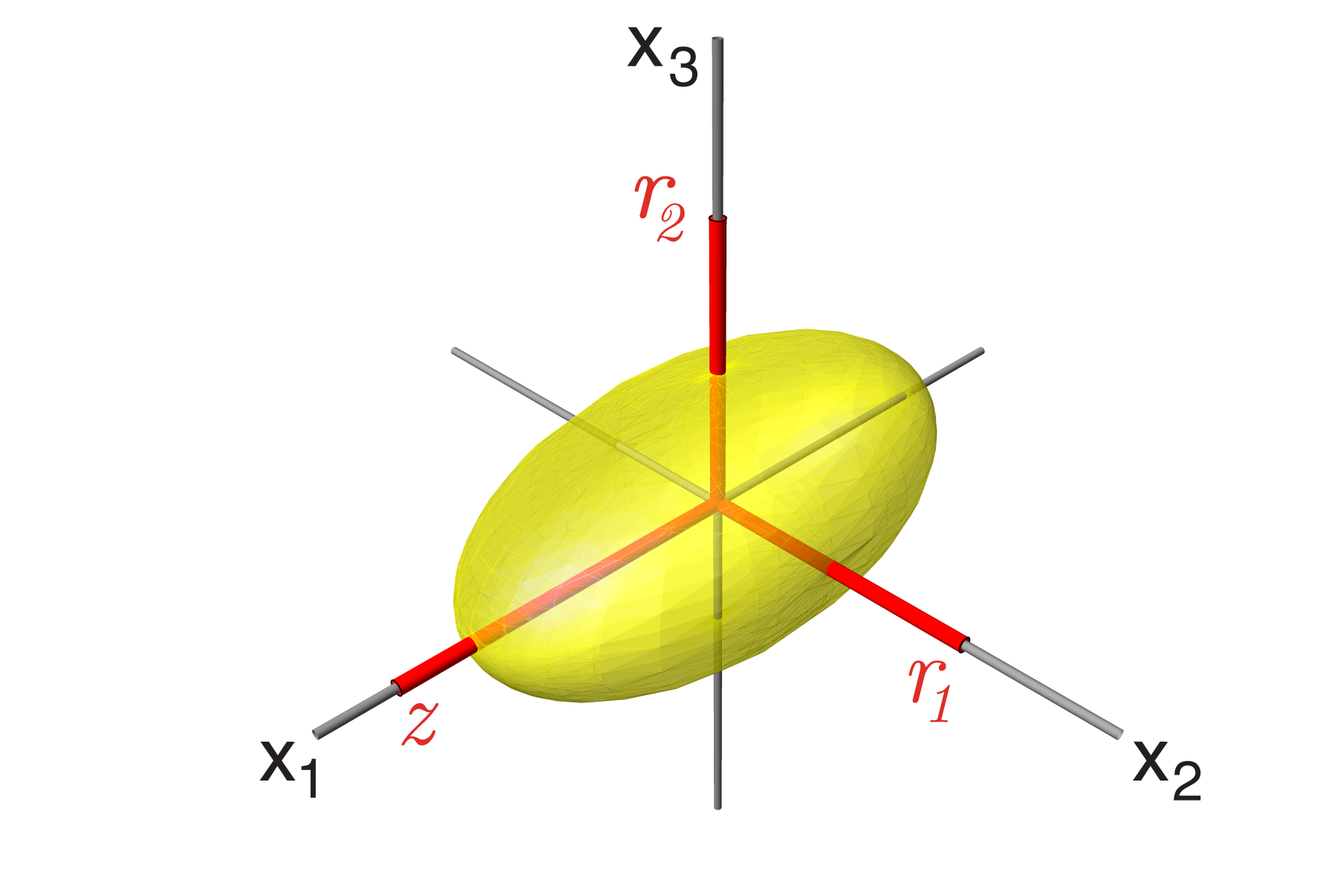}}
\caption{The figure shows the orientations of the body coordinate 
system (in red) of the forcing (ellipsoidal surface in yellow) with respect 
to the laboratory frame (gray), and the conventions we used: (a) when 
the axis of symmetry of the forcing lay close to the $\mathsf{x}_3$
axis of the laboratory frame, and (b) when the axis of symmetry lay
close to $\mathsf{x}_1$ axis of the laboratory frame.  
The coordinate system $(\mathsf{x}_1, \mathsf{x}_2, \mathsf{x}_3)$ 
was fixed in  the laboratory frame, and the coordinate system 
$(r_1, r_2, z)$ was fixed with respect to the symmetry of the forcing.}
\label{fig:forcing_coordinates}
\end{center}
\end{figure}

Figures~\ref{fig:forcing_coordinates}(a) and (b) show the 
coordinate system, $(r_1, r_2, z)$, which was aligned with the symmetry 
axis of the forcing, $z$.  
The two orientations of the forcing coordinate system with respect to
the laboratory frame correspond to the two cases described above.  
Hereafter, `axial' refers to both the direction of the particle
velocity component measured along the axis of symmetry, and, in
discussing two-point statistics, separations along the axis of
symmetry.  
Likewise, `radial' refers to both the direction of the
particle velocity components normal to the axis of symmetry, and
to separations normal to the axis of symmetry.  
Additionally, we refer to two-point quantities whose separation
vector lies along the axis of symmetry as axial quantities, and those
with radial separations as radial quantities.  
For example, a two-point quantity $f(x_1)$ is denoted as $f(z)$ when
the axis of symmetry of the forcing lies along $x_1$, and is called an
axial quantity.  
In keeping with the geometry described above, our `axial'
measurements were in reality about $15^{\circ}$ away from the axis
of symmetry, and the `radial' measurements were about $4^{\circ}$
from its normal.  
Our results do not depend on the angle between the `axial' and
`radial' measurements.  

\section{The measurement protocol}
\label{sec:measurement_protocol}

\begin{figure}
\begin{center}
\subfigure[]{
\includegraphics[scale=0.32]{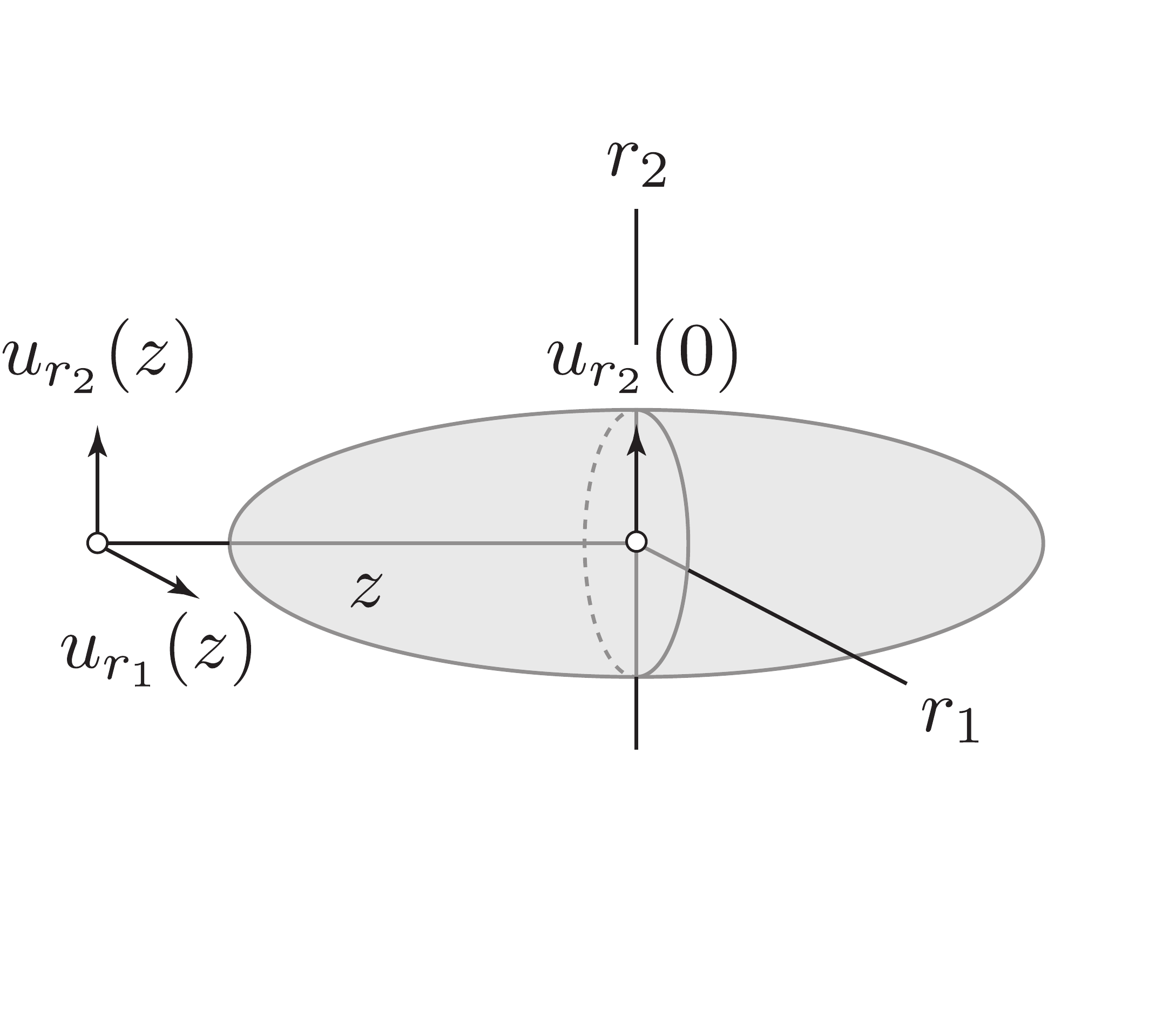}}
\hfill
\subfigure[]{
\includegraphics[scale=0.32]{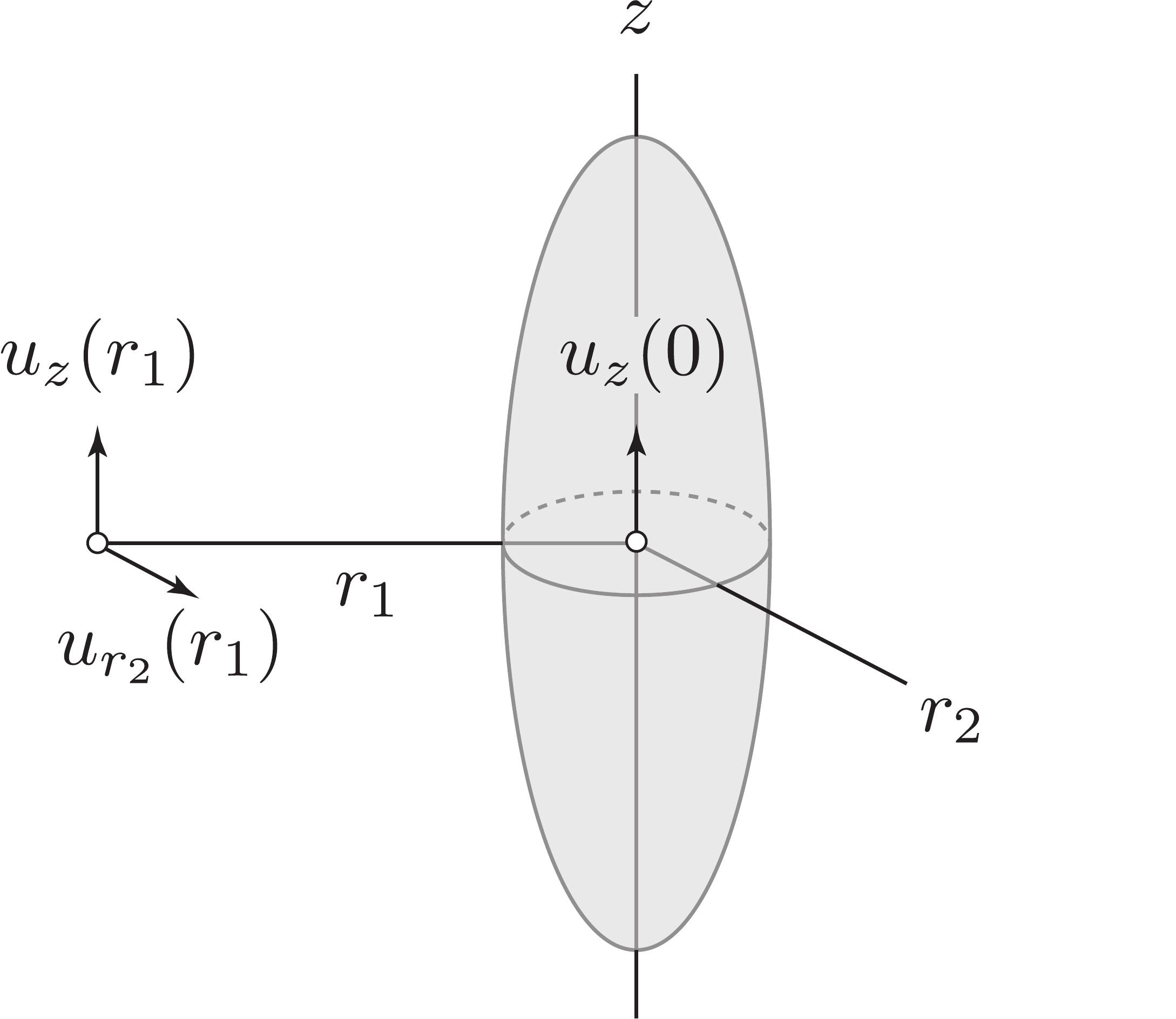}}
\caption{The figure shows the orientations of the measured velocity components with respect to the forcing coordinate system 
when the separation vector lay (a) along the axis of symmetry, or (b) perpendicular to the axis of symmetry.}
\label{fig:velocity_orientations}
\end{center}
\end{figure}
We used the LDV probes in two configurations.  
In one configuration, all three orthogonal components of the fluid
velocities were measured at a single point.  
For these measurements, the probes observed a fixed volume 
(aligned to within 10~$\micro$m with a pin hole) close to the
center of the soccer ball, which is the point $(0,0,0)$
in the laboratory frame shown in figure~\ref{fig:lab_frame}.  
One probe measured a single component of the fluid velocities, 
namely $u_{x_1} (0,0,0)$.  
The second probe measured two orthogonal components of
the fluid velocities, namely $u_{x_2} (0,0,0)$ and $u_{x_3} (0,0,0)$.  
In the second configuration, we aligned the probes in a similar 
way, except that we positioned the two-component probe at different 
points along the $x_1$ axis, using a programmable linear traverse.  
This probe now measured $u_{x_2} (x_1,0,0)$ and $u_{x_3} (x_1,0,0)$.  
In addition, the single-component probe was rotated by $90^{\circ}$ to
measure $u_{x_3} (0,0,0)$, which was coincident with one of the
components measured by the two-component probe when $x_1$ 
equaled zero.  
These velocity components, expressed in the $(r_1, r_2, z)$ coordinate 
system, are shown schematically in figure~\ref{fig:velocity_orientations}.  
The two orientations of the forcing coordinate system relative to the
laboratory frame are described in figure~\ref{fig:lab_frame}.  

\section{Methods}
\label{sec:methods}

We collected two data sets for each of a series of forcing anisotropy, 
$A$, given by equation~\ref{eq:amplitude_ratio}.  
For one data set, the axis of symmetry of the forcing was aligned 
close to $x_3$, and for the other, the symmetry axis was aligned close 
to $x_1$.  
For each value of the forcing anisotropy, and each orientation of the
forcing symmetry axis, data were collected with the two-component LDV
probe positioned at various locations along the $x_1$-axis.  
We varied the forcing anisotropy while fixing the turbulent kinetic
energy in the center of the soccer ball
\begin{equation}
\label{eq:K}
K = \frac{1}{2} \, \biggl[ \langle u_{x_1}^{\prime 2} (0,0,0) \rangle
       + \langle u_{x_2}^{\prime 2} (0,0,0) \rangle 
       + \langle u_{x_3}^{\prime 2} (0,0,0) \rangle \biggr] \,,
\end{equation}
where $u_{x_i}^{\prime}$ is the RMS fluctuations of the fluid velocity in the
$x_i$ direction, and $\langle \cdots \rangle$ denotes temporal
averaging.  
\chapter{The flows}
\label{chap:flows}

In this chapter, we evaluate the quality of the flows generated by the
turbulence apparatus introduced in chapter~\ref{chap:apparatus}.  
Three-dimensional velocity measurements made at single points
in the center region of the flow apparatus show that the anisotropy
of the turbulent velocity RMS fluctuations followed the anisotropy
of the forcing signal.  
Two-point velocity measurements made at points that lay along
a single line in the laboratory frame show that the flows were
approximately homogeneous and axisymmetric.  
For the case of spherically symmetric forcing, with $A=1$ (see 
equation~\ref{eq:amplitude_ratio}), we expected and indeed
found that the turbulence was isotropic.  
In all cases, the mean velocities were less than $0.24$~ms$^{-1}$, 
or less than $23\%$ of the fluctuations and the shear stresses were 
less than $7\%$ of the turbulent kinetic energy, as shown in 
table~\ref{table:flowstat}.  

\section{Anisotropy of the fluctuations}
\label{sec:anisotropy}

\begin{figure}
\begin{center}
\includegraphics[scale=0.25]{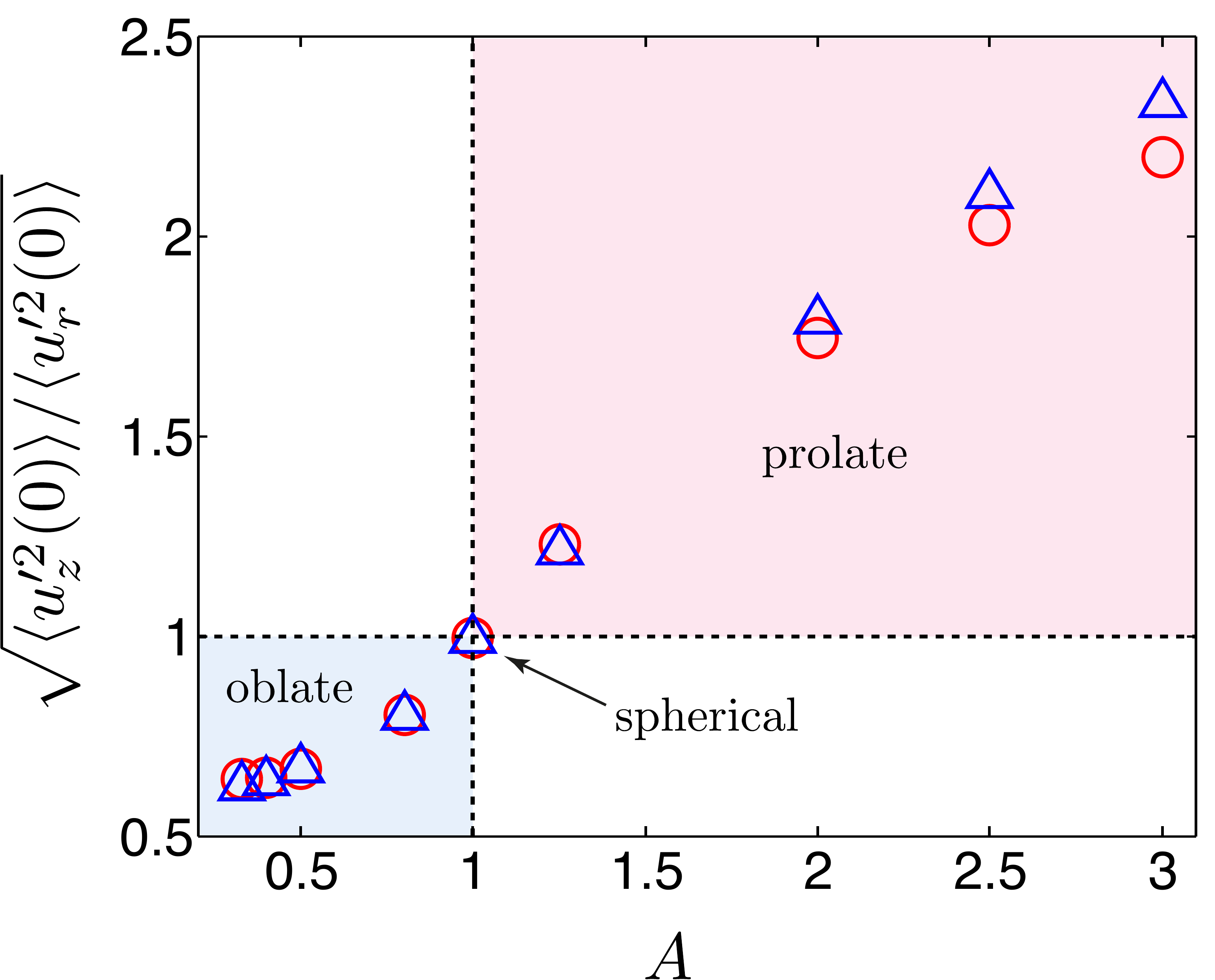}
\caption{The velocity fluctuation ratios as a function of the 
forcing anisotropy, $A$.  The circular symbols ({\color{red} $\ocircle$}) 
are $\sqrt{\langle u^{\prime 2}_{z} (0) \rangle / 
\langle u^{\prime 2}_{r_1} (0) \rangle}$ and the triangular symbols
({\color{blue} $\bigtriangleup$}) are $\sqrt{\langle u^{\prime 2}_{z} (0)
\rangle / \langle u^{\prime 2}_{r_2} (0) \rangle}$.  
Symbols in the region shaded in blue have oblate shape asymetry, 
whereas those in the region shaded in red have prolate shape
asymmetry.  Spherical symmetry lies at the intersection of the two
regions.  Data are collected with the axis of symmetry lying 
along $x_{3}^{\prime}$.  
We obtained similar results (not shown) when the axis of symmetry 
lay along $x_{1}^{\prime}$.}
\label{fig:uratio_aspectratio}
\end{center}
\end{figure}

Figure~\ref{fig:uratio_aspectratio} shows the ratio of axial to
radial velocity fluctuations as a function of the 
forcing anisotropy, $A$.
It is evident that the anisotropy of the turbulence followed
the variation of $A$.  
In addition, the anisotropy was nearly the same whether
measured in the $r_1$ or $r_2$ directions.  
Therefore, the turbulence was close to cylindrical symmetric,
though this was less so at extreme values of $A$.  
The loss of cylindrical symmetry might be explained by the decrease in
the number of loudspeakers driving the turbulence as the value of $A$
moved away from one.  
This is the nature of the forcing algorithm described in
section~\ref{subsec:algorithm}.  
We think that, as the number of loudspeakers effectively decreased, 
the turbulence became more sensitive to mechanical differences 
between the loudspeakers, and to misalignments of the nozzles.  

\section{Axisymmetry of the fluctuations}
\label{sec:axisymmetry}

\begin{figure}
\begin{center}
\subfigure[]{
\includegraphics[scale=0.6]{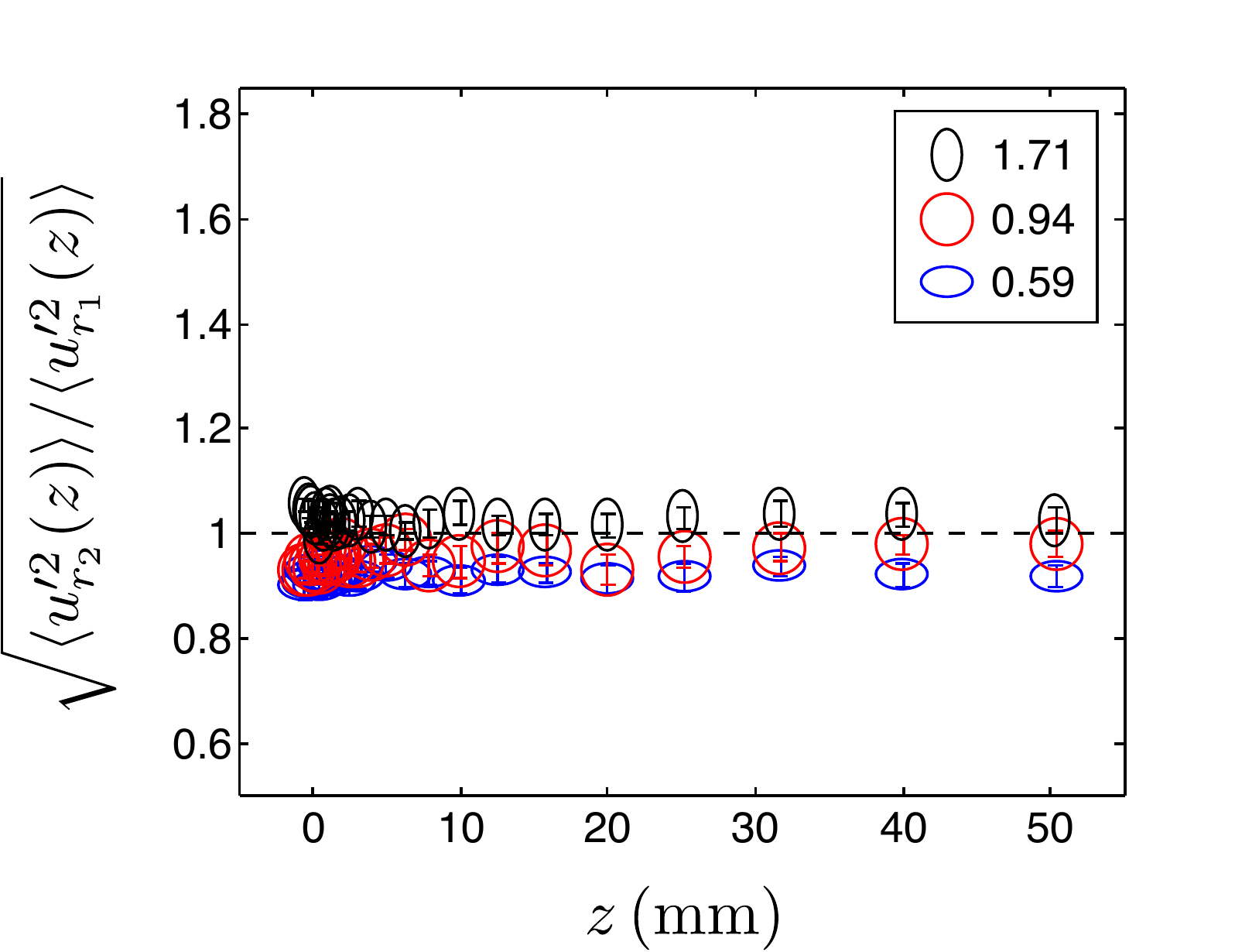}}
\subfigure[]{
\includegraphics[scale=0.6]{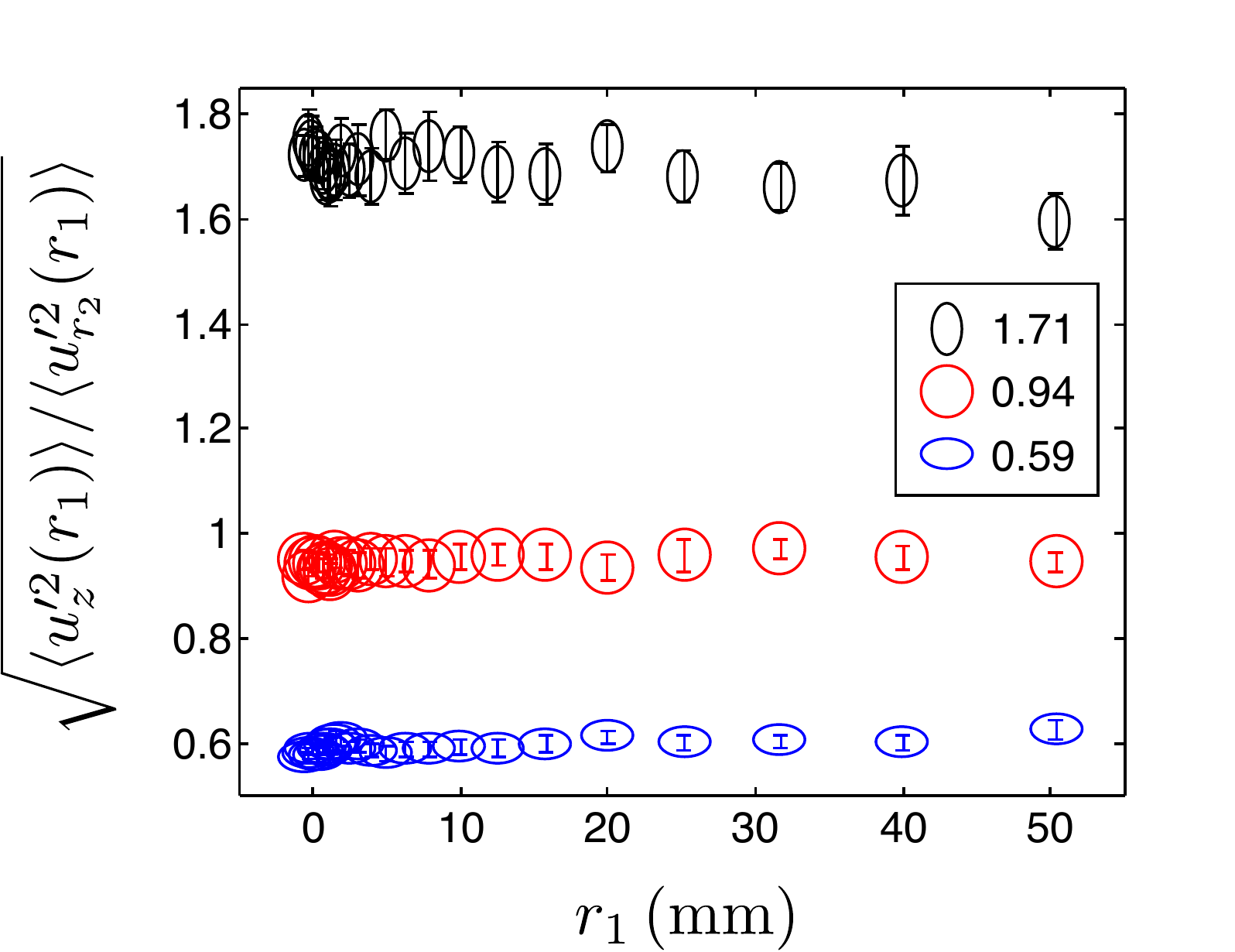}}
\caption{For three values of the anisotropy measured at the centre 
  of the ball, $\sqrt{\langle u^{\prime 2}_{z} (0) 
  \rangle / \langle u^{\prime 2}_{r} (0) \rangle}$, 
  we plot (a) the ratio of the two radial components of the 
  velocity fluctuations 
  $\sqrt{\langle u^{\prime 2}_{r_2} (0,0,z) \rangle / \langle
    u^{\prime 2}_{r_1} (0,0,z) \rangle}$
  and (b) the ratio of the axial fluctuations to the radial
  fluctuations, $\sqrt{\langle u^{\prime 2}_{z} (r_1,0,0) \rangle /
  \langle u^{\prime 2}_{r_2} (r_1,0,0) \rangle}$ 
  at various points moving away from the center of the ball.  
  The values in these legends, and in all others in this thesis, 
  are those of the data in (b) for $r_1 = 0$.  }
\label{fig:vel_anisotropy}
\end{center}
\end{figure}

Here, we characterize the degree to which the axisymmetry was 
spatially uniform.  
In figure~\ref{fig:vel_anisotropy}(a), we examine the ratio of the
two radial fluctuating velocities, $\sqrt{\langle u^{\prime 2}_{r_2}
  (0,0,z) \rangle / \langle u^{\prime 2}_{r_1} (0,0,z) \rangle}$, as a
function of the distance from the center of the soccer ball along the
axial direction for three values of the forcing anisotropy.  
Within $50$~mm from the middle of the flow chamber, the values 
of this ratio deviated by less than $10$\% from the value 1, indicating 
that the turbulence was close to being cylindrically symmetric.  
To gauge the extent of spatial uniformity of the anisotropy, we plot
in figure~\ref{fig:vel_anisotropy}(b) the ratio of the axial
fluctuating velocity to one of the radial fluctuating velocities,
$\sqrt{\langle u^{\prime 2}_{z} (r_1,0,0) \rangle / \langle u^{\prime
    2}_{r_2} (r_1,0,0) \rangle}$, as a function of the distance from
the center of the chamber.  
Here, the measurement points lie on a line nearly normal to the axis of
symmetry, and the values of this ratio at $r_1 = 0$ correspond to the
values shown in figure~\ref{fig:uratio_aspectratio}.  
Again, the velocity fluctuation ratio is approximately
constant within $50$~mm of the center of the chamber.  
Thus, we inferred that the degree of anisotropy was approximately
uniform within a central spherical region with radius $50$~mm.  
In both plots, the error bars were the standard errors of the 
measurements obtained by truncating the original data sets into 
$20$ equal parts.  
They indicate the sampling accuracy and not the measurement 
accuracy.  
These error bars were always less than $\pm 8.5 \%$.  
We concluded that, within the error bars, the ratios of the fluctuating 
velocities at different locations from the center of the soccer ball 
were indistinguishable from each other.  

Since we take axial and radial measurements through the 
turbulence region, we obtained twice the information on the 
radial velocity.  
Thus, the ratio $\sqrt{\langle u^{\prime 2}_{z} (0)
\rangle / \langle u^{\prime 2}_{r} (0) \rangle}$
in our experiment may be calculated in two different ways.  
The first one is the ratio of the two orthogonal 
velocity fluctuations in the limit as $r_1 \to 0$
\begin{equation}
\label{eq:uratio1}
u^{\prime}_z /u^{\prime}_{r_2} =
\lim_{r_1 \to 0}
\sqrt{\langle u^{\prime 2}_{z} (r_1,0,0) \rangle / 
\langle u^{\prime 2}_{r_2} (r_1,0,0) \rangle} \,,
\end{equation}
and the second one is 
\begin{equation}
\label{eq:uratio2}
u^{\prime}_{z} / u^{\prime}_{r_2} = 
\lim_{\substack{r_1 \to 0 \\ z \to 0}}
\sqrt{\langle u^{\prime 2}_{z} (r_1,0,0) \rangle / 
\langle u^{\prime 2}_{r_2} (0,0,z) \rangle} \,.
\end{equation}  

\begin{landscape}
\begin{table}
\begin{center}
\begin{tabular}{*{10}{c}}
\toprule
\addlinespace[8pt]
$A$ & Eq. & 
$0.33$ & $0.4$ & $0.5$ & $0.8$ & $1.0$ & $1.25$ & $2.0$ & $2.5$ \\
\addlinespace[5pt]
\midrule
$u^{\prime}_{z}$ (m~s$^{-1}$) & \ref{eq:uratio1} &
$0.74\pm 0.01$ & $0.75\pm 0.01$ & $0.79\pm 0.01$ & $0.91\pm 0.02$ & $1.08\pm0.02$ & $1.21\pm0.02$ & $1.42\pm 0.04$ & $1.47\pm 0.04$ \\
\addlinespace[8pt]
$u^{\prime}_{r_2}$ (m~s$^{-1}$) & \ref{eq:uratio1} & 
$1.26\pm 0.04$ & $1.24\pm 0.03$ & $1.27\pm 0.03$ & $1.19\pm 0.03$ & $1.15\pm 0.03$ & $1.03\pm 0.03$ & $0.83\pm 0.02$ & $0.75\pm 0.02$ \\
\addlinespace[8pt]
$u^{\prime}_{r_2}$ (m~s$^{-1}$) & \ref{eq:uratio2} &
$1.11\pm 0.02$ & $1.09\pm 0.02$ & $1.12\pm 0.02$ & $1.08\pm 0.02$ & $1.08\pm 0.02$ & $1.04\pm 0.02$ & $0.90\pm 0.02$ & $0.84\pm 0.02$ \\
\addlinespace[8pt]
$u^{\prime}_{z} / u^{\prime}_{r_2}$ & \ref{eq:uratio1} &
$0.59\pm 0.03$ & $0.60\pm 0.02$ & $0.63\pm 0.03$ & $0.77\pm 0.03$ & $0.94\pm 0.04$ & $1.17\pm 0.05$ & $1.71\pm 0.10$ & $1.97\pm 0.10$ \\
\addlinespace[8pt]
$u^{\prime}_{z} / u^{\prime}_{r_2}$ & \ref{eq:uratio2} &
$0.67\pm 0.03$ & $0.69\pm 0.03$ & $0.71\pm 0.03$ & $0.85\pm 0.03$ & $1.00\pm 0.04$ & $1.16\pm 0.03$ & $1.58\pm 0.09$ & $1.76\pm 0.09$ \\
\addlinespace[5pt]
\bottomrule
\end{tabular}
\caption{For forcing anisotropy
ratios $(A)$ ranging from $0.33$ to $2.5$, the table shows the
velocity fluctuation ratio according to the two limits 
discussed in the text (equations~\ref{eq:uratio1} and \ref{eq:uratio2}).}
\label{table:fluctuation_ratio}
\end{center}
\end{table}
\end{landscape}

We calculated the ratio in each case by averaging the RMS
velocity fluctuations at locations within $10$~mm from the center 
of the chamber, and then taking the ratio between the axial
and radial fluctuations.  
Their values are summarized in table~\ref{table:fluctuation_ratio}.  
The variation in the fluctuations in this range was less than $5\%$ 
and we used this variation for estimating the error in the 
velocity fluctuations.  
Our single-point statistics do not depend on this choice of the 
definition for the anisotropy ratio, and we discussed the results
in terms of $u^{\prime}_{z} / u^{\prime}_{r_2}$ defined in \ref{eq:uratio1}.  
The two-point correlation functions, as we will show in 
chapter~\ref{chap:integralscale}, do depend on the anisotropy
of the fluctuations and we discuss the results in terms of
$u^{\prime}_{z} / u^{\prime}_{r_2}$ defined in \ref{eq:uratio2}, 
since these are the velocity fluctuations associated with the radial 
and axial correlation functions.  

\section{Homogeneity of the fluctuations}
\label{sec:homogeneity}

\begin{figure}
\begin{center}
\subfigure[]{
\includegraphics[scale=0.6]{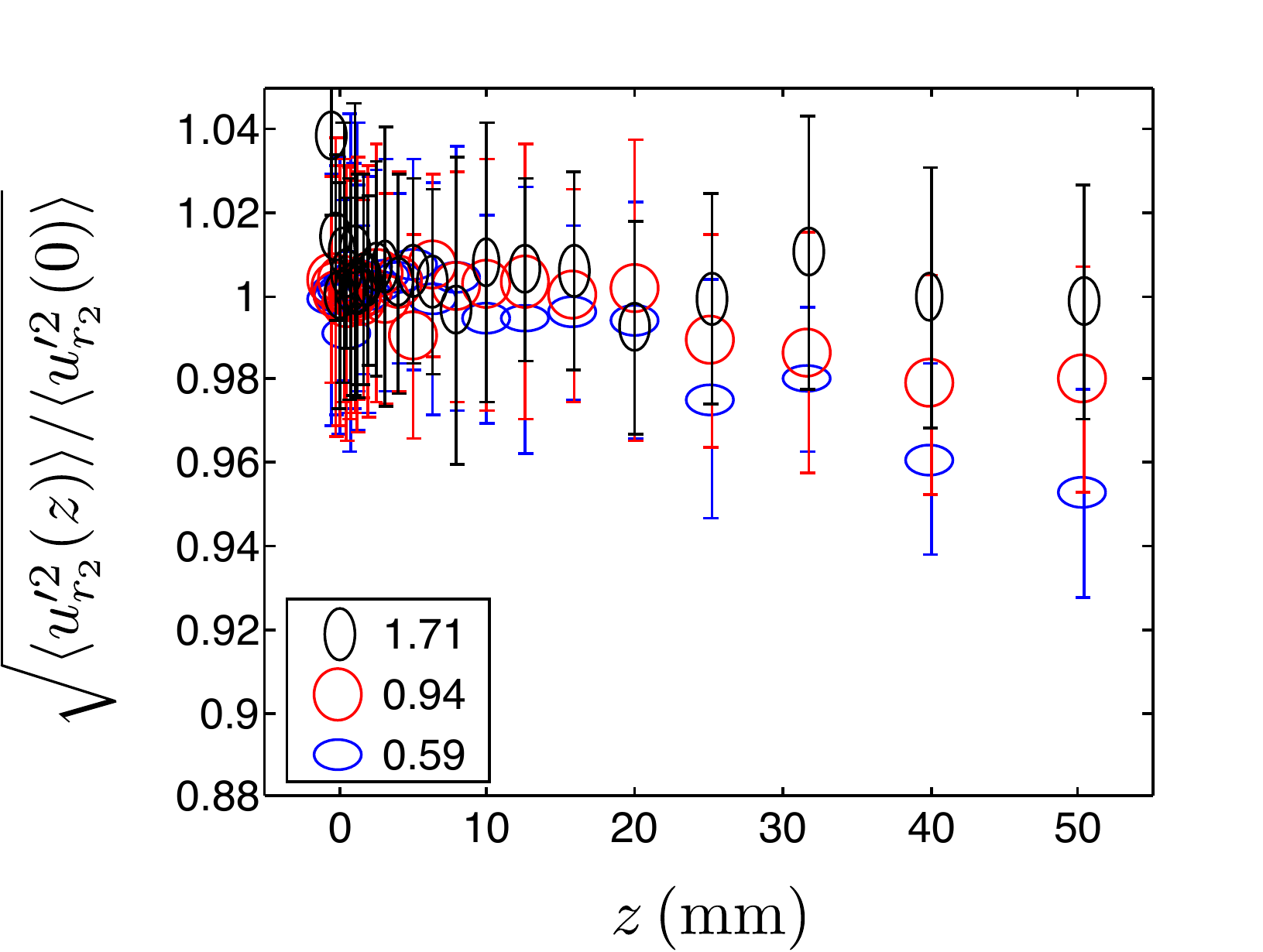}}
\subfigure[]{
\includegraphics[scale=0.6]{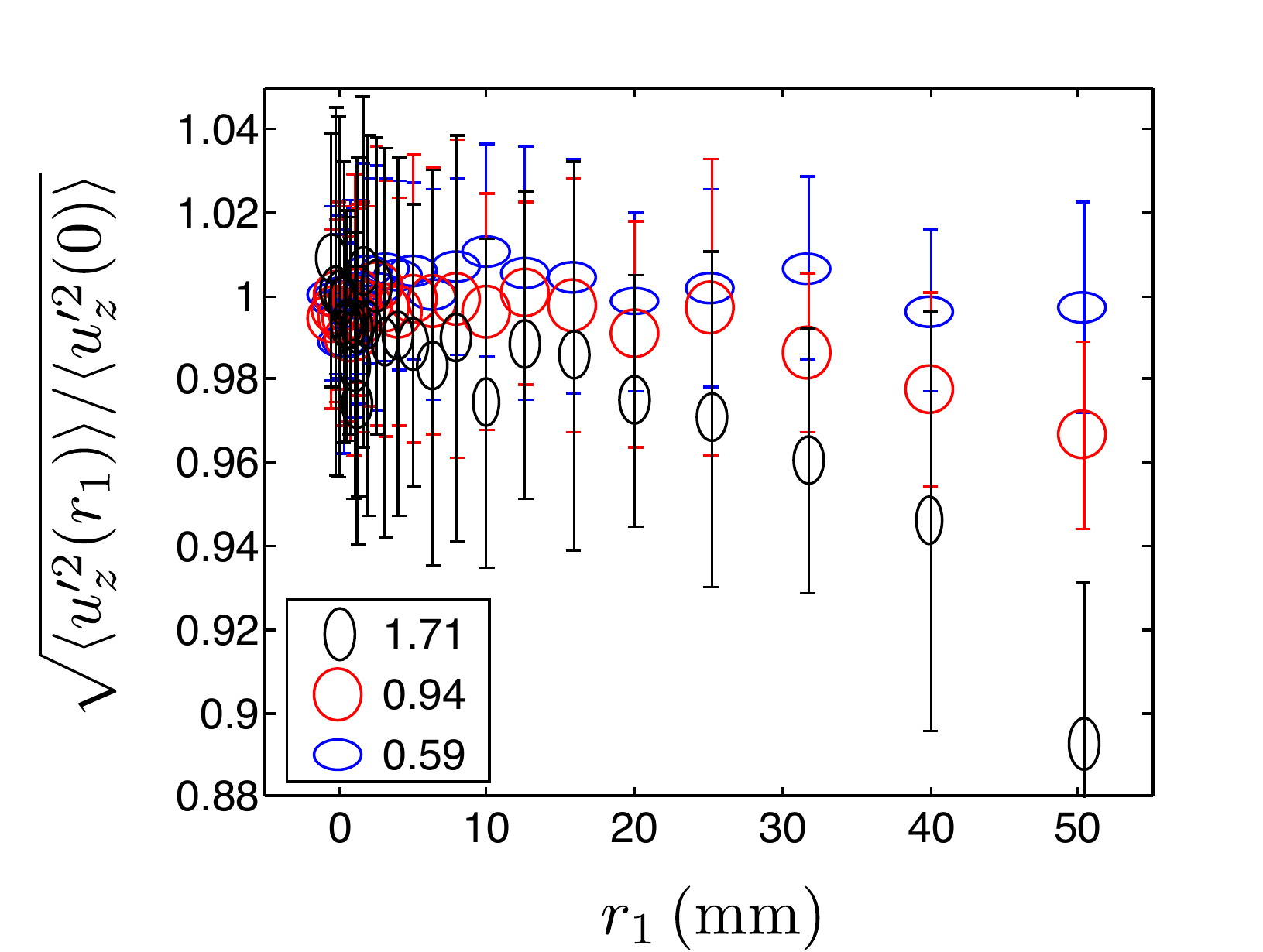}}
\caption{The comparison of fluctuations at various distances 
from the center of the soccer ball to those in the middle.  
(a) Radial fluctuations along axial separations, $\sqrt{\langle u^{\prime
    2}_{r_2} (0,0,z) \rangle / \langle u^{\prime 2}_{r_2} (0,0,0)
  \rangle}$, for anisotropy ratios, $\sqrt{\langle u^{\prime 2}_{z}
  (0) \rangle / \langle u^{\prime 2}_{r} (0) \rangle}$, ranging from 0.59 to
  1.71.  (b) Axial fluctuations along radial separations, 
  $\sqrt{\langle u^{\prime 2}_{z} (r_1,0,0) \rangle / \langle
    u^{\prime 2}_{z} (0,0,0) \rangle}$, for the same range of
  anisotropy.  The anisotropy ratios in the legend are measured at the
  center of the soccer ball.  }
\label{fig:vel_homogeneity}
\end{center}
\end{figure}

To evaluate the homogeneity of the turbulent fluctuations, 
we present measurements of 
$\sqrt{\langle u^{\prime 2}_{r_2} (0,0,z) \rangle / 
\langle u^{\prime 2}_{r_2} (0,0,0) \rangle}$ and 
$\sqrt{\langle u^{\prime 2}_{z} (r_1,0,0) \rangle / 
\langle u^{\prime 2}_{z} (0,0,0) \rangle}$ in 
figures~\ref{fig:vel_homogeneity}.  
These ratios compare the amplitude of the velocity fluctuations at
locations away from the center to those in the center of the chamber.  
The error bars were the standard errors of the measurements
obtained by dividing the original data sets into $20$ equal parts.  
Within a radius of $50$~mm, the difference between $\sqrt{\langle
  u^{\prime 2}_{r_2} (0,0,z) \rangle / \langle u^{\prime 2}_{r_2}
  (0,0,0) \rangle}$ and its value at the origin is within $5$\% of the
value at the origin.  
The inequality also holds for $\sqrt{\langle u^{\prime 2}_{z}
(r_1,0,0) \rangle / \langle u^{\prime 2}_{z} (0,0,0) \rangle}$.  
Thus, we conclude that the RMS fluctuations are homogeneous.  

\section{Anisotropy of the mean flow}
\label{sec:anisotropy_mean_flow}

\begin{figure}
\begin{center}
\subfigure[]{
\includegraphics[scale=0.6]{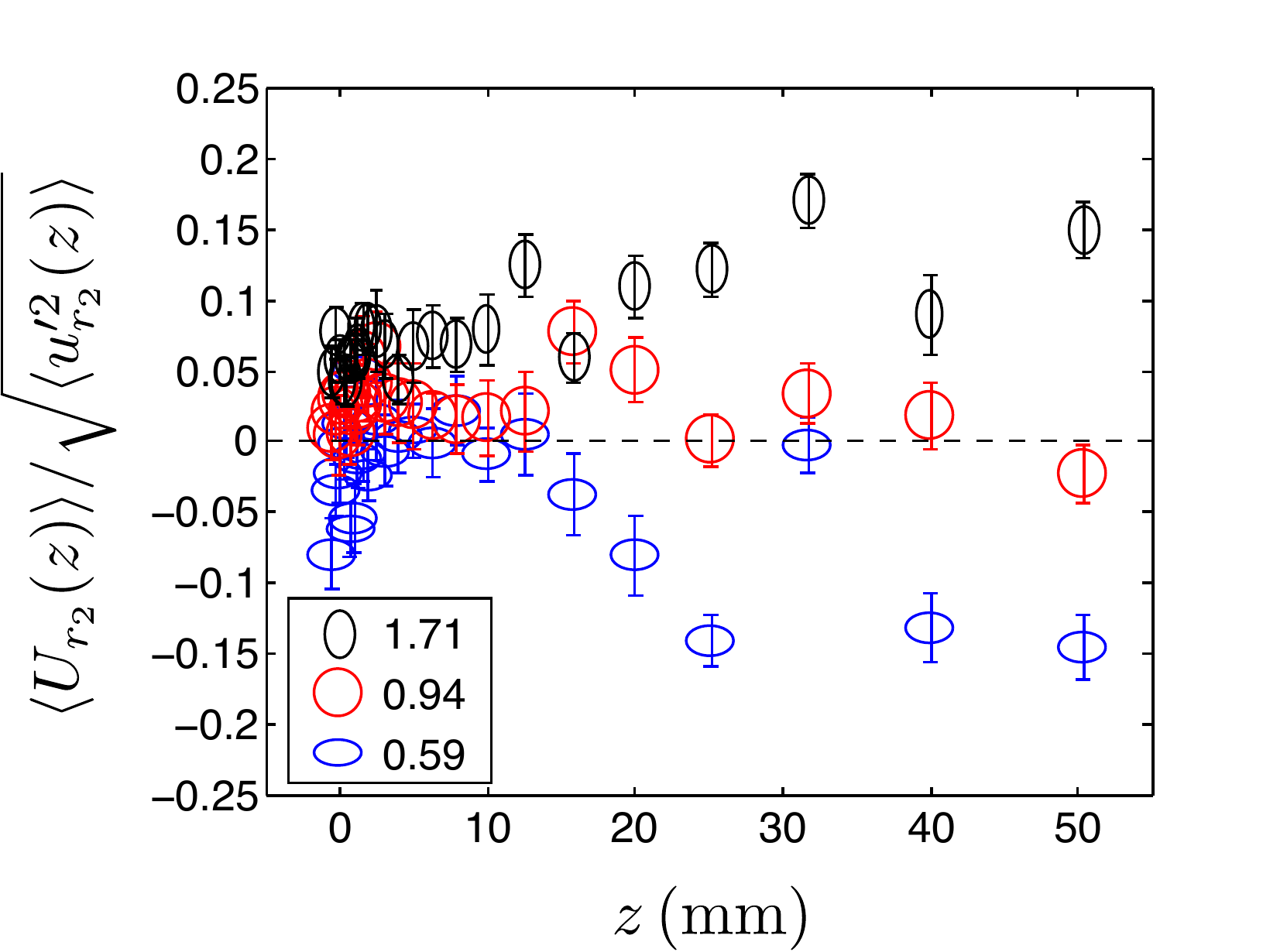}}
\subfigure[]{
\includegraphics[scale=0.6]{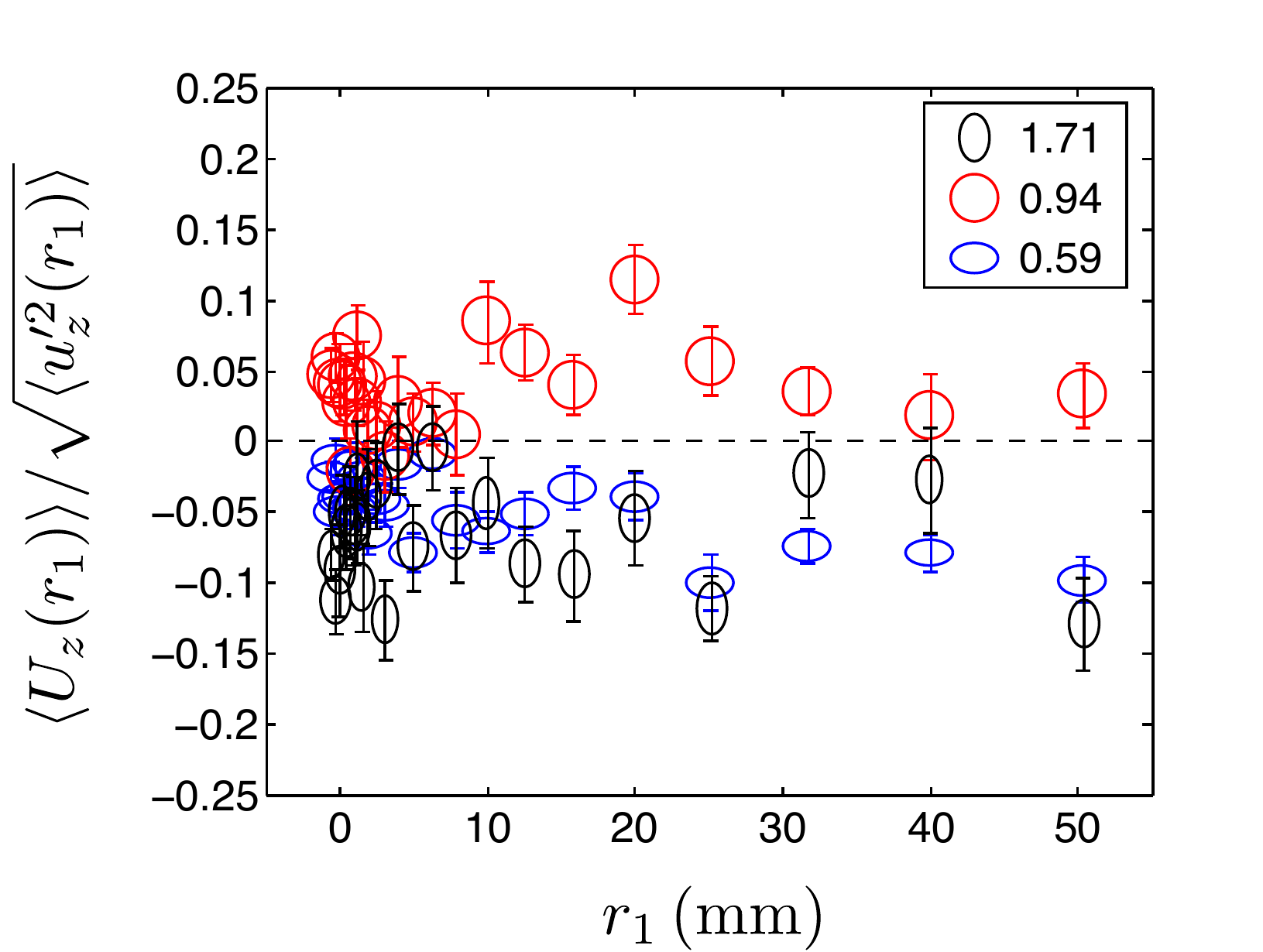}}
\caption{The mean flow as a fraction of the fluctuating velocity
(a) $\langle U_{r_2} (0,0,z) \rangle / 
\sqrt{\langle u^{\prime 2}_{r_2} (0,0,z) \rangle}$, and 
(b) $\langle U_{z} (r_1,0,0) \rangle /
\sqrt{\langle u^{\prime 2}_{z} (r_1,0,0) \rangle}$ for 
axial-to-radial velocity fluctuation ratios ranging from
$0.59$ to $1.71$.  
The values in the legend are the anisotropy measured 
at the centre of the soccer ball.  }
\label{fig:mean_to_fluc_homogeneity}
\end{center}
\end{figure}

Because anisotropy can manifest itself not only in the fluctuations, but also
through spatial variation of the mean velocity, we show in
figure~\ref{fig:mean_to_fluc_homogeneity}(a)  and (b) the ratio of
the mean flow velocity to the velocity fluctuations, $\langle U_{r_2} (0,0,z)
\rangle / \langle u^{\prime 2}_{r_2} (0,0,z) \rangle^{1/2}$ and
$\langle U_{z} (r_1,0,0) \rangle / \langle u^{\prime 2}_{z} (r_1,0,0)
\rangle^{1/2}$.  
It is difficult to predict a trend in the data, but may indicate
that the actual center of the turbulence was displaced 
from the center of our coordinate system.  
We concluded that the variation in the mean velocity was
negligible because it was typically less than $10$\% of the
fluctuations.  
Only at the extreme values of the anisotropy was the mean velocity as
much as $15$\% of the fluctuations.  
This may have been due to the sensitivity of the turbulence to small
differences between the loudspeakers in this range of anisotropies, 
as discussed in section~\ref{sec:anisotropy}.  

\section{Reflectional symmetry}
\label{sec:reflection}

\begin{figure}
\begin{center}
\subfigure[]{
\includegraphics[scale=0.6]{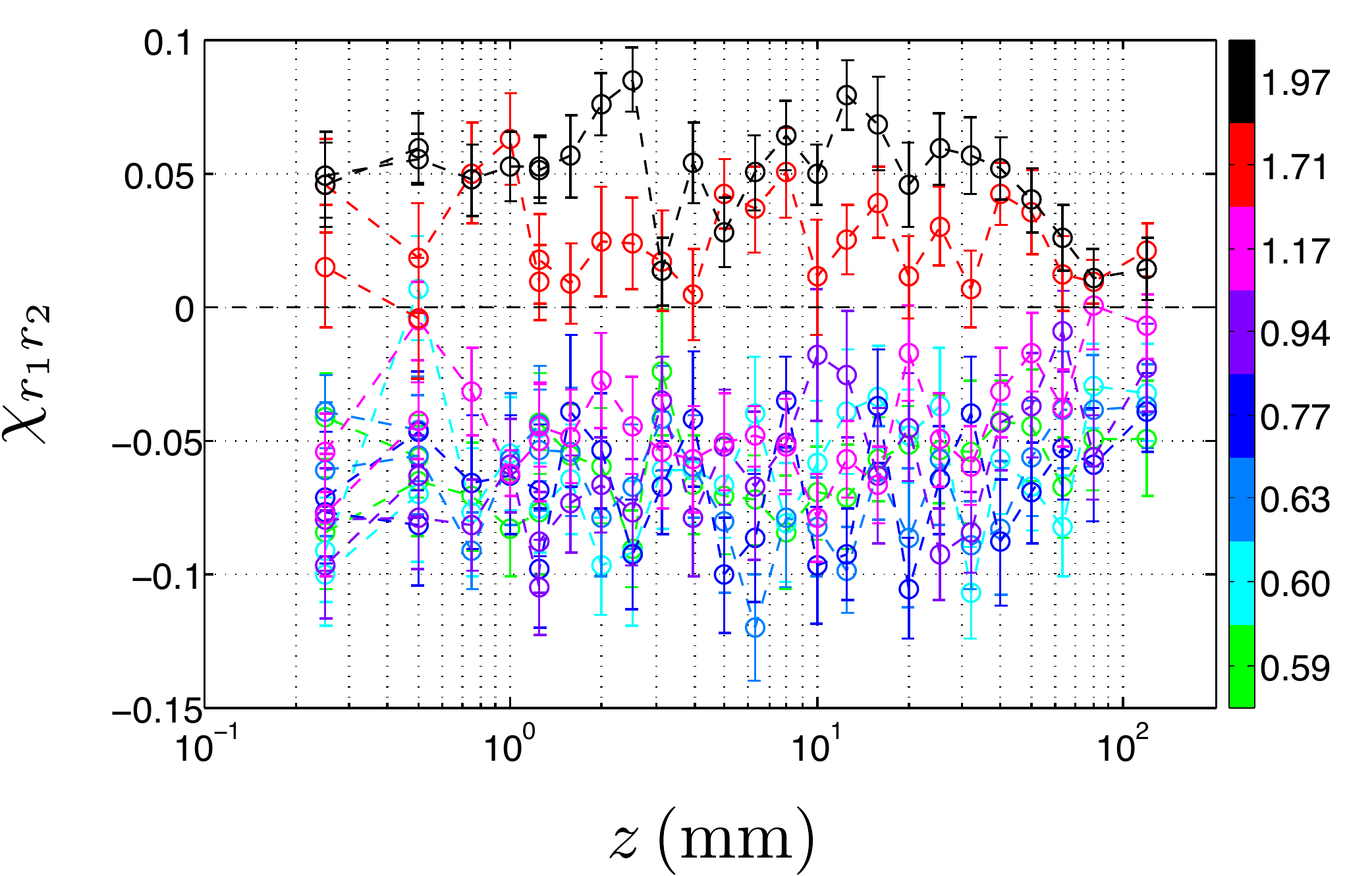}}
\subfigure[]{
\includegraphics[scale=0.6]{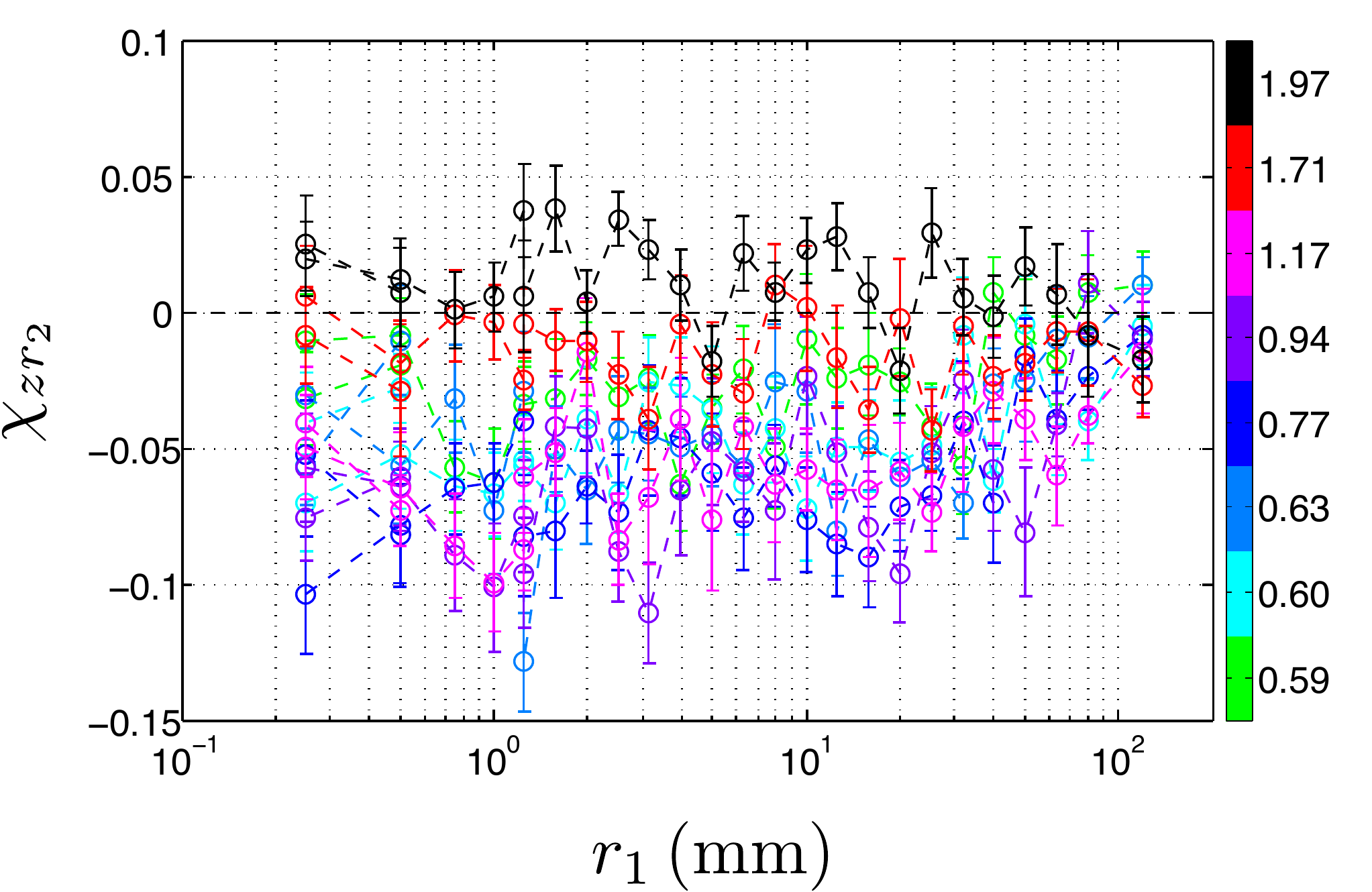}}
\caption{The cross correlations of orthogonal velocity components at various distances from the center of the soccer ball (a) along the axis of symmetry, and (b) along an axis perpendicular to the axis of symmetry of the forcing.  
The values in the color bar are the anisotropy measured at the center of 
the soccer ball for various values of the anisotropy of the forcing.}
\label{fig:crosscorr}
\end{center}
\end{figure}

As described in section~\ref{sec:anisotropic_turbulence}, 
reflectional symmetry in a plane containing the axis of 
symmetry implies that \cite[e.g.][]{lindborg:1995}
\begin{equation}
\chi_{r_1 r_2} = 
\dfrac{\langle u^{\prime}_{r_1} (0,0,z) \, u^{\prime}_{r_2} (0,0,0) \rangle}%
{\sqrt{\langle u^{\prime 2}_{r_1} (0,0,z) \rangle \langle u^{\prime 2}_{r_2} (0,0,0) \rangle}} = 0 \,.
\end{equation}
In addition, reflectional symmetry in a plane normal to the the 
axis of symmetry implies that
\begin{equation}
\chi_{z r_2} = 
\dfrac{\langle u^{\prime}_{z} (r_1,0,0) \, u^{\prime}_{r_2} (0,0,0) \rangle}%
{\sqrt{\langle u^{\prime 2}_{z} (r_1,0,0) \rangle \langle u^{\prime 2}_{r_2} (0,0,0) \rangle}} = 0 \,.
\end{equation}
The measurements of these correlations are shown in 
figure~\ref{fig:crosscorr}.  
The shape of the curves are difficult to interpret, but may suggest 
that the symmetry planes of the turbulence were slightly tilted from 
the plane of measurement.  
In both plots, the error bars were the standard errors of the 
measurements obtained by truncating the original data sets into 
$20$ equal parts.  
These errors were always less than $\pm 3 \%$.  
We do not believe that the trend and the scatter in the data were
significant enough to be the basis of a theory.  
Because the variation in the cross correlations was always less than 
$10\%$ (the mean was no more than $5\%$), we concluded that the 
turbulence possessed remarkable reflectional symmetries

\section{Reynolds shear stresses}
\label{sec:reynolds_shear_stress}

Two observations of turbulent shear flows may be useful.  
First, results obtained in the strongly sheared part of the 
boundary layer by \cite{saddoughi:1994} 
indicated that the width of the Kolmogorov scaling range
of the second order transverse structure function is 
significantly shorter than that of the longitudinal function 
(see figure~\ref{fig:saddoughi1994}).  
In an overview of turbulence spectra obtained in 
various flows, \cite{sreenivasan:1995} noted that shear-flow
turbulence exhibits inertial range scaling in the transverse spectrum 
only at values of $R_{\lambda}$ larger than $10^3$.  
A later study by \cite{noullez:1997} lowered this threshold to $500$.  
Second, there is evidence from experiments 
\cite[e.g.][]{stolovitzky:1993, shen:2002} and numerical simulations 
\cite[e.g.][]{benzi:1996} that the presence of a strong 
shear at low Reynolds numbers may destroy the extended 
self-similarity method (ESS) introduced by \cite{benzi:1993} 
(see section~\ref{sec:higher_order_sf}).  
We surmise from these observations that a delayed
emergence of a scaling region can be avoided by
eliminating shear from the turbulence.  
The process of elimination entails creating a uniform mean profile,
$\langle U \rangle$, and minimizing the Reynolds shear stresses,
$\langle u \, v \rangle$.  
The total shear stress being\footnote{It is customary to use 
$U$, $V$, and $W$ to denote the mean velocities in the $x$, $y$, 
and $z$ directions, respectively; and $u$, $v$, and $w$ their 
corresponding RMS of the velocity fluctuations in the $x$, $y$, 
and $z$ directions.}
\begin{equation}
\tau (y) = \rho \, \nu \dfrac{\partial \langle U \rangle}{\partial y} 
- \rho \langle u \, v \rangle \,,
\end{equation}
by having a uniform mean profile, the gradient term, 
$\partial \langle U \rangle / \partial y$, can be kept small; 
and if the Reynolds shear stress, $\langle u \, v \rangle$, is 
small too, the total shear stress is then negligible.  

It was hoped that randomizing the forcing would suppress
the growth of the Reynolds shear stress term, $\langle u \, v \rangle$.  
We gauged the Reynolds stresses by measuring the normalized 
correlations between orthogonal velocity components, 
\begin{equation}
\tilde{\tau}_{r_1 r_2} = 
\dfrac{\langle u^{\prime}_{r_1} (0,0,z) \, u^{\prime}_{r_2} (0,0,z) \rangle}%
{\sqrt{\langle u^{\prime 2}_{r_1} (0,0,z) \rangle \langle u^{\prime 2}_{r_2} (0,0,z) \rangle}} \,,
\end{equation}
and 
\begin{equation}
\tilde{\tau}_{z r_2} = 
\dfrac{\langle u^{\prime}_{z} (r_1,0,0) \, u^{\prime}_{r_2} (r_1,0,0) \rangle}%
{\sqrt{\langle u^{\prime 2}_{z} (r_1,0,0) \rangle \langle u^{\prime 2}_{r_2} (r_1,0,0) \rangle}} \,,
\end{equation}
at various locations, $z$ and $r_1$, along the axial and radial axes.  
Figure~\ref{fig:crosscorr}(a) and (b) show tests at different 
levels of anisotropy.  
\begin{figure}
\begin{center}
\subfigure[]{
\includegraphics[scale=0.6]{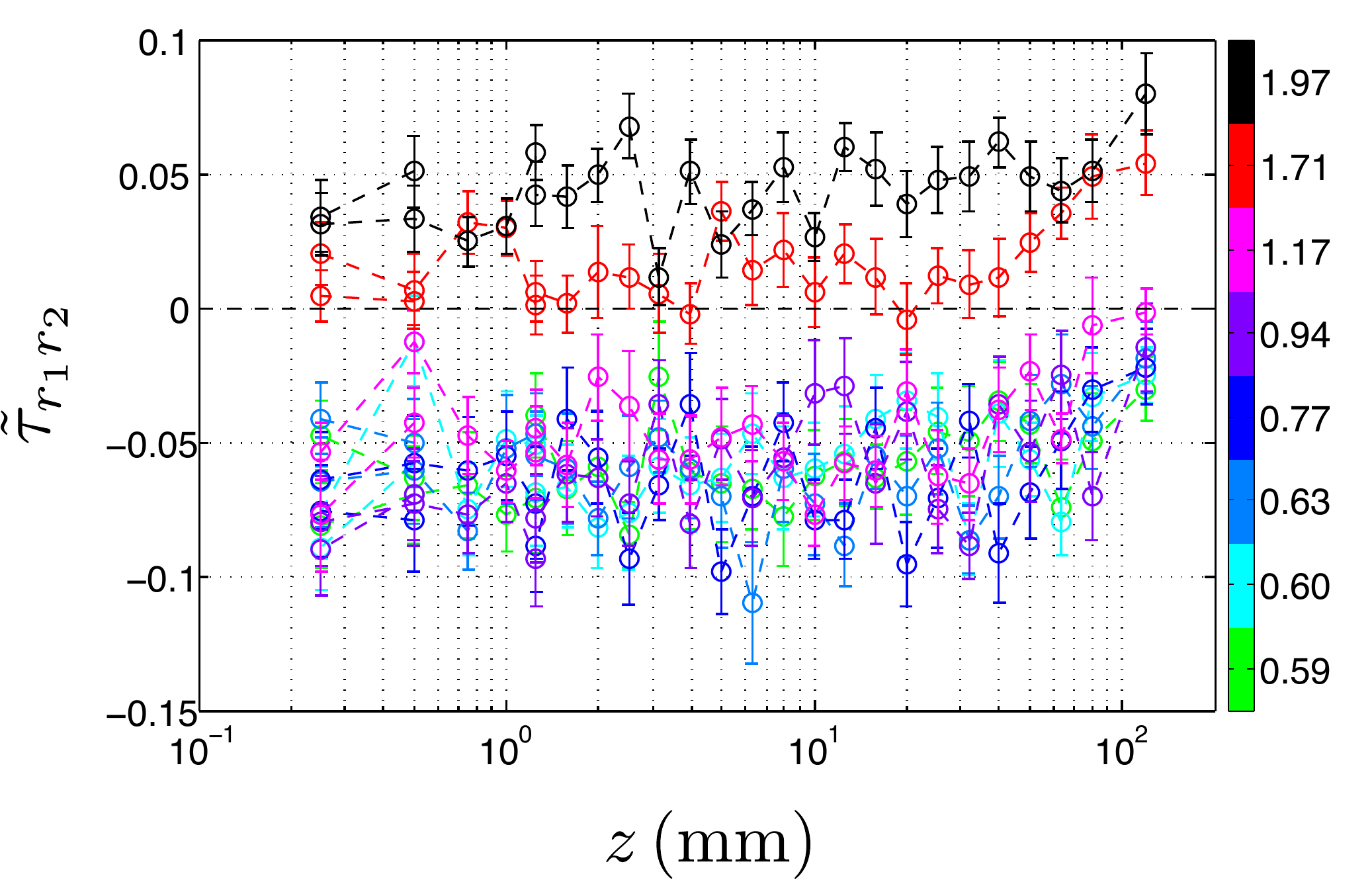}}
\subfigure[]{
\includegraphics[scale=0.6]{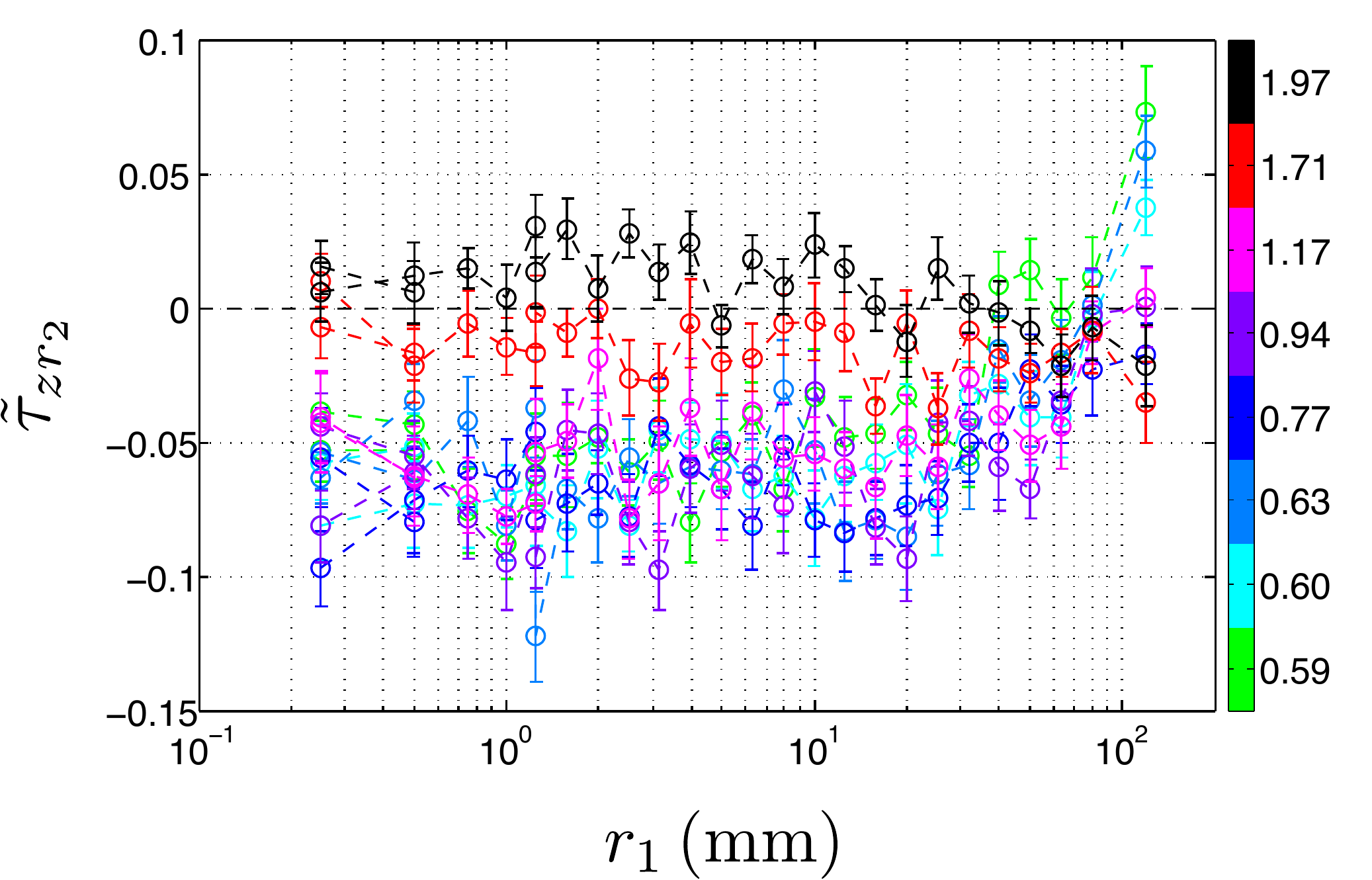}}
\caption{The figure shows the normalized Reynolds stress of orthogonal velocity components at various locations (a) along the axis of symmetry 
of the forcing, and (b) perpendicular to it.  
The values in the color bar are the anisotropy measured at the center 
of the soccer ball for various values of the anisotropy of the forcing.}
\label{fig:reynolds_stress}
\end{center}
\end{figure}
The shear stresses were always closer to zero than $\pm 0.1$ over the 
entire range of distances from the center of the soccer ball; 
\cite{shen:2000} and \cite{shen:2002} have measured the
shear stress in sheared turbulence and found that the value is
about $-0.4$ and is constant throughout the flow.  
We concluded that the turbulence is essentially shearless because
the mean shear stresses were less than $7\%$ of the corresponding 
kinetic energies, $u^{\prime}_{r_1} \, u^{\prime}_{r_2}$ and 
$u^{\prime}_{z} \, u^{\prime}_{r_2}$, respectively.  
The present forcing randomization scheme is a simple method that 
gave remarkably satisfactory shearless turbulence.  
We believe the shear can be reduced further, for example, 
by covering the nozzle with specially designed passive grid
with solidity\footnote{The solidity of a grid, $\sigma$, is defined 
as the projected solid area per unit total area.  
For a biplane grid with square mesh and round rods, the solidity is 
$\sigma = \tfrac{d}{M} \, \big(2 - \tfrac{d}{M}\big)$, where $d$ is 
the rod diameter and $M$ is the distance between rod centerlines, 
or mesh size.  The popular grid solidity $\sigma=0.34$ is a compromise 
between low turbulence intensity, $u^{\prime}/\langle U \rangle$, 
and high flow stability.} $\sigma=0.34$.  
For extreme values of solidity, $\sigma > 0.4$, the resulting turbulence
behind the grid is unstable and non-uniform\footnote{The critical 
solidity $\sigma = 0.4$ limits the turbulence intensity behind the grid, 
since the turbulence intensity is proportional to the drag coefficient, 
and that, in turn, is proportional to the solidity.} \cite[][]{corrsin:1963}.  
Useful design guidelines regarding pressure losses across the grid,
the uniformity of the turbulence behind the grid, and its stability
are collected in \cite{roach:1987}.  

\section{Conclusions}
\label{sec:flow_conclusions}

Our apparatus is unique among other loudspeaker-driven 
flow chambers because it permits a systematic exploration 
of anisotropic turbulence.  
We have shown that we are able to control the large-scale anisotropy
of the turbulence, as summarized by the ratio between axial and radial
components of the root-mean-square velocity, to any value between
$0.6$ and $2.3$, with $1$ being the value for isotropic turbulence.  
Table~\ref{table:flowstat} summarizes the values of various parameters 
for the flows.  
Because the anisotropy, the mean flow, and the strength of the
fluctuations are approximately constant within $50$~mm of the center
of the soccer ball, we call this the region of homogeneity.
\begin{table}
\begin{center}
\begin{tabular}{*{9}{c}}
\toprule
\addlinespace[8pt]
$A$ & $0.33$ & $0.4$ & $0.5$ & $0.8$ & $1.0$ & $1.25$ & $2.0$ & $2.5$ \\
\addlinespace[5pt]
\midrule
\addlinespace[8pt]
$u^{\prime}_z$ (m~s$^{-1}$) & 
$0.74$ & $0.75$ & $0.79$ & $0.91$ & $1.08$ & $1.21$ & $1.42$ & $1.47$ \\
\addlinespace[8pt]
$u^{\prime}_{r_2}$ (m~s$^{-1}$) &
$1.26$ & $1.24$ & $1.27$ & $1.19$ & $1.15$ & $1.03$ & $0.83$ & $0.75$ \\
\addlinespace[8pt]
$u^{\prime}_{z} / u^{\prime}_{r_2}$ & 
$0.59$ & $0.60$ & $0.63$ & $0.77$ & $0.94$ & $1.17$ & $1.71$ & $1.97$ \\
\addlinespace[8pt]
$K$ (m$^2$~s$^{-2}$) &
$1.87$ & $1.83$ & $1.92$ & $1.82$ & $1.91$ & $1.79$ & $1.69$ & $1.65$ \\
\addlinespace[8pt]
$|U_{z} / U_{r_2}|$ & 
$0.31$ & $0.16$ & $0.13$ & $0.15$ & $4.04$ & $2.57$ & $7.58$ & $23.9$ \\
\addlinespace[8pt]
$|U|$ (m~s$^{-1}$) & 
$0.17$ & $0.18$ & $0.13$ & $0.10$ & $0.05$ & $0.07$ & $0.14$ & $0.24$ \\
\addlinespace[8pt]
$\tilde{\tau}_{r_1 r_2}$ &
$-0.07$ & $-0.06$ & $-0.06$ & $-0.07$ & $-0.07$ & $-0.05$ & $0.01$ & $0.04$ \\
\addlinespace[8pt]
$\tilde{\tau}_{z r_2}$ &
$-0.06$ & $-0.07$ & $-0.06$ & $-0.07$ & $-0.07$ & $-0.06$ & $-0.01$ & $0.02$ \\
\addlinespace[8pt]
$\epsilon$ (m$^2$~s$^{-3}$) &
$5.41$ & $4.94$ & $5.43$ & $6.00$ & $6.71$ & $6.98$ & $6.49$ & $6.45$ \\
\addlinespace[8pt]
$\eta$ ($\micro$m) &
$163$ & $167$ & $163$ & $159$ & $155$ & $153$ & $156$ & $156$ \\
\addlinespace[8pt]
$\tau_{\eta}$ (ms) &
$1.7$ & $1.8$ & $1.7$ & $1.6$ & $1.5$ & $1.5$ & $1.6$ & $1.6$ \\
\addlinespace[8pt]
$\lambda$ (mm) & 
$7.4$ & $7.6$ & $7.5$ & $6.9$ & $6.7$ & $6.3$ & $6.4$ & $6.3$ \\
\addlinespace[8pt]
$R_{\lambda}$ &
$525$ & $536$ & $538$ & $485$ & $480$ & $443$ & $434$ & $423$ \\
\addlinespace[5pt]
\bottomrule
\end{tabular}
\caption{The table shows the turbulence 
  parameters, for loudspeaker RMS amplitude ratios ($A$) ranging from 
  0.33 to 2.5.  
  $K$ is the turbulent kinetic
  energy, $\frac{1}{2} (u_{z}^{\prime 2}+2 \, u_{r_2}^{\prime 2})$,
  and $|U|$ is $(U_z^2 + 2 \, U_{r_2}^2)^{1/2}$.  
  $\tilde{\tau}_{r_1 r_2}$ and $\tilde{\tau}_{z r_2}$ are the average
  normalized Reynolds stresses measured along and normal to the
  axis of symmetry, respectively (see section~\ref{sec:reynolds_shear_stress}).  
  $\epsilon$ is the energy dissipation rate estimated according to 
  the procedure prescribed in section~\ref{sec:structure_functions}.  
  $\eta$ and $\tau_\eta$ are the Kolmogorov length and time scales,
  respectively.  
  $\lambda$ is the Taylor length (see section~\ref{sec:c2}) and
  $R_{\lambda}$ is the Taylor scale Reynolds number.  
  Data were collected at the center of the soccer ball 
  when the axis of symmetry of the forcing is horizontal.
  Similar results (not shown) were obtained when the axis of symmetry
  is vertical.  }
\label{table:flowstat}
\end{center}
\end{table}

\chapter{Universality in the structure functions}
\label{chap:universality}

In this chapter, we examine how the inertial scales of 
the turbulence generated by the apparatus introduced in 
chapter~\ref{chap:apparatus} are influenced by the anisotropy 
of the forcing, through the measurements of the second order 
moment of velocity increments.  
We measured the transverse structure functions, $D_{r_2 r_2} (z)
= \langle (u^{\prime}_{r_2} (0,0,z) - u^{\prime}_{r_2} (0,0,0))^2 \rangle$
and $D_{zz} (r_1) = \langle (u^{\prime}_{z} (r_1,0,0) - 
u^{\prime}_{z} (0,0,0))^2 \rangle$, with the separation vector lying 
nearly along the axis of symmetry or perpendicular to it, 
and show that, in an inertial range that spanned nearly a decade,
the Kolmogorov constant and the inertial range scaling exponents
were scalar valued in unsheared anisotropic turbulence, within the
experimental error of 10\%.

\section{Isotropy of the structure functions}
\label{sec:structure_functions}

\begin{figure}
\begin{center}
\subfigure[]{
\includegraphics[scale=0.6]{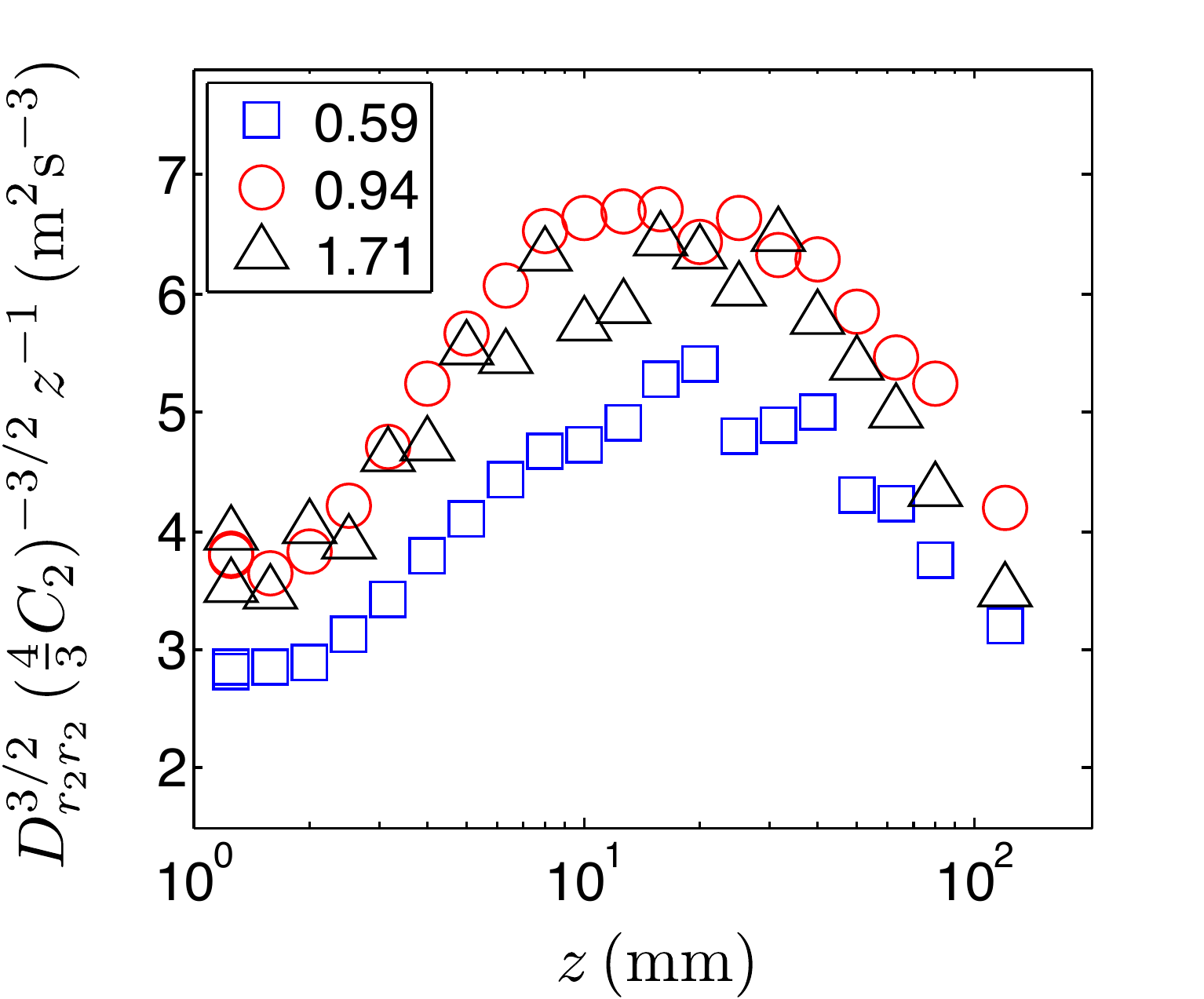}}
\subfigure[]{
\includegraphics[scale=0.6]{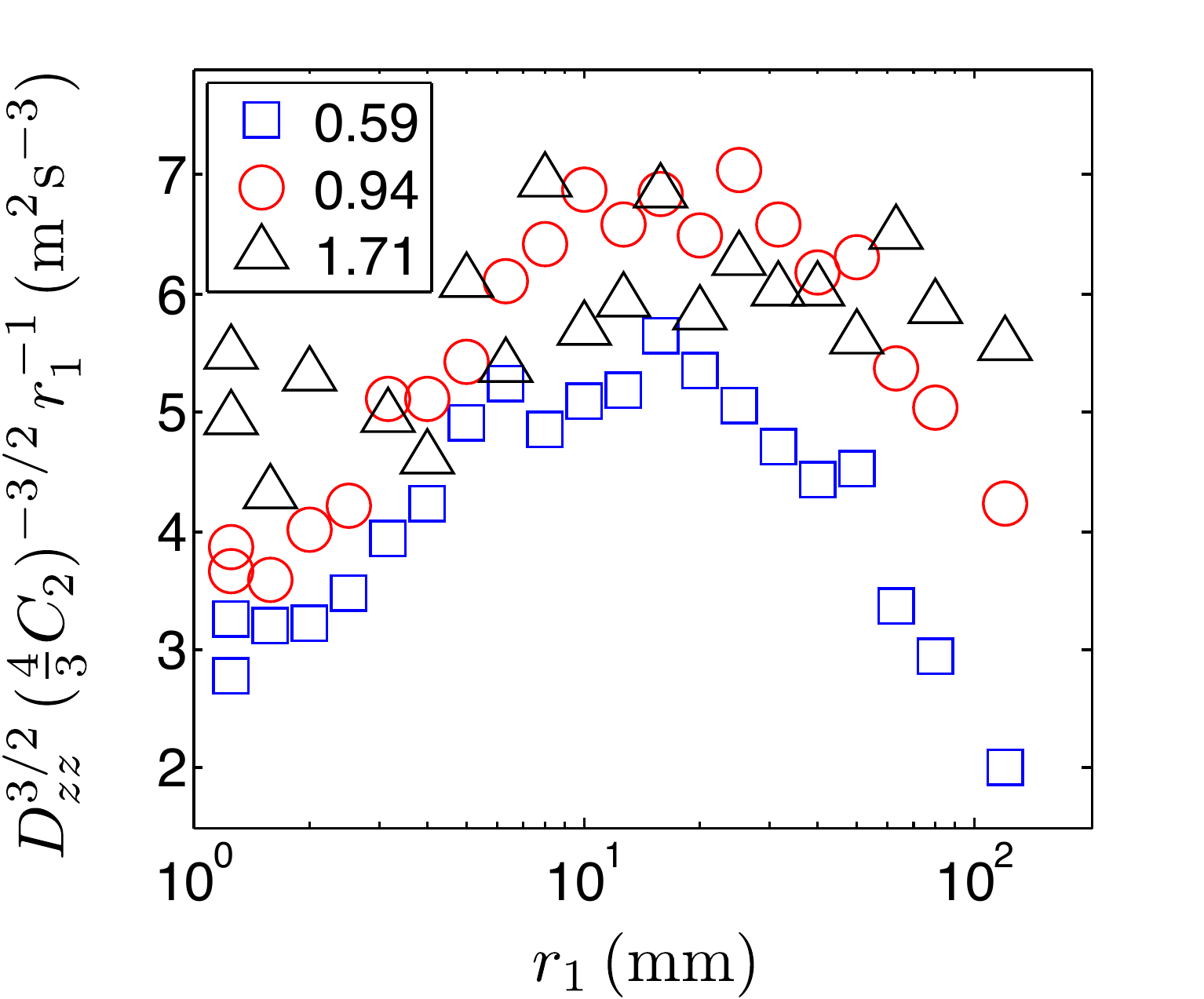}}
\caption{The figure shows the (a) axial structure functions, 
$D_{r_2 r_2} (z)$, and (b) radial structure functions, $D_{zz} (r_1)$.  
In both cases, the structure functions are divided by the inertial
range scaling predicted by Kolmogorov.  
The error bars are smaller than the symbols.  
The values in the legends are the anisotropy measured at the 
center of the soccer ball for various values of the anisotropy 
 of the forcing.}
\label{fig:comp_structure_func}
\end{center}
\end{figure}

To determine the inertial range of  the turbulence for different
anisotropies, we plot in figures~\ref{fig:comp_structure_func}(a)
and (b) the measure for energy per unit time and mass, 
$(D_{r_2 r_2} (z))^{3/2} \, (4 \, C_2 / 3)^{-3/2} \, z^{-1}$
and $(D_{zz}(r_1))^{3/2} \, (4\, C_2 / 3)^{-3/2} \, r_1^{-1}$, 
as functions of $z$ and $r_1$, respectively.  
No noise corrections were applied to the structure functions.  
The value of $C_2$ was taken to be $2.1$ \cite[][]{sreenivasan:1995}.  
We found that the inertial range spanned approximately one decade, 
as indicated by the extent of flat region in 
figures~\ref{fig:comp_structure_func}(a) and (b).  

\begin{figure}
\begin{center}
\includegraphics[scale=0.6,type=pdf,ext=.pdf,read=.pdf]{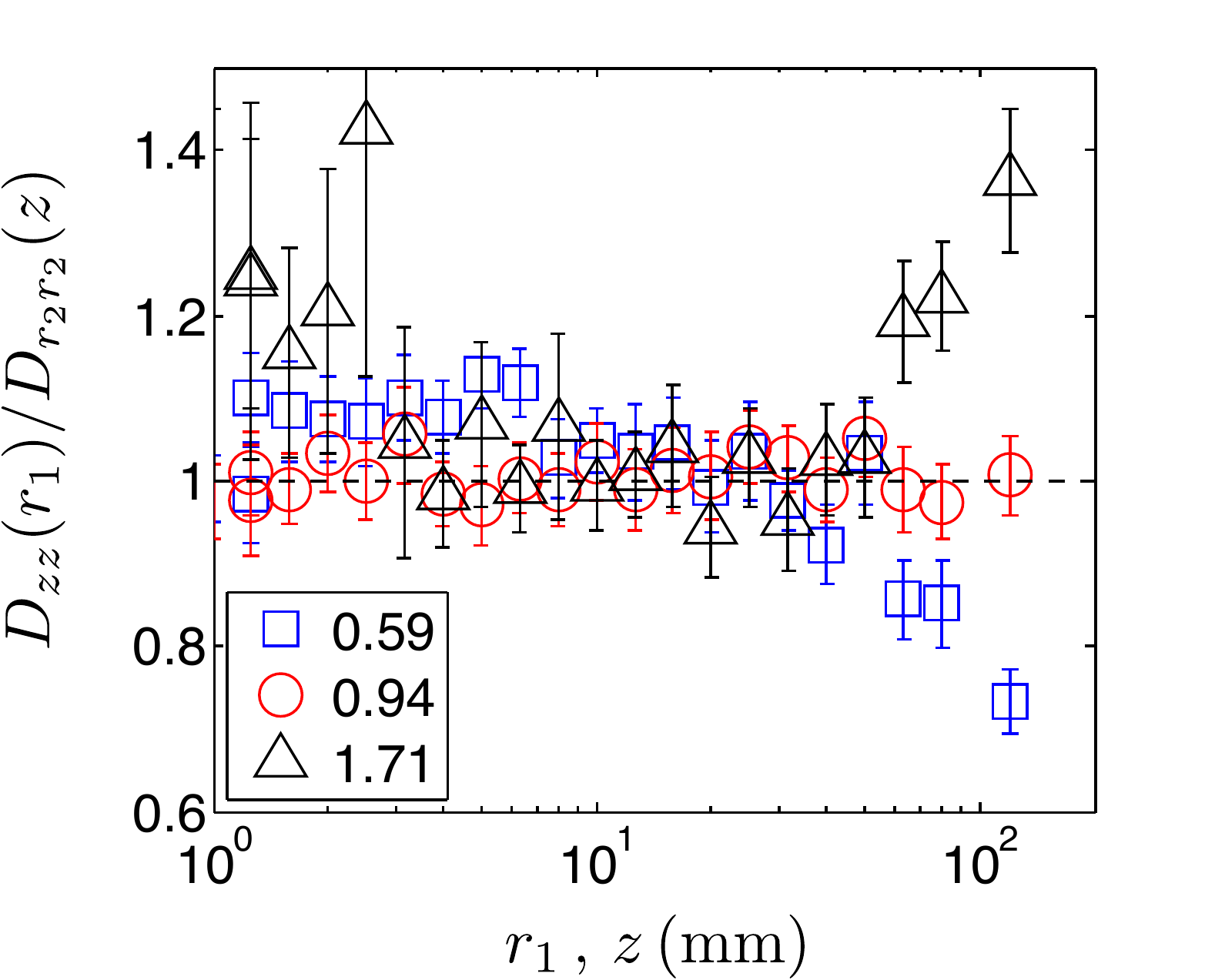}
\caption{The ratios between structure functions in different 
directions but in the same flow, $D_{zz} (r) / D_{r_2 r_2} (z)$, 
approach the isotropic value of one when the separation 
distance, $r_1 = z$, decreases.  
Results for data taken at other values of the large-scale 
anisotropy are consistent with those shown here.  
The values in the legends are the anisotropy measured 
at the center of the soccer ball.}
\label{fig:dzz_drr_ratio_3cases}
\end{center}
\end{figure}

Figure~\ref{fig:dzz_drr_ratio_3cases} shows the scale-dependent
measure of isotropy, namely $D_{zz} (r_1) / D_{r_2 r_2} (z)$, where 
$r_1 = z$.  
In the limit of large separations, the ratio should approach 
$u^{\prime 2}_z / u^{\prime 2}_{r_2}$, since the velocities 
$u^{\prime}_i (\boldsymbol{r})$ and $u^{\prime}_i (0)$ are 
uncorrelated when $\boldsymbol{r}$ is large enough.  
Although we could not resolve this asymptotic limit due to 
experimental limitations, the values of this ratio in anisotropic 
cases did clearly separate from the isotropic value.  
Moving toward small scales, local isotropy requires that
$D_{zz} (r) / D_{r_2 r_2} (z)$ approaches one.  
The figure reveals that within the experimental error we observed
local isotropy for separations smaller than the energy injection
scale.  

\section{Equality of transverse ESS scaling exponents}
\label{sec:scaling_exponents}

\begin{figure}
\begin{center}
\subfigure[]{
\includegraphics[scale=0.6]{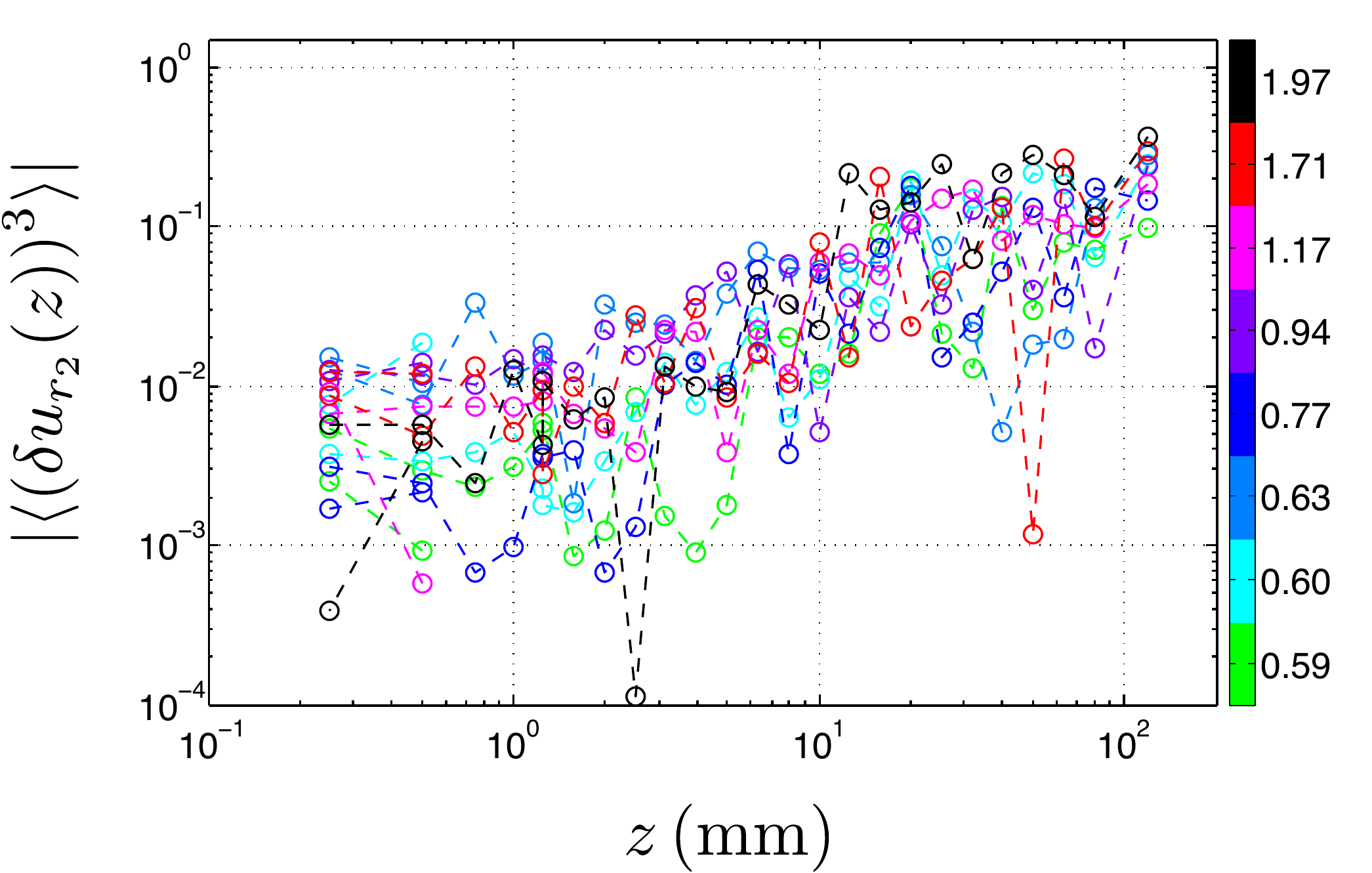}}
\subfigure[]{
\includegraphics[scale=0.6]{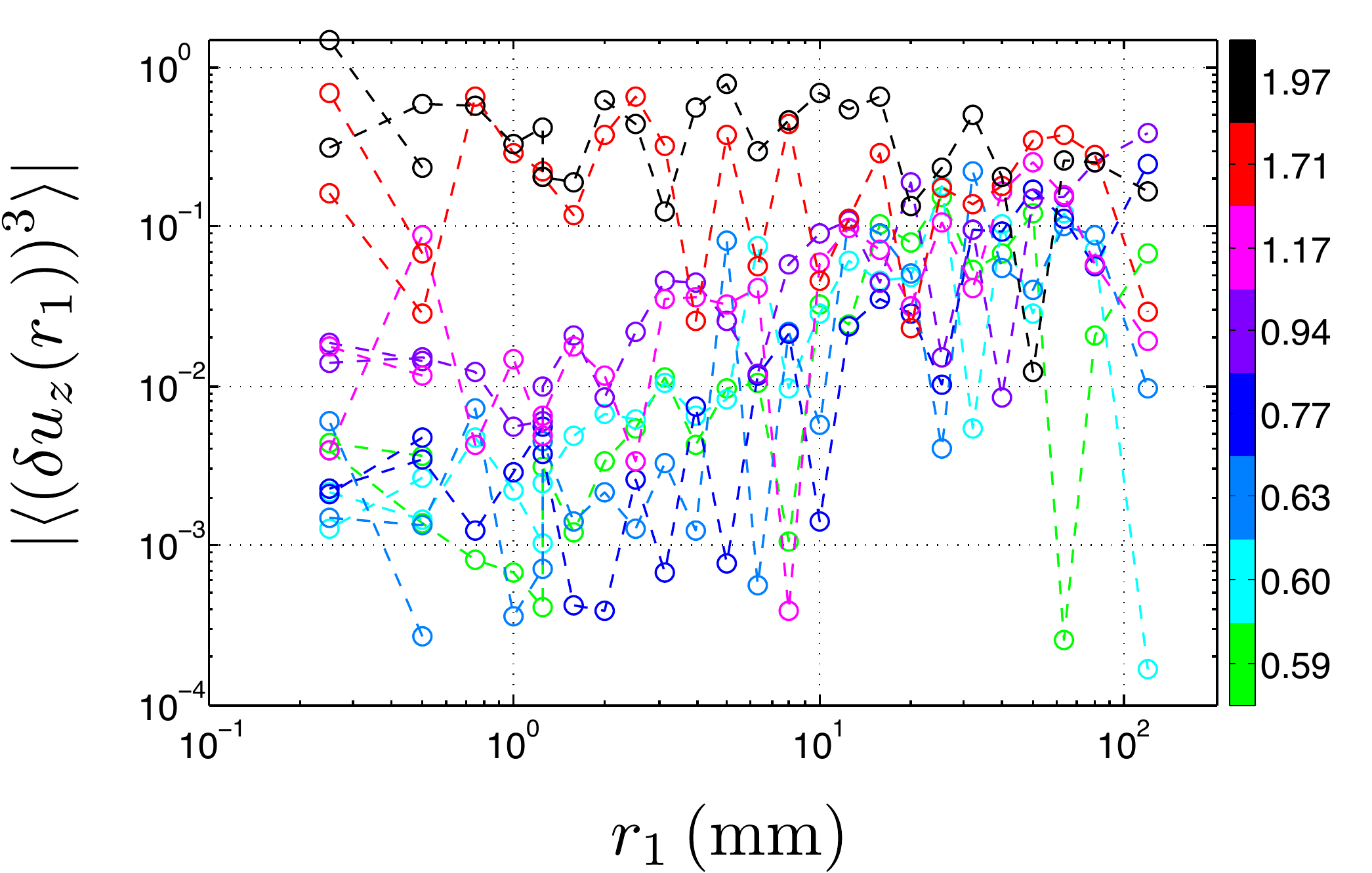}}
\caption{The third-order structure functions with separations in 
the direction (a) along the axis of symmetry, and (b) perpendicular
to the axis of symmetry.
The values in the color bar are the anisotropy measured at the center
of the soccer ball.}
\label{fig:sf_3rd}
\end{center}
\end{figure}

\begin{figure}
\begin{center}
\subfigure[]{
\includegraphics[scale=0.6]{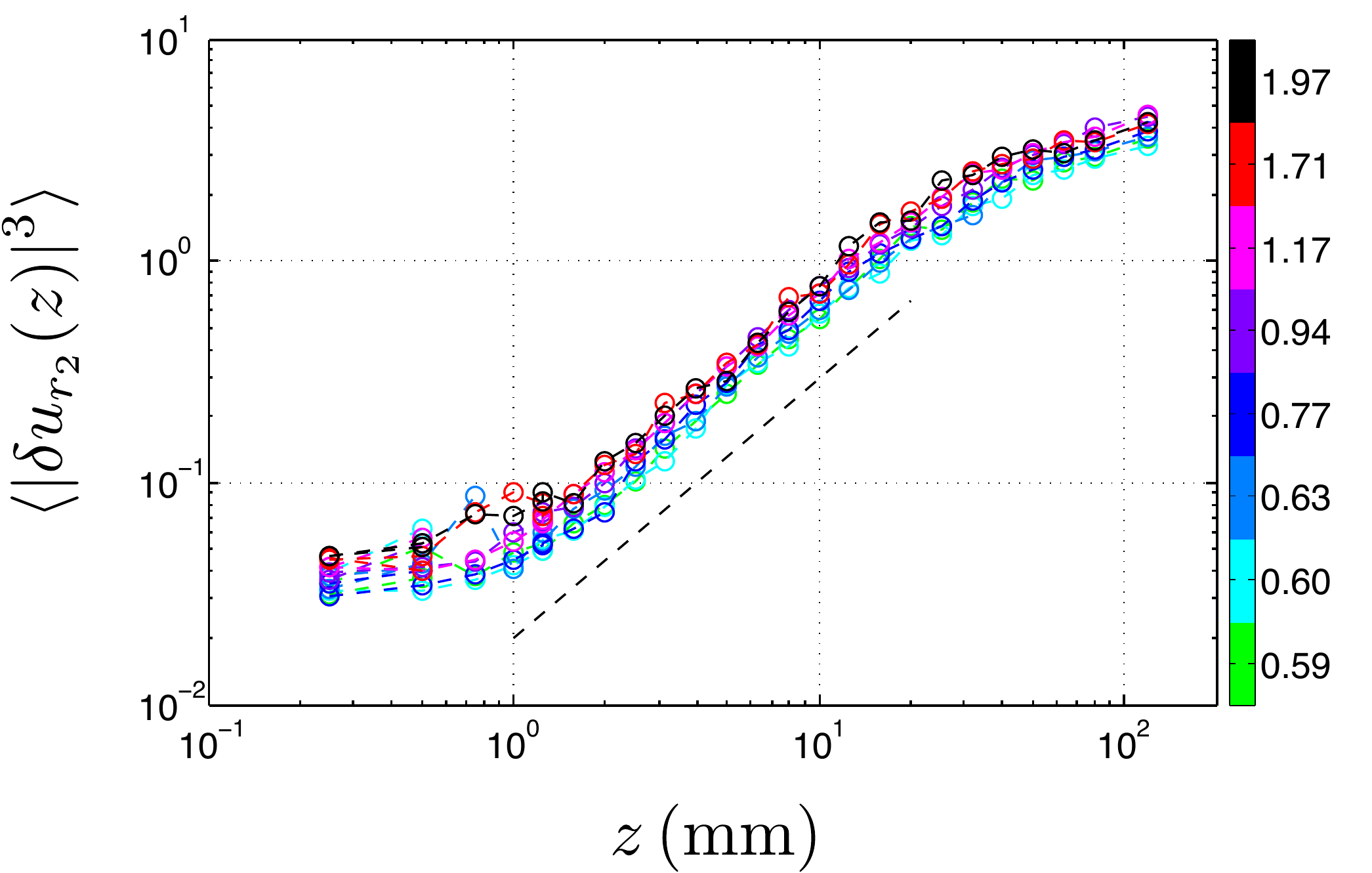}}
\subfigure[]{
\includegraphics[scale=0.6]{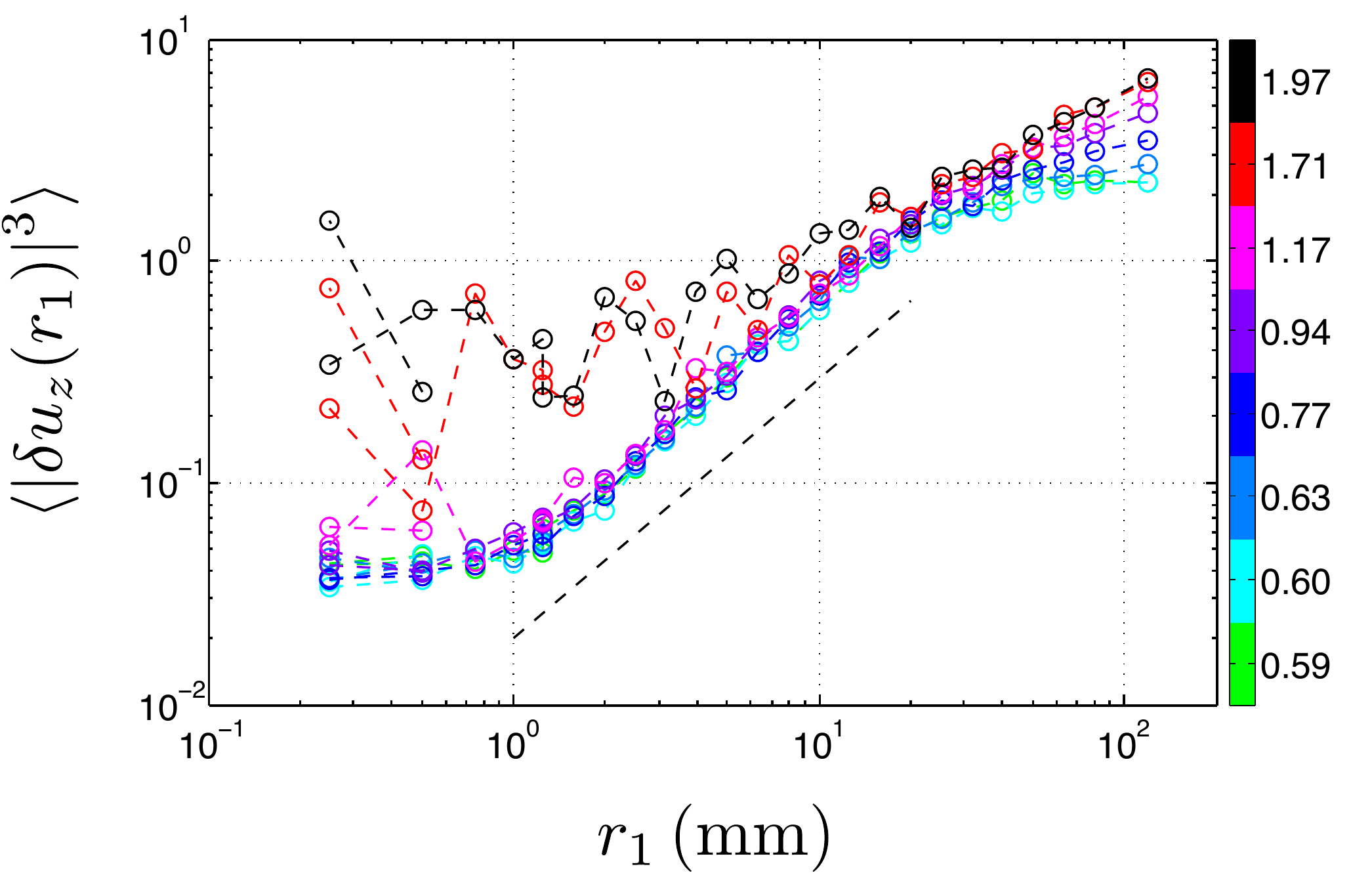}}
\caption{The third-order moments of the absolute values of the velocity increments with separations in the direction (a) along the axis of 
symmetry, and (b) perpendicular to the axis of symmetry.  
The dashed lines are proportional to a power law with exponent $1.17$.  
The values in the color bar are the anisotropy measured at the center
of the soccer ball.}
\label{fig:sf_3ESS}
\end{center}
\end{figure}

Here, we investigate how the scaling exponent measured using 
the ESS method, as described in section~\ref{sec:higher_order_sf}, 
depends on the anisotropy of the forcing.  
We begin with an examination of the third order transverse 
structure function, $\langle (\delta u (x))^3 \rangle$, 
and its ESS variant, $\langle |\delta u (x)|^3 \rangle$.  
Figures~\ref{fig:sf_3rd}(a) and (b) show 
$\langle (\delta u_{r_2} (z))^3 \rangle$ and 
$\langle (\delta u_{z} (r_1))^3 \rangle$ as functions of 
$z$ and $r_1$, respectively.  
No noise corrections were applied to these structure functions.  
It can be seen that $\langle (\delta u_{r_2} (z))^3 \rangle$ 
and $\langle (\delta u_{z} (r_1))^3 \rangle$ were nearly zero 
and varied unsystematically with separations, as all
odd-order moments should if the turbulence is locally
isotropic \cite[][]{shen:2002}.  
This contrasts the predictions in shear flows, where simple
Kolmogorov-type argument assuming that 
$\langle (\delta u (x))^3 \rangle$ is proportional to the shear yields
$\langle (\delta u (x))^3 \rangle \sim x^{4/3}$ \cite[][]{lumley:1967}.  
The absence of a four-thirds scaling reflects the 
absence of shear in our flow.  
Figures~\ref{fig:sf_3ESS}(a) and (b) show the scaling of 
$\langle |\delta u_{r_2} (z)|^3 \rangle$ and 
$\langle |\delta u_{z} (r_1)|^3 \rangle$ with
separations, $z$ and $r_1$.  
It can be seen that the scaling range spans approximately
a decade from 2~mm to 20~mm.  
For reference, it might be noted that the empirical scaling for 
$\langle |\delta u (x)|^3 \rangle$ is 
$\langle |\delta u (x)|^3 \rangle \sim x^{1.17}$ for the 
present measurement, for which the value $1.17$ has been 
obtained by averaging the exponents obtained from straight-line 
fits, $\log \langle |\delta u (x)|^3|\rangle \sim \log x$, to the 
individual curves.  

\begin{figure}
\begin{center}
\includegraphics[scale=0.9,type=pdf,ext=.pdf,read=.pdf]{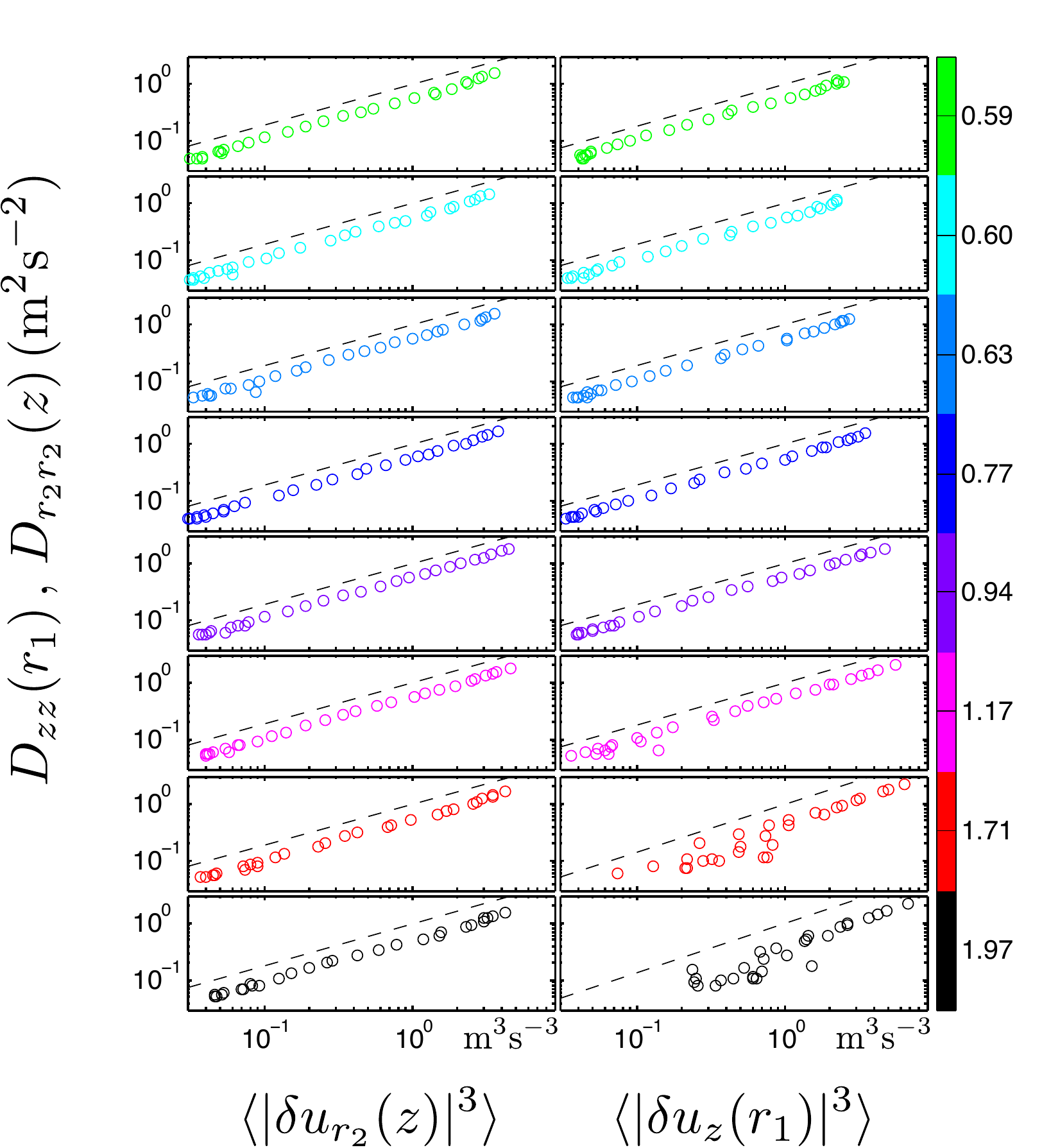}
\caption{The extended-self-similarity plots for the second-order
transverse structure functions.  
The dashed lines are proportional to the power laws obtained from 
least squares fits to data.  
The values in the color bar are the anisotropy measured at the center 
of the soccer ball for various values of the anisotropy of the forcing.  
The error bars are smaller than the symbols.}
\label{fig:sf_ESS}
\end{center}
\end{figure}

\begin{figure}
\begin{center}
\includegraphics[scale=0.7,type=pdf,ext=.pdf,read=.pdf]{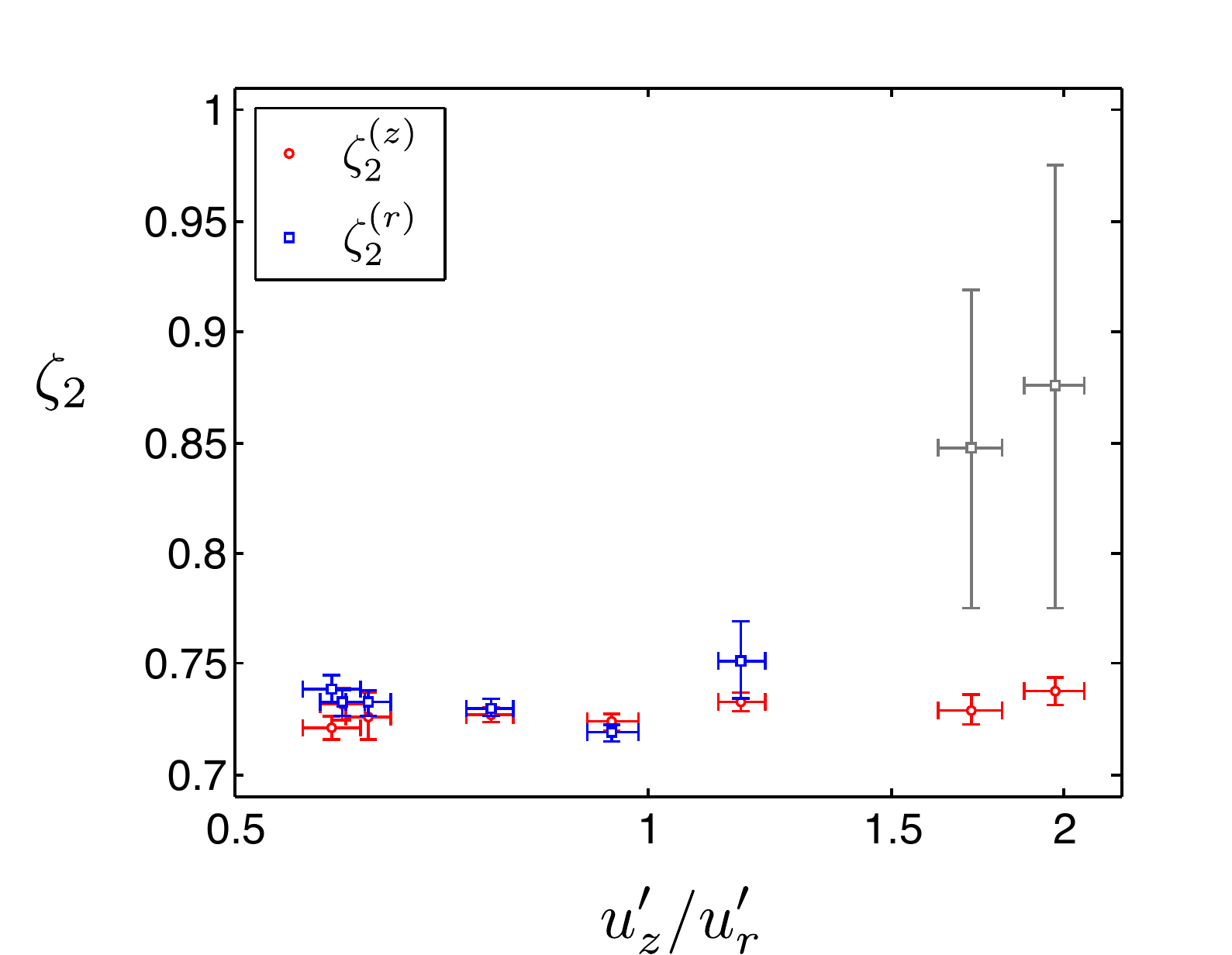}
\caption{The scaling exponents for the transverse
structure functions measured using the extended-self-similarity
method.  
The ${\color{red} \ocircle}$ symbols are the scaling exponents
for $D_{r_2 r_2} (z)$ and the ${\color{blue} \square}$ symbols 
are for $D_{zz} (r_1)$.  
The last two points in the radial data set have large uncertainties
due to spatial resolution.  
They are marked in gray to show the general trend of the 
exponents.}
\label{fig:zeta_all}
\end{center}
\end{figure}

We now consider the ESS plots for $D_{r_2 r_2} (z) \sim 
\langle |\delta u_{r_2} (z)|^3 \rangle^{\zeta_2^{(z)}}$ and $D_{zz}(r_1) 
\sim \langle |\delta u_{z} (r_1)|^3 \rangle^{\zeta_2^{(r_1)}}$ in
figure~\ref{fig:sf_ESS}.  
We obtain two sets of exponents, $\zeta_2^{(z)}$ for the 
axial function, $D_{r_2 r_2} (z)$, and $\zeta_2^{(r)}$ for the 
radial function, $D_{zz}(r_1)$, by calculating the slopes obtained 
from straight-line fits, $\log (D (r)) \sim 
\log \langle |\delta u(r)|^3 \rangle$, to the data.  
The exponents,  $\zeta_2^{(z)}$ and $\zeta_2^{(r)}$, are shown 
in figure~\ref{fig:zeta_all}.  
The uncertainty in the exponents was estimated with the standard
error of the slopes in the straight-line fits 
(see equation~\ref{eq:SEb} in appendix~\ref{app:statistical_tools}).  
Table~\ref{table:zeta} lists the numerical values for 
$\zeta_2^{(z)}$ and $\zeta_2^{(r)}$.  
The scatter in the data sets for $D_{zz} (r_1)$ with anisotropy ratios 
$u^{\prime}_z/u^{\prime}_r = 1.71$ and $1.97$ could be due
to a number of reasons.  
It may have been an artifact of the limitation in instrument spatial
resolution, a residual shear in the turbulence, or an insufficiently 
developed turbulent region.  
For these reasons, we have plotted their exponents with gray symbols
in figure~\ref{fig:zeta_all}.  
The average value for the scaling exponents, excluding the last 
two radial data sets, was $0.73 \pm 0.02$, for which the error 
was the the standard deviation of the values of the exponents.  
The value for the mean exponent is largely consistent with 
values measured using similar transverse ESS method reported in 
the literature at comparable Reynolds numbers, see 
table~\ref{table:zeta_literature}.  
\begin{table}
\begin{center}
\begin{tabular}{cccc}
\toprule
\addlinespace[8pt]
$u^{\prime}_z / u^{\prime}_r$ & $\zeta_2^{(r)}$ & $\zeta_2^{(z)}$ &
$\zeta_2^{(r)} / \zeta_2^{(z)}$ \\
\addlinespace[5pt]
\midrule
\addlinespace[8pt]
$0.59\pm 0.03$ & $0.739\pm 0.007$ & $0.721\pm 0.005$ & $1.024\pm 0.017$ \\
$0.60\pm 0.02$ & $0.733\pm 0.007$ & $0.732\pm 0.008$ & $1.001\pm 0.020$ \\
$0.63\pm 0.03$ & $0.733\pm 0.006$ & $0.726\pm 0.011$ & $1.009\pm 0.024$ \\
$0.77\pm 0.03$ & $0.730\pm 0.004$ & $0.727\pm 0.004$ & $1.004\pm 0.011$ \\
$0.94\pm 0.04$ & $0.719\pm 0.004$ & $0.724\pm 0.004$ & $0.993\pm 0.012$ \\
$1.16\pm 0.05$ & $0.752\pm 0.018$ & $0.733\pm 0.005$ & $1.026\pm 0.031$ \\
$1.71\pm 0.10$ & $0.847\pm 0.072$ & $0.730\pm 0.007$ & $1.161\pm 0.111$ \\
$1.98\pm 0.10$ & $0.875\pm 0.101$ & $0.738\pm 0.007$ & $1.186\pm 0.147$ \\
\addlinespace[5pt]
\bottomrule
\end{tabular}
\caption{The numerical data for radial scaling exponents 
($\zeta_2^{(r)}$), axial scaling exponents ($\zeta_2^{(z)}$), 
and the ratio between the two.}
\label{table:zeta}
\end{center}
\end{table}
\begin{table}
\begin{center}
\begin{tabular}{lll}
\toprule
\addlinespace[8pt]
$\zeta_2$ & $R_{\lambda}$ & Source \\
\addlinespace[5pt]
\midrule
\addlinespace[8pt]
$0.70\pm 0.005$ & $37-82$ & \cite{camussi:1996} \\
$0.70 \pm 0.01$ & $365-605$ & \cite{noullez:1997} \\
$0.71$ & $1000$ & \cite{zhou:2001} \\
$0.71-0.74$ & $40-10^4$ & \cite{pearson:2001} \\
$0.73 \pm 0.02$ & $\myrlambda$ & Present work \\
\addlinespace[5pt]
\bottomrule
\end{tabular}
\caption{The value of the scaling exponent of transverse second 
order structure functions, $\zeta_2$, measured using the ESS method 
for various laboratory flows.}
\label{table:zeta_literature}
\end{center}
\end{table}

\begin{figure}
\begin{center}
\includegraphics[scale=0.7,type=pdf,ext=.pdf,read=.pdf]{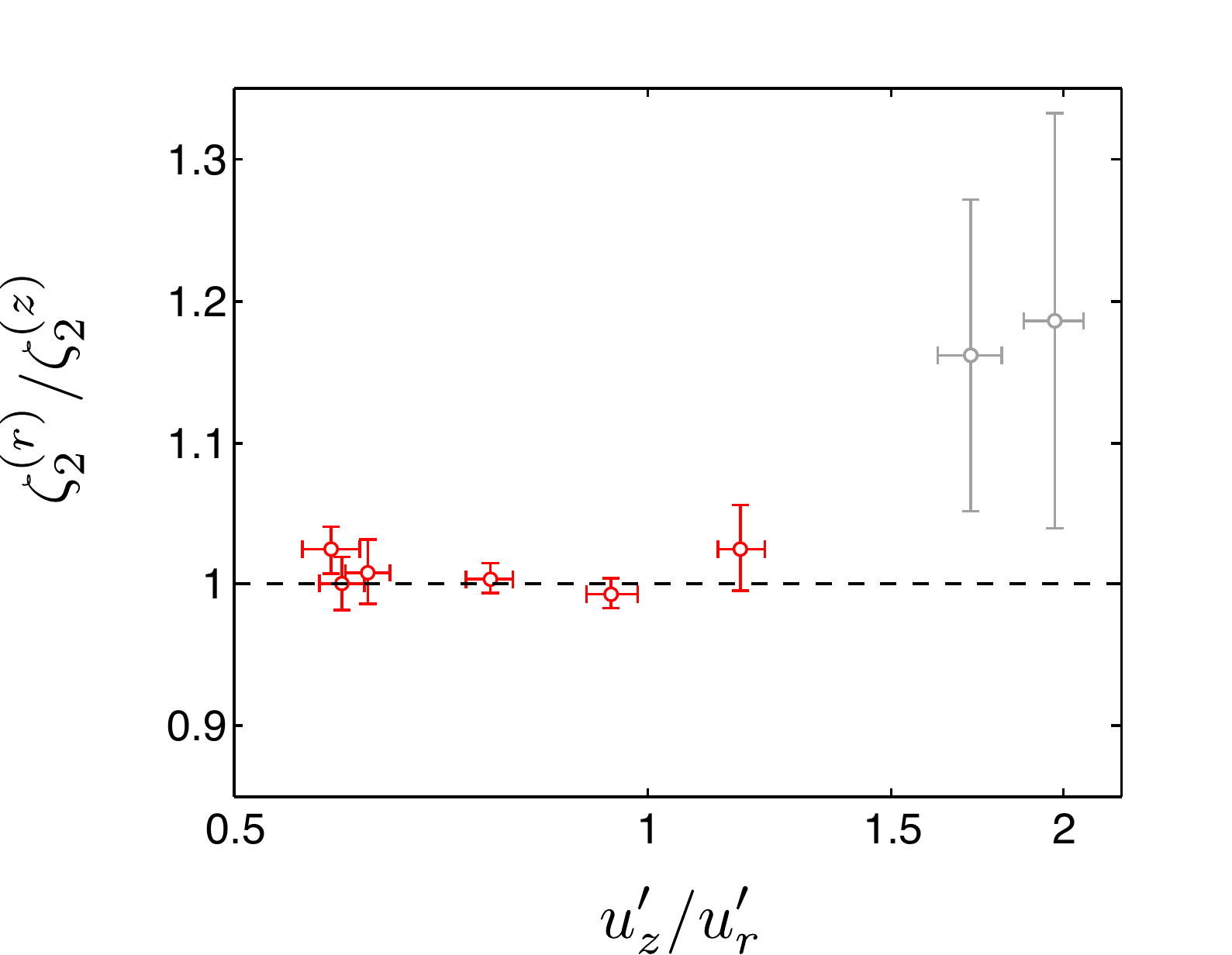}
\caption{The figure shows $\zeta_{2}^{(r)} / \zeta_{2}^{(z)}$, 
the ratio between the scaling exponents of the radial
structure functions and those of the axial structure functions.  
The last two points with large uncertainties due to spatial 
resolution are marked in gray.}
\label{fig:zetaratio}
\end{center}
\end{figure}

We now consider the ratio formed by the two exponents, 
$\zeta_2^{(r)}/\zeta_2^{(z)}$, and observe how it varies
with anisotropy, $u^{\prime}_z /u^{\prime}_r$.  
As shown in figure~\ref{fig:zetaratio}, except for the 
last two values of anisotropy, the ratio is nearly 1.  
Again, noise or residual shear might have affected the
last two radial structure functions, even though the exponents
$\zeta_2^{(z)}$ for the corresponding axial structure functions
were well within the mean of the estimated values.  
Thus, based on our measurements for anisotropy ratios 
$0.59 \leqslant u^{\prime}_z / u^{\prime}_r \leqslant 1.16$, 
we conclude that $\zeta_2^{(z)}$ and $\zeta_2^{(r)}$ are equal.  

\section{Scalar-valuedness of the Kolmogorov constant}
\label{sec:c2}

\begin{figure}
\begin{center}
\includegraphics[scale=0.9,type=pdf,ext=.pdf,read=.pdf]{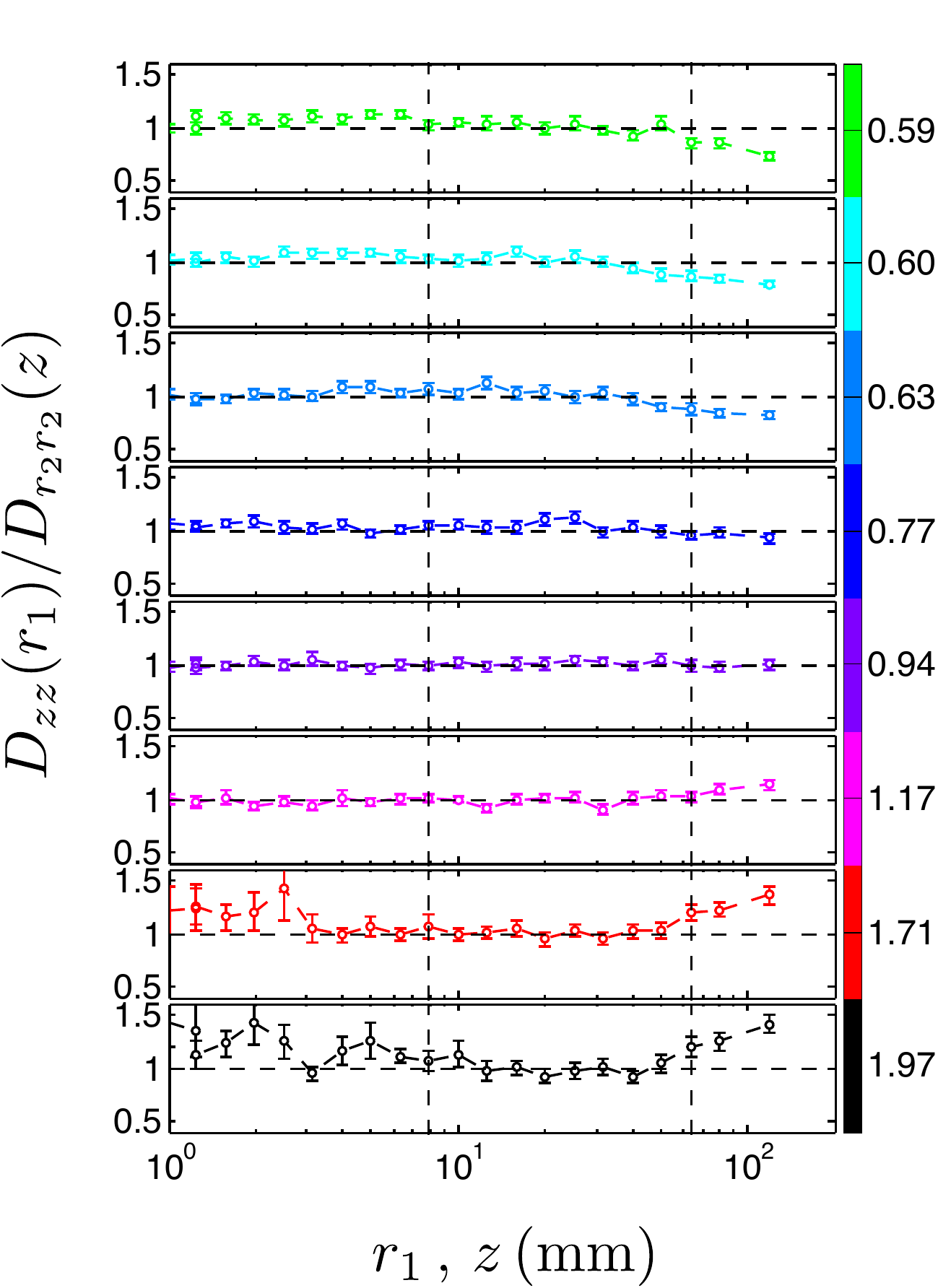}
\caption{The figure shows the ratio between structure functions in 
different directions in the same flow, $D_{zz} (r_1) / D_{r_2 r_2} (z)$, 
for all values of the anisotropy of the forcing.  
The data are the same as in figure~\ref{fig:dzz_drr_ratio_3cases}.  
They are plotted separately to display individual trends.  
The two vertical dashed lines demarcate the range of scales 
between $7$ and $70$~mm where the compensated structure
functions displayed a reasonable plateau.  
The values in the color bar are the fluctuations anisotropy 
measured at the center of the soccer ball for various values 
of the anisotropy of the forcing.}
\label{fig:dzz_drr_ratio_all}
\end{center}
\end{figure}

\begin{figure}
\begin{center}
\includegraphics[scale=0.9,type=pdf,ext=.pdf,read=.pdf]{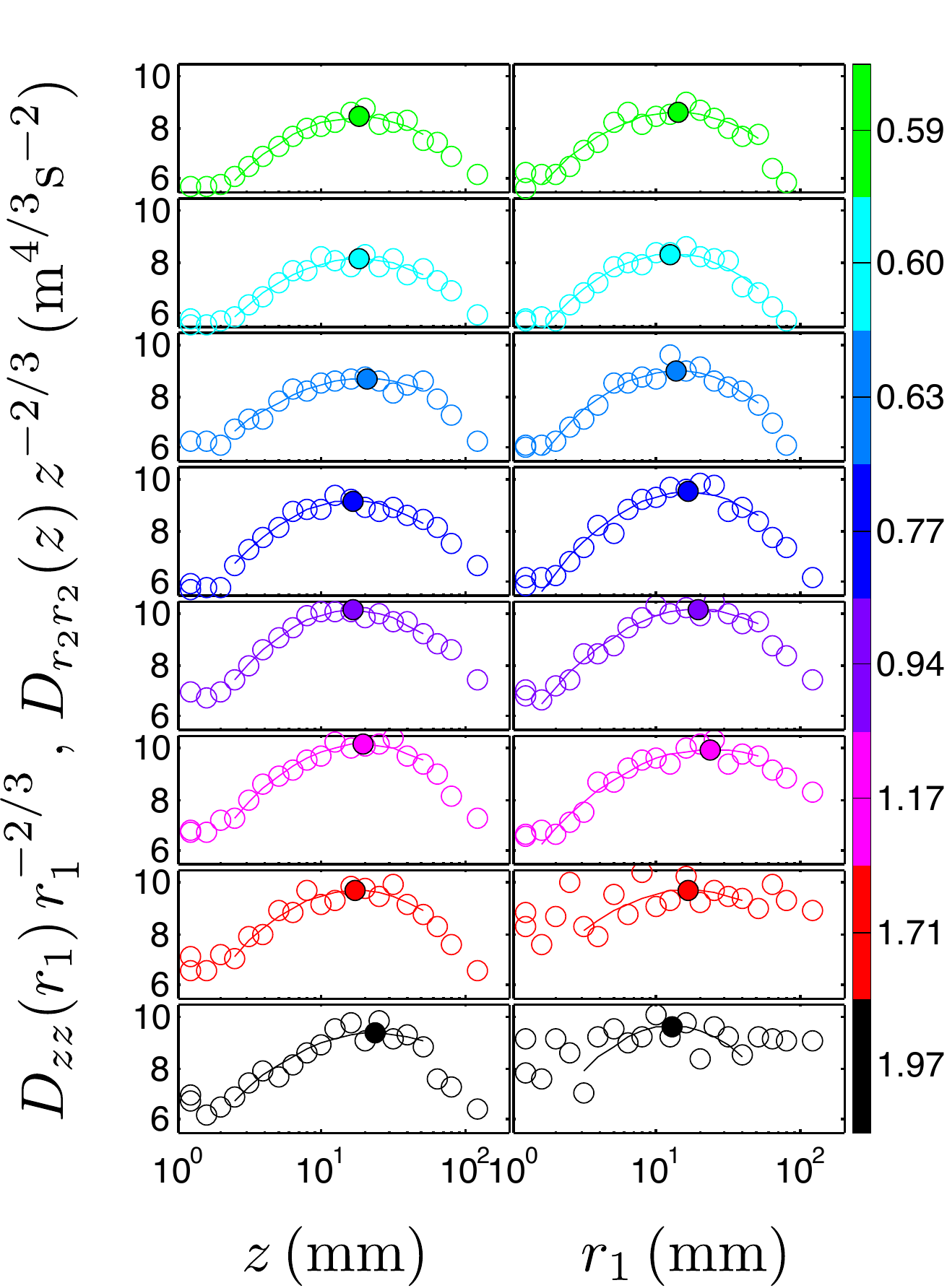}
\caption{The compensated axial and radial transverse structure 
functions for all values of anisotropy.  
Only the data in the range of scales between $2$ and $50$~mm are 
used in the determination of the ratio of the Kolmogorov constant.  
Solid lines are the second-order, least-square, linear-log polynomial 
fits to the structure function data.  
Solid symbols are the maxima of the fitted curves.  
The error bars are smaller than the symbols.  
The values in the color bar are the anisotropy measured at the center 
of the soccer ball for various values of the anisotropy of the forcing.}
\label{fig:s_over_r23}
\end{center}
\end{figure}

Let us now examine the Kolmogorov constant, $C_2$ 
(see equation~\ref{eq:dll_k41} and \ref{eq:dnn_k41}).
Anisotropy of the large scales allows for a family of equations 
in the form of equation~\ref{eq:dnn_k41}, but with different 
constants, $C_2^{(x)}$, for each direction of $\boldsymbol{r}$.  
A measure of inertial range anisotropy is $C_{2}^{(r)} / C_{2}^{(z)}$,
which we calculated in two ways.  
First, in figure~\ref{fig:dzz_drr_ratio_all}, we plot 
$D_{zz} (r_1) / D_{r_2 r_2} (z)$ against the separations $r_1$ and $z$.
These are the same data sets shown in 
figure~\ref{fig:dzz_drr_ratio_3cases}.
For clarity, they are displayed individually.  
We computed an average of the structure function ratio, 
$\langle D_{zz} (r_1) / D_{r_2 r_2} (z) \rangle$, taken over 
the range of separations where the compensated structure
functions displayed a reasonable plateau, or 
$7 \, \mathrm{mm} < r_1 \, , \, z < 70 \, \mathrm{mm}$.  
Since the energy dissipation, $\epsilon$, is a scalar quantity, 
the ratio of structure functions in the inertial range of scales is 
equal to the ratio $C_{2}^{(r)} / C_{2}^{(z)}$.  
To estimate the uncertainty in the measurements, we have used 
the standard error in the measurements of the ratios 
(see appendix~\ref{app:statistical_tools}, equation~\ref{eq:SE}).  
The second method was designed to mimic the 
one usually employed to measure $C_2$ (or $\epsilon$)
when the data are collected in only one direction.  
That is, we estimated $C_2$ from the maxima of the
compensated structure functions, $D (x) \, x^{-2/3}$.  
In order to reduce the influence of noise, we fit a second-order 
polynomial function 
\begin{equation}
y = a + b \, \log (x) + c \, (\log (x))^2 \,,
\end{equation}
to each data set in the range of scales between 
$2$ and $50$~mm, see figure~\ref{fig:s_over_r23}.  
Although this form is decidedly inappropriate since the structure function
should approach a constant for large separations, it proves to be very instructive in analyzing the data.  
We estimated the maximum value of $C_{2}^{(x)} \, \epsilon^{2/3}$
for each curve by calculating the ordinate of the peak of the parabola, 
given analytically in terms of the fitting parameters $a$, $b$, and $c$
\begin{equation}
y_{\rm max} = a - \dfrac{b^2}{4 \, c} \,.
\end{equation}
The ratio between the maxima, 
$\max (D_{zz} (r_1) \, r_1^{-2/3}) / \max (D_{r_2 r_2} (z) \, z^{-2/3})$, 
was then calculated.  
The ratio so formed, the energy dissipation rate being a scalar quantity 
falls out of the ratio, gave a second measure of $C_{2}^{(r)} / C_{2}^{(z)}$.  
To evaluate the fit, we calculated the difference between 
experimental and predicted values, $\xi_i = y_i - \tilde{y}_i$, 
where $y_i$ is the original value from experiment and $\tilde{y}_i$ is 
the value predicted by the parabolic fit.  
The standard deviation, or variability, of $\xi_i$
\begin{equation}
\mathrm{Variability} = \bigg[ \dfrac{1}{N-1} \, \sum_{i=1}^{N} (\xi_i - 
\langle \xi \rangle)^2 \bigg]^{1/2} \,,
\end{equation}
where $N$ is the number of data points used in the fit, 
gave an estimate for the error in the measurements

\begin{figure}
\begin{center}
\includegraphics[scale=0.7,type=pdf,ext=.pdf,read=.pdf]{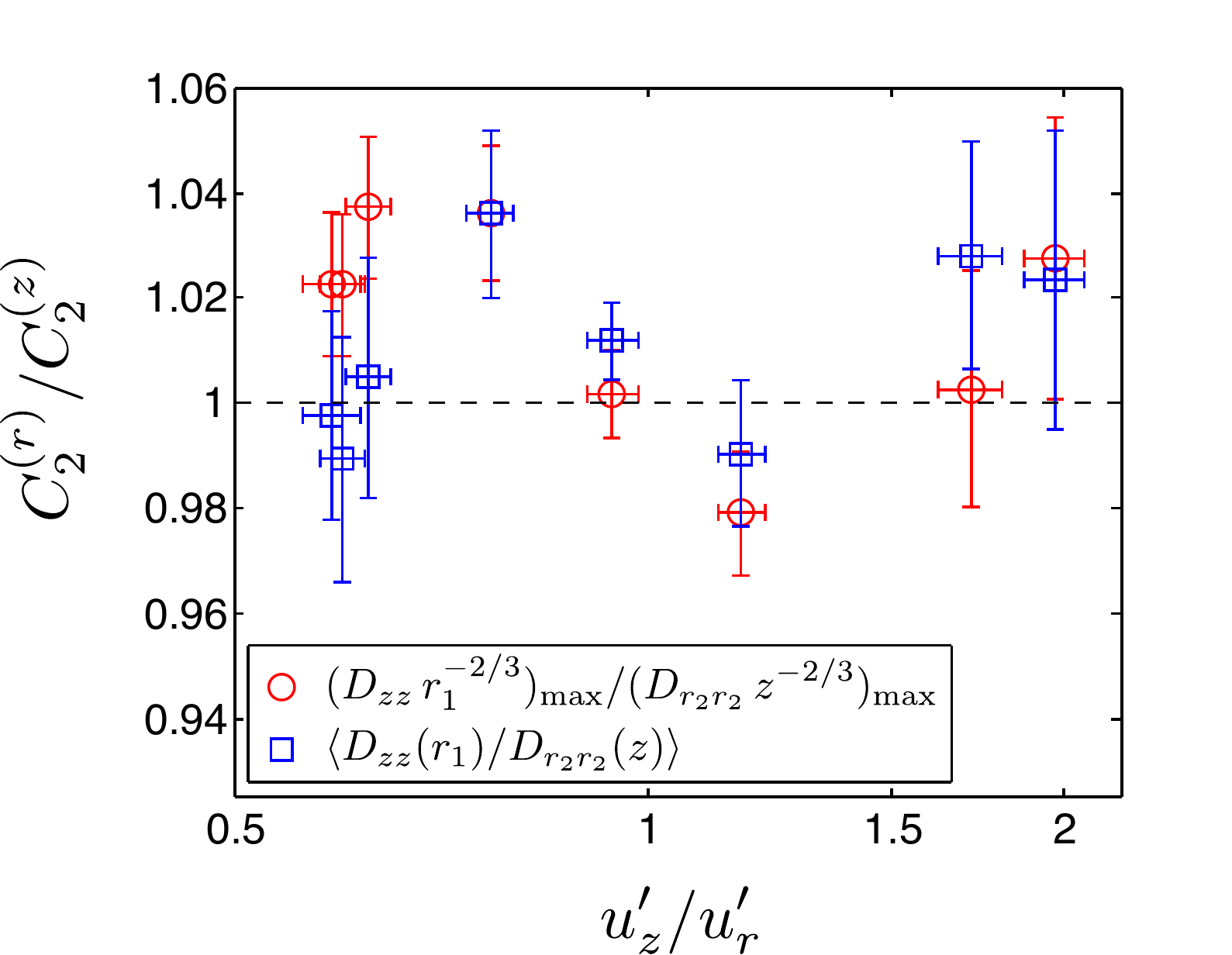}
\caption{The ratio between Kolmogorov constants
measured in different directions $C_{2}^{(r)} / C_{2}^{(z)}$.  
The open circular symbols (${\color{red} \ocircle}$) are the ratio 
between the maximum values of the compensated radial 
and axial structure functions.  
The open square symbols (${\color{blue} \square}$) are 
$\langle D_{zz} (r_1) / D_{r_2 r_2} (z) \rangle$, the ratio of radial 
to axial structure functions averaged over a range of 
scales $7 < r < 70~{\rm mm}$.}
\label{fig:c2ratio}
\end{center}
\end{figure}

The two measures of $C_{2}^{(r)} / C_{2}^{(z)}$ are shown in 
figure~\ref{fig:c2ratio}, and their numerical values are
listed side by side in table~\ref{table:c2ratio}.  
It can be seen that within measurement error, $C_2^{(r)}$ equals
$C_2^{(z)}$; the small downward trend is comparable to the scatter.  
Thus, we conclude that the Kolmogorov constant is scalar valued
over the range of anisotropies studied here.  
\begin{table}
\begin{center}
\begin{tabular}{ccccc}
\toprule
\addlinespace[8pt]
& \multicolumn{4}{c}{$C_{2}^{(r)} / C_{2}^{(z)}$} \\
\addlinespace[5pt]
\cmidrule(l){2-5}
\addlinespace[5pt]
$\dfrac{u^{\prime}_{z}}{u^{\prime}_{r}}$ & 
$\left\langle \dfrac{D_{zz} (r_1)}{D_{r_2 r_2} (z)} \right\rangle$ & Variability &
$\dfrac{\max (D_{zz} / r_1^{2/3})}{\max (D_{r_2 r_2} / z^{2/3})}$ & Variability \\
\addlinespace[5pt]
\midrule
\addlinespace[5pt]
$0.59\pm 0.03$ & $0.998$ & $\pm 0.020$ & $1.023$ & $\pm 0.014$ \\
$0.60\pm 0.02$ & $0.989$ & $\pm 0.024$ & $1.022$ & $\pm 0.014$ \\
$0.63\pm 0.03$ & $1.004$ & $\pm 0.023$ & $1.037$ & $\pm 0.014$ \\
$0.77\pm 0.03$ & $1.036$ & $\pm 0.016$ & $1.036$ & $\pm 0.013$ \\
$0.94\pm 0.04$ & $1.012$ & $\pm 0.007$ & $1.002$ & $\pm 0.009$ \\
$1.16\pm 0.05$ & $0.990$ & $\pm 0.014$ & $0.979$ & $\pm 0.012$ \\
$1.71\pm 0.10$ & $1.028$ & $\pm 0.022$ & $1.003$ & $\pm 0.023$ \\
$1.98\pm 0.10$ & $1.024$ & $\pm 0.029$ & $1.028$ & $\pm 0.027$ \\
\addlinespace[5pt]
\bottomrule
\end{tabular}
\caption{The numerical data for the ratio of the 
Kolmogorov constant for transverse structure functions measured in 
nearly orthogonal directions, computed from averaging the ratio
$D_{zz} (r_1) / D_{r_2 r_2} (z)$ over a range of scales between $7$
and $70$~mm (second column), as well as from taking the ratio
between the maxima of appropriately compensated structure 
functions, $\max(D_{zz} / r^{2/3}_1) / \max (D_{r_2 r_2} / z^{2/3})$ 
(fourth column).  
The values in the third and the last columns are the corresponding 
estimated error.}
\label{table:c2ratio}
\end{center}
\end{table}

Let us conclude with the calculation of the flow parameters.  
We determined $\epsilon$ by taking the peak values of the
compensated structure functions, as plotted in 
figure~\ref{fig:comp_structure_func}, assuming that
the second order structure functions obey Kolmogorov-type
scaling law.  
This assumption may not be valid in view of the anomaly
of the scaling exponent discussed in 
section~\ref{sec:scaling_exponents}, but dimensional
considerations compelled us to adhere to this definition.  
The dissipation rates depended on the anisotropy and 
reached a maximum value of $6.71$~m$^{2}$s$^{-3}$ for 
the case of isotropic turbulence.  
The variation of the dissipation rate with varying anisotropy
of the agitation is most likely the result of our decision to fix the 
total kinetic energy of the turbulence, see equation~\ref{eq:K}.  
Fixing the turbulence kinetic energy does not guarantee 
a constant energy dissipation rate, since the energy dissipation 
rate depends also on the large scales of the flow, which in turn 
depend on the anisotropy.  

Under isotropic forcing, the Taylor scale was 
$\lambda = \sqrt{15 \, \nu \, u^{\prime 2} / \epsilon} = 6.7$~mm, 
where $\nu = 1.568\times 10^{-5}$~m$^2$s$^{-1}$ is the kinematic 
viscosity of air at room temperature, and the Taylor-microscale Reynolds
number was $R_{\lambda} = \lambda \, u^{\prime} / \nu = \myrlambda$.  
The Reynolds numbers for the anisotropic cases were slightly lower.  
The corresponding Kolmogorov length and time scales were $\eta =
(\nu^3 / \epsilon)^{1/4} = 155$~$\micro$m and 
$\tau_{\eta} = \sqrt{\nu / \epsilon} =1.5$~ms.  

\section{Conclusions}

We investigated systematically the influence of the anisotropic
agitation on the inertial scales of turbulence in flows for
which the ratio of axial to radial RMS velocity fluctuations was between
$0.6$ and $2.3$, with Taylor based Reynolds number 
$R_{\lambda}=\myrlambda$.  
For scales smaller than the energy injection scale, 
the second-order transverse velocity structure functions were
independent of anisotropy.  
Within the experimental uncertainty, the inertial range scaling
exponent and the Kolmogorov constant were independent
of anisotropy.  
We expect anisotropic corrections, such as those introduced by
the SO(3) decomposition \cite[][]{biferale:2005}, to be smaller
than the error in our measurements (about 5\%).  
This will be true unless the corrections would cancel in the two
directions we measured; an outcome we consider unlikely.  
Our findings contrast with previous results by \cite{shen:2002},
who found that the scaling exponent for the transverse
structure function in the direction of a mean shear was
different from the scaling exponent in the direction of the
mean wind by about $0.1$; this indicates that the anisotropy
produced by shear may be inherently different from that
produced by turbulent fluctuations.
Our result is of practical importance in providing a means of
unambiguously estimating the parameters of anisotropic flows
without shear, such as the energy dissipation rate and the 
Reynolds number.  
\chapter{Higher order statistics}
\label{chap:higherorderstats}

In this chapter, we extend our investigation of the influence of 
anisotropy on the higher order statistics.  
We provide measurements of the structure functions, 
$\langle |\delta u (x)|^n \rangle$, for $n=4$, $5$, $6$, as described 
in section~\ref{sec:higher_order_sf}, and the nondimensional ratios, 
$\langle |\delta u (x)|^4 \rangle / \langle |\delta u (x)|^2 \rangle^2$ and 
$\langle |\delta u (x)|^6 \rangle / \langle |\delta u (x)|^2 \rangle^3$.  

It has now become common in studies of intermittency to
look at higher order structure functions.  
Measurements of higher order structure functions are very difficult.  
This is because higher order moments are influenced by rare events in the 
tails of the probability distribution (PDF), thus requiring a longer data set.  
We can estimate how much time and storage space it requires to 
collect these rare events.  
Experiments \cite[e.g.][]{anselmet:1984} suggest that the probability
distribution of the velocity increments, $\delta u$, can be approximated 
by an exponential, $P(x) \sim \mathrm{e}^{-x}$, where 
$x = \delta u / \langle |\delta u|^2 \rangle^{1/2}$.  
For good convergence of the $n$th order moment, 
$\langle |\delta u|^n \rangle$, the rule of thumb is to resolve $\delta u$
up to $n \, \langle |\delta u|^2 \rangle^{1/2}$, which corresponds to 
the location of the peak of $x^n \, P(x)$ \cite[e.g.][]{staicu:2003a}.  
Let us consider the sixth order moment.  
From figure~16 of \cite{shen:2000}, we estimate that $P(x=6) \approx 10^{-4}$.  
If $10$ counts are required around $x=6$, then
the total number of samples we need to collect is 
$10 / 10^{-4} = 10^5$.  
This is very close to the number of samples we actually collected
for $\delta u$ in the experiment.  
This amount of data occupied approximately $50$~megabytes of 
disc space in $20$~minutes.  
The total amount of time for $28$ different separation distances 
and $16$ anisotropies ($8$ for scanning along the symmetry axis of the 
forcing, and another $8$ perpendicular to it) was then $6.2$ days and
the total disc space was about $22$~gigabytes.  
Now for the $12$th order moment, $P(x=12) \approx 10^{-9}$ 
(\cite{anselmet:1984}, figure~3).  
Taking $10$ again as a reasonable number of counts,
the total amount of time and disc space required would be about
$1700$~years and $2.1$~petabytes.  
Thus, it can be seen that a precise measurement of the high-order
moments of the velocity increments with LDV may take many 
generations of graduate students to accomplish.  

\section{Higher order structure functions}
\label{sec:higherordersf}

\begin{figure}
\begin{center}
\includegraphics[scale=0.9,type=pdf,ext=.pdf,read=.pdf]{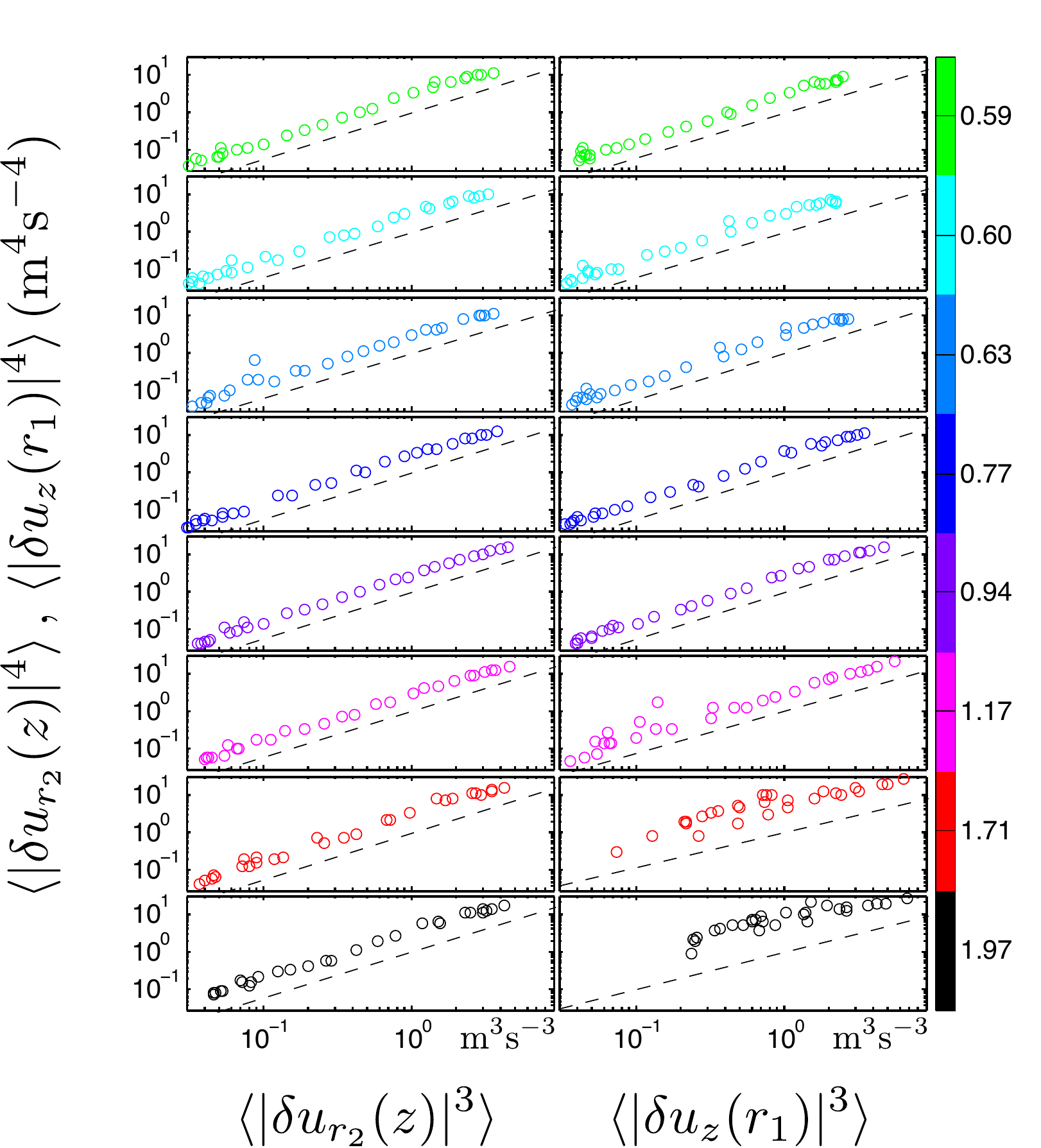}
\caption{The extended-self-similarity plots for the fourth-order
transverse structure functions.  
Dashed lines are proportional to power laws obtained from 
least squares fits to data.  
The values in the color bar are the anisotropy measured at the center 
of the soccer ball.}
\label{fig:sf_ESS_4th}
\end{center}
\end{figure}

\begin{figure}
\begin{center}
\includegraphics[scale=0.9,type=pdf,ext=.pdf,read=.pdf]{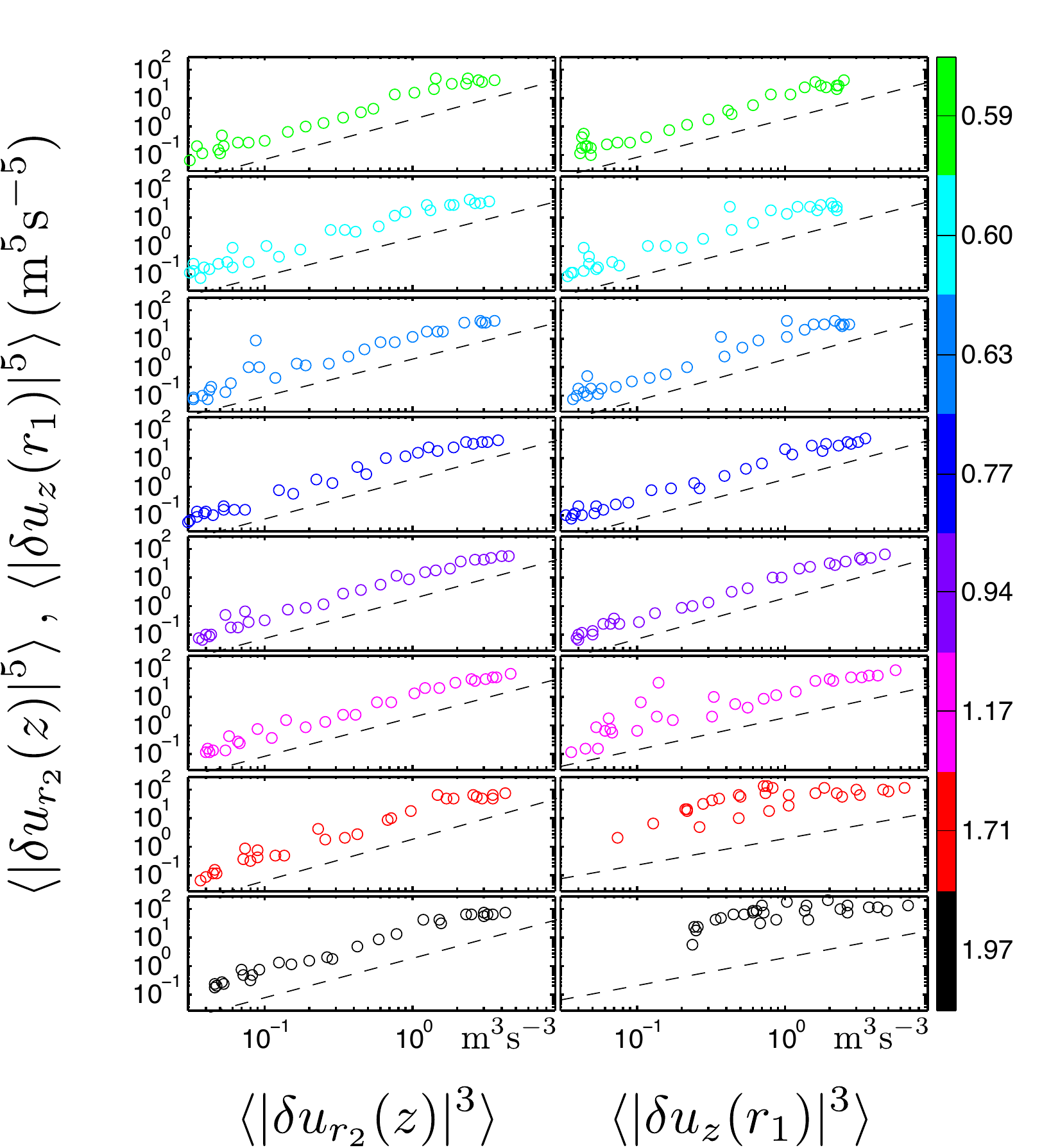}
\caption{The extended-self-similarity plots for the fifth-order
transverse structure functions.  
Dashed lines are proportional to power laws obtained from 
least squares fits to data.  
For legend, see figure~\ref{fig:sf_ESS_4th}.}
\label{fig:sf_ESS_5th}
\end{center}
\end{figure}

\begin{figure}
\begin{center}
\includegraphics[scale=0.9,type=pdf,ext=.pdf,read=.pdf]{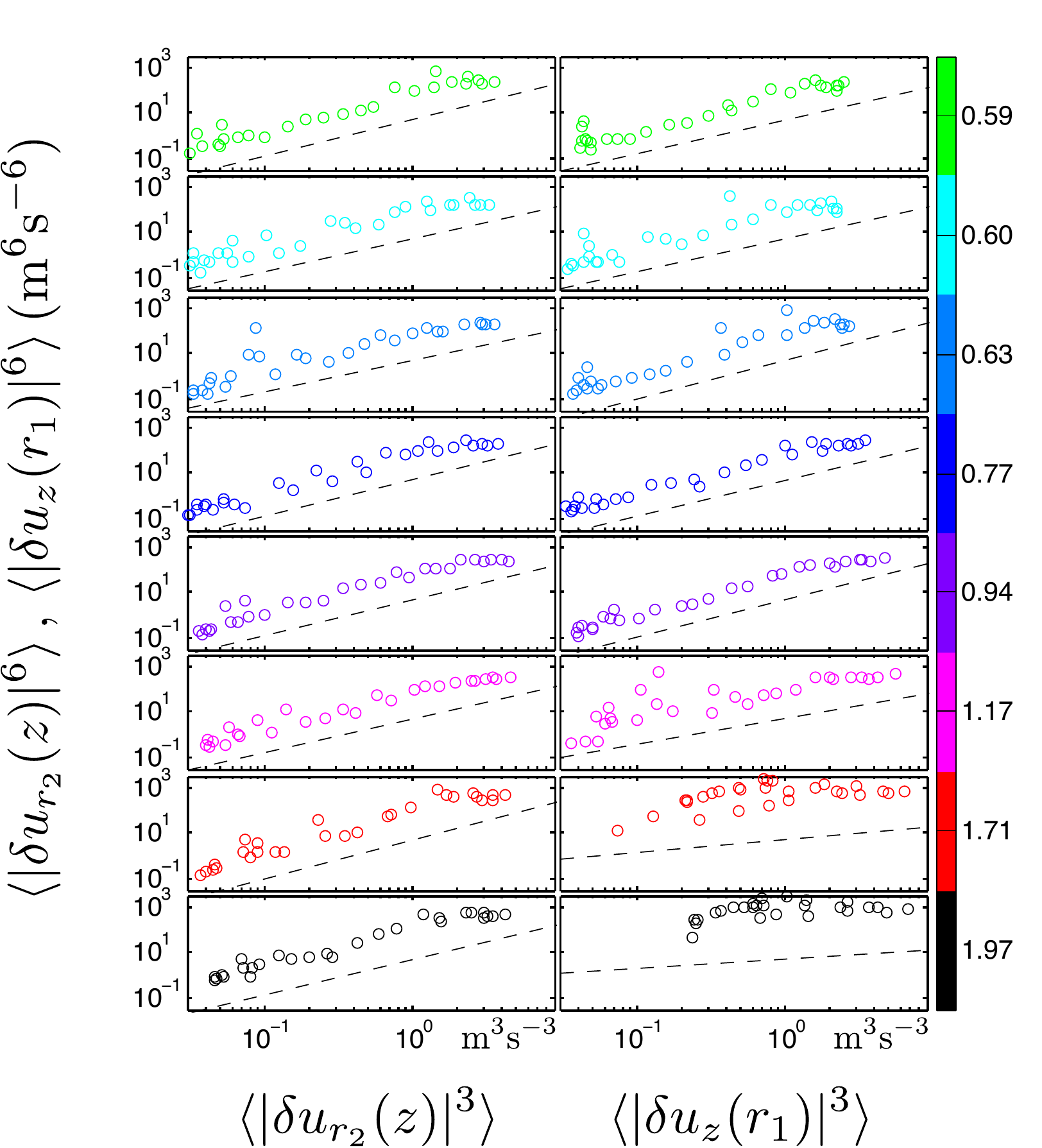}
\caption{The extended-self-similarity plots for the sixth-order
transverse structure functions.  
Dashed lines are proportional to power laws obtained from 
least squares fits to data.  
For legend, see figure~\ref{fig:sf_ESS_4th}.}
\label{fig:sf_ESS_6th}
\end{center}
\end{figure}

\begin{figure}
\begin{center}
\includegraphics[scale=1,type=pdf,ext=.pdf,read=.pdf]{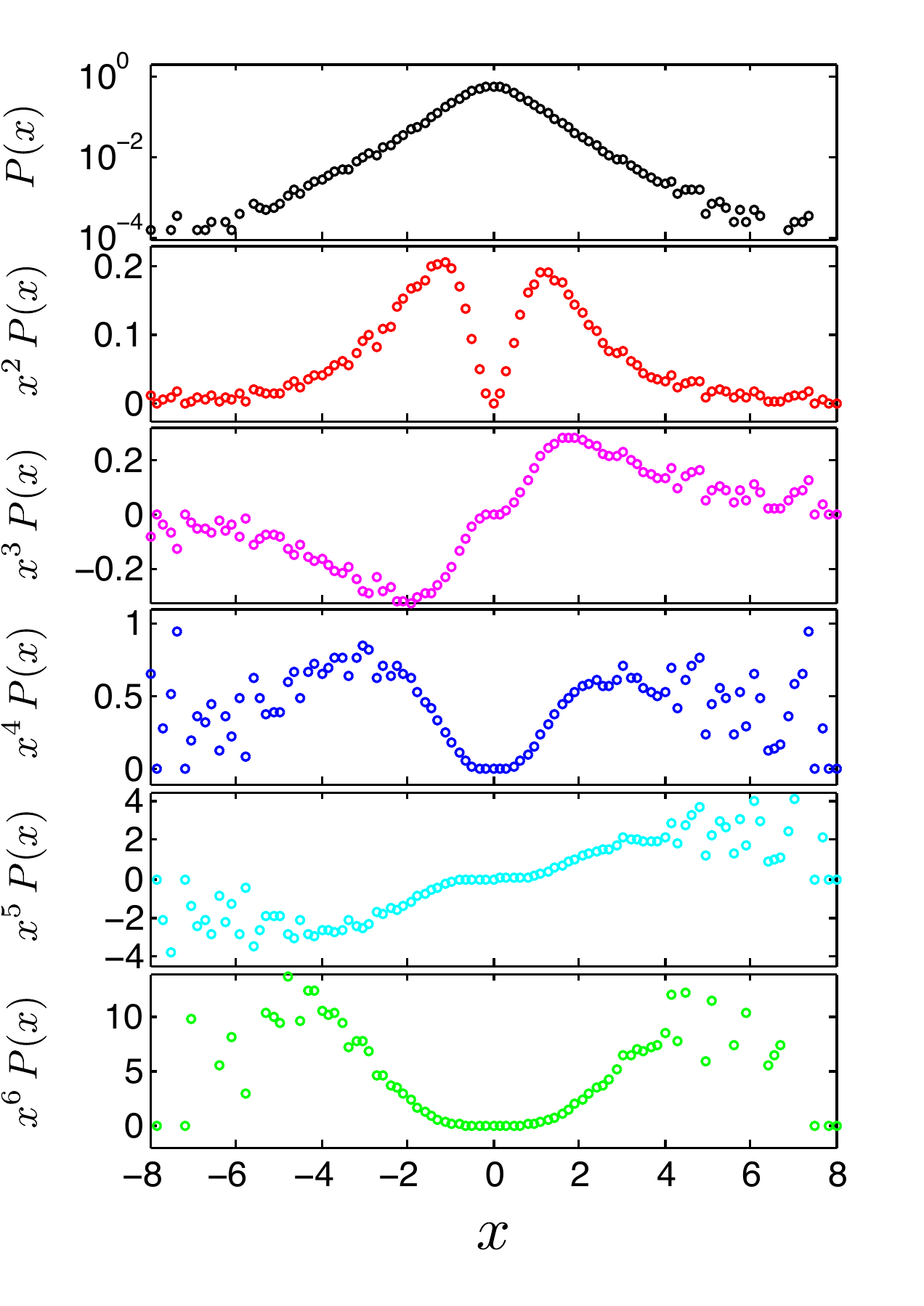}
\caption{The probability density of velocity increments, $\delta u_{r_2}(z)$,
and their moments for an inertial range separation $z/\eta\approx 130$
and spherically symmetric forcing, $u^{\prime}_z / u^{\prime}_r = 0.94$,
at $R_{\lambda} = \myrlambda$ (see table~\ref{table:flowstat}).  
Here $x = \delta u_{r_2} (z) / \langle |\delta u_{r_2}(z)|^2 \rangle^{1/2}$.}
\label{fig:pdf_moments_isotropic}
\end{center}
\end{figure}

In figures~\ref{fig:sf_ESS_4th}, \ref{fig:sf_ESS_5th}, and \ref{fig:sf_ESS_6th},
we show the ESS plots of transverse structure functions up to the sixth
order.  
As can be seen, the quality of data for higher order moments
of velocity increments quickly deteriorates when increasing the order.  
This is especially so for extreme prolate shape anisotropies and
for separations in the dissipative range.  
Let us examine the important issue of convergence.  
Figure~\ref{fig:pdf_moments_isotropic} shows, from top to bottom, 
the normalized probability density of $\delta u_{r_2} (z)$, 
denoted as $P(x)$, where $x = \delta u_{r_2} (z)/\langle |\delta 
u_{r_2} (z)|^2 \rangle^{1/2}$ is the velocity increments normalized 
by their standard deviation, and $P(x)$ multiplied by $x^2$, $x^3$, $x^4$,
$x^5$, and $x^6$ for a nearly spherically symmetric turbulent flow, 
$u^{\prime}_z / u^{\prime}_r = 0.94$, at $R_{\lambda} = \myrlambda$.  
Here, the probe volume spacing was $20$~mm, which corresponded to
about $130 \, \eta$.  
We obtained the PDF from approximately 78000 data points and
resolved the PDF to approximately 8 standard deviations.  
There seemed to be no convergence problems for the second
to fifth moments.  
The tails of these curves tended to flatten out at high $x$.  
The sixth moment was not as well converged as the lower orders; 
yet all moments showed strong asymmetry.  
This is clearly indicated in table~\ref{table:moments_area_isotropic} 
by the difference in the area integrated under the curves.  
The values of the third to the sixth moments were 
$-0.107$, $6.88$, $-0.689$, and $141$.  
The same features can be seen in the inertial range PDF 
obtained by \cite{shen:2000} in a wind tunnel homogeneous 
nonsheared turbulent flow at $R_{\lambda} = 934$ and a probe 
spacing of $153 \, \eta$.  
All their third to sixth moments showed strong asymmetry;
their values were $-0.46$, $5.14$, $-7.81$, and $66.7$.  
Notice that the asymmetry in \cite{shen:2000} is left-right
inverted because their definition of the velocity increments 
is $\delta u (z) = u(0)-u(z)$ whereas we have defined it as
$\delta u(z) = u(z) - u(0)$.  
The difference in signs between our odd order moments and those
of \cite{shen:2000} has been corrected in the above comparison.  

\begin{figure}
\begin{center}
\includegraphics[scale=1,type=pdf,ext=.pdf,read=.pdf]{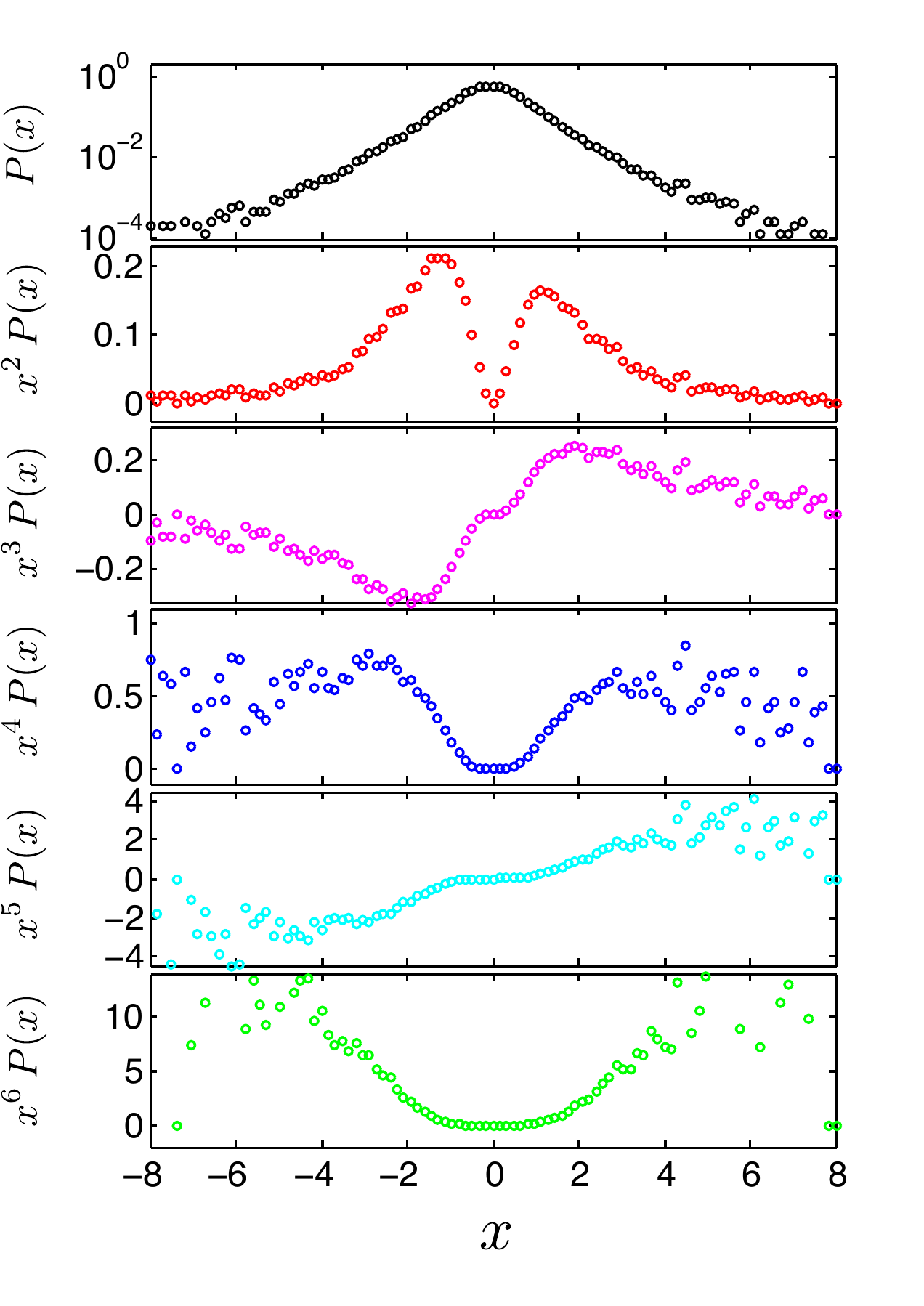}
\caption{The probability density of velocity increments, $\delta u_{r_2}(z)$,
and their moments for an inertial range separation $z/\eta\approx 130$
and oblate shape forcing, $u^{\prime}_z / u^{\prime}_r = 0.60$, at 
$R_{\lambda} = 536$ (see table~\ref{table:flowstat}).  
Here $x = \delta u_{r_2} (z) / \langle |\delta u_{r_2}(z)|^2 \rangle^{1/2}$.}
\label{fig:pdf_moments_anisotropic}
\end{center}
\end{figure}

Figure~\ref{fig:pdf_moments_anisotropic} shows the PDF of 
$\delta u_{r_2} (z)$ for an oblate shape forcing, 
$u^{\prime}_z / u^{\prime}_r = 0.60$, at the same probe separation 
and $R_{\lambda}$ comparable to the spherically symmetric case.  
We obtained about $10^5$ points.  
The improvement brought about by an increase in the sample size 
to the convergence of the higher order moments was marginal; 
yet the asymmetry in the PDF was evident, as indicated by an increase 
in the values of the odd order moments in 
table~\ref{table:moments_area_anisotropic}.  
The values for the third to the sixth moments were $-0.19$, $7.07$,
$-2.11$ and $154$.  
The asymmetry in the PDF of $\delta u$ was highly reproducible from 
flow to flow with different values of anisotropy.  


\begin{table}
\begin{center}
\begin{tabular}{rcccccc}
\toprule
\addlinespace[8pt]
Order of moment ($n$) & $0$ & $2$ & $3$ & $4$ & $5$ & $6$ \\
\addlinespace[5pt]
\midrule
\addlinespace[8pt]
$(-1)^n \, \int_{-\infty}^{0} x^n \, P(x) \, \mathrm{d} x$ & 
$0.512$ & $0.519$ & $1.14$ & $3.57$ & $14.6$ & $71.8$ \\
\addlinespace[5pt]
$\int_{0}^{\infty} x^n \, P(x) \, \mathrm{d} x$ & 
$0.488$ & $0.468$ & $1.03$ & $3.31$ & $13.9$ & $69.6$ \\
\addlinespace[5pt]
$\int_{-\infty}^{\infty} x^n \, P(x) \, \mathrm{d} x$ &
$1.00$ & $0.987$ & $-0.107$ & $6.88$ & $-0.689$ & $141$ \\
\addlinespace[5pt]
\bottomrule
\end{tabular}
\caption{The numerical integral of the PDF and its moments from
the second to the sixth order of an isotropically forced turbulent flow, 
shown in figure~\ref{fig:pdf_moments_isotropic}.  
The isotropy ratio was $u^{\prime}_z / u^{\prime}_r = 0.94$
and the Reynolds number was $R_{\lambda} = \myrlambda$ 
(see table~\ref{table:flowstat}).}
\label{table:moments_area_isotropic}
\end{center}
\end{table}
\begin{table}
\begin{center}
\begin{tabular}{r*{6}{c}}
\toprule
\addlinespace[8pt]
Order of moment ($n$) & $0$ & $2$ & $3$ & $4$ & $5$ & $6$ \\
\addlinespace[5pt]
\midrule
\addlinespace[8pt]
$(-1)^n \, \int_{-\infty}^{0} x^n \, P(x) \, \mathrm{d} x$ & 
$0.541$ & $0.530$ & $1.17$ & $3.79$ & $16.2$ & $83.4$ \\
\addlinespace[5pt]
$\int_{0}^{\infty} x^n \, P(x) \, \mathrm{d} x$ & 
$0.459$ & $0.430$ & $0.98$ & $3.27$ & $14.1$ & $71.5$ \\
\addlinespace[5pt]
$\int_{-\infty}^{\infty} x^n \, P(x) \, \mathrm{d} x$ &
$1.00$ & $0.96$ & $-0.19$ & $7.07$ & $-2.11$ & $154$ \\
\addlinespace[5pt]
\bottomrule
\end{tabular}
\caption{The numerical integral of the PDF and its moments from
the second to the sixth order of figure~\ref{fig:pdf_moments_anisotropic}.
The flow anisotropy was $u^{\prime}_z / u^{\prime}_r = 0.60$ at 
$R_{\lambda} = 536$ (see table~\ref{table:flowstat}).}
\label{table:moments_area_anisotropic}
\end{center}
\end{table}

\section{Kurtosis and hyperkurtosis of structure functions}

We now turn to the normalized structure functions.
When normalized by the second-order structure functions, 
the fourth and sixth order moments are given 
special names, namely the kurtosis
\begin{equation}
K_4 (x) = \dfrac{\langle |\delta u (x)|^4 \rangle}{\langle |\delta u (x)|^2 
\rangle^{2}} \,,
\end{equation}
and the hyper-kurtosis
\begin{equation}
K_6 (x) = \dfrac{\langle |\delta u (x)|^6 \rangle}{\langle |\delta u (x)|^2 
\rangle^{3}} \,.
\end{equation}
At the small scales, $K_4 > 3$ and $K_6 > 15$ because of 
intermittency \cite[][]{frisch:1995}.  
At the large scales the PDF of $\delta u$ follows a normal distribution
and $K_4 = 3$ and $K_6 = 15$.  

\begin{figure}
\begin{center}
\subfigure[]{
\includegraphics[scale=0.6,type=pdf,ext=.pdf,read=.pdf]{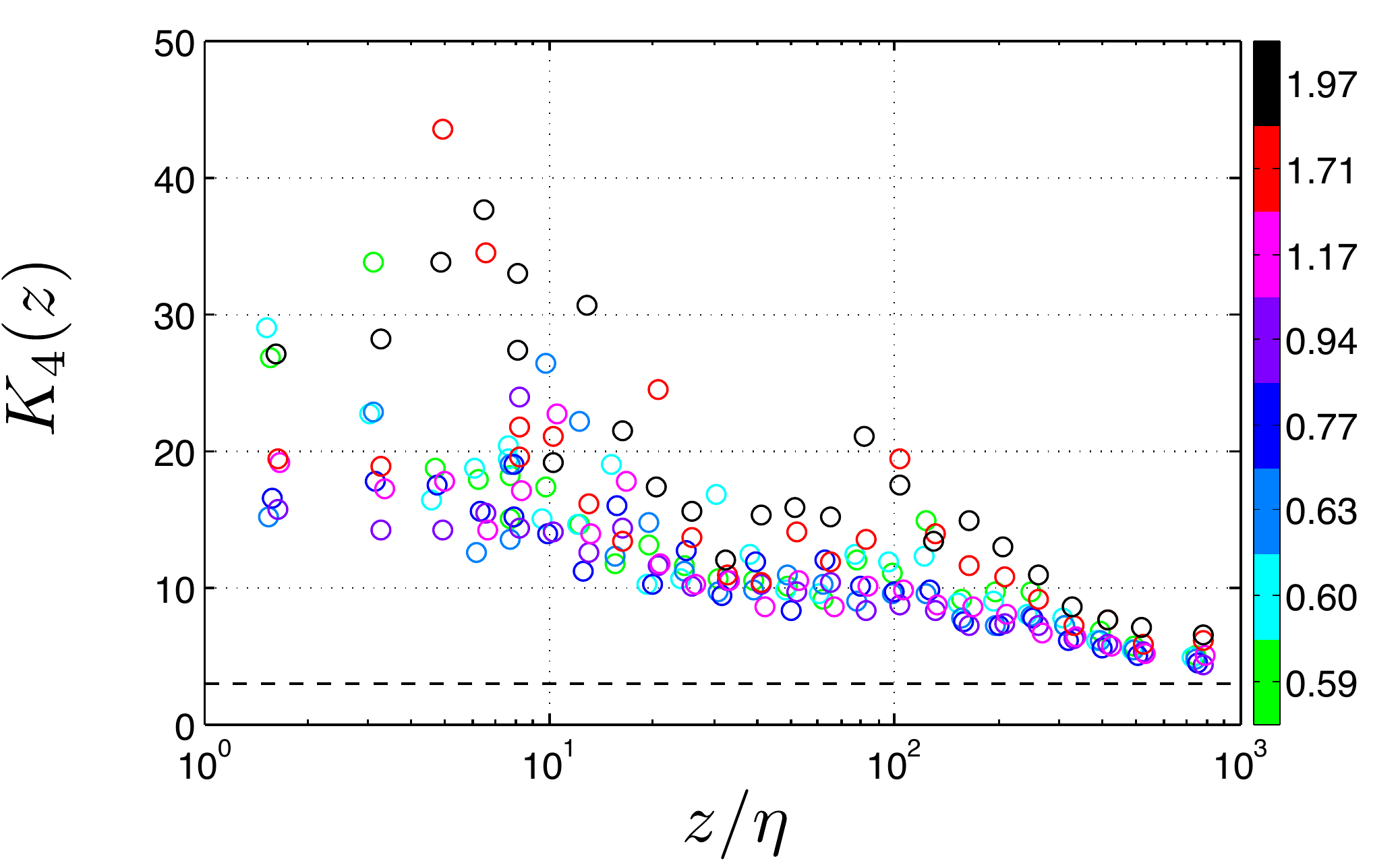}}
\subfigure[]{
\includegraphics[scale=0.6,type=pdf,ext=.pdf,read=.pdf]{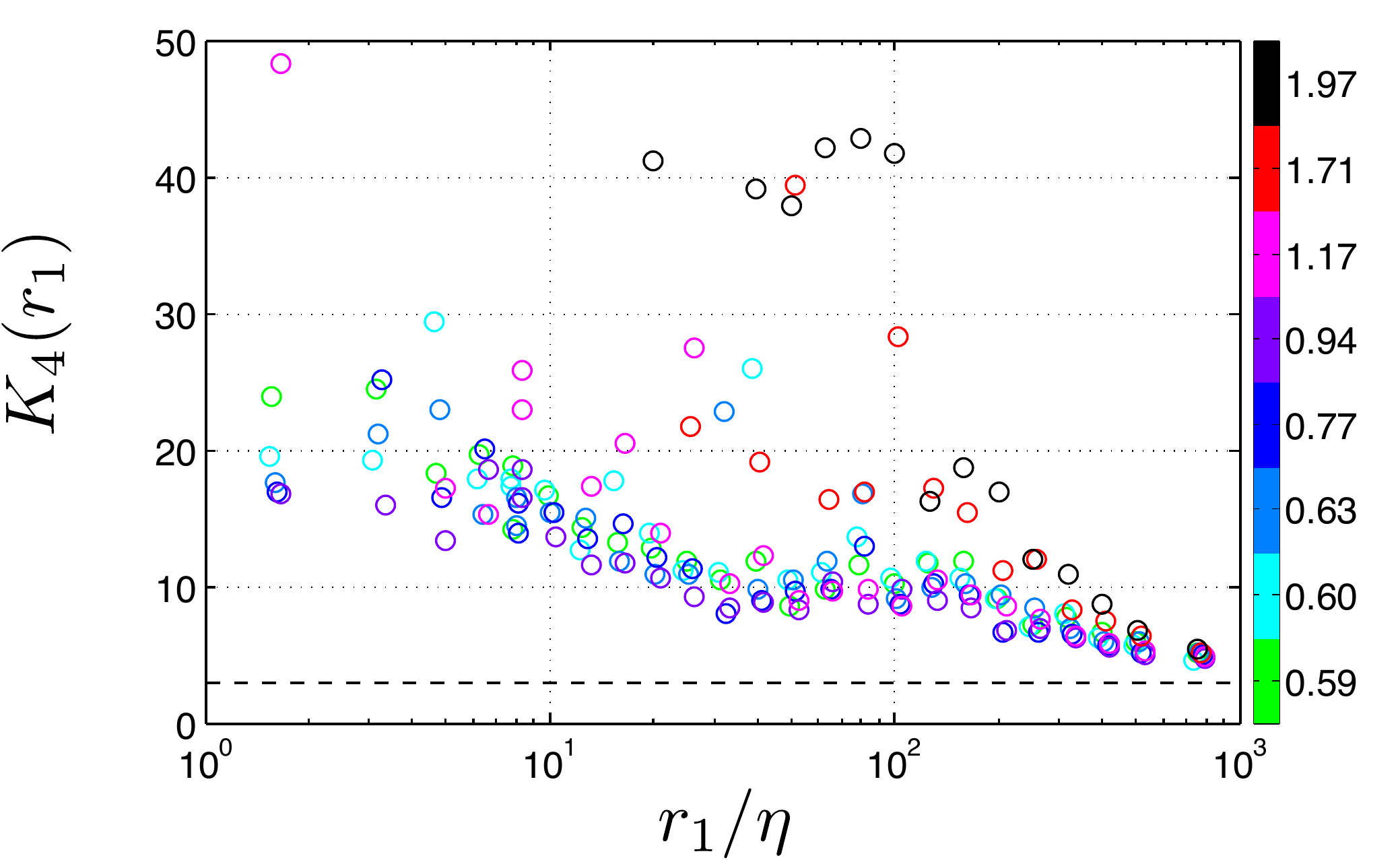}}
\caption{The kurtoses as a function of separations.   
(a) The axial kurtoses, $K_4 (z) = \langle |\delta u_{r_2} (z)|^4 \rangle / \langle |\delta u_{r_2}(z)|^2 \rangle^2$, and (b) 
the radial kurtoses $K_4 (r_1) = \langle |\delta u_{z} (r_1)|^4 \rangle / 
\langle |\delta u_{z} (r_1)|^2 \rangle^2$.  
The dashed lines are at the gaussian value of $3$.  
The values in the color bar are the anisotropy measured at the center
of the soccer ball.  
Error bars (not shown) are smaller than the symbols.}
\label{fig:kurtosis}
\end{center}
\end{figure}

\begin{figure}
\begin{center}
\subfigure[]{
\includegraphics[scale=0.6,type=pdf,ext=.pdf,read=.pdf]{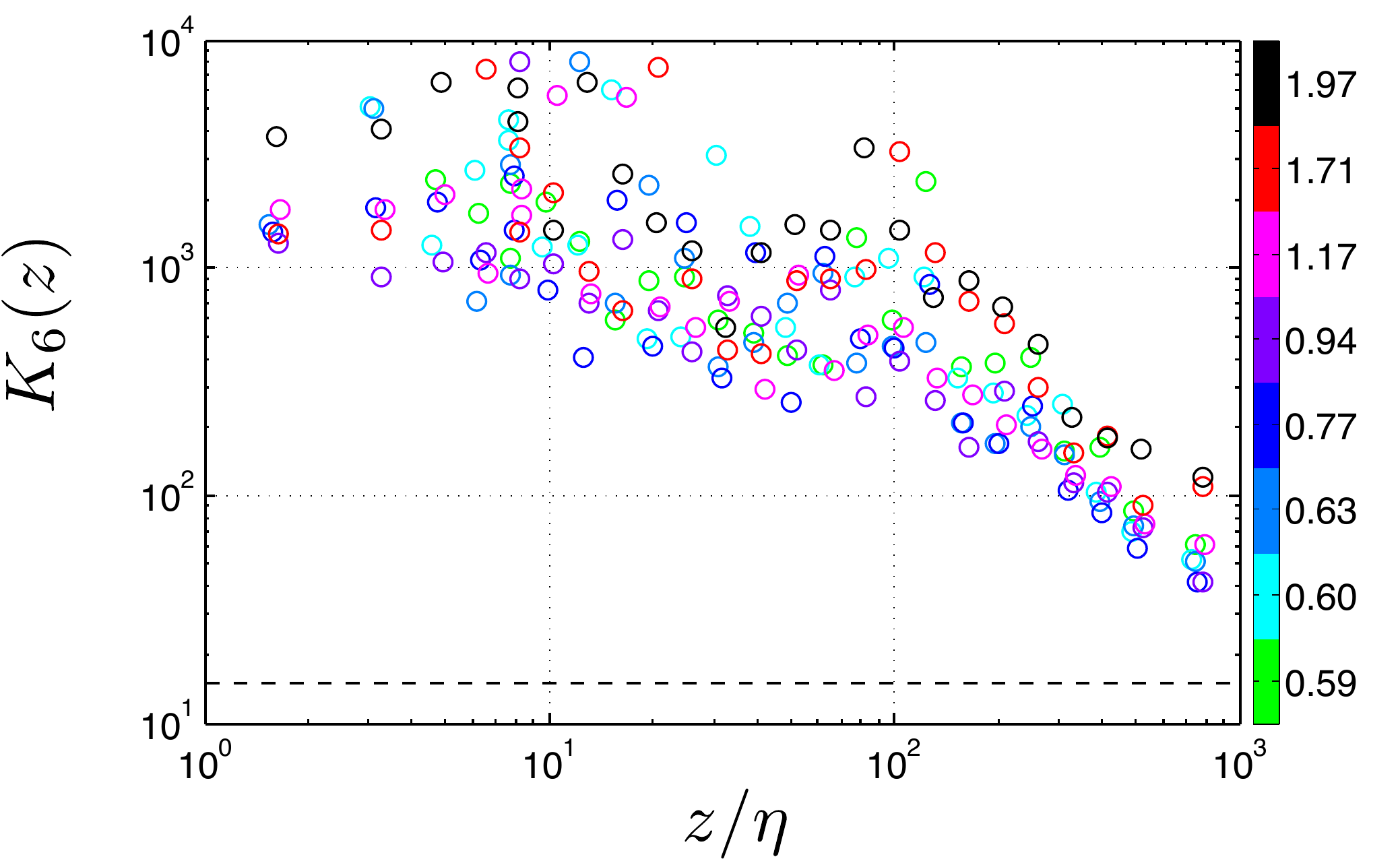}}
\subfigure[]{
\includegraphics[scale=0.6,type=pdf,ext=.pdf,read=.pdf]{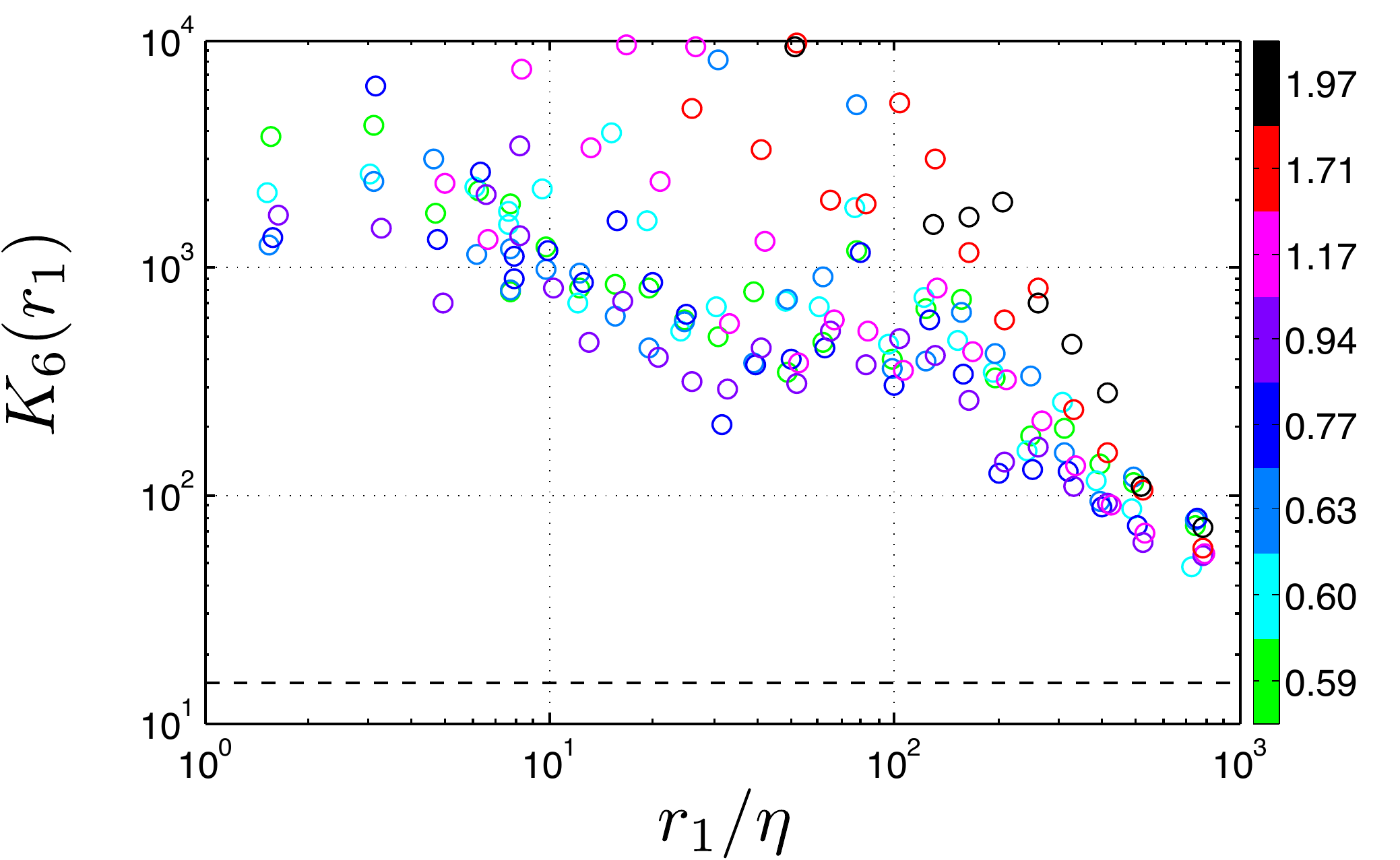}}
\caption{The hyper-kurtoses as a function of separations.  
(a) shows $ K_6 (z) = \langle |\delta u_{r_2} (z)|^6 \rangle / 
\langle |\delta u_{r_2} (z)|^2 \rangle^3$ 
and (b) shows $K_6 (r_1) = \langle |\delta u_{z} (r_1)|^6 \rangle / 
\langle |\delta u_{z}(r_1)|^2 \rangle^3$.  
The dashed lines are at the gaussian value of $15$.  
The values in the color bar are the anisotropy measured at the center
of the soccer ball.  
Error bars (not shown) are smaller than the symbols.}
\label{fig:hyperkurtosis}
\end{center}
\end{figure}

The measurements for the normalized axial and radial structure functions, 
$K_4 (z)$, $K_4 (r_1)$, $K_6 (z)$, and $K_6 (r_1)$, are shown in 
figures~\ref{fig:kurtosis} and \ref{fig:hyperkurtosis}.  
It is evident that our flows were highly intermittent at the small scales.  
Both $K_4$ and $K_6$ showed trends of decreasing to the gaussian 
values of $3$ and $15$, respectively.  
\cite{shen:2000} and \cite{garg:1998} have shown that $K_4$ and 
$K_6$ in sheared and nonsheared turbulent flows asymptote to the 
same limits at the integral scale, approximately at $2000 \, \eta$.  
If we can extend our measurement that far out, $K_4$ and $K_6$ in 
our experiment may continue to decrease and eventually reach 
their respective gaussian values.  
This is not physically possible in our experiment because 
$2000 \, \eta$ would correspond to a distance of about $0.3$~m, 
which is very close to the boundary of the apparatus where the 
anisotropy of the flow is heavily influenced by the confining walls 
rather than by the anisotropy of the agitation.  
What we gain by generating a stationary turbulent flow 
with negligible mean within confining walls we lose by having
a finite extend in our turbulent region.  
Nevertheless, it is remarkable that $K_4$ and $K_6$ in our flow
show downward trends toward their gaussian values
similar to those observed in the wind tunnel.  
\cite{shen:2000} also noted the presence of a small bump at 
$200 \, \eta$ and attributed this to the scatter in the data.  
This bump was also noticeable in our measurements of $K_4$ and $K_6$,
albeit at $100 \, \eta$.  
The magnitude of this bump was as large as the scatter and 
may have evaded detection in experiments.  
We are only able to find another instance of the appearance 
of this bump at approximately $150 \, \eta$ in the measurement 
of \cite{siebert:2010} in the turbulent region of stratocumulus clouds 
at $R_{\lambda} \approx 5000$.  
We note that the scale at which the bump occurs approximately 
coincides with the Taylor scales, $\lambda = 15^{1/4} \, 
R_{\lambda}^{1/2} \, \eta$ (see e.g. \cite{pope:2000}, page~200, 
for a derivation of $\lambda$), which were $40 \, \eta$, $60 \, \eta$, 
and $140 \, \eta$ for $R_{\lambda} = \myrlambda$, $934$, and 
$5000$, respectively.  
From the numerical experiments of \cite{donzis:2010}, we infer
that the second order structure functions at $R_{\lambda} = 140 - 1000$ 
reach a plateau at around the Taylor scales $\lambda = 20 \, \eta - 
60 \, \eta$.  
Thus, we speculate that the reason for the occurrence of this bump 
may be due to transition range dynamics.  

\section{Higher order scaling exponents}

Here, we document the inertial range scaling exponents
of higher order structure functions.  
The method for obtaining the higher order scaling exponents is 
the same as in section~\ref{sec:scaling_exponents}.  
We calculated the slopes obtained from straight-line fits,
$\log\langle |\delta u (x)|^n \rangle \sim 
\log \langle |\delta u (x)|^3 \rangle$, to the data shown
in figures~\ref{fig:sf_ESS_4th}, \ref{fig:sf_ESS_5th}, and
\ref{fig:sf_ESS_6th}.  
Tables~\ref{table:zeta4}, \ref{table:zeta5}, and \ref{table:zeta6}
list the values for the axial and radial scaling exponents, $\zeta_p^{(z)}$
and $\zeta_p^{(r)}$, for $p = 4$, $5$, and $6$.  
The radial structure functions for $u^{\prime}_z / u^{\prime}_r = 1.71$
and $1.98$ were very noisy and the scatter in these two data sets might 
have been due to a residual shear in the turbulence or an insufficiently
developed turbulent region.  
The errors in all the exponents were estimated with the standard
error of the slopes in the straight-line fits 
(see appendix~\ref{app:statistical_tools}).  
Within the experimental error, we found that the values of the exponents 
are in fair agreement with those found in the literature 
(see section~\ref{sec:anomalous_scaling}).  

\begin{figure}
\begin{center}
\includegraphics[scale=0.75,type=pdf,ext=.pdf,read=.pdf]{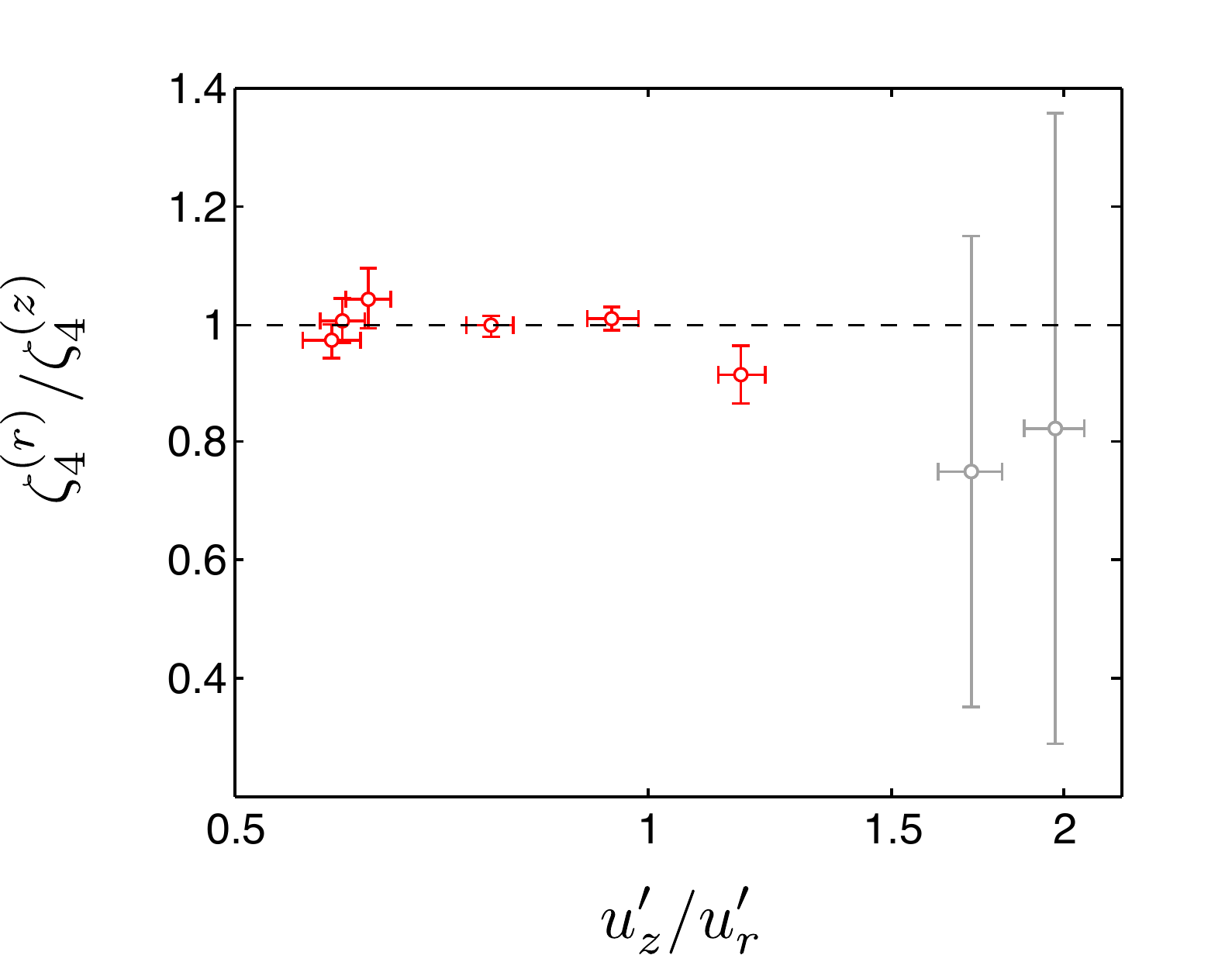}
\caption{The figure shows $\zeta_{4}^{(r)} / \zeta_{4}^{(z)}$, 
the ratio between the fourth-order scaling exponents of the radial
structure functions and those of the axial structure functions.  
The last two points with large uncertainties due to spatial 
resolution are marked in gray.}
\label{fig:zetaratio_ess4}
\end{center}
\end{figure}

\begin{figure}
\begin{center}
\includegraphics[scale=0.75,type=pdf,ext=.pdf,read=.pdf]{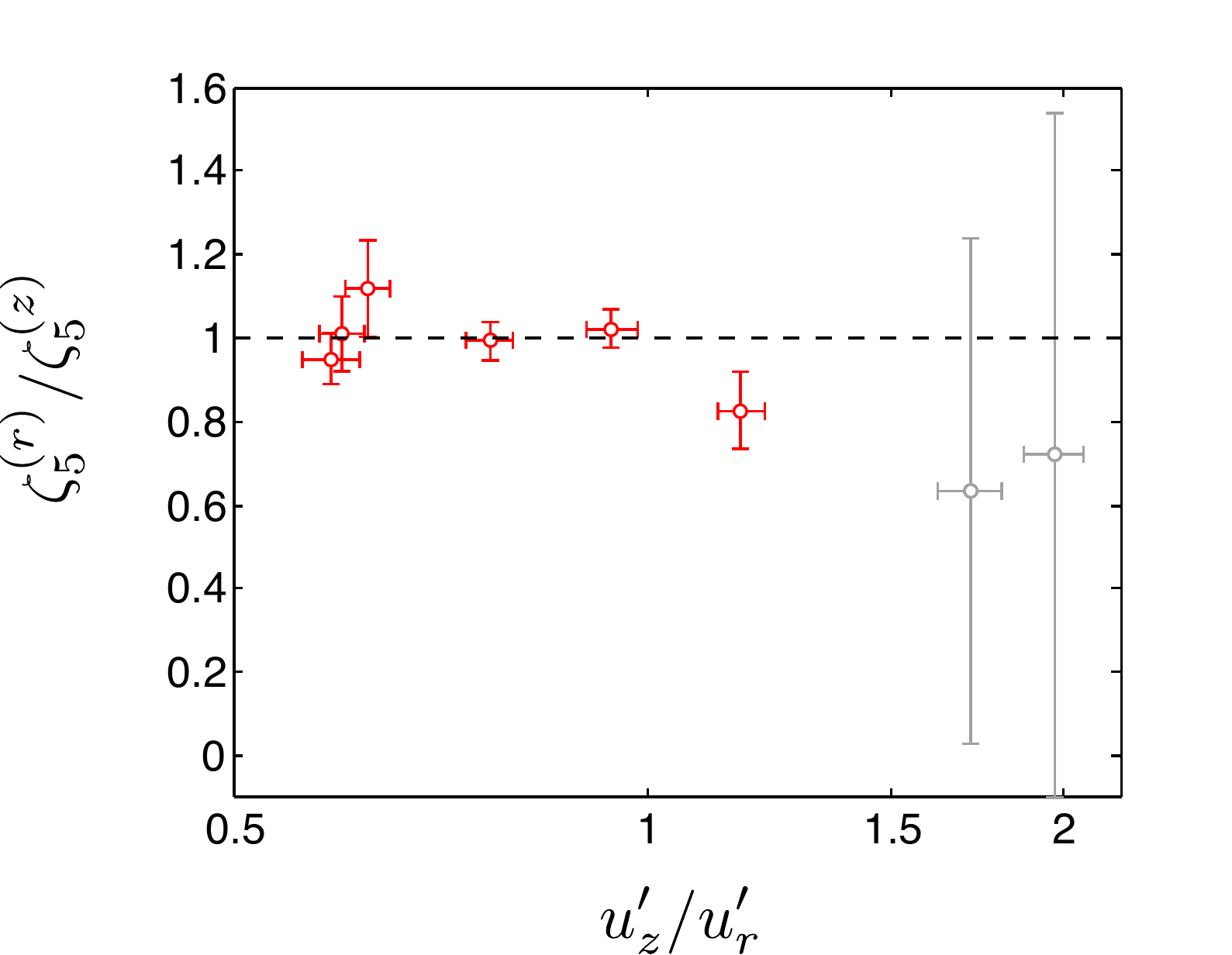}
\caption{The figure shows $\zeta_{5}^{(r)} / \zeta_{5}^{(z)}$, 
the ratio between the fifth-order scaling exponents of the radial
structure functions and those of the axial structure functions.  
The last two points with large uncertainties due to spatial 
resolution are marked in gray.}
\label{fig:zetaratio_ess5}
\end{center}
\end{figure}

We now examine the ratio of radial to axial exponents for order
$4$ and $5$.  
In figures~\ref{fig:zetaratio_ess4} and \ref{fig:zetaratio_ess5}, 
we plot $\zeta_4^{(r)}/\zeta_4^{(z)}$ and 
$\zeta_5^{(r)}/\zeta_5^{(z)}$ against anisotropy, 
$u^{\prime}_z / u^{\prime}_r$.  
Because the majority of the ratios between radial and axial 
exponents are close to one, we conjecture that 
$\zeta_p^{(r)}$ and $\zeta_p^{(z)}$ are equal in anisotropic 
turbulence.  
We note that \cite{ouellette:2006} measured the Lagrangian
scaling exponents up to the tenth order in a von K\'{a}rm\'{a}n 
counter-rotating flow and found that the axial and radial exponents 
are the same to within experimental accuracy.  
Thus, our investigation here concerning the equality of Eulerian 
scaling exponents forms an extension to the above work.  
\begin{table}
\begin{center}
\begin{tabular}{cccc}
\toprule
\addlinespace[8pt]
$u^{\prime}_z / u^{\prime}_r$ & $\zeta_4^{(r)}$ & $\zeta_4^{(z)}$ &
$\zeta_4^{(r)} / \zeta_4^{(z)}$ \\
\addlinespace[5pt]
\midrule
\addlinespace[8pt]
$0.59\pm 0.03$ & $1.186\pm 0.021$ & $1.220\pm 0.017$ & $0.972\pm 0.031$ \\
$0.60\pm 0.02$ & $1.189\pm 0.027$ & $1.184\pm 0.020$ & $1.005\pm 0.040$ \\
$0.63\pm 0.03$ & $1.223\pm 0.023$ & $1.173\pm 0.038$ & $1.043\pm 0.053$ \\
$0.77\pm 0.03$ & $1.213\pm 0.012$ & $1.216\pm 0.013$ & $0.998\pm 0.021$ \\
$0.94\pm 0.04$ & $1.227\pm 0.011$ & $1.215\pm 0.015$ & $1.010\pm 0.021$ \\
$1.16\pm 0.05$ & $1.098\pm 0.049$ & $1.199\pm 0.013$ & $0.916\pm 0.051$ \\
$1.71\pm 0.10$ & $0.923\pm 0.486$ & $1.231\pm 0.022$ & $0.750\pm 0.400$ \\
$1.98\pm 0.10$ & $0.989\pm 0.629$ & $1.199\pm 0.016$ & $0.824\pm 0.535$ \\
\addlinespace[5pt]
\bottomrule
\end{tabular}
\caption{The values of the fourth order radial and axial scaling exponents, 
$\zeta_4^{(r)}$ and $\zeta_4^{(z)}$, measured in our experiment
using ESS.  The last column shows the ratio between the two.}
\label{table:zeta4}
\end{center}
\end{table}

\begin{table}
\begin{center}
\begin{tabular}{cccc}
\toprule
\addlinespace[8pt]
$u^{\prime}_z / u^{\prime}_r$ & $\zeta_5^{(r)}$ & $\zeta_5^{(z)}$ &
$\zeta_5^{(r)} / \zeta_5^{(z)}$ \\
\addlinespace[5pt]
\midrule
\addlinespace[8pt]
$0.59\pm 0.03$ & $1.338\pm 0.052$ & $1.407\pm 0.043$ & $0.951\pm 0.066$ \\
$0.60\pm 0.02$ & $1.327\pm 0.072$ & $1.312\pm 0.049$ & $1.011\pm 0.092$ \\
$0.63\pm 0.03$ & $1.441\pm 0.061$ & $1.290\pm 0.082$ & $1.117\pm 0.118$ \\
$0.77\pm 0.03$ & $1.394\pm 0.032$ & $1.403\pm 0.036$ & $0.993\pm 0.048$ \\
$0.94\pm 0.04$ & $1.429\pm 0.029$ & $1.396\pm 0.039$ & $1.023\pm 0.050$ \\
$1.16\pm 0.05$ & $1.118\pm 0.099$ & $1.355\pm 0.036$ & $0.825\pm 0.095$ \\
$1.71\pm 0.10$ & $0.923\pm 0.852$ & $1.458\pm 0.054$ & $0.633\pm 0.608$ \\
$1.98\pm 0.10$ & $0.989\pm 1.102$ & $1.373\pm 0.038$ & $0.720\pm 0.822$ \\
\addlinespace[5pt]
\bottomrule
\end{tabular}
\caption{The numerical data for the fifth order radial and axial 
scaling exponents, $\zeta_5^{(r)}$, $\zeta_5^{(z)}$, 
and the ratio between the two.}
\label{table:zeta5}
\end{center}
\end{table}

\begin{table}
\begin{center}
\begin{tabular}{cccc}
\toprule
\addlinespace[8pt]
$u^{\prime}_z / u^{\prime}_r$ & $\zeta_6^{(r)}$ & $\zeta_6^{(z)}$ &
$\zeta_6^{(r)} / \zeta_6^{(z)}$ \\
\addlinespace[5pt]
\midrule
\addlinespace[8pt]
$0.59\pm 0.03$ & $1.490\pm 0.085$ & $1.584\pm 0.076$ & $0.941\pm 0.099$ \\
$0.60\pm 0.02$ & $1.447\pm 0.120$ & $1.424\pm 0.082$ & $1.017\pm 0.143$ \\
$0.63\pm 0.03$ & $1.668\pm 0.104$ & $1.395\pm 0.127$ & $1.196\pm 0.183$ \\
$0.77\pm 0.03$ & $1.562\pm 0.055$ & $1.584\pm 0.067$ & $0.986\pm 0.076$ \\
$0.94\pm 0.04$ & $1.630\pm 0.051$ & $1.570\pm 0.069$ & $1.038\pm 0.078$ \\
$1.16\pm 0.05$ & $1.131\pm 0.144$ & $1.499\pm 0.063$ & $0.754\pm 0.128$ \\
$1.71\pm 0.10$ & $0.583\pm 0.170$ & $1.696\pm 0.090$ & $0.344\pm 0.119$ \\
$1.98\pm 0.10$ & $0.426\pm 0.175$ & $1.544\pm 0.062$ & $0.276\pm 0.124$ \\
\addlinespace[5pt]
\bottomrule
\end{tabular}
\caption{The numerical data for the sixth order radial and axial 
scaling exponents, $\zeta_6^{(r)}$, $\zeta_6^{(z)}$, 
and the ratio between the two.}
\label{table:zeta6}
\end{center}
\end{table}

\begin{figure}
\begin{center}
\includegraphics[scale=0.75,type=pdf,ext=.pdf,read=.pdf]{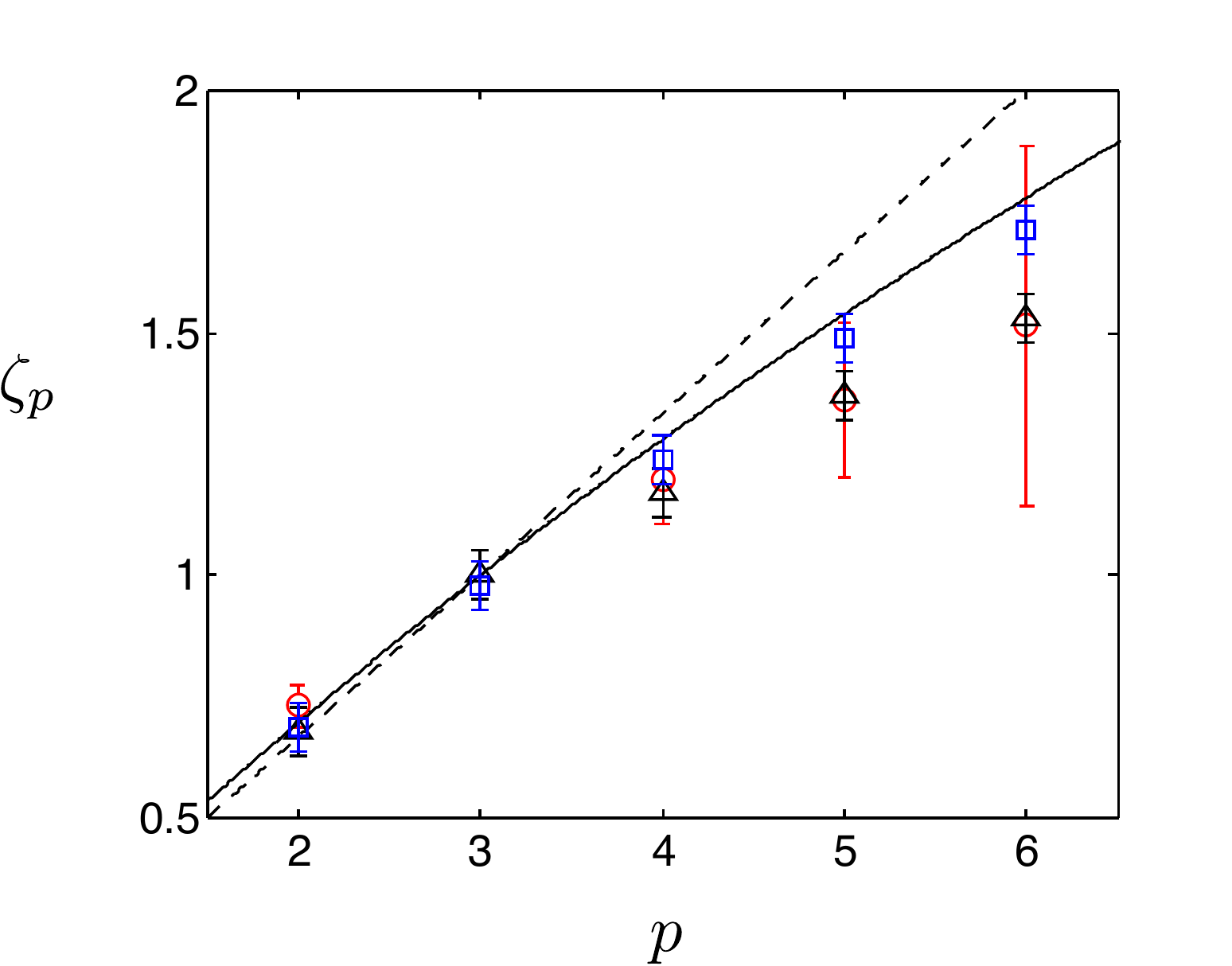}
\caption{The circular symbols ({\color{red} $\ocircle$}) are our 
measurements of the inertial range scaling exponents, $\zeta_p$, 
of the transverse structure functions using ESS method.  
The triangular symbols ($\bigtriangleup$) are the exponents
measured by \cite{shen:2002} in an unsheared wind tunnel homogeneous 
and isotropic turbulent flow at an $R_{\lambda}$ of $134$, and 
the square symbols (${\color{blue} \square}$) are those measured by 
them at an $R_{\lambda}$ of $863$.  
The dashed line is the Kolmogorov prediction (equation~\ref{eq:zetap_k41}).  
The solid line is the She-Leveque prediction (equation~\ref{eq:zetap_she_leveque}).}
\label{fig:intermittency_plot}
\end{center}
\end{figure}

\begin{table}
\begin{center}
\begin{tabular}{*{4}{c}}
\toprule
\addlinespace[8pt]
\multirow{2}{*}{Order $p$} & \multicolumn{3}{c}{$R_{\lambda}$} \\
\addlinespace[5pt]
\cmidrule(l){2-4}
\addlinespace[5pt]
& $\myrlambda$ & $134$ & $863$ \\
& Present work & S\&W (2002) & S\&W (2002) \\
\addlinespace[5pt]
\midrule
\addlinespace[8pt]
$2$ & $0.731 \pm 0.017$ & $0.68 \pm 0.05$ & $0.69 \pm 0.05$ \\
\addlinespace[8pt]
$3$ & $1$ & $1.00 \pm 0.05$ & $0.98 \pm 0.05$ \\
\addlinespace[8pt]
$4$ & $1.198 \pm 0.088$ & $1.17 \pm 0.05$ & $1.24 \pm 0.05$ \\
\addlinespace[8pt]
$5$ & $1.360 \pm 0.167$ & $1.37 \pm 0.05$ & $1.49 \pm 0.05$ \\
\addlinespace[8pt]
$6$ & $1.516 \pm 0.096$ & $1.53 \pm 0.05$ & $1.71 \pm 0.05$ \\
\addlinespace[5pt]
\bottomrule
\end{tabular}
\caption{The inertial range transverse scaling exponents, $\zeta_p$,
measured in the current experiment using the ESS method and those
measured by \cite{shen:2002}, denoted as S\&W (2002) above, in an
unsheared wind tunnel homogeneous and isotropic turbulent flow at 
two different $R_{\lambda}$ of $134$ and $863$.}
\label{table:measured_exponents}
\end{center}
\end{table}

\section{Anomalous scaling}
\label{sec:anomalous_scaling}

We now turn to the issue of how the transverse scaling exponents, 
$\zeta_p$, vary with order, $p$.  
In the absence of intermittency, Kolmogorov scaling predicts a linear
relationship between $\zeta_p$ and $p$, see equation~\ref{eq:zetap_k41}.  
Figure~\ref{fig:intermittency_plot} and table~\ref{table:measured_exponents}
show the mean of $\zeta_p$ in tables~\ref{table:zeta}, \ref{table:zeta4},
\ref{table:zeta5}, and \ref{table:zeta6} as a function of $p$.  
The error bars were the standard deviation of the values of $\zeta_p$ in
each table.  
Our observation is that, within this error, the exponents show an anomalous
scaling (departure from Kolmogorov prediction), similar to what is observed 
for the scaling exponents measured in a nonsheared wind tunnel
homogeneous and isotropic turbulent flow at an $R_{\lambda}$ of $134$ 
\cite[][]{shen:2002}.  
The values of both sets of exponents are below the values obtained 
by \cite{shen:2002} at an $R_{\lambda}$ of $863$.  
The difference is less than $6\%$ at the fourth order and increases
monotonically up to about $13\%$ at the sixth order.  
We suspect that the reason for the difference is a combination of 
Reynolds number and large scale anisotropy effects.  
We believe that if we could achieve higher Reynolds numbers, the 
values of the scaling exponents would increase.  

\section{Conclusions}

We studied the influence of the large scale anisotropy on the 
structure functions of order 4, 5, and 6.  
We found indications that anisotropy in the velocity field
intensifies the asymmetry of the probability density of the
velocity increments.  
We have shown that the small scales of the flows are highly
intermittent, as indicated by the measurements of the kurtosis
and hyper-kurtosis, and that both kurtosis and hyper-kurtosis 
asymptote to their gaussian limits at the large scales.  
Within the experimental uncertainty, we found evidence that the 
scaling exponents measured using the ESS method are 
independent of anisotropy.  
The exponents showed a departure from the Kolmogorov prediction
and their values were below those measured in a nonsheared wind 
tunnel homogeneous turbulent flow at an $R_{\lambda}$ of $863$ 
\cite[][]{shen:2002}.  
This is a subtle difference and we believe the difference would become
smaller if we increase the Reynolds number.  
\chapter{The integral scales of turbulence}
\label{chap:integralscale}

Recently, there is increasing evidence to suggest that 
macroscopic properties of a turbulent flow are linked to
the turbulence energy spectrum.  
Some of the macroscopic properties considered
include the cross-sectional mean velocity of gravity-driven 
open channel flows \cite[][]{gioia:2001}, the depth of  turbulent 
cauldron \cite[][]{gioia:2005}, the friction factor \cite[][]{gioia:2006, 
guttenberg:2009, tran:2010}, and the mean-velocity profile of 
pipe flows \cite[][]{gioia:2010}.  
The authors established their findings based on the assumption that
Kolmogorov's theory for homogeneous and isotropic turbulence 
is also applicable to anisotropic and inhomogeneous flows
\cite[][]{knight:1990, moser:1994, lundgren:2003}.  
They explain their findings in terms of the varying habits of 
momentum transfer with varying sizes of turbulent eddies.  
Here, we study the physical space equivalent of the spectrum --
the correlation of velocity fluctuations separated in space.  
Our results highlight the interesting feature that correlation 
functions measured in different flow directions collapse onto 
one self-similar curve when appropriately scaled with the 
anisotropy that drives the turbulence at the large scale.  
The scaling form allows us to express the integral length,
which is representative of the length scale of the most
energetic eddies (\cite{batchelor:1956}, chapter 6), 
as a power law of its characteristic velocity, with an exponent 
related to the inertial range scaling exponent.  
Our work complements the existing studies and suggests that 
self-similarity and Kolmogorov theory are relevant to the 
phenomenological description of the large scale of anisotropic
turbulent flows.  

\section{The integral scale and correlation function}
\label{sec:integralscale}

We begin with a brief introduction to the integral length of 
the correlation function in order to facilitate the discussion that follows.  

As described in the introductory chapter, an important tenet in the 
Richardson picture of the turbulence energy cascade is that the 
energy dissipation rate per unit mass in a turbulent flow is finite 
(see e.g. \cite{frisch:1995}, 5.2).  
This is summarized in the following empirical law 
\cite[][]{taylor:1935}
\begin{equation}
\label{eq:dissipation_law}
\epsilon = \dfrac{A \, u^{\prime 3}}{L} \,.
\end{equation}
Here, $A$ is a number of order unity, $u^{\prime}$ is a 
velocity scale characteristic of the most energetic eddies, 
and $L$ their characteristic length scales.  
This order of magnitude assumption has found support from 
a theoretical consideration by \cite{lohse:1994}.  
Starting from an energy balance equation derived from
the Navier-Stokes equation with a mean field approach, 
\cite{lohse:1994} showed that $A \approx (a/C_2)^{3/2}$, 
where $C_2$ is the Kolmogorov constant for the second-order
structure function (see equations~\ref{eq:dll_k41} and \ref{eq:dnn_k41})
and $a \approx 1.25$.  
Recent updates on the empirical law of energy dissipation rate
can be found in \cite{sreenivasan:1984} and \cite{pearson:2002}.  
A simple interpretation of the above empirical law is that
the decay time of the turbulence is a few characteristic
periods of the most energetic eddies (\cite{batchelor:1956}, 6.1).  
Identifying $u^{\prime}$ with the RMS velocity fluctuation, 
$u^{\prime 2} / \epsilon$ can be regarded as the time scale
of the decay of energy, and $L / u^{\prime}$ can be 
regarded as the characteristic period of the most energetic 
eddies.  

The above empirical relation has found wide use in estimating
the energy dissipation rate of turbulent flows.  
Measuring $u^{\prime}$ is relatively simple.  
Single-point measurements of the turbulent velocity
fluctuation suffice to yield $u^{\prime}$.  
In contrast to $u^{\prime}$, measuring $L$ is not so
straightforward because it is not directly accessible 
(see e.g. \cite{batchelor:1956}, appendix to 6.1).  
If we consider the three-dimensional energy spectrum, 
$\Phi (k)$, it is permissible to think of the part of the function 
around the maximum, at wavenumber $\kappa_{0}$, as responsible 
for supplying energy into the flow (the most energetic eddies).  
The reciprocal of the wavenumber at which this maximum
occurs, $L_{E} = 1 / \kappa_0$, then characterizes the 
principal energy bearing part of the spectrum (the most energetic
eddies).  
On the other hand, if the contribution from higher wavenumbers
is negligible, we may approximate $L$ with the transverse 
integral length (see e.g. \cite{batchelor:1956})
\begin{equation}
L_{g} = \dfrac{3 \, \pi}{8} \, \dfrac{\int_{0}^{\infty} \kappa^{-1} \, 
\Phi (\kappa) \, \mathrm{d}\kappa}{\int_{0}^{\infty} \Phi (\kappa) \,
\mathrm{d}\kappa} \,,
\end{equation}
which, for isotropic turbulence, is related to the one-dimentional 
transverse correlation function in the following way
\begin{equation}
L_{g} = \int_{0}^{\infty} g (r) \, \mathrm{d}r \,.
\end{equation}
Here, $g(r)$ is the diagonal element of the much more general
correlation tensor
\begin{equation}
g_{ij} (\boldsymbol{r}) = \dfrac{\langle u^{\prime}_{i} (\boldsymbol{r}) \, u^{\prime}_{j} (0) \rangle}{\sqrt{\langle u^{\prime 2}_{i} (\boldsymbol{r}) \rangle \, \langle u^{\prime 2}_{j} (0) \rangle}} \,,
\end{equation}
which is the velocity correlation tensor introduced in 
section~\ref{sec:kolmogorov_theory} normalized by the fluctuations.  
The careful reader may note that, in general, the velocity fluctuations
are functions of space and time.  
The time dependence of the fluctuations, and consequently $g_{ij} 
(\boldsymbol{r})$, has been suppressed for the simplicity
of discussion.  
If the velocity field satisfies the homogeneity and isotropy conditions,
then the correlation tensor can be expressed in terms of two scalar 
functions (\cite{batchelor:1956}, 3.4).  
With reference to experimental work, we usually make the choice 
$\boldsymbol{r} = r \boldsymbol{e}_1$ and introduce the
longitudinal correlation function
\begin{equation}
f(r) = g_{11} (r) \,,
\end{equation}
and the transverse correlation functions
\begin{equation}
g(r) = g_{22} (r) = g_{33} (r) \,.
\end{equation}
In this way, we obtain the form for the correlation tensor first
derived by \cite{karman:1938} and systematically formulated by
\cite{robertson:1940}
\begin{equation}
g_{ij} (\boldsymbol{r}) = 
g (r) \, \delta_{ij} + [f(r) - g(r)] \, \dfrac{r_i \, r_j}{r^2} \,.
\end{equation}
$f(r)$ and $g(r)$ are related by the continuity equation in much the 
same way as the longitudinal and transverse structure functions are
\cite[e.g.][]{pope:2000}.  
Relevant to our experiments are the following two transverse 
correlation functions
\begin{align}
g_{r_2 r_2} (0,0,z) &= \dfrac{\langle u_{r_2}^{\prime} (0,0,z) \, u_{r_2}^{\prime} (0,0,0) \rangle}{\sqrt{\langle u_{r_2}^{\prime 2} (0,0,z) \rangle 
\langle u_{r_2}^{\prime 2} (0,0,0) \rangle}} \,, \\
g_{zz} (r_1,0,0) &= \dfrac{\langle u_{z}^{\prime} (r_1,0,0) \, u_{z}^{\prime} (0,0,0) \rangle}{\sqrt{\langle u_{z}^{\prime 2} (r_1,0,0) \rangle 
\langle u_{z}^{\prime 2} (0,0,0) \rangle}} \,.
\end{align}
They are related simply to the structure functions by expanding 
the product in equation~\ref{eq:sf_correlation}
\begin{equation}
g (x) = \dfrac{1}{2} \, \bigg[ \sqrt{\dfrac{\langle u^{\prime 2} (x) \rangle}{\langle u^{\prime 2} (0) \rangle}} + \sqrt{\dfrac{\langle u^{\prime 2} (0) \rangle}{\langle u^{\prime 2} (x) \rangle}} - \dfrac{D(x)}{\sqrt{\langle u^{\prime 2} (x) \rangle \langle u^{\prime 2} (0) \rangle}} \bigg] \,,
\end{equation}
which, under the assumption of homogeneity $\langle u^{\prime 2} (x)
\rangle = \langle u^{\prime 2} (0) \rangle = u^{\prime 2}$, reduces to
\begin{equation}
g(x) = 1 - \dfrac{D(x)}{2 \, u^{\prime 2}} \,.
\label{eq:g_D}
\end{equation}
Here, $x$ is understood to represent $z$ or $r_1$.  
Given that $L_g$ is an important control parameter in aerodynamic 
stability analyses of bridges with sharp edges \cite[e.g.][]{fransos:2010}, 
as well as being a key parameter in the design of gas turbine engines 
\cite[e.g.][]{vanfossen:1994a, vanfossen:1995, barrett:2001, carullo:2011}, 
a more fundamental understanding of $L_g$, and thus of $g(r)$, is needed.  

\section{Scaling and self-similarity}

There is no rigorous theory that yields the shape for 
the longitudinal and transverse correlation functions in
fully developed turbulence, but we have a rough picture 
of how these curves should look like from experiments 
\cite[e.g.][]{comtebellot:1971} and numerical simulations 
\cite[e.g.][]{kim:1987}.  
\cite{batchelor:1948b} studied the decay of turbulence and
showed that in the final period of decay of the turbulence, 
when the effects of inertial forces are negligible, 
the longitudinal correlation function asymptotes to a self-similar 
function
\begin{equation}
f(r,t) = \exp\bigg(-\dfrac{r^2}{8 \, \nu \, t} \bigg) \,,
\end{equation}
corroborating the hypothesis made by 
\cite{karman:1938} (for more details on the self-preservation 
hypothesis, see e.g. \cite{monin:1975}, 16).  
Temporally self-similar correlation functions were also derived 
analytically in decay problems of Burgers turbulence 
\cite[][]{gurbatov:1997} and passive scalar in Kraichnan's 
model \cite[][]{eyink:2000}.  
\cite{ewing:2007} examined the correlation function of 
streamwise velocity component in the far field of an 
axisymmetric jet and showed that the similarity solutions
of the governing equations for the correlation functions 
depend only on the separation distance between 
the points in the streamwise similarity coordinate, 
namely $\psi^{\prime} - \psi$, where the similarity coordinate
$\psi = \ln x$ is measured from a virtual origin in space $x$.  
The spatial self-similarity of correlation functions in anisotropic 
turbulence considered here complements the existing works.  

The similarity argument to be presented here is a special
case of Widom's scaling (see e.g. \cite{huang:1987}, 16.5) with the 
dimension of homogeneity $d$ equals zero \cite[][]{eyink:2011}, 
and is in the same vein as \cite{goldenfeld:2006}.  
Two notable features of this approach are the universal
scaling function that governs the correlation function 
and the reduction of the number of variables in the function.  
The scaling form we propose for the transverse correlation
function is
\begin{equation}
g(x, u^{\prime}) = \mathscr{G} (x \, u^{\prime \beta}) \,,
\end{equation}
where $\mathscr{G} (\xi)$ is a universal scaling function of 
a single variable $\xi$, and $\beta$ is an exponent to be 
determined.  
To determine $\beta$, we connect the scaling function 
$\mathscr{G} (\xi)$ to that in the inertial range.  
We note that $g(x, u^{\prime})$ scales as
\begin{equation}
g(x, u^{\prime}) \propto \dfrac{x^\zeta}{u^{\prime 2}} \,,
\end{equation}
in the inertial range.  
This can be seen by substituting the inertial range 
scaling of the structure function, 
$D(x) \propto x^{\zeta}$, into equation~\ref{eq:g_D}.  
This requires that $\mathscr{G} (\xi) \propto \xi^{\zeta}$, 
in the inertial range, and therefore $\beta = -2/\zeta$.  
Thus, the scaling form for the correlation function is
\begin{equation}
g(x, u^{\prime}) = \mathscr{G} (x \, u^{\prime - 2 / \zeta}) \,.
\label{eq:g_scaling_form}
\end{equation}
The scaling form predicts that correlation functions will
collapse onto a single curve when plotted as $g$ against 
$x \, u^{\prime -2/\zeta}$.  

The scaling form in \ref{eq:g_scaling_form} implies a relationship
between the integral length and the large scale velocity fluctuation.  
By changing the variable of integration to $\xi = x \, u^{\prime -2/\zeta}$, 
we obtain
\begin{equation}
L = u^{\prime 2/\zeta} \, \mathscr{I} \,,
\end{equation}
where $\mathscr{I} = \int_{0}^{\infty} \mathscr{G} (\xi) \, 
\mathrm{d} \xi$ is a constant which may depend on some
other parameters of the flow, e.g. $R_{\lambda}$ or $\zeta$.  
This dependence can be eliminated if we form the ratio of two
integral scales obtained from two different correlation functions
\begin{equation}
\dfrac{L_1}{L_2} = 
\bigg( \dfrac{u^{\prime}_1}{u^{\prime}_2} \bigg)^{2/\zeta} \,.
\label{eq:u_L_scaling}
\end{equation}
We cannot test these predictions in isotropic turbulence
because the correlation functions and RMS velocity fluctuation 
are isotropic by definition.  
If we can drive the flow away from isotropy, as we can in our
flow generator, we can test these predictions.  

\section{Integral length scaling}

\begin{figure}
\begin{center}
\includegraphics[scale=1]{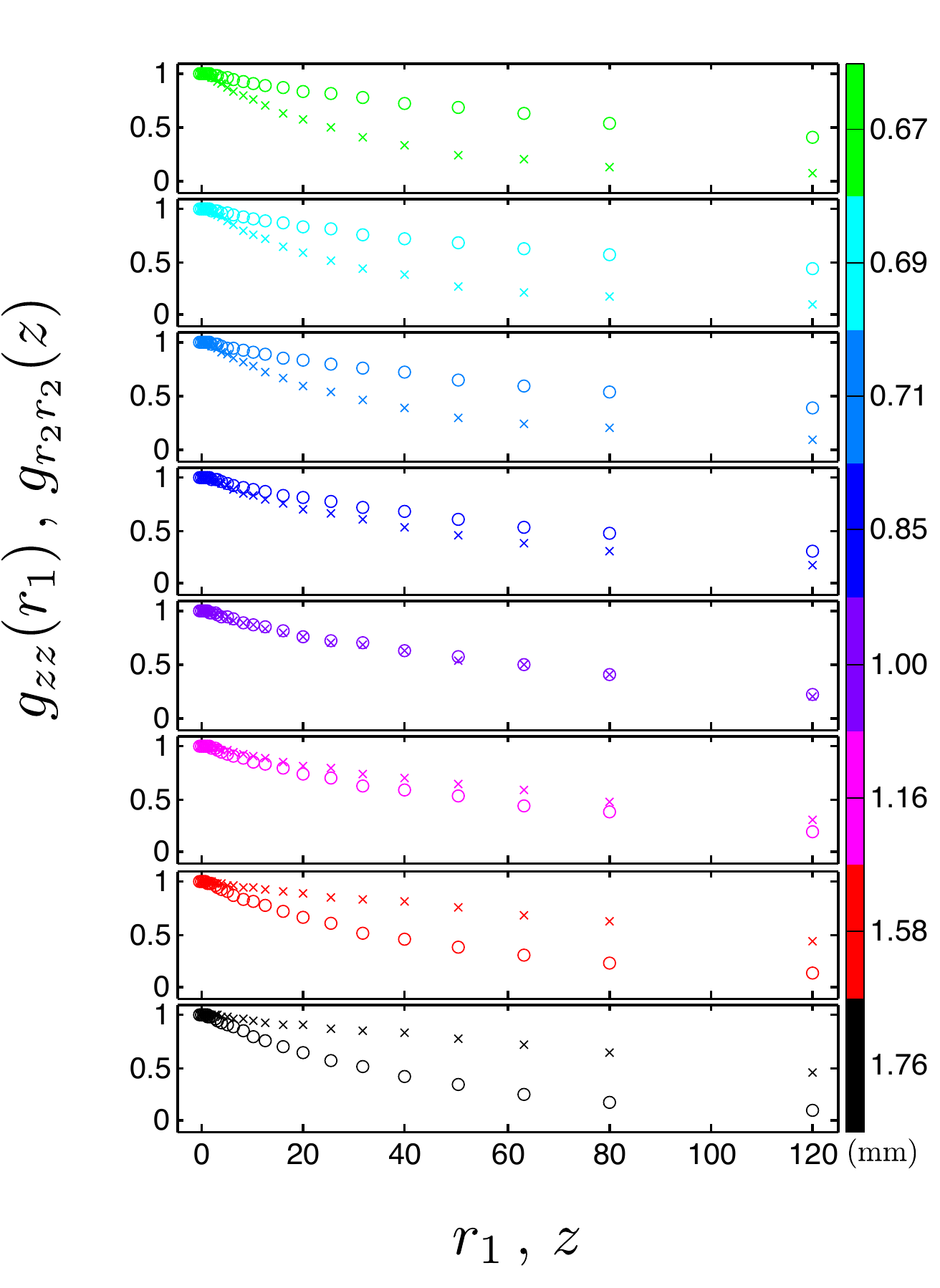}
\caption{%
The figure shows the 8 pairs of transverse correlation functions
obtained in 8 different flows with varying degree of large scale
anisotropy.  
Circles (${\ocircle}$) are for $g_{r_2 r_2} (z)$ and crosses 
($\times$) are $g_{zz} (r_1)$.  
The error bars are smaller than the symbols.  
The values in the color bar are the fluctuations anisotropy 
measured at the center of the soccer ball for various values 
of the anisotropy of the forcing.}
\label{fig:corrfunc}
\end{center}
\end{figure}

Figure~\ref{fig:corrfunc} shows 8 pairs of correlation functions, 
$g_{r_2 r_2} (z)$ and $g_{zz} (r_1)$, obtained in 8 different flows 
with different large scale anisotropies, but with approximately 
the same energy dissipation rate and Reynolds number.  
We found that correlation functions measured in different directions 
in anisotropic flows were different when plotted against separation.   
Note that we normalized the correlation functions by the 
velocity fluctuations at both points, $0$ and $r_1$ (or $z$), 
although homogeneity would allow us to set them equal, 
because it compensated for the small inhomogeneity of our flow.  
In addition, we divided the correlation functions by their values 
at zero separation, that is, when the distance between probe 
volumes was zero.  
The values of the correlations at zero separation were not equal 
to one because of noise in the signals, but were approximately 
equal to $0.98$, for all correlation functions.  

\begin{figure}
\begin{center}
\includegraphics[scale=0.75]{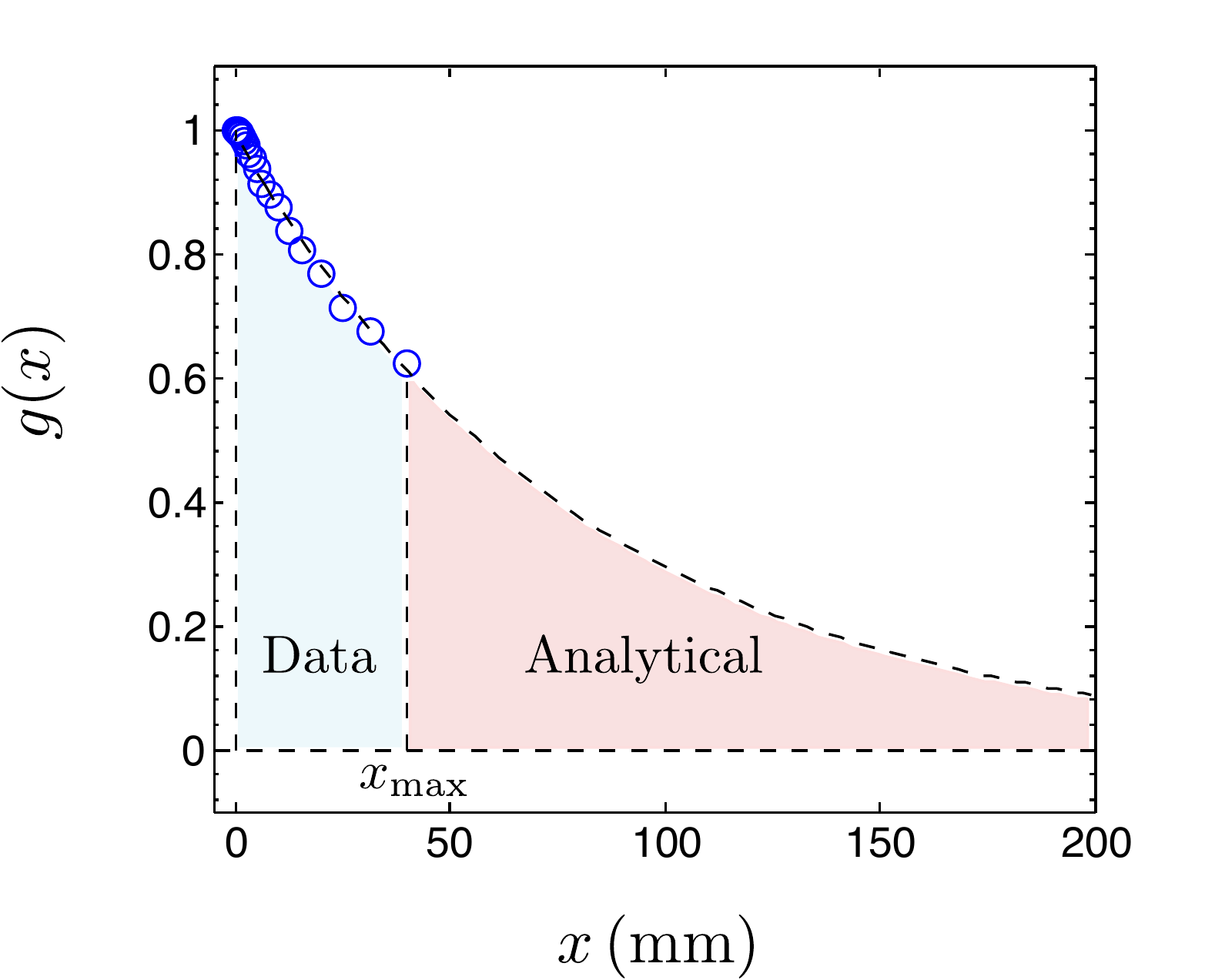}
\caption{The extrapolation of the partially measured correlation function.  
The blue circular symbols (${\color{blue} \ocircle}$) are the experimental
data.  The dashed line is the exponential extrapolated from the data.  
The area in the region shaded in blue is numerically integrated and 
the area in the region shaded in pink is analytically calculated.  
The sum of the two yields the integral length.}
\label{fig:fit_exponential}
\end{center}
\end{figure}

\begin{figure}
\begin{center}
\includegraphics[scale=0.75]{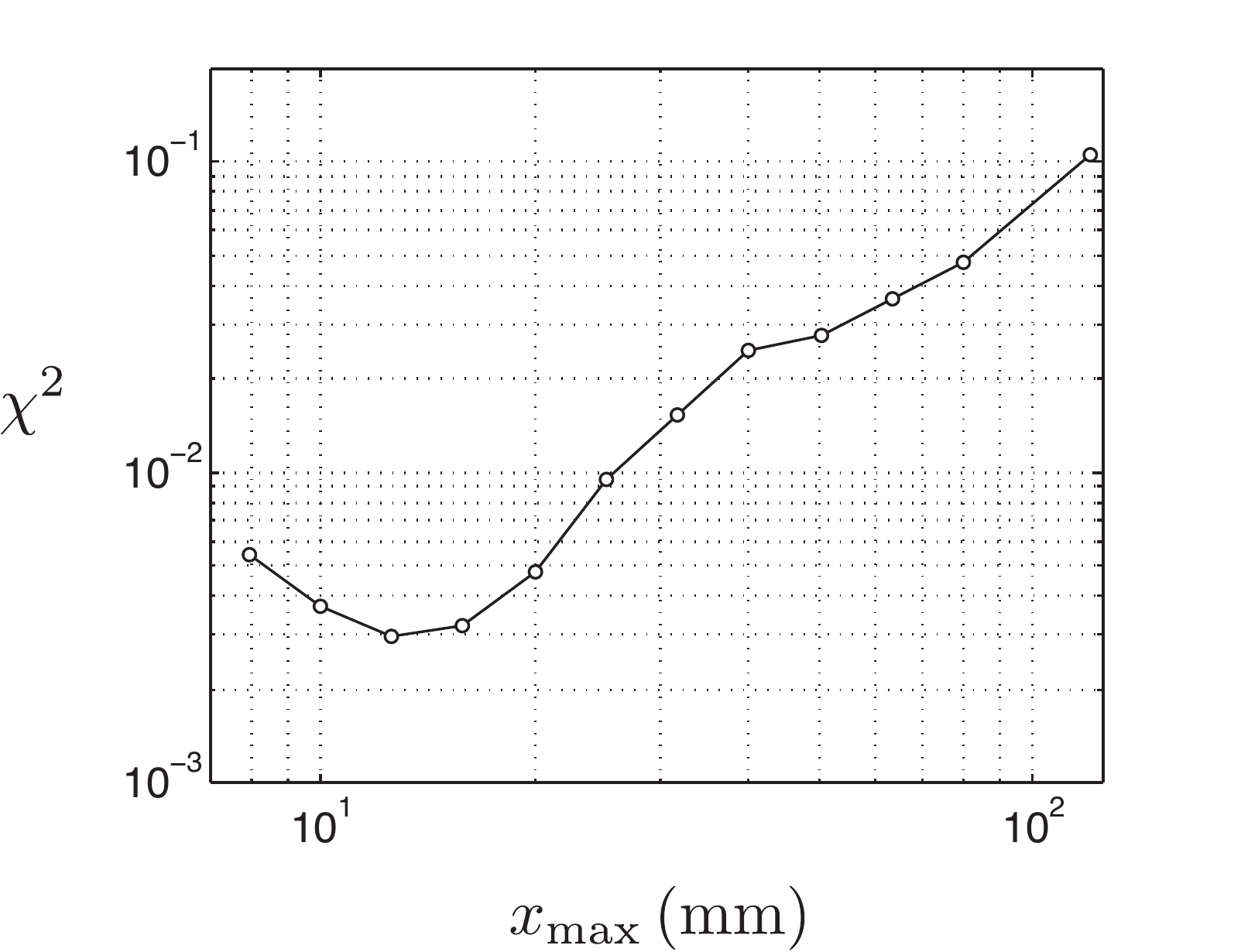}
\caption{The sum of squares of residuals in the determination 
of $x_{\mathrm{max}}$ that yields the exponential from a 
least squares fit of an exponential tail to the correlation function data.}
\label{fig:chi_sqr}
\end{center}
\end{figure}

In order to test equation~\ref{eq:u_L_scaling}, we need to 
estimate the integral scale.  
\cite{batchelor:1948a} observed that transverse correlation functions 
cross zero before they return to zero asymptotically, but it is not
unusual that experimental measurements do not resolve this limit 
because of the finite size of experiments 
\cite[e.g.][]{hwang:2004, dejong:2009, siebert:2010}.  
In fact, the first zero crossing would occur outside the region of 
our flow that is approximately homogeneous.  
On the other hand, \cite{comtebellot:1971} argued that integral 
scales in finite flows should strictly be zero, if data could be 
collected that extended far enough.  
This can be seen by considering the case $k=0$ in the 
Wiener-Khinchin relation between the one-dimensional transverse 
spectrum, $E_{22} (k)$, and its correlation function, $R_{22} (r)$
(see equations~\ref{eq:rtoe} and \ref{eq:etor}), which yields
\begin{equation}
L_g = \dfrac{\pi \, E_{22} (0)}{2 \, \langle u^{\prime 2} \rangle} \,.
\end{equation}
If we consider, however, the spectrum given by
$E_{22}(k) = \tilde{u} (k) \, \tilde{u}^{*} (k)$, where the Fourier coefficients 
of the velocity field averaged over a distance $L$ in physical space are 
$\tilde{u} (k) = \tfrac{1}{L} \, \int_{0}^{L} u^{\prime} (r) \, 
\mathrm{e}^{-i k r} \, \mathrm{d} r$, given that the mean velocity 
$\tfrac{1}{L} \int_{0}^{L} u^{\prime} (r) \, \mathrm{d}r$ is zero, it follows 
that the coefficient of the zeroth Fourier mode is zero, $\tilde{u} (0) = 0$.  
It can be immediately seen that $E_{22}(0) = 0$, and thus $L_g = 0$.  
Only hypothetical infinite flows could have nonzero integral scales.  
\cite{comtebellot:1971} note that the only way to construct 
a nonzero integral scale from real data is to make `simple extrapolations'
of partially measured correlation functions in a reasonable way.  
We chose to extrapolate the correlation functions with 
exponential functions \cite[e.g.][]{vanfossen:1994a, 
vanfossen:1994b, dejong:2009}.  
We are aware that physically plausible extrapolations of the 
transverse correlation functions must include zero crossing, but 
\cite{lenschow:1986} found that, following the suggestion of 
\cite{comtebellot:1971}, approximating the one-dimensional 
longitudinal spectrum (see equation~\ref{eq:r11_e11}) with a 
downward parabola with zero slope at zero wavenumber gives 
a correlation function with oscillatory tail.  
Integrating this correlation function up to its first zero-crossing
yields a value that overestimates the integral length by no more
than 5\%.  
We found empirically that using more elaborate functions, such
as the modified Bessel function, for the extrapolations 
did not alter our results for the integral length scaling exponent.  
The difference between the exponents calculated using different
fitting functions is less than $10\%$.  
We suggest that this is because we only consider ratios of integral 
scales, not the integral scales themselves, and the influence of the 
two neglected negative parts of the correlation functions cancel 
each other.  
Figure~\ref{fig:fit_exponential} shows how a typical extrapolation is
done.  
For each correlation function, we fit an exponential using a least 
squares algorithm \cite[e.g.][]{draper:1998} to the last 15 data points 
collected between $0 \leqslant x \leqslant x_{\mathrm{max}}$, 
where the maximum separation, $x_{\mathrm{max}}$, ranged from
$8$~mm to $120$~mm.  
We then chose the value of the maximum separation, 
$x_{\mathrm{max}}$, that minimized $\chi^2$, the sum of squares of
the vertical differences between the experimental, $g(x_i)$, and the 
fitted values of the correlation function $\tilde{g} (x_i)$
\begin{equation}
\chi^2 = \sum_{i=1}^{N} (g(x_i) - \tilde{g} (x_i))^2 \,,
\end{equation}
where $N$ is the total number of samples used in the fit.  
Figure~\ref{fig:chi_sqr} shows the variation of $\chi^2$ with 
$x_{\mathrm{max}}$.  
It can be inferred that $x_{\mathrm{max}} = 13$~mm yields the fit
with the smallest sum of squares of residuals.  
Data for separations $x>x_{\mathrm{max}}$ were extrapolated
with exponential tails, whose integral we added to the numerical 
integration of the data, to yield the integral length.  
The values of the integral scales did depend on $x_{\mathrm{max}}$, 
leading to a variation in the value of the exponent 
(see figure~\ref{fig:exp_xmax}).  
We then used the standard error (see appendix~\ref{app:statistical_tools})
of value of the integral scale calculated with different 
$x_{\mathrm{max}}$ to estimate the error of our integral scale 
measurement.  

We note that, since $u^{\prime}_i$ is a transverse velocity, the 
subscript in $u^{\prime}_i$ indicates the direction of the velocity 
and the subscript in $L_i$ indicates the direction of the separation 
over which the integral length is calculated.   
Therefore, the expected scaling is a power law with a 
negative exponent
\begin{equation}
\dfrac{L_z}{L_r} = 
\bigg( \dfrac{u^{\prime}_z}{u^{\prime}_r} \bigg)^{-2/\zeta} \,,
\end{equation}
in conformity with our definition of axial and radial directions.  
We test this relationship in figure~\ref{fig:uratio_Lratio}.  
Note that we have used the definition given in equation~\ref{eq:uratio2}
for $u^{\prime}_z / u^{\prime}_r$, in compliance with the definition
of $u^{\prime}$ in \ref{eq:u_L_scaling}.  
The integral length scaling exponent obtained from a least squares 
straight-line fit
\begin{equation}
\dfrac{L_z}{L_r} = 
\bigg( \dfrac{u^{\prime}_z}{u^{\prime}_r} \bigg)^{\beta} \,,
\end{equation}
to the data is $\beta = -2.56 \pm 0.07$, for which 
the uncertainty was estimated with the standard error of the slope 
in the straight-line fit (see appendix~\ref{app:statistical_tools}).  
To check for consistency, we calculated the value for $\beta$ with
the mean value of the inertial range scaling exponents measured 
using ESS method, $\zeta_2$ (see section~\ref{sec:scaling_exponents}), 
for which we find $\beta_2 = -2/\zeta_2 = -2.74 \pm 0.06$, assuming 
that the percentage error remains the same ($\Delta \beta_2 / 
\beta_2 = \Delta \zeta_2 / \zeta_2$).  
The value for $\beta_2$ is no more than $7\%$ greater than $\beta$.  
Minor differences between the two scaling exponents may arise 
from the difference between the exponential and the real correlation
curve, which is at present unknown.  
A function that takes into account the transition range between
the inertial scales and the large scales will be able to account
for the difference.  
Nevertheless, even with these simple methods, we find
excellent agreement between the two exponents.  

In addition, we note that the value of the exponent showed a departure
from the Kolmogorov prediction, which has a value of $-3$ for $\beta$
if the value of $\zeta$ is $2/3$.  
The difference is very likely due to a Reynolds number effect.  
We suspect that the difference would become smaller if we increase
the Reynolds number.  
\begin{figure}
\begin{center}
\includegraphics[scale=0.75]{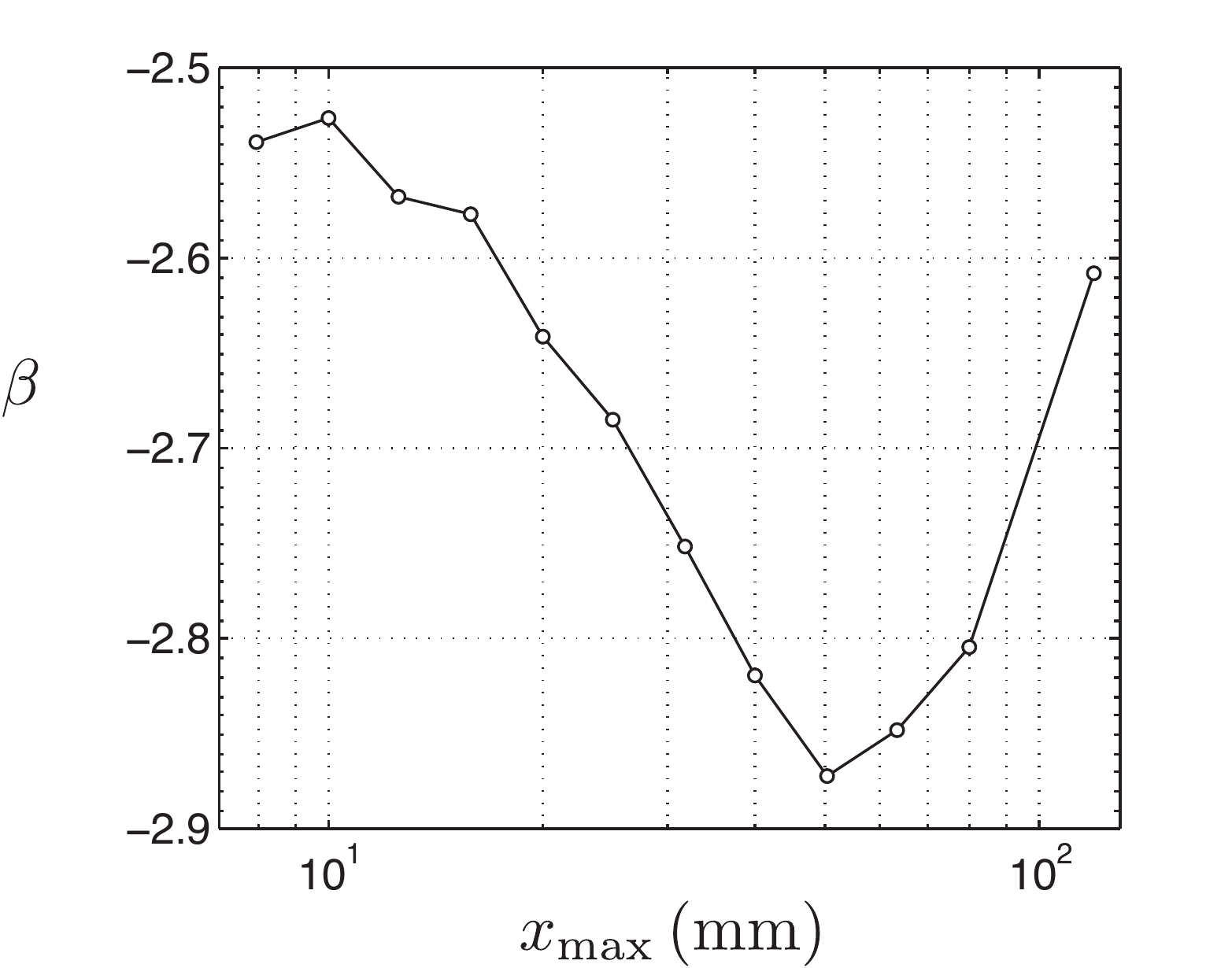}
\caption{The variation in the value of the exponent with
the maximum separation in the least squares fit of exponential
tails to the data.  }
\label{fig:exp_xmax}
\end{center}
\end{figure}
\begin{figure}
\begin{center}
\includegraphics[scale=0.75,type=pdf,ext=.pdf,read=.pdf]{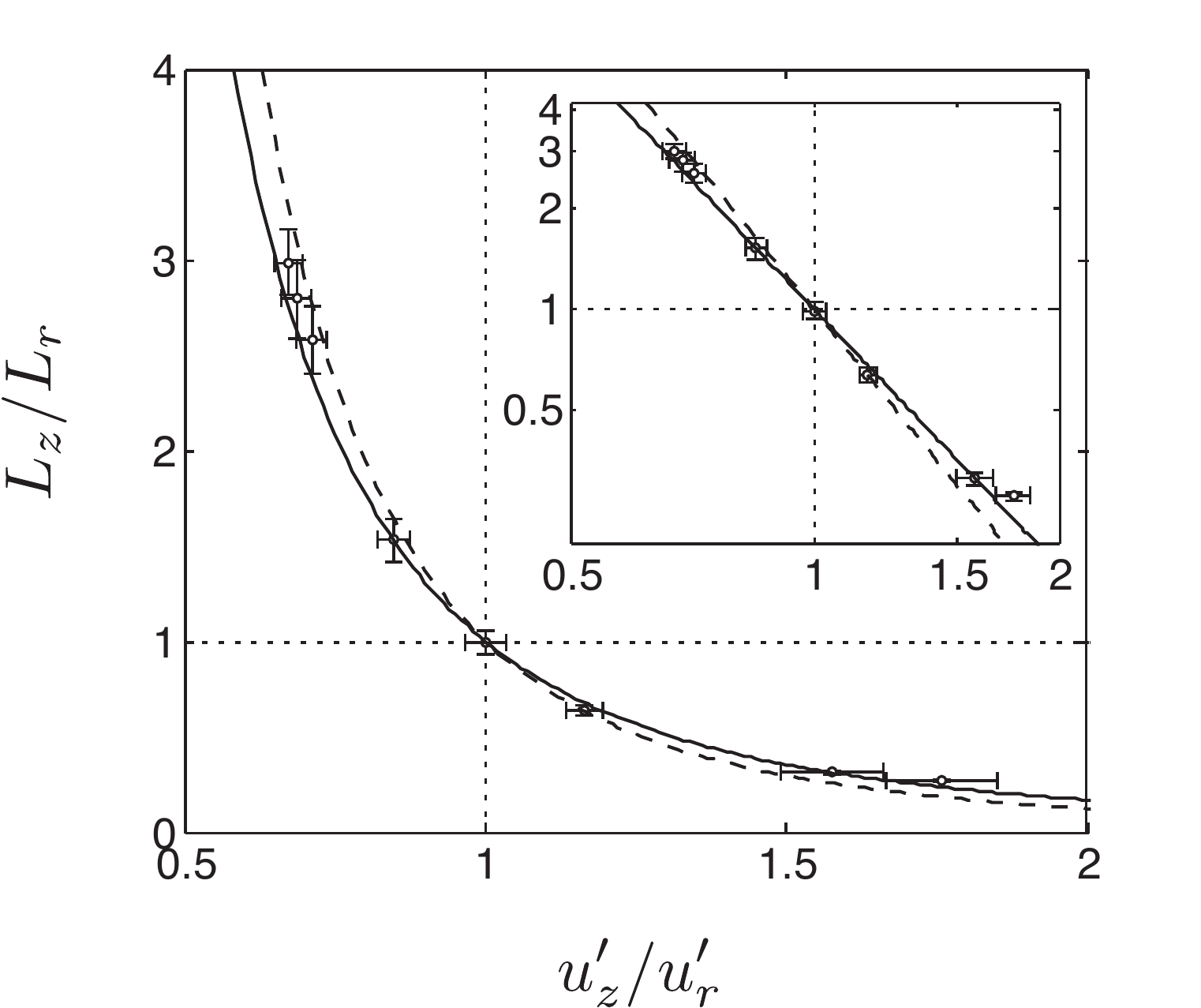}
\caption{The main figure shows the ratio of integral scales, 
$L_z / L_r$, as a function of the anisotropy in the fluctuating 
velocities, $u^{\prime}_z / u^{\prime}_r$.  
The solid line is the least squares fit to the data and the dashed
line is the power law with exponent $-3$ predicted by the 
Kolmogorov inertial range scaling argument.  
The inset shows the same data on a logarithmic scale.
The straight line in the inset has a slope of $-2.56$.}
\label{fig:uratio_Lratio}
\end{center}
\end{figure}
\begin{table}
\begin{center}
\begin{tabular}{r@{$\, \pm \, $}lr@{$\, \pm \, $}lr@{$\, \pm \, $}lr@{$\, \pm\, $}l}
\toprule
\addlinespace[8pt]
\multicolumn{2}{c}{$u^{\prime}_z / u^{\prime}_r$} & 
\multicolumn{2}{c}{$L_z$ (mm)} & 
\multicolumn{2}{c}{$L_r$ (mm)} & 
\multicolumn{2}{c}{$L_z / L_r$} \\
\addlinespace[5pt]
\midrule
\addlinespace[8pt]
$0.67$ & $0.03$ & $100$  & $4.3$  & $33.5$ & $0.6$ & $2.99$ & $0.18$ \\
$0.69$ & $0.03$ & $95.6$ & $4.8$ & $34.2$ & $0.9$ & $2.80$ & $0.21$ \\
$0.71$ & $0.03$ & $93.0$ & $4.1$ & $36.1$ & $1.0$ & $2.58$ & $0.18$ \\
$0.84$ & $0.03$ & $76.9$ & $3.3$ & $50.2$ & $1.9$ & $1.53$ & $0.12$ \\
$1.00$ & $0.04$ & $67.5$ & $2.5$ & $68.0$ & $2.1$ & $0.99$ & $0.07$ \\
$1.16$ & $0.03$ & $61.0$ & $1.9$ & $95.4$ & $2.6$ & $0.64$ & $0.04$ \\
$1.58$ & $0.09$ & $45.1$ & $1.0$ & $144$  & $3.9$ & $0.31$ & $0.02$ \\
$1.76$ & $0.09$ & $42.6$ & $0.5$ & $155$  & $4.5$ & $0.28$ & $0.01$ \\
\addlinespace[5pt]
\bottomrule
\end{tabular}
\caption{The numerical data for axial integral lengths 
($L_z$), radial integral lengths ($L_r$), 
and the ratio between the two.}
\label{table:integral_length}
\end{center}
\end{table}

\section{Correlation functions collapse}

\begin{figure}
\begin{center}
\includegraphics[scale=1,type=pdf,ext=.pdf,read=.pdf]{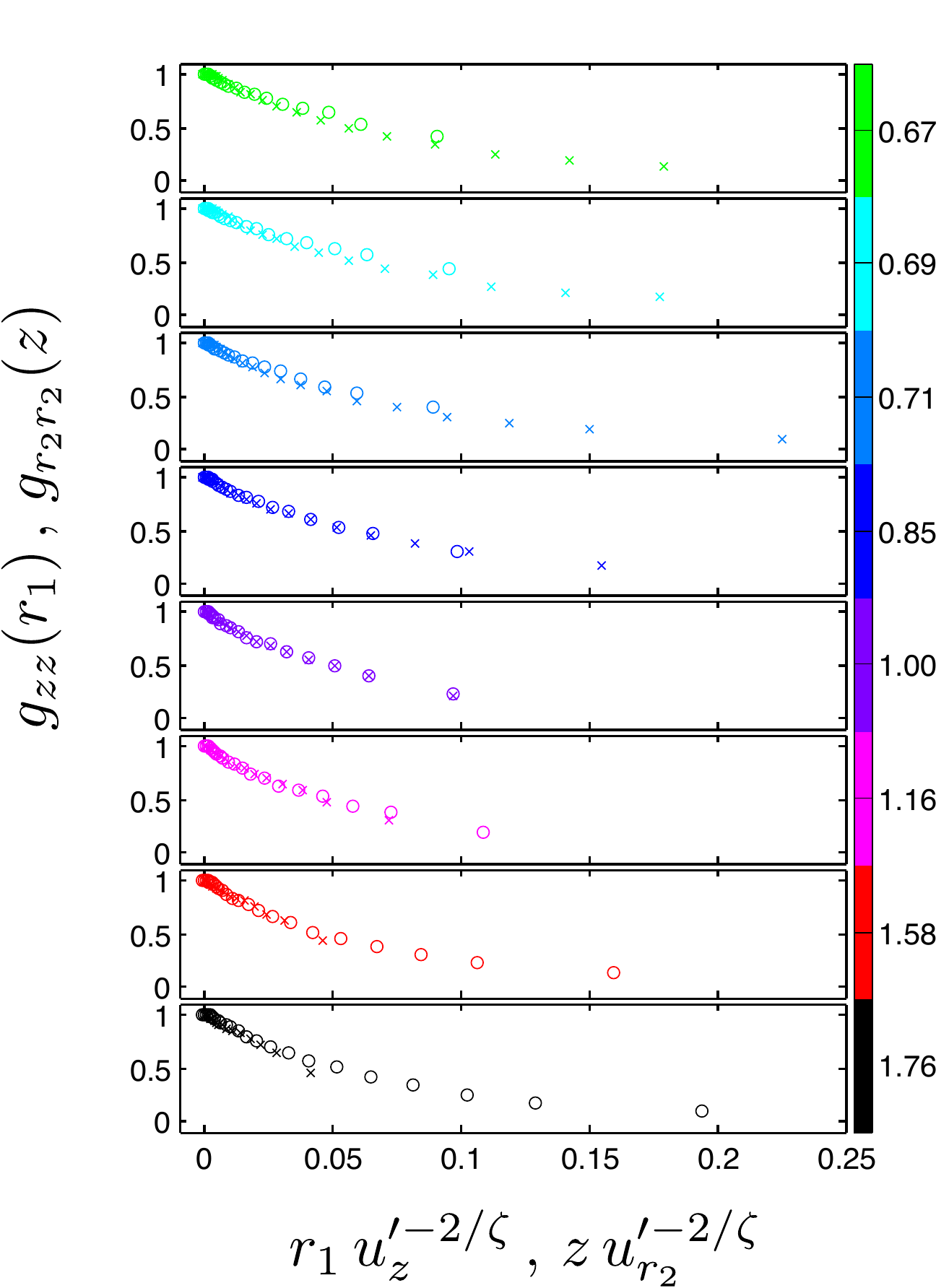}
\caption{The figure shows the same correlation functions from 
figure~\ref{fig:corrfunc} plotted against the reduced variables, 
$r_1 \, u_{z}^{\prime - 2/\zeta}$ and $z \, u_{r_2}^{\prime - 2/\zeta}$.  
Here, $\zeta = 0.73$.  
Circles (${\ocircle}$) are for $g_{r_2 r_2} (z)$ and crosses ($\times$) 
are for $g_{zz} (r_1)$.  
The values in the color bar are the fluctuations anisotropy 
measured at the center of the soccer ball.}
\label{fig:corrfunc_collapse}
\end{center}
\end{figure}

The test of data collapse is shown in figure~\ref{fig:corrfunc_collapse},
The value for the inertial range scaling exponent is $\zeta = 0.73$, 
obtained using the ESS method (see section~\ref{sec:scaling_exponents}).  
It can be seen that correlation functions stretching out in different
directions on the plane in figure~\ref{fig:corrfunc} collapse 
onto one single curve.  
Small deviations from the data collapse are visible, especially at 
large separations, which may be a reflection of the inhomogeneity
occurring outside the turbulent boundary, where the effects of 
individual jets are felt.  

\section{Conclusions}

We have presented, in the foregoing pages, a similarity
argument applicable to the correlation functions whose
main predictions is the power-law scaling of the ratio of 
integral lengths with the large scale anisotropy ratio, with
an exponent governed by the inertial range scaling 
exponent.  
Within experimental accuracy, the scaling exponent
determined by direct measurement of the ratio
of integral lengths and the large scale anisotropy ratio
was found to be in excellent agreement with the value derived 
from the inertial range scaling exponent determined with ESS 
method.  
Further, we presented evidence to support the claim that
correlation functions can be made to collapse when separations
were rescaled with a scaling factor derived from the similarity
hypothesis.  

\chapter{Summary and Outlook}
\label{chap:summary}

In this dissertation, we have succeeded in isolating the effect
of shear from the large scale anisotropy.  
We saw that anisotropy in the velocity field has difference 
effects on the inertial and integral scales of turbulence.  

In chapter~\ref{chap:flows} we demonstrated that asymmetry 
of forcing leads to asymmetry of turbulence in a controllable way.  
We saw that the flows are approximately homogeneous and
axisymmetric, have negligible mean and shear.  

In chapter~\ref{chap:universality}, we investigated second-order 
moments of velocity increments measured in different directions in 
anisotropic turbulent flows.  
We ruled out the possibility that the scaling exponent and 
the Kolmogorov constant might depend on the direction
in which the structure function is measured.  
This is in contrast with previously published results and suggests
that the anisotropy produced by shear may be inherently 
different from that produced only in the fluctuations.  

In chapter~\ref{chap:higherorderstats}, we examined moments of 
the velocity increments up to the sixth order.  
We found evidence that the asymmetry in the probability
density of the velocity increments is enhanced when the large
scale anisotropy is present.  
In addition, we found indications that the inertial range scaling 
exponents are very likely to be independent of anisotropy.  

In chapter~\ref{chap:integralscale}, we presented a scaling 
argument to describe the anisotropy observed at the large scale.  
We saw that correlation functions measured in different
directions of the flow collapse onto a single curve when 
appropriately scaled.  
The scaling implies a power-law relationship between
the ratio of integral lengths and velocity fluctuation anisotropy,
whose exponent is closely linked to the inertial range 
scaling exponent.  
We found that the exponent measured from the power-law
relationship is consistent with the value estimated from
the inertial range scaling exponent with extended-self-similarity
method.  
This power-law relationship suggests that self-similarity and
Kolmogorov theory are relevant to the large-scale 
phenomenology of turbulence.  

It is important that the reader is aware of some of the difficulties 
and shortcomings of the scaling argument presented.  
A point of vagueness is the complete neglect of the dissipation 
range; correlation function in this range has been shown to follow 
an analytic scaling law.  
The second point is the observed anomaly and intermittency in 
the inertial range scaling exponent.  
These questions will not be settled by the present measurements,
but these measurements will provide more information on 
which to base a future study.  

Further insight into these questions may be provided with tools 
like the particle image velocimetry \cite[e.g.][]{raffel:2007} 
and the Lagrangian particle tracking \cite[e.g.][]{ouellette:2006}.  
These measurement systems would allow us to probe the small
scales with even higher precision at higher data rate.  
If further experiments were to be conducted, then we should 
investigate the angular dependence of the longitudinal 
structure functions and correlation functions.  
Do they follow the same scaling law?  
Or are they intrinsically different from transverse functions, 
as measurements in shear flows suggest?
In addition, we should examine the angular dependence of 
higher-order structure functions with longer measurement time, 
derivative structure functions, and the Lagrangian structure 
functions in our system.  

To conclude, let us discuss a fascinating extension of the 
present work to turbulence in dimensions other than three.  
A well-founded scientific proof must explore and exhaust all
the parameter space in the problem.  
A parameter we have not been able to vary is the inertial 
range scaling exponent, $\zeta_2$, because our system is inherently 
three-dimensional and it obeys the three-dimensional forced 
Navier-Stokes equation.  
Two-dimensional turbulence \cite[e.g.][]{kraichnan:1980} 
has captured the fascination of the researchers because 
many geophysical and magneto-hydrodynamical 
phenomena can be formulated as a two-dimensional problem.  
In the inertial range of two-dimensional turbulence, 
Kolmogorov-type dimensional reasoning gave $r^2$ scaling 
for the second-order velocity structure functions
\cite[e.g.][]{kellay:2002}.  
The question would then be, does the ratio of integral lengths
scale with the fluctuations anisotropy ratio following a power-law 
with a different inertial range scaling exponent?
Measurements in thin layers of conducting fluid \cite[e.g.][]{xia:2011} 
and in soap films \cite[e.g.][]{kellay:2002} will provide the answer to 
this question.  

We have presented in this thesis a flow apparatus 
that permits a systematic exploration of turbulence away from 
isotropy and described some unique features of this system
through measurements of structure functions and correlation
functions.  
We believe that understanding the large scale phenomenology
of turbulence may elucidate the underlying mechanism 
of turbulent motions at the small scales.  
Any useful theory of turbulence must incorporate some degree of
anisotropy, and the purpose of our system is to provide a test bed.  

\appendix
\chapter{Evaluating the dimensionless integral $A_{\gamma}$}
\label{app:agamma}

The integral
\begin{equation}
A_{\gamma} = \int_{-\infty}^{\infty} (1 - \mathrm{e}^{i x}) |x|^{-(\gamma + 1)} \, \mathrm{d} x \quad , \quad (0 < \gamma < 2) \,,
\end{equation}
may be written as a sum of two integrals
\begin{equation}
A_{\gamma} = \int_{-\infty}^{0} (1 - \mathrm{e}^{i x}) |x|^{-(\gamma + 1)} \, \mathrm{d} x + \int_{0}^{\infty} (1 - \mathrm{e}^{i x}) |x|^{-(\gamma + 1)} \, \mathrm{d} x \,.
\end{equation}
By making a substitution of variables $y=-x$ in the first integral and employing the trigonometric identity $\cos (x) = \tfrac{1}{2} (\mathrm{e}^{ix} + \mathrm{e}^{-ix})$, we can combine the two integrals to obtain
\begin{equation}
A_{\gamma} = 2 \, \int_{0}^{\infty} (1 - \cos (x)) \, x^{-(\gamma + 1)} \, \mathrm{d} x \,,
\end{equation}

\begin{figure}
\begin{center}
\includegraphics[scale=0.3]{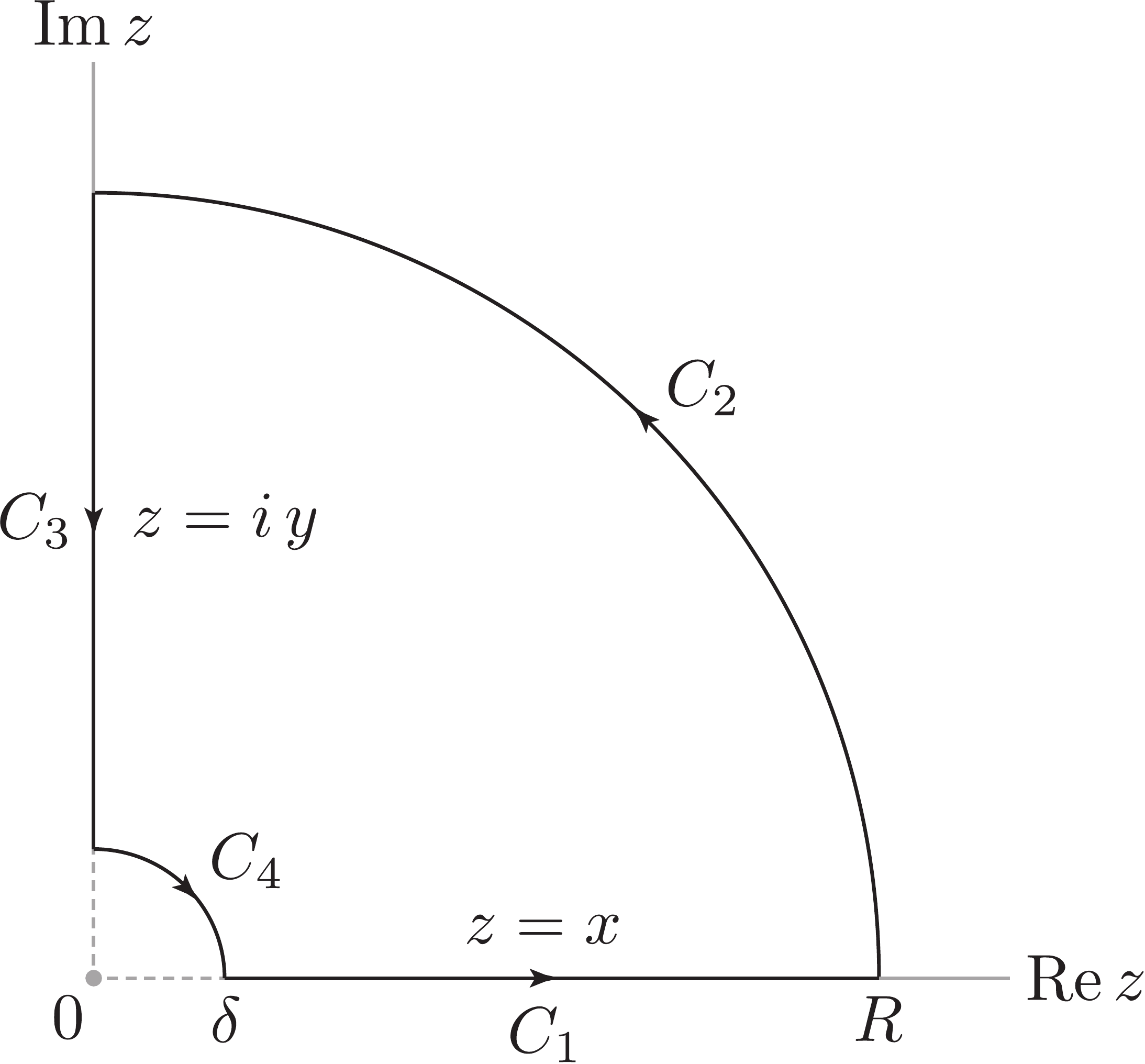}
\caption{Contour for the integral \ref{eq:complex_integral}.  
The pole at zero has been avoided by following the small circular quadrant near the origin.  
The integrals over the two circular quadrants tend to zero as 
$R \rightarrow \infty$ and $\delta \rightarrow 0$.  
Only the integrals over the line segments contribute to the integral 
\ref{eq:complex_integral}.}
\label{fig:contour_quadrant}
\end{center}
\end{figure}

Integrating the above integral by parts, the resulting boundary term 
can be shown to vanish at the boundary and we are left with
\begin{equation}
\label{eq:agamma_sine}
A_{\gamma} = \dfrac{2}{\gamma} \, \int_{0}^{\infty} x^{-\gamma} \, \sin (x) \, \mathrm{d} x \,.
\end{equation}
This integral can be evaluated by considering a complex integral of the form
\begin{equation}
\label{eq:complex_integral}
\textgoth{A}= \oint z^{-\gamma} \, \mathrm{e}^{i z} \, \mathrm{d} z \,,
\end{equation}
and by applying Cauchy's theorem to the closed curve consisting of two line segments along positive $\mathrm{Re} \, z$ and $\mathrm{Im} \, z$ axes and two quadrants in the upper half plane; one very large with radius $R$ and one very small with radius $\delta$, as shown in figure~\ref{fig:contour_quadrant}.  
The point $z=0$ has to be avoided; we do this by following the small circular quadrant of radius $\delta$.  
The integral over the big circular quadrant $C_2$ tends to 
$0$ as $R \rightarrow \infty$.  
The integral over the small circular quadrant near the origin $C_4$ can 
also be seen to approach $0$ as $\delta \rightarrow 0$.  
Only the integrals over the line segments remain and the complex 
integral reduces to
\begin{equation}
\textgoth{A} = \int_{0}^{\infty} x^{-\gamma} \, 
\mathrm{e}^{ix} \, \mathrm{d} x 
- \mathrm{e}^{i(1-\gamma)\pi/2} \, \int_{0}^{\infty} y^{-\gamma} \,
\mathrm{e}^{-y} \, \mathrm{d} y \,.
\end{equation}

Because we have avoided the pole at the origin, the integral \textgoth{A} equals $0$ by Cauchy's theorem.  
After a substitution of $\Gamma (1-\gamma) = \int_{0}^{\infty} y^{-\gamma} \, \mathrm{e}^{-y} \, \mathrm{d} y$, the imaginary part of \textgoth{A} yields
\begin{equation}
\int_{0}^{\infty} x^{-\gamma} \, \sin (x) \, \mathrm{d} x = \Gamma (1-\gamma) \cos \Bigl( \dfrac{\gamma \, \pi}{2} \Bigr) \,.
\end{equation}
From Euler's reflection formula, $\Gamma (1 - \gamma) \, \Gamma (\gamma) = \pi / \sin (\gamma \, \pi)$, 
and the double-angle formula, $\sin (2 \, \theta) = 2 \, \sin \theta \, \cos \theta$, the above integral can be written in the form
\begin{equation}
\int_{0}^{\infty} x^{-\gamma} \, \sin (x) \, \mathrm{d} x =
\dfrac{\pi}{2 \, \Gamma (\gamma) \, \sin (\gamma \, \pi / 2)} \,.
\end{equation}
Finally, with a substitution of the above in equation~\ref{eq:agamma_sine} and an explicit use of the recurrence relation $\Gamma (\gamma + 1) = \gamma \, \Gamma (\gamma)$, we find
\begin{equation}
A_{\gamma} = 
\dfrac{\pi}{\Gamma (\gamma + 1) \, \sin (\gamma \, \pi / 2)} \,.
\end{equation}

\cite{gradshteyn:2007} (3.761, equation~4) give
\begin{equation}
\int_{0}^{\infty} x^{\mu - 1} \, \sin (a\, x) \, \mathrm{d} x = \dfrac{\Gamma (\mu)}{a^{\mu}} \, \sin \Bigl( \dfrac{\mu \, \pi}{2} \Bigr) \quad [a > 0; \quad 0 < |\mathrm{Re} \, \mu| < 1] \,.
\end{equation}
The integral in equation~\ref{eq:agamma_sine} is a special case of that found in \cite{gradshteyn:2007} with $a = 1$ and $\mu = 1 - \gamma$.  
\chapter{Statistical tools}
\label{app:statistical_tools}

In this thesis, unless otherwise stated, the following 
statistical estimators are used.  
For a sample (data set) with sample size $N$ and observed values
$x_1, x_2, \ldots , x_N$, the sample mean, $\langle x \rangle$, 
is the sum of the observations divided by the sample size
\begin{equation}
\langle x \rangle = \dfrac{1}{N} \, \sum_{i=1}^{N} x_i \,.
\end{equation}
The measure of spread, or variability, in the sample is the sample 
standard deviation
\begin{equation}
\sigma_x = \bigg[ \dfrac{1}{N-1} \, 
\sum_{i=1}^{N} (x_i - \langle x \rangle)^2 \bigg]^{1/2} \,.
\end{equation}
The measure of spread, or variability, in the sample mean is
the standard error (the central limit theorem)
\begin{equation}
\label{eq:SE}
\mathrm{SE} = \dfrac{\sigma_x}{\sqrt{N}} \,.
\end{equation}

For a sample with $N$ pairs of independent ($x_i$) and dependent
variables ($y_i$), if a straight line, $\tilde{y}_i = a + b \, x_i$, is 
appropriate for the range of values studied, the linear 
regression coefficients, $a$ and $b$, are those for which the sum 
of squared residuals, $\sum_{i}^{N} (y_i - \tilde{y}_i)^2$, is minimized.  
Defining the following sums of squares
\begin{align}
S_{xx} &= \sum_{i=1}^{N} (x_i - \langle x \rangle)^2 \, , \\
S_{yy} &= \sum_{i=1}^{N} (y_i - \langle y \rangle)^2 \, , \\
S_{xy} &= \sum_{i=1}^{N} (x_i - \langle x \rangle)
(y_i - \langle y \rangle) \, ,
\end{align}
the slope, $b$, and intercept, $a$, are given by
\begin{align}
b &= \dfrac{S_{xy}}{S_{xx}} \, , \\
a &= \langle y \rangle - b \, \langle x \rangle \,.
\end{align}
Let $\tilde{y}_i = a + b \, x_i$ be the value predicted by the 
least-squares fit, then the difference between the original value
and the predicted value is given by $\xi_i = y_i - \tilde{y}_i$.  
The measure for the spread in $\xi_i$ is the sum of squares of 
residuals
\begin{equation}
\label{eq:residuals_squares}
s_{\xi} = \bigg[ \dfrac{1}{N-2} \, 
\sum_{i=1}^{N} \xi_i^2 \bigg]^{1/2} \,,
\end{equation}
or it can be rewritten in terms of the sums of squares
\begin{equation}
s_{\xi} = \bigg[ \dfrac{S_{xx} \, S_{yy} - 
S_{xy}^2}{(N-2) \, S_{xx}} \bigg]^{1/2} \,.
\end{equation}
The standard errors for the intercept, $a$, and the slope, $b$, are
\begin{align}
\label{eq:SEa}
\mathrm{SE} (a) &= s_{\xi} \, \bigg[ 
\dfrac{1}{N} + \dfrac{\langle x \rangle^2}{S_{xx}} \bigg]^{1/2} \, , \\
\mathrm{SE} (b) &= \dfrac{s_{\xi}}{\sqrt{S_{xx}}} \,.
\label{eq:SEb}
\end{align}
For a reference on linear regression, see e.g. \cite{acton:1966}.  
\chapter{Generating exponentially correlated colored noise}
\label{app:foxnoise}

\begin{lstlisting}
%%%%%%%%%%%%%%%%%%%%%%%%%%%%%%%%%%%%%%%%%%%%%%%%%%%%%%%%%%%%%%%%%
% This Matlab script calculates the exponentially				%
% correlated colored noise using the algorithm by				%
% Fox et al. (PRA, 38, 5938-5940, 1988).     					%
%%%%%%%%%%%%%%%%%%%%%%%%%%%%%%%%%%%%%%%%%%%%%%%%%%%%%%%%%%%%%%%%%

% Setting parameters for the algorithm
a = 0;              % Initializing random number
b = 0;              % Initializing random number
lambda = 1./(0.1);  % 1/Correlation time (1/seconds)
D = 1./lambda;      % Correlation time (seconds), D*lambda = 1
N = 2.^17;          % Number of samples
f = 50;             % Carrier frequency (Hz)
Fs = 3000;          % Sampling frequency
dt = 1./Fs;         % Sample time
t = dt.*[0:1:(N-1)]';   % Time array
E = exp(-lambda.*dt);

% Here begins the algorithm
e = zeros(N,1);         % Initializing noise array
for i=1:(N-1)
    a = rand(1);
    b = rand(1);
    h = sqrt(-2.*D.*lambda.*(1-E.^2).*log(a)).*cos(2.*pi.*b);
    e(i+1) = e(i).*E + h;
end

% Modulate the amplitude of the carrier wave
amp = e.*sin(2.*pi.*f.*t);

% Display the signal
figure;
plot(t, amp, 'k-');
xlabel('Time (s)');
ylabel('Voltage (V)');

% Take the Fourier transform of the signal
Nfft = 2^nextpow2(N);
famp = fft(amp,Nfft)/N;
freq = (Fs./2).*linspace(0,1,Nfft/2+1);
y    = abs(famp(1:Nfft/2+1)).^2;

% Display the spectrum
figure;
loglog(freq, y, 'ko');
xlabel('Frequency (Hz)');
ylabel('|Amplitude|^2 (A.U.)');

% Get rid of noise in the spectrum
windowsize = 30;
yf = filter(ones(1,windowsize)/windowsize,1,y);

% Display the filtered spectrum
figure;
loglog(freq, yf, 'ko');
xlabel('Frequency (Hz)');
ylabel('|Amplitude|^2 (A.U.)');
\end{lstlisting}
\chapter{Calculating statistics with inter-arrival 
time weighting}
\label{app:interarrival}

\begin{lstlisting}
%%%%%%%%%%%%%%%%%%%%%%%%%%%%%%%%%%%%%%%%%%%%%%%%%%%%%%%%%%%%%
% This Matlab function calculates statistics				%
% according to inter-arrival time weighting					%
%															%
%   [out, sigma, num] = InterArrival(t, v);					%
%															%
% Each vector is weighted by the time past since previous	%
% measurement.												%
% Measurements made after 5 times of the mean arrival time	%
% are rejected.												%
%%%%%%%%%%%%%%%%%%%%%%%%%%%%%%%%%%%%%%%%%%%%%%%%%%%%%%%%%%%%%

function [out, varargout] = InterArrival(t, v)

% Check for error in input arguments
error(nargchk(1,2,nargin));
if (nargin < 2) || isempty(v)
    error('Input argument must contain two vectors');
end

if length(t) ~= length(v)
    error('Vectors must have the same length');
end

% Specify the criterion for rejection
limit = 5;

% Calculate the inter-arrival time
dt 	  = diff(t);
v     = v(2:length(v));

% Reject measurements with dt > 5*mean(dt)
accept = logical(dt < limit*mean(dt, 1));

dt = dt(accept);
v  = v(accept);

% Calculate the weighted sum
out = sum(v.*dt) / sum(dt);

% Provide the standard deviation and the number of statistics
if (nargout > 1)
    varargout(1) = {std(v)};
end
if (nargout == 3)
    varargout(2) = {length(v)};
end
\end{lstlisting}
\chapter{Resampling LDV velocity signals}
\label{app:resample}

\begin{lstlisting}
%%%%%%%%%%%%%%%%%%%%%%%%%%%%%%%%%%%%%%%%%%%%%%%%%%%%%%%%%%%%%
%   This Matlab function resamples LDV velocity signal		%
%   tin will be rounded up to the nearest multiples			%
%   of tsep.												%
%   If there are more than one velocity within a particular	%
%   tin, an average of the velocities is calculated.		%
%															%
%   syntax:													%
%       [vout, tout, nout] = LDVResample(vin, tin, tsep);	%
%   inputs:													%
%       vin  -- velocity vector								%
%       tin  -- time vector									%
%       tsep -- separation time								%
%   outputs:												%
%       vout -- resampled velocity vector					%
%       tout -- resampled time vector in multiples of tsep	%
%       nout -- number of elements within a bin				%
%%%%%%%%%%%%%%%%%%%%%%%%%%%%%%%%%%%%%%%%%%%%%%%%%%%%%%%%%%%%%

function [vout, tout, nout] = LDVResample(vin, tin, tsep)

% Check for error in input arguments
error(nargchk(1,3,nargin));
if nargin < 2 || isempty(tin)
    error('Time array is missing!');
end

if nargin < 3 || isempty(tsep)
    error('Specify separation time.');
end

if ischar(vin) || ischar(tin) || ischar(tsep)
    error('Input arguments must be numeric.');
end

% Initialize vectors
vout = zeros(1+round(max(tin)/tsep), 1);
nout = vout;

% Take the sum of velocities within a bin
for i = 1:length(tin)
    j 		= 1 + round(tin(i)/tsep);
    vout(j) = vout(j) + vin(i);
    nout(j) = nout(j) + 1;
end

% Calculate average
nout = nout + (nout==0);
vout = vout ./ nout;
tout = tsep*[0:round(max(tin)/tsep)]';
\end{lstlisting}
\chapter{Calculating velocity autocorrelation}
\label{app:ldvcorr}

\begin{lstlisting}
%%%%%%%%%%%%%%%%%%%%%%%%%%%%%%%%%%%%%%%%%%%%%%%%%%%%%%%%%%%%%
% This Matlab function calculates velocity correlation		%
% for equally spaced data									%
%															%
%   syntax:													%
%       [fout, stat] = LDVCorr(v1, v2, t1, t2);				%
%   inputs:													%
%       v1, v2  -- velocity vectors							%
%       t1, t2  -- time vectors, equally spaced				%
%   outputs:												%
%       fout    -- velocity correlation function			%
%       stat    -- [v1_RMS, v2_RMS, v1_mean, v2_mean, 		%
%                   nfout, nv1_mean, nv2_mean]				%
%       RMS     -- RMS of coincident velocities				%
%       mean    -- mean of nonzero velocities.				%
%       n's     -- number of statistics						%
%%%%%%%%%%%%%%%%%%%%%%%%%%%%%%%%%%%%%%%%%%%%%%%%%%%%%%%%%%%%%

function [fout, varargout] = LDVCorr(v1, v2, t1, t2)

% Check for error in input arguments
error(nargchk(1,4,nargin));
if nargin < 2 || isempty(v2)
    error('v2 is missing!');
end
if nargin < 3 || isempty(t1)
    error('t1 for v1 is missing!');
end
if nargin < 4 || isempty(t2)
    error('t2 for v2 is missing!');
end

% Locate nonzero elements
cutoff = min([length(t1) length(t2)]);
nonzero = find(v1(1:cutoff).*v2(1:cutoff));

% Calculate the correlation
foo = (v1(nonzero)-mean(v1(find(v1)))).*(v2(nonzero)-mean(v2(find(v2)))) / ...
    (std(v1(nonzero)-mean(v1(find(v1))))*std(v2(nonzero)-mean(v2(find(v2)))));
fout = sum(foo)/length(nonzero);

% Provide the RMS, the mean and the number of statistics
if (nargout > 1)
    stat    = zeros(1,7);
    RMS1    = std(v1(nonzero)-mean(v1(find(v1))));
    RMS2    = std(v2(nonzero)-mean(v2(find(v2))));
    MEAN1   = mean(v1(find(v1)));
    MEAN2   = mean(v2(find(v2)));
    NFOUT   = length(nonzero);
    NMEAN1  = length(find(v1));
    NMEAN2  = length(find(v2));
    stat(:) = [RMS1 RMS2 MEAN1 MEAN2 NFOUT NMEAN1 NMEAN2];
    varargout(1) = {stat};
end
\end{lstlisting}
\chapter{Calculating least-squares fit coefficients}
\label{app:linearfit}

For an introduction to the theory of linear regression, the reader could consult \cite{draper:1998}.
\begin{lstlisting}
%%%%%%%%%%%%%%%%%%%%%%%%%%%%%%%%%%%%%%%%%%%%%%%%%%%%%%%%%%%%%%%%%%%%%
% This Matlab function calculates the least 						%
% square fit coefficients											%
% Usage:															%
% [B] = LinearMultiVarFit(Y, X1, X2, ..., XN, flag);				%
% Input:															%
% Y - fit function													%
% X1, X2, ... XN - variables (2D arrays)							%
% flag - 'zero' for regression through the origin.					%
%        default is no crossing at origin.							%
% Output:															%
% B - an array of fit coefficients									%
%																	%
% LinearMultiVarFit calculates the fit coefficients of 				%
% the multivariate function											%
% Y = B0 + B1*X1 + B2*X2 + ... + BN*XN								%
% with matrix left division in the least square sense.				%
% If X's and Y are M by N matrices, B(:,j) is a 					%
% column vector [B0 B1 B2 ... BN] of fit coefficients 				%
% for each set of Y(:,j), X1(:,j), X2(:,j), ... XN(:,j)				%
%%%%%%%%%%%%%%%%%%%%%%%%%%%%%%%%%%%%%%%%%%%%%%%%%%%%%%%%%%%%%%%%%%%%%

function B = LinearMultiVarFit(varargin)

% Check for error in input argument
if (nargin < 2)
    error('Input argument must contain at least two arrays');
end

% Check for the number of input arrays
if isstr(varargin{iend})
    nin      = nargin-1;
    zeroflag = varargin{iend};
else
    nin      = nargin;
    zeroflag = 'boobs';
end

% Form an ND array
for i=1:nin
    X(:,:,i) = varargin{i};
end

[M, N, P] = size(X);

% Initialize array
if (M==1)
    X = permute(X, [2 1 3]);
    [NX, NY, NZ] = size(X);
    B = zeros(NZ,1);
else
    if (N==1)
        [NX, NY, NZ] = size(X);
        B = zeros(NZ,1);
    else
        [NX, NY, NZ] = size(X);
        B = zeros(NZ,NY);
    end
end

% Compute fit coefficients
if (strcmpi(zeroflag,'zero'))
    B(1,:) = 0;
    Mat = zeros(NX,NZ-1);
    for i=1:NY
        Mat(:,:)   = X(:,i,2:iend);
        out        = Mat\X(:,i,1);
        B(2:iend,i) = out;
        clear out
    end
else
    Mat = zeros(NX,NZ);
    Mat(:,1) = ones(NX,1);
    for i=1:NY
        Mat(:,2:iend) = X(:,i,2:iend);
        out          = Mat\X(:,i,1);
        B(:,i)       = out;
        clear out
    end
end
\end{lstlisting}

\cleardoublepage
\addcontentsline{toc}{chapter}{Bibliography}
\bibliography{thesisbib}

\begin{thebibliography}{222}
\providecommand{\natexlab}[1]{#1}
\providecommand{\url}[1]{\texttt{#1}}
\expandafter\ifx\csname urlstyle\endcsname\relax
  \providecommand{\doi}[1]{doi: #1}\else
  \providecommand{\doi}{doi: \begingroup \urlstyle{rm}\Url}\fi

\bibitem[Acton(1966)]{acton:1966}
F.~S. Acton.
\newblock \emph{Analysis of Straight-Line Data}.
\newblock Dover, New York, 1966.

\bibitem[Adrian and Yao(1987)]{adrian:1987}
R.~J. Adrian and C.~S. Yao.
\newblock Power spectra of fluid velocities measured by laser doppler
  velocimetry.
\newblock \emph{Experiments in Fluids}, 5:\penalty0 17--28, 1987.

\bibitem[Albrecht et~al.(2003)Albrecht, Borys, Damaschke, and
  Tropea]{albrecht:2003}
H.-E. Albrecht, M.~Borys, N.~Damaschke, and C.~Tropea.
\newblock \emph{Laser Doppler and phase Doppler measurement techniques}.
\newblock Springer-Verlag, Berlin Heidelberg New York, 2003.

\bibitem[Anfossi et~al.(2000)Anfossi, Degrazia, Ferrero, Gryning, Morselli, and
  Castelli]{anfossi:2000}
D.~Anfossi, G.~Degrazia, E.~Ferrero, S.~E. Gryning, M.~G. Morselli, and S.~T.
  Castelli.
\newblock Estimation of the {L}agrangian structure function constant ${C}_0$
  from surface-layer wind data.
\newblock \emph{Boundary-Layer Meteorology}, 95:\penalty0 249--270, 2000.

\bibitem[Anselmet et~al.(1984)Anselmet, Gagne, Hopfinger, and
  Antonia]{anselmet:1984}
F.~Anselmet, Y.~Gagne, E.~J. Hopfinger, and R.~A. Antonia.
\newblock High-order velocity structure functions in turbulent shear flows.
\newblock \emph{Journal of Fluid Mechanics}, 140:\penalty0 63--89, 1984.

\bibitem[Antonia and Pearson(1999)]{antonia:1999}
R.~A. Antonia and B.~R. Pearson.
\newblock Low-order velocity structure functions in relatively high {R}eynolds
  number turbulence.
\newblock \emph{Europhysics Letters}, 48\penalty0 (2):\penalty0 163--169, 1999.

\bibitem[Antonia et~al.(2002)Antonia, Zhou, and Romano]{antonia:2002}
R.~A. Antonia, T.~Zhou, and G.~P. Romano.
\newblock Small-scale turbulence characteristics of two-dimensional bluff body
  wakes.
\newblock \emph{Journal of Fluid Mechanics}, 459:\penalty0 67--92, 2002.

\bibitem[Arad et~al.(1998)Arad, Dhruva, Kurien, {L'vov}, Procaccia, and
  Sreenivasan]{arad:1998}
I.~Arad, B.~Dhruva, S.~Kurien, V.~S. {L'vov}, I.~Procaccia, and K.~R.
  Sreenivasan.
\newblock The extraction of anisotropic contributions in turbulent flows.
\newblock \emph{Physical Review Letters}, 81\penalty0 (24):\penalty0
  5330--5333, 1998.

\bibitem[Arad et~al.(1999)Arad, Biferale, Mzzitelli, and Procaccia]{arad:1999a}
I.~Arad, L.~Biferale, I.~Mzzitelli, and I.~Procaccia.
\newblock Disentangling scaling properties in anisotropic and inhomogeneous
  turbulence.
\newblock \emph{Physical Review Letters}, 82\penalty0 (25):\penalty0
  5040--5043, 1999.

\bibitem[Barenblatt and Goldenfeld(1995)]{barenblatt:1995}
G.~I. Barenblatt and N.~Goldenfeld.
\newblock Does fully developed turbulence exist? {R}eynolds number independence
  versus asymptotic covariance.
\newblock \emph{Physics of Fluids}, 7:\penalty0 3078--3082, 1995.

\bibitem[Barrett and Hollingsworth(2001)]{barrett:2001}
M.~J. Barrett and D.~K. Hollingsworth.
\newblock On the calculation of length scales for turbulent heat transfer
  correlation.
\newblock \emph{Journal of Heat Transfer}, 123:\penalty0 878--883, 2001.

\bibitem[Batchelor(1946{\natexlab{a}})]{batchelor:1946a}
G.~K. Batchelor.
\newblock The {T}heory of {A}xisymmetric {T}urbulence.
\newblock \emph{Proc. R. Soc. Lond. A}, 186\penalty0 (1007):\penalty0 480--502,
  1946{\natexlab{a}}.

\bibitem[Batchelor(1946{\natexlab{b}})]{batchelor:1946b}
G.~K. Batchelor.
\newblock Double velocity correlation function in turbulent motion.
\newblock \emph{Nature}, 158:\penalty0 883--884, 1946{\natexlab{b}}.

\bibitem[Batchelor(1947)]{batchelor:1947}
G.~K. Batchelor.
\newblock Kolmogoroff's theory of locally isotropic turbulence.
\newblock \emph{Proc. Camb. Phil. Soc.}, 43:\penalty0 533--559, 1947.

\bibitem[Batchelor(1948)]{batchelor:1948a}
G.~K. Batchelor.
\newblock Decay of isotropic turbulence in the initial period.
\newblock \emph{Proc. R. Soc. Lond. A}, 193:\penalty0 539--558, 1948.

\bibitem[Batchelor(1951)]{batchelor:1951}
G.~K. Batchelor.
\newblock Pressure fluctuations in isotropic turbulence.
\newblock \emph{Mathematical Proceedings of the Cambridge Philosophical
  Society}, 47:\penalty0 359--374, 1951.

\bibitem[Batchelor(1952)]{batchelor:1952}
G.~K. Batchelor.
\newblock Diffusion in a field of homogeneous turbulence. ii. {T}he relative
  motion of particles.
\newblock \emph{Proc. Camb. Phil. Soc.}, 48:\penalty0 345--362, 1952.

\bibitem[Batchelor(1956)]{batchelor:1956}
G.~K. Batchelor.
\newblock \emph{The theory of homogeneous turbulence}.
\newblock Cambridge University Press, Cambridge, U. K., 1956.

\bibitem[Batchelor and Townsend(1948)]{batchelor:1948b}
G.~K. Batchelor and A.~A. Townsend.
\newblock Decay of turbulence in the final period.
\newblock \emph{Proc. R. Soc. Lond. A}, 194\penalty0 (1039):\penalty0 527--543,
  1948.

\bibitem[Benzi et~al.(1993)Benzi, Ciliberto, Tripiccione, Baudet, Massaioli,
  and Succi]{benzi:1993}
R.~Benzi, S.~Ciliberto, R.~Tripiccione, C.~Baudet, F.~Massaioli, and S.~Succi.
\newblock Extended self-similarity in turbulent flows.
\newblock \emph{Physical Review E}, 48:\penalty0 R29--R32, 1993.

\bibitem[Benzi et~al.(1996)Benzi, Biferale, Ciliberto, Struglia, and
  Tripiccione]{benzi:1996}
R.~Benzi, L.~Biferale, S.~Ciliberto, M.~V. Struglia, and R.~Tripiccione.
\newblock Generalized scaling in fully developed turbulence.
\newblock \emph{Physica D}, 96:\penalty0 162--181, 1996.

\bibitem[Bershadskii(2008)]{bershadskii:2008}
A.~Bershadskii.
\newblock Near-dissipation range in nonlocal turbulence.
\newblock \emph{Physics of Fluids}, 20\penalty0 (8):\penalty0 085103, 2008.

\bibitem[Bewley et~al.(2008)Bewley, Sreenivasan, and Lathrop]{bewley:2008}
G.~P. Bewley, K.~R. Sreenivasan, and D.~P. Lathrop.
\newblock Particles for tracing turbulent liquid helium.
\newblock \emph{Experiments in Fluids}, 44:\penalty0 887--896, 2008.

\bibitem[Biferale and Procaccia(2005)]{biferale:2005}
L.~Biferale and I.~Procaccia.
\newblock Anisotropy in turbulent flows and in turbulent transport.
\newblock \emph{Physics Reports}, 414:\penalty0 43--164, 2005.

\bibitem[Biferale and Toschi(2001)]{biferale:2001}
L.~Biferale and F.~Toschi.
\newblock Anisotropic homogeneous turbulence: hierarchi adn intermittency of
  scaling exponents in the anisotropic sectors.
\newblock \emph{Physical Review Letters}, 86\penalty0 (21):\penalty0
  4831--4834, 2001.

\bibitem[Boratav and Pelz(1997)]{boratav:1997}
O.~N. Boratav and R.~B. Pelz.
\newblock Structures and structure functions in the inertial range of
  turbulence.
\newblock \emph{Physics of Fluids}, 9:\penalty0 1400--1415, 1997.

\bibitem[Borgas and Sawford(1991)]{borgas:1991}
M.~S. Borgas and B.~L. Sawford.
\newblock The small-scale structure of acceleration correlations and its role
  in the statistical theory of turbulent dispersion.
\newblock \emph{Journal of Fluid Mechanics}, 228:\penalty0 69--99, 1991.

\bibitem[Buchhave(1975)]{buchhave:1975}
P.~Buchhave.
\newblock Biasing errors in individual particle measurements.
\newblock In \emph{Proc. LDA Symp. Copenhagen}, pages 258--278, Tonsbakken
  16-18, 2740 Skovlunde, Denmark, 1975.

\bibitem[Buchhave et~al.(1979)Buchhave, {Jr. George}, and
  Lumley]{buchhave:1979}
P.~Buchhave, W.~K. {Jr. George}, and J.~L. Lumley.
\newblock The measurement of turbulence with the laser-dopper anemometer.
\newblock \emph{Annu. Rev. Fluid Mech.}, 11:\penalty0 443--504, 1979.

\bibitem[Camussi and Benzi(1997)]{camussi:1997}
R.~Camussi and R.~Benzi.
\newblock Hierarchy of transverse structure functions.
\newblock \emph{Physics of Fluids}, 9:\penalty0 257--259, 1997.

\bibitem[Camussi et~al.(1996)Camussi, Barbagallo, Guj, and
  Stella]{camussi:1996}
R.~Camussi, D.~Barbagallo, G.~Guj, and F.~Stella.
\newblock Transverse and longitudinal scaling laws in non-homogeneous low {R}e
  turbulence.
\newblock \emph{Physics of Fluids}, 8:\penalty0 1181--1191, 1996.

\bibitem[Carullo et~al.(2011)Carullo, Nasir, Cress, Ng, Thole, Zhang, and
  Moon]{carullo:2011}
J.~S. Carullo, S.~Nasir, R.~D. Cress, W.~F. Ng, K.~A. Thole, L.~J. Zhang, and
  H.~K. Moon.
\newblock The effects of freestream turbulence, turbulence lengt scale, and
  exit {R}eynolds number on turbine blade heat transfer in a transonic cascade.
\newblock \emph{Journal of Turbomachinery}, 133\penalty0 (011030):\penalty0
  1--11, 2011.

\bibitem[Chakraborty et~al.(2010)Chakraborty, Frisch, and
  Ray]{chakraborty:2010}
S.~Chakraborty, U.~Frisch, and S.~S. Ray.
\newblock Extended self-similarity works for the {B}urgers equation and why.
\newblock \emph{Journal of Fluid Mechanics}, 649:\penalty0 275--285, 2010.

\bibitem[Chandrasekhar(1950)]{chandrasekhar:1950}
S.~Chandrasekhar.
\newblock The {T}heory of {A}xisymmetric {T}urbulence.
\newblock \emph{Phil. Trans. R. Soc. Lond. A}, 242:\penalty0 557--577, 1950.

\bibitem[Chen et~al.(1997)Chen, Sreenivasan, Nelkin, and Cao]{chen:1997}
S.~Chen, K.~R. Sreenivasan, M.~Nelkin, and N.~Cao.
\newblock A refined similarity hypothesis for transverse structure functions.
\newblock \emph{Physical Review Letters}, 79:\penalty0 2253--2256, 1997.

\bibitem[{Comte-Bellot} and Corrsin(1966)]{comtebellot:1966}
G.~{Comte-Bellot} and S.~Corrsin.
\newblock The use of a contraction to improve the isotropy of {grid-generated}
  turbulence.
\newblock \emph{Journal of Fluid Mechanics}, 25:\penalty0 657--682, 1966.

\bibitem[{Comte-Bellot} and Corrsin(1971)]{comtebellot:1971}
G.~{Comte-Bellot} and S.~Corrsin.
\newblock Simple {E}ulerian time correlation of full- and narrow-band velocity
  signals in grid-generated, `isotropic' turbulence.
\newblock \emph{Journal of Fluid Mechanics}, 48 part 2:\penalty0 273--337,
  1971.

\bibitem[Corrsin(1951)]{corrsin:1951}
S.~Corrsin.
\newblock On the spectrum of isotropic temperature fluctuations in isotropic
  turbulence.
\newblock \emph{J. Appl. Phys.}, 22:\penalty0 469--473, 1951.

\bibitem[Corrsin(1963{\natexlab{a}})]{corrsin:1963}
S.~Corrsin.
\newblock Estimates of the relations between {E}ulerian and {L}agrangian scales
  in large {R}eynolds number turbulence.
\newblock \emph{Journal of the Atmospheric Sciences}, 20:\penalty0 115--119,
  1963{\natexlab{a}}.

\bibitem[Corrsin(1963{\natexlab{b}})]{corrsin:1963b}
S.~Corrsin.
\newblock \emph{Turbulence: Experimental Methods, in Handbuch der Physik},
  volume VIII/2, chapter~4, pages 524--590.
\newblock Springer-Verlag, 1963{\natexlab{b}}.

\bibitem[Corrsin and Lumley(1956)]{corrsin:1956}
S.~Corrsin and J.~L. Lumley.
\newblock On the equation of motion for a particle in turbulent fluid.
\newblock \emph{Applied Scientific Research}, 6A:\penalty0 114--116, 1956.

\bibitem[Counihan(1975)]{counihan:1975}
J.~Counihan.
\newblock Adiabatic atmospheric boundary layers: a review and analysis of data
  from the period 1880-1972.
\newblock \emph{Atmos. Environ.}, 19:\penalty0 871--905, 1975.

\bibitem[Cramer(1959)]{cramer:1959}
H.~E. Cramer.
\newblock Measurements of turbulence structure near the ground within the
  frequency range from 0.5 to 0.01 cycles sec$^{-1}$.
\newblock \emph{Adv. Geophys.}, 6:\penalty0 75--96, 1959.

\bibitem[{de Jong} et~al.(2009){de Jong}, Cao, Woodward, Salazar, Collins, and
  Meng]{dejong:2009}
J.~{de Jong}, L.~Cao, S.~H. Woodward, J.~P. L.~C. Salazar, L.~R. Collins, and
  H.~Meng.
\newblock Dissipation rate estimation from {PIV} in zero-mean isotropic
  turbulence.
\newblock \emph{Experiments in Fluids}, 46:\penalty0 499--515, 2009.

\bibitem[Degrazia et~al.(2008)Degrazia, Welter, Wittwer, Carvalho, Roberti,
  Acevedo, Moraes, and Velho]{degrazia:2008}
G.~A. Degrazia, G.~S. Welter, A.~R. Wittwer, J.~C. Carvalho, D.~R. Roberti,
  O.~C. Acevedo, O.~L.~L. Moraes, and H.~F.~C. Velho.
\newblock Estimation of the {L}agrangian {K}olmogorov constant from {E}ulerian
  measurements for distinct {R}eynolds number with application to pollution
  dispersion model.
\newblock \emph{Atmospheric Environment}, 42:\penalty0 2415--2423, 2008.

\bibitem[{del \'{A}lamo} and Jim\'{e}nez(2009)]{delalamo:2009}
J.~{del \'{A}lamo} and J.~Jim\'{e}nez.
\newblock Estimation of turbulent convection velocities and corrections to
  {T}aylor's approximation.
\newblock \emph{Journal of Fluid Mechanics}, 640:\penalty0 5--26, 2009.

\bibitem[Dhruva et~al.(1997)Dhruva, Tsuji, and Sreenivasan]{dhruva:1997}
B.~Dhruva, Y.~Tsuji, and K.~R. Sreenivasan.
\newblock Transverse structure functions in high-{R}eynolds-number turbulence.
\newblock \emph{Physical Review E}, 56\penalty0 (5):\penalty0 R4928--R4930,
  1997.

\bibitem[Dobler et~al.(2003)Dobler, Haugen, Yousef, and
  Brandenburg]{dobler:2003}
W.~Dobler, N.~E.~L. Haugen, T.~A. Yousef, and A.~Brandenburg.
\newblock Bottleneck effect in three-dimensional turbulence simulations.
\newblock \emph{Physical Review E}, 68\penalty0 (2):\penalty0 026304, 2003.

\bibitem[Donzis and Sreenivasan(2010)]{donzis:2010}
D.~A. Donzis and K.~R. Sreenivasan.
\newblock The bottleneck effect and the {K}olmogorov constant in isotropic
  turbulence.
\newblock \emph{Journal of Fluid Mechanics}, 657:\penalty0 171--188, 2010.

\bibitem[Draper and Smith(1998)]{draper:1998}
N.~R. Draper and H.~Smith.
\newblock \emph{Applied regression analysis}.
\newblock Wiley-Interscience, New York, 1998.

\bibitem[Dryden et~al.(1937)Dryden, Schubauer, Mock, and
  Skramstad]{dryden:1937}
H.~L. Dryden, G.~B. Schubauer, W.~C. Mock, and H.~K. Skramstad.
\newblock Measurements of intensity and scale of wind tunnel turbulence and
  their relaiton to the critical {R}eynolds number of spheres.
\newblock Technical Report 581, Nat. Adv. Com. Aeronaut., 1937.

\bibitem[Ducet et~al.(2000)Ducet, {Le-Traon}, and Reverdin]{ducet:2000}
N.~Ducet, P.~Y. {Le-Traon}, and G.~Reverdin.
\newblock Global high-resolution mapping of ocean circulation from
  {TOPEX/Poseidon} and {ERS-1} and {-2}.
\newblock \emph{Journal of Geophysical Research}, 105\penalty0 (C8):\penalty0
  19477--19498, 2000.

\bibitem[Elghobashi(1994)]{elghobashi:1994}
S.~Elghobashi.
\newblock On predicting particle-laden turbulent flows.
\newblock \emph{Applied Scientific Research}, 52:\penalty0 309--329, 1994.

\bibitem[Elghobashi and Truesdell(1993)]{elghobashi:1993}
S.~Elghobashi and G.~C. Truesdell.
\newblock On the two-way interaction between homogeneous turbulence and
  dispersed solid particles. {I}: {T}urbulence modification.
\newblock \emph{Physics of Fluids A}, 5\penalty0 (7):\penalty0 1790--1801,
  1993.

\bibitem[Ewing et~al.(2007)Ewing, Frohnapfel, George, Pedersen, and
  Westerweel]{ewing:2007}
D.~Ewing, B.~Frohnapfel, W.~K. George, J.~M. Pedersen, and J.~Westerweel.
\newblock Two-point similarity in the round jet.
\newblock \emph{Journal of Fluid Mechanics}, 577:\penalty0 309--330, 2007.

\bibitem[Eyink(2011)]{eyink:2011}
G.~Eyink.
\newblock Personal communication, 15 March 2011.

\bibitem[Eyink and Xin(2000)]{eyink:2000}
G.~L. Eyink and J.~Xin.
\newblock Self-similar decay in the {K}raichnan model of a passive scalar.
\newblock \emph{Journal of Statistical Physics}, 100\penalty0 (3/4):\penalty0
  679--741, 2000.

\bibitem[Fairhall et~al.(1996)Fairhall, Gat, {L'vov}, and
  Procaccia]{fairhall:1996}
A.~L. Fairhall, O.~Gat, V.~{L'vov}, and I.~Procaccia.
\newblock Anomalous scaling in a model of passive scalar advection: {E}xact
  results.
\newblock \emph{Physical Review E}, 53\penalty0 (4):\penalty0 3518--3535, 1996.

\bibitem[Falkovich(1994)]{falkovich:1994}
G.~Falkovich.
\newblock Bottleneck phenomenon in developed turbulence.
\newblock \emph{Physics of Fluids}, 6:\penalty0 1411--1414, 1994.

\bibitem[Falkovich and {L'vov}(1995)]{falkovich:1995}
G.~Falkovich and V.~S. {L'vov}.
\newblock Isotropic and anisotropic turbulence in {C}lebsch variables.
\newblock \emph{Chaos, Solitons \& Fractals}, 5:\penalty0 1855--1869, 1995.

\bibitem[Feynman et~al.(1963)Feynman, Leighton, and Sands]{feynman:1963}
R.~P. Feynman, R.~B. Leighton, and M.~Sands.
\newblock \emph{The {F}eynman {L}ectures on {P}hysics}, volume~1.
\newblock Addison-Wesley, Reading MA, 1963.

\bibitem[Finnigan(2000)]{finnigan:2000}
J.~Finnigan.
\newblock Turbulence in plant canopies.
\newblock \emph{Annu. Rev. Fluid Mech.}, 32:\penalty0 519--571, 2000.

\bibitem[Fox et~al.(1988)Fox, Gatland, Roy, and Vemuri]{fox:1988}
R.~F. Fox, I.~R. Gatland, R.~Roy, and G.~Vemuri.
\newblock Fast, accurate algorithm for numerical simulation of exponentially
  correlated colored noise.
\newblock \emph{Physical Review A}, 38\penalty0 (11):\penalty0 5938--5940,
  1988.

\bibitem[Fransos and Bruno(2010)]{fransos:2010}
D.~Fransos and L.~Bruno.
\newblock Edge degree-of-sharpness and free-stream turbulence scale effects on
  the aerodynamics of a bridge deck.
\newblock \emph{Journal of Wind Engineering and Industrial Aerodynamics},
  98:\penalty0 661--671, 2010.

\bibitem[Franzese and Cassiani(2007)]{franzese:2007}
P.~Franzese and M.~Cassiani.
\newblock A statistical theory of turbulent relative dispersion.
\newblock \emph{Journal of Fluid Mechanics}, 571:\penalty0 391--417, 2007.

\bibitem[Friehe et~al.(1991)Friehe, Shaw, Rogers, Davidson, Large, Stage,
  Crescenti, Khalsa, Greenhut, and Li]{friehe:1991}
C.~A. Friehe, W.~J. Shaw, D.~P. Rogers, K.~L. Davidson, W.~G. Large, S.~A.
  Stage, G.~H. Crescenti, S.~J.~S. Khalsa, G.~K. Greenhut, and F.~Li.
\newblock Air-sea fluxes and surface layer turbulence around a sea surface
  temperature front.
\newblock \emph{Journal of Geophysical Research}, 96\penalty0 (C5):\penalty0
  8593--8609, 1991.

\bibitem[Frisch(1995)]{frisch:1995}
U.~Frisch.
\newblock \emph{Turbulence: {T}he {L}egacy of {A}. {N}. {K}olmogorov}.
\newblock Cambridge University Press, Cambridge, U. K., 1995.

\bibitem[Frisch et~al.(2008)Frisch, Kurien, Pandit, Pauls, Ray, Wirth, and
  Zhu]{frisch:2008}
U.~Frisch, S.~Kurien, R.~Pandit, W.~Pauls, S.~S. Ray, A.~Wirth, and J.~Zhu.
\newblock Hyperviscosity, {G}alerkin truncation, and bottlenecks in turbulence.
\newblock \emph{Physical Review Letters}, 101\penalty0 (14):\penalty0 144501,
  2008.

\bibitem[Garg and Warhaft(1998)]{garg:1998}
S.~Garg and Z.~Warhaft.
\newblock On the small scale structure of simple shear flow.
\newblock \emph{Physics of Fluids}, 10\penalty0 (3):\penalty0 662--673, 1998.

\bibitem[Gioia and Bombardelli(2001)]{gioia:2001}
G.~Gioia and F.~A. Bombardelli.
\newblock Scaling and similarity in rough channel flows.
\newblock \emph{Physical Review Letters}, 88\penalty0 (014501), 2001.

\bibitem[Gioia and Bombardelli(2005)]{gioia:2005}
G.~Gioia and F.~A. Bombardelli.
\newblock Localized turbulent flows on scouring granular beds.
\newblock \emph{Physical Review Letters}, 95\penalty0 (014501), 2005.

\bibitem[Gioia and Chakraborty(2006)]{gioia:2006}
G.~Gioia and P.~Chakraborty.
\newblock Turbulent friction in rough pipes and the energy spectrum of the
  phenomenological theory.
\newblock \emph{Physical Review Letters}, 96\penalty0 (044502), 2006.

\bibitem[Gioia et~al.(2010)Gioia, Guttenberg, Goldenfeld, and
  Chakraborty]{gioia:2010}
G.~Gioia, N.~Guttenberg, N.~Goldenfeld, and P.~Chakraborty.
\newblock Spectral theory of the turbulent mean-velocity profile.
\newblock \emph{Physical Review Letters}, 105\penalty0 (184501), 2010.

\bibitem[Gledzer et~al.(1996)Gledzer, Villermaux, Kahalerras, and
  Gagne]{gledzer:1996}
E.~Gledzer, E.~Villermaux, H.~Kahalerras, and Y.~Gagne.
\newblock On the log-{P}oisson statistics of the energy dissipation field and
  related problems of developed turbulence.
\newblock \emph{Physics of Fluids}, 8\penalty0 (12):\penalty0 3367--3378, 1996.

\bibitem[Glezer and Amitay(2002)]{glezer:2002}
A.~Glezer and M.~Amitay.
\newblock Synthetic jets.
\newblock \emph{Annu. Rev. Fluid Mech.}, 34:\penalty0 503--529, 2002.

\bibitem[G\"{o}decke(1935)]{goedecke:1935}
K.~G\"{o}decke.
\newblock Messungen der atmosph\"{a}rischen {T}urbulenz in {B}odenn\"{a}he mit
  einer {H}itzdrahtmethode.
\newblock \emph{Ann. Hydrogr.}, {-- }\penalty0 (10):\penalty0 400--410, 1935.

\bibitem[Goepfert et~al.(2010)Goepfert, Mari{\'e}, Chareyron, and
  Lance]{goepfert:2010}
C.~Goepfert, J.-L. Mari{\'e}, D.~Chareyron, and M.~Lance.
\newblock Characterization fo a system generating a homogeneous isotropic
  turbulence field by free synthetic jets.
\newblock \emph{Experiments in Fluids}, 48:\penalty0 809--822, 2010.

\bibitem[Goldenfeld(2006)]{goldenfeld:2006}
N.~Goldenfeld.
\newblock Roughness-induced critical phenomena in a turbulent flow.
\newblock \emph{Physical Review Letters}, 96\penalty0 (044503), 2006.

\bibitem[Gotoh et~al.(2002)Gotoh, Fukayama, and Nakano]{gotoh:2002}
G.~Gotoh, D.~Fukayama, and T.~Nakano.
\newblock Velocity field statistics in homogeneous steady turbulence obtained
  using a high-resolution direct numerical simulation.
\newblock \emph{Physics of Fluids}, 14\penalty0 (3):\penalty0 1065--1081, 2002.

\bibitem[Gradshteyn and Ryzhik(2007)]{gradshteyn:2007}
I.~S. Gradshteyn and I.~M. Ryzhik.
\newblock \emph{Table of integrals, series, and products}.
\newblock Academic Press, 7th edition edition, 2007.

\bibitem[Grant et~al.(1962)Grant, Stewart, and Moilliet]{grant:1962}
H.~L. Grant, R.~W. Stewart, and A.~Moilliet.
\newblock Turbulence spectra from a tidal channel.
\newblock \emph{Journal of Fluid Mechanics}, 12:\penalty0 241--268, 1962.

\bibitem[Grossmann et~al.(1994)Grossmann, Lohse, {L'vov}, and
  Procaccia]{grossmann:1994}
S.~Grossmann, D.~Lohse, V.~{L'vov}, and I.~Procaccia.
\newblock Finite size corrections to scaling in high {R}eynolds number
  turbulence.
\newblock \emph{Physical Review Letters}, 73\penalty0 (3):\penalty0 432--435,
  1994.

\bibitem[Grossmann et~al.(1997)Grossmann, Lohse, and Reeh]{grossmann:1997}
S.~Grossmann, D.~Lohse, and A.~Reeh.
\newblock Different intermittency for longitudinal and transversal turbulent
  fluctuations.
\newblock \emph{Physics of Fluids}, 9\penalty0 (12):\penalty0 3817--3825, 1997.

\bibitem[Gurbatov et~al.(1997)Gurbatov, Simdyankin, Aurell, Frisch, and
  T\'{o}th]{gurbatov:1997}
S.~N. Gurbatov, S.~I. Simdyankin, E.~Aurell, U.~Frisch, and G.~T\'{o}th.
\newblock On the decay of {B}urgers turbulence.
\newblock \emph{Journal of Fluid Mechanics}, 344:\penalty0 339--374, 1997.

\bibitem[Guttenberg and Goldenfeld(2009)]{guttenberg:2009}
N.~Guttenberg and N.~Goldenfeld.
\newblock Friction factor of two-dimensional rough-boundary turbulent soap film
  flows.
\newblock \emph{Physical Review E}, 79\penalty0 (065306), 2009.

\bibitem[Hao et~al.(2008)Hao, Zhou, Zhou, and Mi]{hao:2008}
Z.~Hao, T.~Zhou, Y.~Zhou, and J.~Mi.
\newblock Reynolds number dependence of the inertial range scaling of energy
  dissipation rate and enstrophy in a cylinder wake.
\newblock \emph{Experiments in Fluids}, 44:\penalty0 279--289, 2008.

\bibitem[He et~al.(1999)He, Doolen, and Chen]{he:1999}
G.~He, G.~D. Doolen, and S.~Chen.
\newblock Calculations of longitudinal and transverse velocity structure
  functions using a vortex model of isotropic turbulence.
\newblock \emph{Physics of Fluids}, 11\penalty0 (12):\penalty0 3743--3748,
  1999.

\bibitem[Heisenberg(1948)]{heisenberg:1948}
W.~Heisenberg.
\newblock Zur statistischen {T}heorie der {T}urbulenz.
\newblock \emph{Zeit. f. Phys.}, 124:\penalty0 628--657, 1948.

\bibitem[Herweijer and {Van de Water}(1995)]{herweijer:1995}
J.~A. Herweijer and W.~{Van de Water}.
\newblock Transverse structure functions of turbulence.
\newblock In \emph{Advances in Turbulence V}, pages 210--216. Kluwer, 1995.

\bibitem[Hestroni(1989)]{hestroni:1989}
G.~Hestroni.
\newblock Particles turbulence interaction.
\newblock \emph{International Journal of Multiphase Flow}, 15:\penalty0
  735--746, 1989.

\bibitem[Huang(1987)]{huang:1987}
K.~Huang.
\newblock \emph{Statistical Mechanics}.
\newblock John Wiley \& Sons, 1987.

\bibitem[Hwang and Eaton(2004)]{hwang:2004}
W.~Hwang and J.~K. Eaton.
\newblock Creating homogeneous and isotropic turbulence without a mean flow.
\newblock \emph{Experiments in Fluids}, 36:\penalty0 444--454, 2004.

\bibitem[Inoue(1951)]{inoue:1951}
E.~Inoue.
\newblock On the {L}agrangian correlation coefficient for the turbulent
  diffusion and its application to the atmospheric diffusion phenomena.
\newblock Unpublished report, {G}eophys. Inst., Univ. of Tokyo, 1951.

\bibitem[Isaza et~al.(2009)Isaza, Warhaft, and Collins]{isaza:2009}
J.~C. Isaza, Z.~Warhaft, and L.~R. Collins.
\newblock Experimental investigation of the large-scale velocity statistics in
  homogeneous turbulent shear flow.
\newblock \emph{Physics of Fluids}, 21\penalty0 (065105), 2009.

\bibitem[Ivanov and Stratonovich(1963)]{ivanov:1963}
V.~N. Ivanov and R.~L. Stratonovich.
\newblock On the {L}agrange characteristics of turbulence.
\newblock \emph{Izv. Akad. Nauk. SSSR, Ser. Geofiz.}, {-- }\penalty0
  (10):\penalty0 1581--1593, 1963.

\bibitem[Kahalerras et~al.(1996)Kahalerras, Malecot, and
  Gagne]{kahalerras:1996}
H.~Kahalerras, Y.~Malecot, and Y.~Gagne.
\newblock Transverse velocity structure functions in developed turbulence.
\newblock In \emph{Advances in Turbulence VI}, pages 235--238, Dordrecht, 1996.
  Kluwer.

\bibitem[Kahalerras et~al.(1998)Kahalerras, Mal\'{e}cot, Gagne, and
  Castaing]{kahalerras:1998}
H.~Kahalerras, Y.~Mal\'{e}cot, Y.~Gagne, and B.~Castaing.
\newblock Intermittency and {R}eynolds number.
\newblock \emph{Physics of Fluids}, 10\penalty0 (4):\penalty0 910--921, 1998.

\bibitem[Kaneda et~al.(2003)Kaneda, Ishihara, Yokokawa, Itakura, and
  Uno]{kaneda:2003}
Y.~Kaneda, T.~Ishihara, M.~Yokokawa, K.~Itakura, and A.~Uno.
\newblock Energy dissipation rate and energy spectrum in high resolution direct
  numerical simulations of turbulence in a periodic box.
\newblock \emph{Physics of Fluids}, 15\penalty0 (2):\penalty0 L21--L24, 2003.

\bibitem[K\'{a}rm\'{a}n and Howarth(1938)]{karman:1938}
T.~K\'{a}rm\'{a}n and L.~Howarth.
\newblock On the statistical theory of isotropic turbulence.
\newblock \emph{Proc. R. Soc. Lond. A}, 164:\penalty0 192--215, 1938.

\bibitem[Kellay and Goldburg(2002)]{kellay:2002}
H.~Kellay and W.~I. Goldburg.
\newblock Two-dimensional turbulence: a review of some recent experiments.
\newblock \emph{Rep. Prog. Phys.}, 65:\penalty0 845--894, 2002.

\bibitem[Kim et~al.(1987)Kim, Moin, and Moser]{kim:1987}
J.~Kim, P.~Moin, and R.~Moser.
\newblock Turbulence statistics in fully developed channel flow at low
  {R}eynolds number.
\newblock \emph{Journal of Fluid Mechanics}, 177:\penalty0 133--166, 1987.

\bibitem[Kistler and Vrebalovich(1966)]{kistler:1966}
A.~L. Kistler and T.~Vrebalovich.
\newblock Grid turbulence at large {R}eynolds numbers.
\newblock \emph{Journal of Fluid Mechanics}, 26:\penalty0 37--47, 1966.

\bibitem[Knight and Sirovich(1990)]{knight:1990}
B.~Knight and L.~Sirovich.
\newblock Kolmogorov inertial range for inhomogeneous turbulent flows.
\newblock \emph{Physical Review Letters}, 65\penalty0 (11):\penalty0
  1356--1359, 1990.

\bibitem[Kolmogorov(1940{\natexlab{a}})]{kolmogorov:1940a}
A.~N. Kolmogorov.
\newblock Curves in {H}ilbert space which are invariant with respect to
  {one-parameter} group motion.
\newblock \emph{Dokl. Akad. Nauk. SSSR}, 26\penalty0 (1):\penalty0 6--9,
  1940{\natexlab{a}}.

\bibitem[Kolmogorov(1940{\natexlab{b}})]{kolmogorov:1940b}
A.~N. Kolmogorov.
\newblock Wiener's spiral and some interesting curves in {H}ilbert space.
\newblock \emph{Dokl. Akad. Nauk. SSSR}, 26\penalty0 (2):\penalty0 115--118,
  1940{\natexlab{b}}.

\bibitem[Kolmogorov(1941{\natexlab{a}})]{kolmogorov:1941}
A.~N. Kolmogorov.
\newblock The local structure of turbulence in incompressible viscous fluid for
  very large {R}eynolds numbers.
\newblock \emph{Dokl. Akad. Nauk. SSSR}, 30:\penalty0 299--303,
  1941{\natexlab{a}}.

\bibitem[Kolmogorov(1941{\natexlab{b}})]{kolmogorov:1941b}
A.~N. Kolmogorov.
\newblock Dissipation of energy in locally isotorpic turbulence.
\newblock \emph{Dokl. Akad. Nauk. SSSR}, 32:\penalty0 19--21,
  1941{\natexlab{b}}.

\bibitem[Kolmogorov(1991)]{kolmogorov:1991}
A.~N. Kolmogorov.
\newblock Dissipation of energy in locally isotorpic turbulence.
\newblock \emph{Proc. R. Soc. Lond. A}, 434:\penalty0 15--17, 1991.

\bibitem[Kraichnan(1966)]{kraichnan:1966}
R.~H. Kraichnan.
\newblock Isotropic turbulence and inertial-range structure.
\newblock \emph{Physics of Fluids}, 9:\penalty0 1728--1752, 1966.

\bibitem[Kraichnan and Montgomery(1980)]{kraichnan:1980}
R.~H. Kraichnan and D.~Montgomery.
\newblock Two-dimensional turbulence.
\newblock \emph{Rep. Prog. Phys.}, 43:\penalty0 547--619, 1980.

\bibitem[Kurien and Sreenivasan(2000)]{kurien:2000b}
S.~Kurien and K.~R. Sreenivasan.
\newblock Anisotropic scaling contributions to high-order structure functions
  in high-{R}eynolds-number turbulence.
\newblock \emph{Physical Review E}, 62\penalty0 (2):\penalty0 2206--2212, 2000.

\bibitem[Kurien et~al.(2000)Kurien, {L'vov}, Procaccia, and
  Sreenivasan]{kurien:2000a}
S.~Kurien, V.~S. {L'vov}, I.~Procaccia, and K.~R. Sreenivasan.
\newblock Scaling structure of the velocity statistics in atmospheric boundary
  layers.
\newblock \emph{Physical Review E}, 61\penalty0 (1):\penalty0 407--421, 2000.

\bibitem[Kurien et~al.(2004)Kurien, Taylor, and Matsumoto]{kurien:2004}
S.~Kurien, M.~A. Taylor, and T.~Matsumoto.
\newblock Cascade time scales for energy and helicity in homogeneous isotropic
  turbulence.
\newblock \emph{Physical Review E}, 69\penalty0 (6):\penalty0 066313, 2004.

\bibitem[{La Porta} et~al.(2000){La Porta}, Voth, Crawford, Alexander, and
  Bodenschatz]{laporta:2000}
A.~{La Porta}, G.~A. Voth, A.~M. Crawford, J.~Alexander, and E.~Bodenschatz.
\newblock Fluid particle accelerations in fully developed turbulence.
\newblock \emph{Nature}, 409:\penalty0 1017--1019, 2000.

\bibitem[Landau and Lifshitz(1959)]{landau:1959}
L.~D. Landau and E.~M. Lifshitz.
\newblock \emph{Fluid Mechanics}.
\newblock Addison-Wesley, Reading MA, 1959.

\bibitem[Lenschow and Stankov(1986)]{lenschow:1986}
D.~H. Lenschow and B.~B. Stankov.
\newblock Length scales in the convective boundary layer.
\newblock \emph{Journal of the Atmospheric Sciences}, 43\penalty0
  (12):\penalty0 1198--1209, 1986.

\bibitem[Lighthill(1978)]{lighthill:1978}
J.~Lighthill.
\newblock Acoustic streaming.
\newblock \emph{Journal of Sound and Vibration}, 61\penalty0 (3):\penalty0
  391--418, 1978.

\bibitem[Lin(1953)]{lin:1953}
C.~C. Lin.
\newblock On {T}aylor's hypothesis and the acceleration terms in the
  {N}avier-{S}tokes equations.
\newblock \emph{Quarterly of Applied Mathematics}, 10\penalty0 (4):\penalty0
  295--306, 1953.

\bibitem[Lin(1960)]{lin:1960}
C.~C. Lin.
\newblock On a theory of dispersion by continuous movements.
\newblock \emph{Proc. Natl. Acad. Sci. USA}, 46:\penalty0 566--570, 1960.

\bibitem[Lindborg(1995)]{lindborg:1995}
E.~Lindborg.
\newblock Kinematics of homogeneous axisymmetric turbulence.
\newblock \emph{Journal of Fluid Mechanics}, 302:\penalty0 179--201, 1995.

\bibitem[Lohse(1994)]{lohse:1994}
D.~Lohse.
\newblock Cross-over from high to low reynolds number turbulence.
\newblock \emph{Physical Review Letters}, 73\penalty0 (24):\penalty0
  3223--3226, 1994.

\bibitem[Lohse and {M\"{u}ller-Groeling}(1995)]{lohse:1995}
D.~Lohse and A.~{M\"{u}ller-Groeling}.
\newblock Bottleneck effects in turbulence-scaling phenomena in $r$-space
  versus $p$-space.
\newblock \emph{Physical Review Letters}, 74\penalty0 (10):\penalty0
  1747--1750, 1995.

\bibitem[Lu et~al.(2008)Lu, Fugal, Nordsiek, Saw, Shaw, and Yang]{lu:2008}
J.~Lu, J.~P. Fugal, H.~Nordsiek, E.~W. Saw, R.~A. Shaw, and W.~Yang.
\newblock Lagrangian particle tracking in three dimensions via single-camera
  in-line digital holography.
\newblock \emph{New Journal of Physics}, 10\penalty0 (125013), 2008.

\bibitem[Lumley(1957)]{lumley:1957}
J.~L. Lumley.
\newblock \emph{Some problems connected with the motion of small particles in
  turbulent fluid}.
\newblock PhD thesis, The Johns Hopkins University, Baltimore, Maryland, 1957.

\bibitem[Lumley(1962)]{lumley:1962}
J.~L. Lumley.
\newblock An approach to the {E}ulerian-{L}agrangian problem.
\newblock \emph{J. Math. Phys.}, 3:\penalty0 309--312, 1962.

\bibitem[Lumley(1967)]{lumley:1967}
J.~L. Lumley.
\newblock Similarity and the turbulent energy spectrum.
\newblock \emph{Physics of Fluids}, 10\penalty0 (4):\penalty0 855--858, 1967.

\bibitem[Lumley(1976)]{lumley:1976}
J.~L. Lumley.
\newblock \emph{Two phase and non-{N}ewtonian flows. In Topics in Applied
  Physics: Turbulence}, volume~12, pages 290--324.
\newblock Springer-Verlag, Berlin, 1976.

\bibitem[Lumley and Yaglom(2001)]{lumley:2001}
J.~L. Lumley and A.~M. Yaglom.
\newblock A {C}entury of {T}urbulence.
\newblock \emph{Flow, Turbulence and Combustion}, 66:\penalty0 241--286, 2001.

\bibitem[Lundgren(2003)]{lundgren:2003}
T.~S. Lundgren.
\newblock Kolmogorov turbulence by matched asymptotic expansions.
\newblock \emph{Physics of Fluids}, 15\penalty0 (4):\penalty0 1074--1081, 2003.

\bibitem[{L'vov} and Procaccia(1996)]{lvov:1996}
V.~{L'vov} and I.~Procaccia.
\newblock The universal scaling exponents of anisotropy in turbulence and their
  measurement.
\newblock \emph{Physics of Fluids}, 8\penalty0 (10):\penalty0 2565--2567, 1996.

\bibitem[Mac{C}ready(1953)]{maccready:1953}
P.~B. Mac{C}ready.
\newblock Atmospheric turbulence measurements and analysis.
\newblock \emph{J. Meteor.}, 10\penalty0 (4):\penalty0 325--337, 1953.

\bibitem[Manton(1977)]{manton:1977}
M.~J. Manton.
\newblock The equation of motion for a small aerosol in a continuum.
\newblock \emph{Pure and Applied Geophysics}, 115:\penalty0 547--559, 1977.

\bibitem[Martinez et~al.(1997)Martinez, Chen, Doolen, Kraichnan, Wang, and
  Zhou]{martinez:1997}
D.~O. Martinez, S.~Chen, G.~D. Doolen, R.~H. Kraichnan, L.~P. Wang, and
  Y.~Zhou.
\newblock Energy spectrum in the dissipation range of fluid turbulence.
\newblock \emph{Journal of Plasma Physics}, 57:\penalty0 195--201, 1997.

\bibitem[Maxey and Riley(1983)]{maxey:1983}
M.~R. Maxey and J.~J. Riley.
\newblock Equation of motion for a small rigid sphere in a nonuniform flow.
\newblock \emph{Physics of Fluids}, 26:\penalty0 883--889, 1983.

\bibitem[Mayo et~al.(1974)Mayo, Shay, and Ritter]{mayo:1974}
W.~T.~J. Mayo, M.~T. Shay, and S.~Ritter.
\newblock Digital estimation of turbulence power spectra from burst counter ldv
  data.
\newblock In \emph{Proceedings of 2nd International Workshop on Laser
  Velocimetry}, pages 16--26, Purdue University, 1974.

\bibitem[McDougall(1980)]{mcdougall:1980}
T.~J. McDougall.
\newblock Bias correction for individual realisation of lda measurements.
\newblock \emph{Journal of Physics E: Scientific Instruments}, 13:\penalty0
  53--60, 1980.

\bibitem[{McLaughlin} and Tiederman(1973)]{mclaughlin:1973}
D.~K. {McLaughlin} and W.~G. Tiederman.
\newblock Biasing correction for individual realisation of laser anemometer
  measurements in turbulent flows.
\newblock \emph{Physics of Fluids}, 15:\penalty0 2082--2088, 1973.

\bibitem[Mei(1996)]{mei:1996}
R.~Mei.
\newblock Velocity fidelity of flow tracer particles.
\newblock \emph{Experiments in Fluids}, 22:\penalty0 1--13, 1996.

\bibitem[Mininni et~al.(2008)Mininni, Alexakis, and Pouquet]{mininni:2008}
P.~D. Mininni, A.~Alexakis, and A.~Pouquet.
\newblock Nonlocal interactions in hydrodynamic turbulence at high {R}eynolds
  numbers: the slow emergence of scaling laws.
\newblock \emph{Physical Review E}, 77:\penalty0 036306, 2008.

\bibitem[Monin and Yaglom(1975)]{monin:1975}
A.~S. Monin and A.~M. Yaglom.
\newblock \emph{Statistical {F}luid {M}echanics}, volume~2.
\newblock MIT Press, Cambridge MA, USA, 1975.

\bibitem[Moser(1994)]{moser:1994}
R.~D. Moser.
\newblock Kolmogorov inertial range spectra for inhomogeneous turbulence.
\newblock \emph{Physics of Fluids}, 6\penalty0 (2):\penalty0 794--801, 1994.

\bibitem[Mydlarski and Warhaft(1998)]{mydlarski:1998}
L.~Mydlarski and Z.~Warhaft.
\newblock Passive scalar statistics in high-{P}\'{e}clet-number grid
  turbulence.
\newblock \emph{Journal of Fluid Mechanics}, 358:\penalty0 135--175, 1998.

\bibitem[Nelkin(1992)]{nelkin:1992}
M.~Nelkin.
\newblock In {W}hat {S}ense {I}s {T}urbulence an {U}nsolved {P}roblem?
\newblock \emph{Science}, 255:\penalty0 566--570, 1992.

\bibitem[Nelkin(1994)]{nelkin:1994}
M.~Nelkin.
\newblock Universality and scaling in fully devloped turbulence.
\newblock \emph{Advances in Physics}, 43\penalty0 (2):\penalty0 143--181, 1994.

\bibitem[Noullez et~al.(1997)Noullez, Wallace, Lempert, Miles, and
  Frisch]{noullez:1997}
A.~Noullez, G.~Wallace, W.~Lempert, R.~B. Miles, and U.~Frisch.
\newblock Transverse velocity increments in turbulent flow using the {RELIEF}
  technique.
\newblock \emph{Journal of Fluid Mechanics}, 339:\penalty0 287--307, 1997.

\bibitem[Novikov(1963)]{novikov:1963}
E.~A. Novikov.
\newblock Random force method in turbulence theory.
\newblock \emph{Sov. Phys., J. Exp. Theor. Phys.}, 17:\penalty0 1449--1454,
  1963.

\bibitem[Obukhov(1941{\natexlab{a}})]{obukhov:1941a}
A.~M. Obukhov.
\newblock On the distribution of energy in the spectrum of turbulent flow.
\newblock \emph{Dokl. Akad. Nauk. SSSR}, 32\penalty0 (1):\penalty0 22--24,
  1941{\natexlab{a}}.

\bibitem[Obukhov(1941{\natexlab{b}})]{obukhov:1941b}
A.~M. Obukhov.
\newblock Spectral energy distribution in a turbulent flow.
\newblock \emph{Izv. Akad. Nauk. SSSR, Ser. Geogr. Geofiz.}, 5\penalty0
  (4-5):\penalty0 453--466, 1941{\natexlab{b}}.

\bibitem[Obukhov(1942)]{obukhov:1942}
A.~M. Obukhov.
\newblock On the theory of atmospheric turbulence.
\newblock \emph{Izv. Akad. Nauk. SSSR, Ser. Fiz.}, 6\penalty0 (1-2):\penalty0
  59--63, 1942.

\bibitem[Obukhov(1949{\natexlab{a}})]{obukhov:1949a}
A.~M. Obukhov.
\newblock Local structure of atmospheric turbulence.
\newblock \emph{Dokl. Akad. Nauk. SSSR}, 67\penalty0 (4):\penalty0 643--646,
  1949{\natexlab{a}}.

\bibitem[Obukhov(1949{\natexlab{b}})]{obukhov:1949b}
A.~M. Obukhov.
\newblock Structure of the temperature field in turbulent flows.
\newblock \emph{Izv. Akad. Nauk. SSSR, Ser. Geogr. Geofiz.}, 13:\penalty0
  58--69, 1949{\natexlab{b}}.

\bibitem[Onsager(1949)]{onsager:1949}
L.~Onsager.
\newblock Statistical hydrodynamics.
\newblock \emph{Nuovo Cimento}, 6\penalty0 (2):\penalty0 279--287, 1949.

\bibitem[Ott and Mann(2000)]{ott:2000}
S.~Ott and J.~Mann.
\newblock An experimental investigation of the relative diffusion of particle
  pairs in three-dimensional turbulent flow.
\newblock \emph{Journal of Fluid Mechanics}, 422:\penalty0 207--223, 2000.

\bibitem[Ouellette(2006)]{ouellette:2006}
N.~T. Ouellette.
\newblock \emph{Probing the statistical structure of turbulence with
  measurements of tracer particle tracks}.
\newblock PhD thesis, Cornell University, Ithaca, New York, 2006.

\bibitem[Ouellette et~al.(2006)Ouellette, Xu, Bourgoin, and
  Bodenschatz]{ouellette:2006b}
N.~T. Ouellette, H.~Xu, M.~Bourgoin, and E.~Bodenschatz.
\newblock Experimental study of turbulent relative dispersion models.
\newblock \emph{New Journal of Physics}, 8\penalty0 (109), 2006.

\bibitem[Paul and Jackson(1971)]{paul:1971}
D.~M. Paul and D.~A. Jackson.
\newblock Rapid velocity sensor using astatic confocal fabry-perot
  interferometer and a single argon laser.
\newblock \emph{Journal of Physics E: Scientific Instruments}, 4:\penalty0
  170--172, 1971.

\bibitem[Pearson and Antonia(2001)]{pearson:2001}
B.~R. Pearson and R.~A. Antonia.
\newblock Reynolds-number dependence of turbulent velocity and pressure
  increments.
\newblock \emph{Journal of Fluid Mechanics}, 444:\penalty0 343--382, 2001.

\bibitem[Pearson et~al.(2002)Pearson, \r{A}. Krogstad, and {Van de
  Water}]{pearson:2002}
B.~R. Pearson, P.~\r{A}. Krogstad, and W.~{Van de Water}.
\newblock Measurements of the turbulent energy dissipation.
\newblock \emph{Physics of Fluids}, 14\penalty0 (3):\penalty0 1288--1290, 2002.

\bibitem[Poggi et~al.(2008)Poggi, Katul, and Cassiani]{poggi:2008}
D.~Poggi, G.~G. Katul, and M.~Cassiani.
\newblock On the anomalous behavior fo the {L}agrangian structure function
  similarity constant inside dense canopies.
\newblock \emph{Atmospheric Environment}, 42\penalty0 (4212-4231), 2008.

\bibitem[Pope(2000)]{pope:2000}
S.~B. Pope.
\newblock \emph{Turbulent Flows}.
\newblock Cambridge University Press, Cambridge, U. K., 2000.

\bibitem[Prandtl(1945)]{prandtl:1945}
L.~Prandtl.
\newblock \"{U}ber die {R}olle der {Z}\"{a}higkeit im {M}echanismus der
  ausgebildete {T}urbulenz.
\newblock GOAR 3712, DLR Archive, 1945.

\bibitem[Praskovsky and Oncley(1994)]{praskovsky:1994}
A.~Praskovsky and S.~Oncley.
\newblock Measurements of the {K}olmogorov constant and intermittency exponent
  at very high {R}eynolds numbers.
\newblock \emph{Physics of Fluids}, 6\penalty0 (9):\penalty0 2886--2888, 1994.

\bibitem[Press et~al.(2007)Press, Teukolsky, Vetterling, and
  Flannery]{press:2007}
W.~H. Press, S.~A. Teukolsky, W.~T. Vetterling, and B.~P. Flannery.
\newblock \emph{Numerical Recipes}.
\newblock Cambridge University Press, Cambridge, U. K., 2007.

\bibitem[Qian(1984)]{qian:1984}
J.~Qian.
\newblock Universal equilibrium range of turbulence.
\newblock \emph{Physics of Fluids}, 27:\penalty0 2229--2233, 1984.

\bibitem[Raffel et~al.(2007)Raffel, Willert, Wereley, and
  Kompenhans]{raffel:2007}
M.~Raffel, C.~E. Willert, S.~T. Wereley, and J.~Kompenhans.
\newblock \emph{Particle Image Velocimetry}.
\newblock Springer-Verlag, 2nd edition, 2007.

\bibitem[Richardson(1922)]{richardson:1922}
L.~F. Richardson.
\newblock \emph{Weather Prediction by Numerical Process}.
\newblock Cambridge University Press, Cambridge, U. K., 1922.

\bibitem[Richardson(1926)]{richardson:1926}
L.~F. Richardson.
\newblock Atmospheric diffusion shown on a distance-neighbour graph.
\newblock \emph{Proc. R. Soc. Lond. A}, 110:\penalty0 709--737, 1926.

\bibitem[Roach(1987)]{roach:1987}
P.~E. Roach.
\newblock The generation of nearly isotropic turbulence by means of grids.
\newblock \emph{International Journal of Heat and Fluid Flow}, 8:\penalty0
  82--92, 1987.

\bibitem[Robertson(1940)]{robertson:1940}
H.~P. Robertson.
\newblock The invariant theory of isotropic turbulence.
\newblock \emph{Proc. Camb. Phil. Soc.}, 36:\penalty0 209--223, 1940.

\bibitem[Rodean(1991)]{rodean:1991}
H.~C. Rodean.
\newblock The universal constant for the {L}agrangian structure function.
\newblock \emph{Physics of Fluids A}, 3\penalty0 (6):\penalty0 1479--1480,
  1991.

\bibitem[Romano and Antonia(2001)]{romano:2001}
G.~P. Romano and R.~A. Antonia.
\newblock Longitudinal and transverse structure functions in a turbulent round
  jet: effect of initial conditions and {R}eynolds number.
\newblock \emph{Journal of Fluid Mechanics}, 436:\penalty0 231--248, 2001.

\bibitem[Roth(2000)]{roth:2000}
M.~Roth.
\newblock Review of atmospheric turbulence over cities.
\newblock \emph{Q. J. R. Meteorol. Soc.}, 126:\penalty0 941--990, 2000.

\bibitem[Saddoughi and Veeravalli(1994)]{saddoughi:1994}
S.~G. Saddoughi and S.~V. Veeravalli.
\newblock Local isotropy in turbulent boundary layers at high {R}eynolds
  number.
\newblock \emph{J.\ Fluid Mech.}, 268:\penalty0 333--372, 1994.

\bibitem[Salazar and Collins(2009)]{salazar:2009}
P.~L.~C. Salazar and L.~R. Collins.
\newblock Two-particle dispersion in isotropic turbulent flows.
\newblock \emph{Annu. Rev. Fluid Mech.}, 41:\penalty0 405--432, 2009.

\bibitem[Sawford(1991)]{sawford:1991}
B.~L. Sawford.
\newblock Reynolds number effects in {L}agrangian stochastic models of
  turbulent dispersion.
\newblock \emph{Physics of Fluids A}, 3\penalty0 (6):\penalty0 1577--1586,
  1991.

\bibitem[She and Leveque(1994)]{she:1994}
Z.~S. She and E.~Leveque.
\newblock Universal scaling laws in fully developed turbulence.
\newblock \emph{Physical Review Letters}, 72:\penalty0 336--339, 1994.

\bibitem[Shen and Warhaft(2000)]{shen:2000}
X.~Shen and Z.~Warhaft.
\newblock The anisotropy of the small scale structure in high {R}eynolds number
  (${R}_{\lambda} \sim 1000$) turbulent shear flow.
\newblock \emph{Physics of Fluids}, 12\penalty0 (11):\penalty0 2976--2989,
  2000.

\bibitem[Shen and Warhaft(2002)]{shen:2002}
X.~Shen and Z.~Warhaft.
\newblock Longitudinal and transverse structure functions in sheared and
  unsheared wind-tunnel turbulence.
\newblock \emph{Physics of Fluids}, 14\penalty0 (1):\penalty0 370--381, 2002.

\bibitem[Shiotani(1955)]{shiotani:1955}
M.~Shiotani.
\newblock On the fluctuation of the temperature and turbulent structure near
  the ground.
\newblock \emph{J. Meteor. Soc. Japan}, 33\penalty0 (3):\penalty0 117--123,
  1955.

\bibitem[Shy et~al.(1997)Shy, Tang, and Fann]{shy:1997}
S.~S. Shy, C.~Y. Tang, and S.~Y. Fann.
\newblock A nearly isotropic turbulence generated by a pair of vibrating grids.
\newblock \emph{Experimental Thermal and Fluid Science}, 14:\penalty0 251--262,
  1997.

\bibitem[Siebert et~al.(2010)Siebert, Shaw, and Warhaft]{siebert:2010}
H.~Siebert, R.~A. Shaw, and Z.~Warhaft.
\newblock Statistics of small-scale velocity fluctuations an internal
  intermittency in marine stratocumulus clouds.
\newblock \emph{Journal of the Atmospheric Sciences}, 67:\penalty0 262--273,
  2010.

\bibitem[Sreenivasan(1984)]{sreenivasan:1984}
K.~R. Sreenivasan.
\newblock On the scaling of the turbulence energy dissipation rate.
\newblock \emph{Physics of Fluids}, 27\penalty0 (5):\penalty0 1048--1051, 1984.

\bibitem[Sreenivasan(1995)]{sreenivasan:1995}
K.~R. Sreenivasan.
\newblock On the universality of the {K}olmogorov constant.
\newblock \emph{Physics of Fluids}, 7:\penalty0 2778--2784, 1995.

\bibitem[Sreenivasan(1996)]{sreenivasan:1996}
K.~R. Sreenivasan.
\newblock The passive scalar spectrum and the {O}bukhov-{C}orrsin constant.
\newblock \emph{Physics of Fluids}, 8\penalty0 (1):\penalty0 189--196, 1996.

\bibitem[Sreenivasan and Dhruva(1998)]{sreenivasan:1998}
K.~R. Sreenivasan and B.~Dhruva.
\newblock Is there scaling in high-{R}eynolds-number turbulence?
\newblock \emph{Progress of Theoretical Physics Supplement}, -\penalty0
  (130):\penalty0 103--120, 1998.

\bibitem[Sreenivasan and Narasimha(1978)]{sreenivasan:1978}
K.~R. Sreenivasan and R.~Narasimha.
\newblock Rapid distortion of axisymmetric turbulence.
\newblock \emph{Journal of Fluid Mechanics}, 84:\penalty0 497--516, 1978.

\bibitem[Sreenivasan et~al.(1996)Sreenivasan, Vainshtein, Bhiladvala, {San
  Gil}, Chen, and Cao]{sreenivasan:1996b}
K.~R. Sreenivasan, S.~I. Vainshtein, R.~Bhiladvala, I.~{San Gil}, S.~Chen, and
  N.~Cao.
\newblock Asymmetry of velocity increments in fully developed turbulence and
  the scaling of low-order moments.
\newblock \emph{Physical Review Letters}, 77\penalty0 (8):\penalty0 1488--1491,
  1996.

\bibitem[Staicu and {Van de Water}(2003)]{staicu:2003a}
A.~Staicu and W.~{Van de Water}.
\newblock Small scale velocity jumps in shear turbulence.
\newblock \emph{Physical Review Letters}, 90\penalty0 (094501):\penalty0 1--4,
  2003.

\bibitem[Staicu et~al.(2003)Staicu, Vorselaars, and {Van der
  Water}]{staicu:2003b}
A.~Staicu, B.~Vorselaars, and W.~{Van der Water}.
\newblock Turbulence anisotropy and the {SO(3)} description.
\newblock \emph{Physical Review E}, 68:\penalty0 046303, 2003.

\bibitem[Stolovitzky and Sreenivasan(1993)]{stolovitzky:1993}
S.~Stolovitzky and K.~R. Sreenivasan.
\newblock Scaling of structure functions.
\newblock \emph{Physical Review E}, 48:\penalty0 R33--R36, 1993.

\bibitem[Sullivan et~al.(2000)Sullivan, Greeley, Kraft, Wilson, Golombek,
  Herkenhoff, Murphy, and Smith]{sullivan:2000}
R.~Sullivan, R.~Greeley, M.~Kraft, G.~Wilson, M.~Golombek, K.~Herkenhoff,
  J.~Murphy, and P.~Smith.
\newblock Results of the {I}mager for {M}ars {P}athfinder windsock experiment.
\newblock \emph{Journal of Geophysical Research}, 105\penalty0 (E10):\penalty0
  24547--24562, 2000.

\bibitem[Taylor(1935)]{taylor:1935}
G.~I. Taylor.
\newblock Statistical theory of turbulence.
\newblock \emph{Proc. R. Soc. Lond. A}, 151:\penalty0 421--478, 1935.

\bibitem[Taylor(1938)]{taylor:1938}
G.~I. Taylor.
\newblock The spectrum of turbulence.
\newblock \emph{Proc. R. Soc. Lond. A}, 164\penalty0 (919):\penalty0 476--490,
  1938.

\bibitem[Taylor(1955)]{taylor:1955}
R.~J. Taylor.
\newblock Some observations of wind velocity autocorrelations in the lowest
  layers of the atmosphere.
\newblock \emph{Austr. J. Phys.}, 8\penalty0 (4):\penalty0 535--544, 1955.

\bibitem[Thomson(1996)]{thomson:1996}
D.~J. Thomson.
\newblock The second-order moment structure of dispersing plumes and puffs.
\newblock \emph{Journal of Fluid Mechanics}, 320:\penalty0 305--329, 1996.

\bibitem[Toschi and Bodenschatz(2009)]{toschi:2009}
F.~Toschi and E.~Bodenschatz.
\newblock Lagrangian properties of particles in turbulence.
\newblock \emph{Annu. Rev. Fluid Mech.}, 41:\penalty0 375--404, 2009.

\bibitem[Townsend(1948)]{townsend:1948}
A.~A. Townsend.
\newblock Experimental evidence for the theory of local isotropy.
\newblock \emph{Proc. Camb. Phil. Soc.}, 44\penalty0 (4):\penalty0 560--565,
  1948.

\bibitem[Tran et~al.(2010)Tran, Chakraborty, Guttenberg, Prescott, Kellay,
  Goldburg, Goldenfeld, and Gioia]{tran:2010}
T.~Tran, P.~Chakraborty, N.~Guttenberg, A.~Prescott, H.~Kellay, W.~Goldburg,
  N.~Goldenfeld, and G.~Gioia.
\newblock Macroscopic effects of the spectral structure in turbulent flows.
\newblock \emph{Nature Physics}, 6:\penalty0 438--441, 2010.

\bibitem[Vainshtein et~al.(1994)Vainshtein, Sreenivasan, Pierrehumbert,
  Kashyap, and Juneja]{vainshtein:1994}
S.~I. Vainshtein, K.~R. Sreenivasan, R.~T. Pierrehumbert, V.~Kashyap, and
  A.~Juneja.
\newblock Scaling exponents for turbulence and other random processes and their
  relationships with multifractal structure.
\newblock \emph{Physical Review E}, 50\penalty0 (3):\penalty0 1823--1835, 1994.

\bibitem[{Van Atta} and Chen(1970)]{vanatta:1970}
C.~W. {Van Atta} and W.~Y. Chen.
\newblock Structure functions of turbulence in the atmospheric boundary layer
  over the ocean.
\newblock \emph{Journal of Fluid Mechanics}, 44\penalty0 (1):\penalty0
  145--159, 1970.

\bibitem[{Van de Water} and Herweijer(1999)]{vandewater:1999}
W.~{Van de Water} and J.~A. Herweijer.
\newblock High-order structure functions of turbulence.
\newblock \emph{Journal of Fluid Mechanics}, 387:\penalty0 3--37, 1999.

\bibitem[{Van Fossen} and Ching(1994)]{vanfossen:1994a}
G.~J. {Van Fossen} and C.~Y. Ching.
\newblock Measurements of the influence of integral length scale on stagnation
  region heat transfer.
\newblock Technical Memorandum 106503, NASA, 1994.

\bibitem[{Van Fossen} et~al.(1994){Van Fossen}, Simoneau, and
  Ching]{vanfossen:1994b}
G.~J. {Van Fossen}, R.~J. Simoneau, and C.~Y. Ching.
\newblock Influence of turbulence parameters, {R}eynolds number, and body shape
  on stagnation-region heat transfer.
\newblock Technical Paper 3487, NASA, 1994.

\bibitem[{Van Fossen} et~al.(1995){Van Fossen}, Simoneau, and
  Ching]{vanfossen:1995}
G.~J. {Van Fossen}, R.~J. Simoneau, and C.~Y. Ching.
\newblock Influence of turbulence parameters, {R}eynolds number, and body shape
  on stagnation-region heat transfer.
\newblock \emph{Journal of Heat Transfer}, 117:\penalty0 597--603, 1995.

\bibitem[Verma and Donzis(2007)]{verma:2007}
M.~K. Verma and D.~Donzis.
\newblock Energy transfer and bottleneck effect in turbulence.
\newblock \emph{J. Phys. A}, 40\penalty0 (16):\penalty0 4401--4412, 2007.

\bibitem[Voth et~al.(2002)Voth, {La Porta}, Crawford, Alexander, and
  Bodenschatz]{voth:2002}
G.~A. Voth, A.~{La Porta}, A.~M. Crawford, J.~Alexander, and E.~Bodenschatz.
\newblock Measurement of particle accelerations in fully developed turbulence.
\newblock \emph{Journal of Fluid Mechanics}, 469:\penalty0 121--160, 2002.

\bibitem[Warhaft(2002)]{warhaft:2002a}
Z.~Warhaft.
\newblock Turbulence in nature and in the laboratory.
\newblock \emph{Proceedings of the National Academy of Science}, 99\penalty0
  (1):\penalty0 2481--2486, 2002.

\bibitem[Warhaft and Shen(2002)]{warhaft:2002b}
Z.~Warhaft and X.~Shen.
\newblock On the higher order mixed structure functions in laboratory shear
  flow.
\newblock \emph{Physics of Fluids}, 14\penalty0 (7):\penalty0 2432--2438, 2002.

\bibitem[Warnaars et~al.(2006)Warnaars, Hondzo, and Carper]{warnaars:2006}
T.~A. Warnaars, M.~Hondzo, and M.~A. Carper.
\newblock A desktop apparatus for studying interactions between microorganisms
  and small-scale fluid motion.
\newblock \emph{Hydrobiologia}, 563:\penalty0 431--443, 2006.

\bibitem[Webster et~al.(2004)Webster, Brathwaite, and Yen]{webster:2004}
D.~R. Webster, A.~Brathwaite, and J.~Yen.
\newblock A novel laboratory apparatus for simulating isotropic oceanic
  turbulence at low reynolds number.
\newblock \emph{Limnology and Oceanography: Methods}, 2:\penalty0 1--12, 2004.

\bibitem[Weinman and Klimenko(2000)]{weinman:2000}
K.~A. Weinman and A.~Y. Klimenko.
\newblock Estimation of the {K}olmogorov constant ${C}_0$ by direct numerical
  simulation of a continuous scalar.
\newblock \emph{Physics of Fluids}, 12\penalty0 (12):\penalty0 3205--3220,
  2000.

\bibitem[Weizs\"{a}cker(1948)]{weizsaecker:1948}
C.~F.~v. Weizs\"{a}cker.
\newblock Das {S}pektrum der {T}urbulenz bei gro\ss en {R}eynoldschen {Z}ahlen.
\newblock \emph{Zeit. f. Phys.}, 124:\penalty0 614--627, 1948.

\bibitem[Welter et~al.(2009)Welter, Wittwer, Degrazia, Acevedo, {Leal de
  Moraes}, and Anfossi]{welter:2009}
G.~S. Welter, A.~R. Wittwer, G.~A. Degrazia, O.~C. Acevedo, O.~L. {Leal de
  Moraes}, and D.~Anfossi.
\newblock Measurements of the {K}olmogorov constant from laboratory and
  geophysical wind data.
\newblock \emph{Physica A}, 388\penalty0 (18):\penalty0 3745--3751, 2009.

\bibitem[Xia et~al.(2011)Xia, Byrne, Falkovich, and Shats]{xia:2011}
H.~Xia, D.~Byrne, G.~Falkovich, and M.~Shats.
\newblock Upscale energy transfer in thick turbulent fluid layers.
\newblock \emph{Nature Physics}, 7:\penalty0 321--324, 2011.

\bibitem[Yakhot and Zakharov(1993)]{yakhot:1993}
V.~Yakhot and V.~Zakharov.
\newblock Hidden conservation-laws in hydrodynamics energy and dissipation rate
  fluctuation spectra in strong turbulence.
\newblock \emph{Physica D}, 64\penalty0 (4):\penalty0 379--394, 1993.

\bibitem[Yeung and Brasseur(1991)]{yeung:1991}
P.~K. Yeung and J.~G. Brasseur.
\newblock The response of isotropic turbulence to isotropic and anisotropic
  forcing at the large scales.
\newblock \emph{Physics of Fluids A}, 3:\penalty0 884--897, 1991.

\bibitem[Yeung and Zhou(1997)]{yeung:1997}
P.~K. Yeung and Y.~Zhou.
\newblock Universality of the {K}olmogorov constant in numerical simulations of
  turbulence.
\newblock \emph{Physical Review E}, 56:\penalty0 1746--1752, 1997.

\bibitem[Zagarola and Smits(1998)]{zagarola:1998}
M.~V. Zagarola and A.~J. Smits.
\newblock Mean-flow scaling of turbulent pipe flow.
\newblock \emph{Journal of Fluid Mechanics}, 373:\penalty0 33--79, 1998.

\bibitem[Zhou and Antonia(2000)]{zhou:2000}
T.~Zhou and R.~A. Antonia.
\newblock Reynolds number dependence of the small-scale structure of grid
  turbulence.
\newblock \emph{Journal of Fluid Mechanics}, 406:\penalty0 81--107, 2000.

\bibitem[Zhou et~al.(2001)Zhou, Pearson, and Antonia]{zhou:2001}
T.~Zhou, B.~R. Pearson, and R.~A. Antonia.
\newblock Comparison between temporal and spatial transverse velocity
  increments in a turbulent plane jet.
\newblock \emph{Fluid Dynamics Research}, 28:\penalty0 127--138, 2001.

\bibitem[Zhou et~al.(2005)Zhou, Hao, Chua, and Yu]{zhou:2005}
T.~Zhou, Z.~Hao, L.~P. Chua, and S.~C.~M. Yu.
\newblock Scaling of longitudinal and transverse velocity increments in a
  cylinder wake.
\newblock \emph{Physical Review E}, 71:\penalty0 066307, 2005.

\bibitem[Zimmermann et~al.(2010)Zimmermann, Xu, Gasteuil, Bourgoin, Volk,
  Pinton, and Bodenschatz]{zimmermann:2010}
R.~Zimmermann, H.~Xu, Y.~Gasteuil, M.~Bourgoin, R.~Volk, J.-F. Pinton, and
  E.~Bodenschatz.
\newblock The lagrangian exploration module: An apparatus for the study of
  statistically homogeneous and isotropic turbulence.
\newblock \emph{Review of Scientific Instruments}, 81\penalty0 (055112), 2010.

\end{thebibliography}

\end{document}